\documentclass[12pt,twoside,a4paper]{book}

\usepackage{bm} 
\usepackage{amsmath}
\usepackage{amssymb}
\usepackage{epsfig}
\usepackage{xr}
\usepackage{colordvi}
\usepackage{float}
\usepackage{minitoc}
\usepackage{footnpag}
\usepackage{fancyheadings}
\usepackage{rotating}
\usepackage{picinpar}
\addtocounter{secnumdepth}{1} 

\newcommand{\clearemptydoublepage}{\newpage{\pagestyle{empty}\cleardoublepage}}
\begin{document}
\ifx\href\undefined\else\hypersetup{linktocpage=true}\fi
\dominitoc

\renewcommand{\thepage}{\roman{page}}
\pagestyle{empty}

\begin{center}
\textsc{\bf \hskip 2cm \large UNIVERSIDADE T\'ECNICA DE LISBOA \\
\vskip 0.5cm
\hskip 2cm INSTITUTO SUPERIOR T\'ECNICO}
\end{center}


\begin{center}
\vskip 3cm

\hskip 2cm \LARGE \bf Quasinormal Modes \\
\hskip 2cm and\\
\hskip 2cm Gravitational Radiation\\
\hskip 2cm in\\ 
\hskip 2cm Black Hole Spacetimes

\vspace{2cm}
{\hskip 2cm \LARGE \bf Vitor Cardoso} \\
\vskip 0.5cm
{\hskip 2cm \large(Licenciado)}

\vskip 1cm
{\hskip 2cm \large Disserta\c c\~ao para obten\c c\~ao do Grau de Doutor em F\'{\i}sica}


\normalsize\rm
\vfill
\begin{tabular}{ll}
Orientador: & Doutor Jos\'e Pizarro de Sande e Lemos\\
&\\
Presidente: & Reitor da Universidade T\'ecnica\\
Vogais:&Doutor Jos\'e Tito da Luz Mendon\c ca\\
&Doutor Jorge Venceslau Comprido Dias de Deus\\
&Doutor Alfredo Barbosa Henriques\\
&Doutor Kostas Kokkotas\\
&Doutor Jos\'e Pizarro de Sande e Lemos\\
&Doutor Lu\'{\i}s Filipe Lopes Bento\\
\end{tabular}
\vfill

\vfill
{\hskip 2cm \large Dezembro de 2003}

\end{center}

\thispagestyle{empty}

\author{\textsc{Vitor Cardoso}} \title{ \Huge{
\textsc
{\bf
Quasinormal Modes\\
and\\
Gravitational Radiation
\\
in Black Hole Spacetimes
}}}
\voffset -1.5cm
\maketitle

\renewcommand{\thefootnote}{\alph{footnote}}

\pagestyle{fancyplain}
\addtolength{\headwidth}{\marginparsep}
\addtolength{\headwidth}{\marginparwidth}
\renewcommand{\chaptermark}[1]{\markboth{#1}{}}
\renewcommand{\sectionmark}[1]{\markright{\thesection\ #1}}
\lhead[]{\fancyplain{}{\bfseries\leftmark}}
\rhead[\fancyplain{}{\bfseries \rightmark}]{}
\lfoot[\fancyplain{}{\bfseries\thepage}]{}
\rfoot[]{\fancyplain{}{\bfseries\thepage}}
\cfoot{}

\setlength{\headrulewidth}{1pt}

\thispagestyle{empty} 
\chapter*{Resumo}
Buracos negros desempenham um papel fundamental na f\'{\i}sica actual.
Os buracos negros
possuem modos de vibra\c c\~ao pr\'oprios, semelhantes aos de um sino
ou de uma guitarra, a que chamamos modos quasinormais.
Investiga\c c\~oes passadas mostram que os modos quasinormais s\~ao
importantes no entendimento da vibra\c c\~ao de buracos negros 
astrof\'{\i}sicos.
Investiga\c c\~oes recentes mostram que s\~ao importantes
na conex\~ao entre gravita\c c\~ao e mec\^anica qu\^antica.
O estudo de modos quasinormais \'e assim de extrema import\^ancia.

Estes modos controlam todo e qualquer processo
involvendo buracos negros, em particular a emiss\~ao de ondas
gravitacionais geradas, por exemplo, na colis\~ao de dois buracos
negros.  Existe a possibilidade de se criarem buracos negros em
aceleradores, segundo conjecturas que invocam a exist\^encia
de dimens\~oes extra no universo. Torna-se portanto
imperativo fazer um estudo detalhado da radia\c c\~ao gravitacional
emitida na colis\~ao de buracos negros em v\'arias dimens\~oes,
bem assim como o estudo de ondas gravitacionais em
dimens\~oes mais altas.

Na presente disserta\c{c}{\~a}o apresenta-se um estudo aprofundado dos
modos quasinormais de buracos negros, em v\'arios espa\c cos-tempo de fundo.  
Investiga-se tamb\'em a radia\c c\~ao gra\-vitacional
gerada na colis\~ao de buracos negros com part\'{\i}culas e na
colis\~ao entre dois buracos negros.
Finalmente, apresenta-se o estudo de ondas gravitacionais em dimens\~oes
gerais onde, por exemplo, se generaliza a f\'ormula do quadrupolo de
Einstein. \\

\vspace{4cm} {\normalfont\sffamily\bfseries%
\noindent PALAVRAS-CHAVE:} \\
\noindent Buracos negros; Modos quasinormais;
Radia\c c\~ao gravitacional; Colis\~oes entre buracos negros; Propaga\c c\~ao de ondas; Dimens\~oes extra;

\thispagestyle{empty} 
\chapter*{Abstract}
Black holes play a fundamental role in modern physics.
They have characteristic oscillation modes, called quasinormal modes.
Past studies have shown that these modes are important to our understanding
of the dynamics of astrophysical black holes.
Recent studies indicate that they are important as a link between
gravitation and quantum mechanics.
Thus, the investigation of these modes is a timeliness topic.

Quasinormal modes dominate almost every process involving black holes,
in particular gravitational wave emission during, for example, the collision
between two black holes.
It may be possible to create black holes at future accelerators,
according to recent theories advocating the existence of extra dimensions
in our universe. 
It is therefore important to study in depth the gravitational radiation
emitted in high energy collision between two black holes in several dimensions, and also
to make a theoretical study of gravitational waves in higher dimensions.

In this thesis we shall make a thorough study of the quasinormal modes
of black holes in several kinds of background spacetimes.  We shall
investigate the gravitational radiation given away when highly
energetic particles collide with black holes, and also when two black
holes collide with each other.  Finally, we shall study the properties
of gravitational waves in higher dimensions, for instance, we
generalize Einstein's quadrupole formula.\\

\vspace{4cm} {\normalfont\sffamily\bfseries%
\noindent KEY-WORDS:} \\
\noindent Black holes; Quasinormal modes; Gravitational radiation;
High energy black hole collision; Wave propagation; Extra dimensions.

\thispagestyle{empty} 
\chapter*{}
\vspace*{12cm} \noindent Este trabalho foi financiado pela Funda\c{c}{\~a}o Para a Ci{\^e}ncia e a Tecnologia,\\
sob o contrato Praxis XXI/BD/21284/99 (01/01/2000-31/12/2003).

\vspace*{2cm} \noindent This work was
supported by Funda\c{c}{\~a}o Para a Ci{\^e}ncia e a Tecnologia,\\
under the grant Praxis XXI/BD/21284/99 (01/01/2000-31/12/2003).

\thispagestyle{empty} 
\chapter*{Acknowledgements}

I am extremely grateful to my supervisor Jos\'e Lemos, for all the things
he taught me and for all the physics we have done together. It was a
pleasure and an honor to work with him, and i'm certain I could not
have asked for a better adviser. Thanks Jos\'e!

I am also very grateful to \'Oscar Dias, my best and long-dated
friend, for such wonderful moments we had together thinking about
physics and life. His sharp mind and gentle personality has made a
better physicist and human being of me.  

I am very thankful to M\'ario
Pimenta, for showing me how a teacher is supposed to teach, and of
course also for the great meals we had at his place.

It is my pleasure to thank all the elements of CENTRA, 
specially Jorge Dias de Deus, Barbosa Henriques and Ana Mour\~ao, 
for all the support and encouragement during
the completion of this thesis, and also Dulce Concei\c c\~ao and
Paula Espada, for their infinite patience, and for having the
gift to make apparently complex things turn simple.

I am indebted to Emanuele Berti, Kostas Kokkotas, Roman Konoplya, Jos\'e Lages,
Lubos Motl, Jos\'e Nat\'ario, Andrew Neitzke, Hisashi Onozawa, 
Tito Mendon\c ca, Ricardo Schiappa and Shijun Yoshida for very 
interesting discussions and collaborations, 
in which part of this work was done. I have learned a lot from you.

To my family, where I always find unconditional support, I cannot
thank enough in words. I love you all! Mother, I am trying. 

Carlota, thank you! This thesis is half yours.

Of course, peace of mind is important when doing research.
Only special people are able to put up with my continuous 
non-sense, and still smile for me. 
Thanks Pedro (Ching-Cheng) and Ana (Bacaninha).
I love you.

Est\'er, I'll see you some day.

\clearemptydoublepage
\lhead[]
      {\fancyplain{}{\bfseries Table of Contents}}
\rhead[\fancyplain{}{\bfseries Table of Contents}]
      {}
\tableofcontents
%
\normalsize
\nocite{*}

\clearemptydoublepage
\newpage
\pagestyle{fancyplain}
\thispagestyle{empty}

\lhead[]
      {\fancyplain{}{\bfseries Preface}}
\rhead[\fancyplain{}{\bfseries Preface}]
      {}
\setlength{\headrulewidth}{1pt}
\nocite{*}
\thispagestyle{empty}
\addcontentsline{toc}{section}{\numberline{}{\bf \hskip -0.8cm Preface}}

\begin{center}{\Large \bf Preface}
\end{center}
The research included in this thesis has been carried out at Centro
Multidisciplinar de Astrof\'{\i}sica (CENTRA) in the Physics
Department of Instituto Superior T\'ecnico.  I declare that this
thesis is not substantially the same as any that I have submitted for
a degree or diploma or other qualification at any other University and
that no part of it has already been or is being concurrently submitted
for any such degree or diploma or any other qualification.

Chapters \ref{chap:qnmbtz}, \ref{chap:qnmtoro}, \ref{chap:qnmds},
\ref{chap:rads}, \ref{chap:radkerr} and \ref{chap:radads} are the 
outcome of collaborations with Professor Jos\'e Lemos.
Chapter \ref{chap:qnmsads} was done in collaboration with
Professor Jos\'e Lemos and Roman Konoplya.
Chapter \ref{chap:qnmkerr} was done in collaboration with Professor
Kostas Kokkotas, Dr. Emanuele Berti and Dr. Hisashi Onozawa.
Chapter \ref{chap:qnms} is the outcome of a collaboration with Professor
Jos\'e Lemos and Dr. Shijun Yoshida.
Chapter \ref{chap:radddim} is the outcome of a collaboration with 
Professor Jos\'e Lemos and \'Oscar Dias.
Chapter \ref{chap:Tails} is the outcome of a
collaboration with Professor Jos\'e Lemos, Dr. Shijun Yoshida and \'Oscar Dias.
Most of these chapters have been published, 
and some are being submitted for publication.

Chapter \ref{chap:rads} referring to encounters of black holes with particles
in an anti-de Sitter background, was submitted with some modifications, 
under the title ``Numerical Analysis of Partial Differential Equations
in General Relativity and String Theory: Applications to Gravitational Radiation
and to the AdS/CFT Conjecture'', and won the Gulbenkian Prize of Scientific Investigation 2001
(Pr\'emio Gulbenkian de Est\'{\i}mulo \`a Investiga\c c\~ao Cient\'{\i}fica, 2001).

A list of the works published included in this thesis is included below.

\noindent{\Large.} V. Cardoso, J. P. S. Lemos, 
``Scalar, electromagnetic and Weyl perturbations of BTZ black holes: 
quasi normal modes'', {\it Phys. Rev. D} {\bf 63}, 
124015 (2001); gr-qc/0101052. (Chapter \ref{chap:qnmbtz})
\vspace{0.15 cm}

\noindent{\Large.} V. Cardoso, J. P. S. Lemos,  
``Quasi-normal modes of Schwarzschild black holes in anti-de 
Sitter spacetimes: electromagnetic and gravitational perturbations,
{\it Phys. Rev. D} {\bf 64}, 084017 (2001); gr-qc/0105183. (Chapter \ref{chap:qnmsads})
\vspace{0.15 cm}

\noindent{\Large.} V. Cardoso, R. Konoplya, J. P. S. Lemos, 
``Quasinormal frequencies of Schwar\-zschild black holes in 
anti-de Sitter spacetimes: A complete study on the asymptotic 
behavior'', 
{\it Phys. Rev. D} {\bf 68}, 044024 (2003); gr-qc/0305037. (Chapter \ref{chap:qnmsads})
\vspace{0.15 cm}

\noindent{\Large.} V. Cardoso, J. P. S. Lemos,
 ``Quasi-normal modes of toroidal, cylindrical 
and planar black holes in anti-de Sitter
spacetimes'', {\it Class. Quantum Grav.},{\bf 18}, 5257 (2001);
gr-qc/0107098. (Chapter \ref{chap:qnmtoro})
\vspace{0.15 cm}

\noindent{\Large.} V. Cardoso, J. P. S. Lemos,  
``Quasinormal modes of the near extremal
Schwa\-rzschild-de Sitter black hole'', 
{\it Phys. Rev. D} {\bf 67}, 084020 (2003); gr-qc/0301078. (Chapter \ref{chap:qnmds})
\vspace{0.15 cm}

\noindent{\Large.} V. Cardoso, J. P. S. Lemos, S. Yoshida,
``Quasinormal modes of Schwarzschild black holes in four 
and higher dimensions'', 
{\it Phys. Rev. D}, in press (2003), 
gr-qc/0309112; (Chapter \ref{chap:qnms}) 
\vspace{0.15 cm}

\noindent{\Large.} E. Berti, V. Cardoso, K. Kokkotas and H. Onozawa,
``Highly damped quasinormal modes of Kerr black holes'', 
{\it Phys. Rev. D} {\bf }, in press (2003); hep-th/0307013. (Chapter \ref{chap:qnmkerr})
\vspace{0.15 cm}

\noindent{\Large.} V. Cardoso, J. P. S. Lemos,
 ``Gravitational radiation from collisions at the speed of light:
a massless particle falling into a Schwarzschild black hole'', 
{\it Phys. Lett. B} {\bf 538}, 1 (2002);
gr-qc/0202019. (Chapter \ref{chap:rads})
\vspace{0.15 cm}

\noindent{\Large.} V. Cardoso, J. P. S. Lemos,
 ``The radial infall of a highly relativistic point 
particle into a Kerr black hole along the symmetry axis'', 
{\it Gen. Rel. Gravitation} {\bf 35}, L327-333 (2003); gr-qc/0207009. (Chapter \ref{chap:radkerr})
\vspace{0.15 cm}

\noindent{\Large.} V. Cardoso, J. P. S. Lemos,
``Gravitational radiation from the radial infall of highly
relativistic point particles into Kerr black holes'', 
{\it Phys. Rev. D} {\bf 67}, 084005 (2003); gr-qc/0211094. (Chapter \ref{chap:radkerr})
\vspace{0.15 cm}

\noindent{\Large.} V. Cardoso, J. P. S. Lemos, S. Yoshida, 
``Electromagnetic radiation from collisions at almost the
speed of light: an extremely relativistic charged particle 
falling into a Schwarzschild black hole'', {
\it Phys. Rev. D} {\bf }, in press (2003); gr-qc/0307104. (mentioned in Chapter \ref{chap:radkerr})
\vspace{0.15 cm}

\noindent{\Large.} V. Cardoso, J. P. S. Lemos,
 ``Black hole collision with a scalar particle 
in three dimensional anti-de Sitter spacetime'', 
{\it Phys. Rev. D} {\bf 65}, 104032 (2002);
hep-th/0112254. (Chapter \ref{chap:radads})
\vspace{0.15 cm}

\noindent{\Large.} V. Cardoso, J. P. S. Lemos,
 ``Black hole collision with a scalar particle in four, 
five and seven dimensional anti-de Sitter 
spacetimes: ringing and radiation'', 
{\it Phys. Rev. D} {\bf 66}, 064006 (2002); hep-th/0206084. (Chapter \ref{chap:radads})
\vspace{0.15 cm}

\noindent{\Large.} V. Cardoso, J. P. S. Lemos,
 ``Scalar synchrotron radiation 
in the Schwarzsch\-ild-anti-de Sitter Geometry'',
{\it Phys. Rev. D} {\bf 65}, 104033 (2002);
hep-th/0201162. (Chapter \ref{chap:radads})
\vspace{0.15 cm}

\noindent{\Large.} V. Cardoso, \'O. J. C. Dias, J. P. S. Lemos,  
``Gravitational radiation in D-dimensional spacetimes'', 
{\it Phys. Rev. D} {\bf 67}, 064026 (2003); hep-th/0212168. (Chapter \ref{chap:radddim})
\vspace{0.15 cm}

\noindent{\Large.} J. T. Mendon\c ca, V. Cardoso,
``Gravitational optics: Self-phase modulation and harmonic cascades'', 
{\it Phys. Rev. D} {\bf 66}, 104009 (2002); gr-qc/0209076. (mentioned in Chapter \ref{chap:radddim})
\vspace{0.15 cm}

\noindent{\Large.} J. T. Mendon\c ca, V. Cardoso, M. Servin, 
M. Marklund and G. Brodin, 
``Self-phase modulation of spherical gravitational waves'',
{\it Phys. Rev. D} {\bf }, in press (2003); gr-qc/0307031. (mentioned in Chapter \ref{chap:radddim})
\vspace{0.15 cm}

\noindent{\Large.} M. Servin, M. Marklund, G. Brodin, 
J. T. Mendon\c ca and V. Cardoso, 
``Nonlinear self-interaction of plane gravitational waves'',
{\it Phys. Rev. D} {\bf 67}, 087501 (2003) ; astro-ph/0303412. (mentioned in Chapter \ref{chap:radddim})
\vspace{0.15 cm}

\noindent{\Large.} V. Cardoso, S. Yoshida, O. J. C. Dias, J. P. S. Lemos,
``Late-Time Tails of Wave Propagation in Higher Dimensional Spacetimes'', 
{\it Phys. Rev. D} {\bf 68}, 061503(R)[Rapid Communications](2003); 
hep-th/0307122. (Chapter \ref{chap:Tails})
\vspace{0.15 cm}

\clearemptydoublepage
\newpage

\pagestyle{fancyplain}
\thispagestyle{empty}

\lhead[]
      {\fancyplain{}{\bfseries Brief general introduction}}
\rhead[\fancyplain{}{\bfseries Brief general introduction}]
      {}
\setlength{\headrulewidth}{1pt}
\nocite{*}
\thispagestyle{empty}
\addcontentsline{toc}{section}{\numberline{}{\bf \hskip -0.8cm Brief general introduction}}

\begin{center}{\Large \bf Brief general introduction}
\end{center}
This thesis is devoted to the study of (i) the characteristic
vibration modes of black holes, called quasinormal modes (QNMs),  (ii)
radiative processes in non-asymptotically flat spacetimes and
gravitational waves from processes involving black holes, specially
high energy collisions, and (iii) gravitational waves in higher dimensions. 
This thesis has accordingly been divided into three parts: Part
I deals with QNMs of black holes, Part II deals with
gravitational radiation from encounters of black holes with particles, and
Part III deals with gravitational wave physics in higher dimensions.

A few years ago, the very concept of QNMs of black holes was closely
linked to the gravitational radiation emitted by the object, and one
studied quasinormal modes, specially the lowest lying modes, in order
to get a better understanding of the properties of the gravitational
wave signal. For example, the decay time of the gravitational signal
is directly connected to the imaginary part of the associated
quasinormal frequency, and the characteristic frequency of the signal
is related to the real part of the lowest quasinormal frequency. There
are now a couple of significant breakthroughs in fundamental
physics that separated the direct direct dependence of QNMs on gravitational 
radiation, and changed the way we
faced gravitational radiation, as something important only in an
astrophysical set.  First supergravity brought a whole new
interest to anti-de Sitter spacetimes and the AdS/CFT conjecture
gave a new special meaning
to the notion of quasinormal modes of black holes living in anti-de Sitter
spacetimes.
 According to it, the imaginary part of the lowest
quasinormal frequency should describe the typical timescale of
approach to thermal equilibrium in a dual conformal field theory.
Very recently quasinormal modes have made their way to a completely
unsuspected territory: quantum gravity.  Some evidence has been
accumulating that highly damped quasinormal modes may be important
in semi-classical attempts to quantize the black hole area.
This has been confirmed,
rather surprisingly, in Loop Quantum Gravity, an alternative theory to 
quantum gravity.  Second, the recent
attempts to solve the hierarchy problem have put forward the
interesting possibility that there may be large extra dimensions in our
Universe, besides the usual four. This is the so called TeV-scale
gravity scenario.  When the consequences of such a scenario are
carefully studied, one comes to the conclusion that gravitational
waves may be indirectly detected at accelerators such as LHC at
CERN!  

In Part I the quasinormal modes of black holes, both in asymptotically
flat and in non-asymptotically flat spacetimes, will be thoroughly
studied.  In Part II we will study
gravitational wave generation from the collision of black holes
with particles.  We begin by study high energy
collisions between point particles and black holes, and black
hole-black hole collisions at near the speed of light.  This is a
subject of the utmost importance, since one may well produce black
holes at the LHC. We will go on by considering scalar radiation in
asymptotically anti-de Sitter spacetime.
In Part III we shall conclude
this thesis by extending what we know about linearized gravitational
waves in four dimensions to a general $D$-dimensional spacetime.
\renewcommand{\thepage}{\arabic{page}}
\setcounter{page}{1}
\part{Quasinormal modes in four and  higher dimensions}

\thispagestyle{empty} \setcounter{minitocdepth}{1}
\chapter[Quasinormal modes: an introduction]{
Quasinormal modes: an introduction} \label{chap:Intro}
\lhead[]{\fancyplain{}{\bfseries Chapter \thechapter. \leftmark}}
\rhead[\fancyplain{}{\bfseries \rightmark}]{}
\minitoc \thispagestyle{empty}
\renewcommand{\thepage}{\arabic{page}}
\section{Quasinormal modes: an introduction}

\subsection{Motivation}
\begin{figure}
\centerline{\includegraphics[width=9 cm,height=9 cm]
{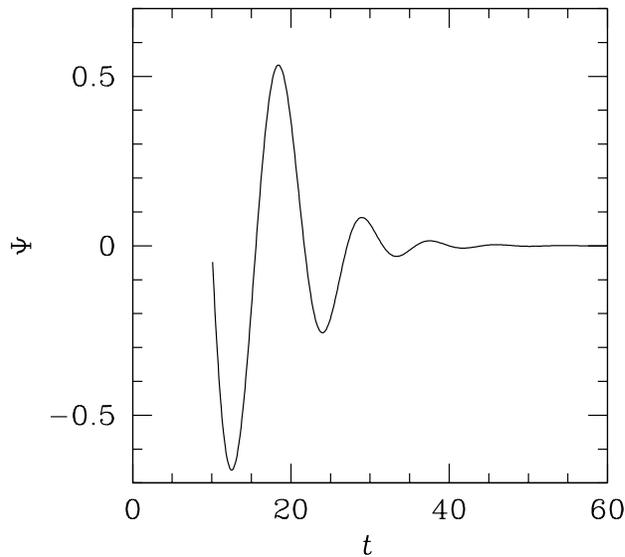}}
\caption{Evolution of a gravitational gaussian wavepacket in the 
neighborhood of a Schwarzschild black hole. One can see that the waveform is dominated
by a characteristic ringing. We have chosen a gaussian wavepacket 
$\Psi \sim e^{-(\frac{v-v_0}{\sigma})^2}$, with $\sigma=3$ and $v_c=10$, but the 
ringing down frequency and damping time is independent of these parameters.
Also, the coordinate $v=t+r_*$ is an advanced time coordinate.}
\label{fig:Ev}
\end{figure}

\begin{figure}
\centerline{\includegraphics[width=9 cm,height=9 cm]
{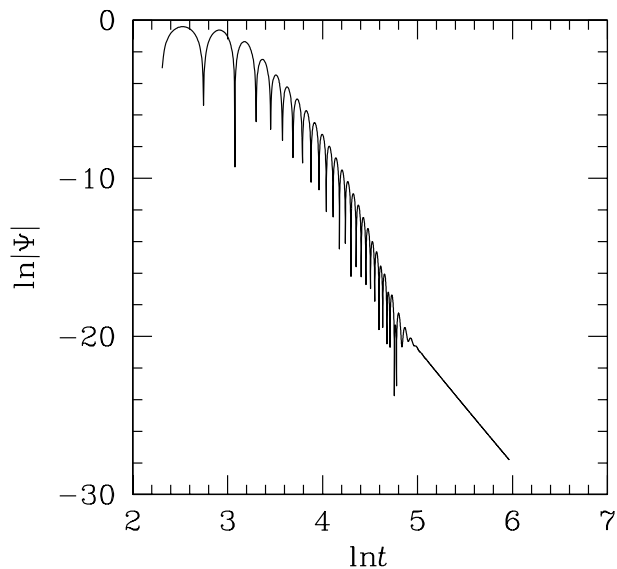}}
\caption{Evolution of a gravitational gaussian wavepacket in the 
neighborhood of a Schwarzschild black hole, here shown in a log-log plot.
The dominance of quasinormal ringing is clear, and at very late times,
a power-law falloff appears. 
}
\label{fig:LogEv}
\end{figure}

We are all familiar with the fact that the ringing of a bell or
the strum of a guitar invariably produces a ``characteristic
sound''. Such systems respond to any excitation by selecting
a set of natural real frequencies, the normal frequencies, and
their response is given as a superposition of stationary modes,
the normal modes.
Black holes have a characteristic sound as well.
As a simple demonstration we may let a gaussian gravitational wavepacket
evolve on the Schwarzschild geometry. The results, at the linearized level,
are shown in Figs. \ref{fig:Ev}-\ref{fig:LogEv}.

A striking feature, first uncovered by Vishveshwara \cite{vish} and
clearly apparent in Figs.\ref{fig:Ev}-\ref{fig:LogEv}, 
is that the signal is dominated, 
during a certain time, by damped single frequency oscillations.  The
frequency and damping of these oscillations depend only on the
parameters characterizing the black hole, which in the Schwarzschild
case is its mass.  They are completely independent of the particular
initial configuration that caused the excitation of such vibrations.
That such characteristic oscillations always appear and dominate the
signal at intermediate times, in any event involving black holes, has
been tested time and time again.  It has been verified at the linearized
level \cite{vish,davis,davis2}, in which the fields are treated as a
perturbation in the single black hole spacetime, but also for example,
on fully numerical simulations of black hole-black hole collision
processes \cite{anninos,gleiser}, or stellar collapse \cite{seidell,smarrl}.  
They are therefore a characteristic of
black holes. These characteristic oscillations have been termed
``quasinormal modes'' and the associated frequencies ``quasinormal
frequencies'' \cite{press}.  The ``normal'' part derives from the
obvious similarity between these and normal mode systems. There are
however important differences between these two systems, which
justifies the ``quasi'':
first, quasinormal modes are not stationary modes, since they are
exponentially damped.  This is merely reflecting the fact that the
black hole spacetime is radiating energy away to infinity, through the
form of gravitational (or any massless field) waves.  When describing
a resonant system, such as a bell, a guitar or a star, one often makes
the convenient and sometimes accurate
assumption, that there is no damping. This then leads to a normal
mode, complete expansion of the field.  The stationary mode expansion
is just reflecting the no energy loss assumption.  As soon as one
turns to the more realistic situation, by allowing a dissipation
mechanism, one expects no such naive stationary normal mode expansion
to exist.  Black hole oscillations occupy a very special place here.
It is impossible, even in principle, even in the most idealized
scenario, to turn off the dissipation mechanism. In fact, black holes
are made from the fabric of spacetime, and any spacetime oscillation implies
the generation of gravitational waves, carrying energy away to
infinity. Indeed the very equations describing black hole oscillations
are nothing more than a description of gravitational waves.
Quasinormal modes were first discussed in the black hole context, but
from the above discussion one anticipates they will also appear in
other dissipative physical systems, such as a vibrating string coupled to a
surrounding medium (and therefore transmitting energy to infinity
through it), laser cavities or a star
when one does not neglect gravitational radiation.
Another important difference between normal and quasinormal modes is
the completeness issue, which is a rather subtle one,
mathematically. The response of a normal mode system can be given for
all times as a superposition of normal modes. However, quasinormal
modes seem to appear only over a limited time interval; this is also
shown in Fig. \ref{fig:LogEv}, where it is seen that at very late times quasinormal
ringing gives way to a power-law falloff.  A thorough account of
quasinormal modes in asymptotically flat spacetimes, their properties,
a thorough comparison between normal and quasinormal mode systems, and
a discussion about the incompleteness of quasinormal modes can be found
in the classical reviews by Kokkotas and Schmidt \cite{kokkotas} and
Nollert \cite{nollert}, and references therein.
From now on we shall refer to quasinormal modes as ``QNMs'' and 
we shall refer to quasinormal
frequencies as ``QN frequencies''.  
\subsection{Definition of quasinormal modes}
\begin{figure}
\centerline{\includegraphics[width=9 cm,height=9 cm]
{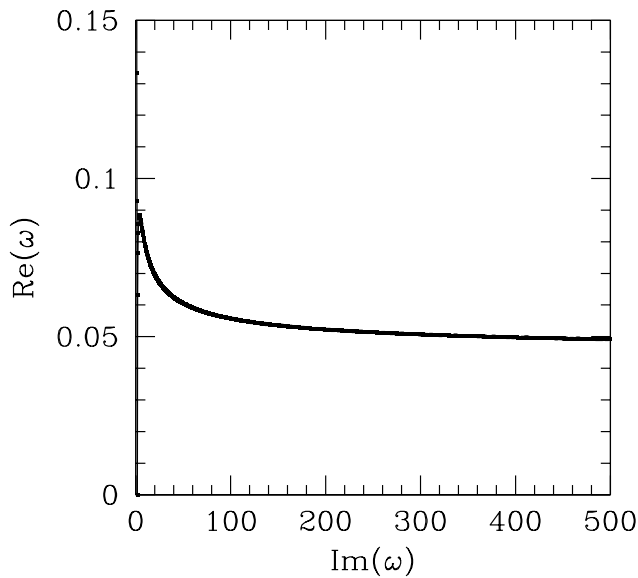}}
\caption{Some of the higher overtone QN frequencies for $l=2$
gravitational perturbations. Notice that the real part seems to go
to a constant, whereas the imaginary part grows without bound.
In fact, one can show numerically 
that $M \omega _n =0.0437123-\frac{i}{4}(n+1/2)+{\mathcal O}(\frac{1}{\sqrt n})$}.
\label{fig:asymptintro}
\end{figure}

Most of the problems concerning wave propagation in 
black hole spacetimes can be reduced to a second order partial 
differential equation of the form
\begin{equation}
\frac{\partial ^2 }{\partial x^2}\Psi-\frac{\partial ^2 }{\partial t^2}\Psi-V \Psi=0.
\label{waveeqmod}
\end{equation}
Here $x$ is a spatial variable, usually but not always ranging 
(in a special coordinate system) from
$-\infty$ to $+\infty$. When dealing with black hole spacetimes the
horizon is usually at $-\infty$, and for the rest of this discussion
we shall assume so.  Also, $V$ is an $x$-dependent potential.  To
define in a phenomenological way what a QNM is, we shall
proceed in the usual way by assuming a time dependence
\begin{equation}
\Psi(t,x)=e^{-i\omega t} \phi(x).
\label{fourierintro}
\end{equation}
Inserting this in (\ref{waveeqmod}) we get an ordinary differential 
equation in the spatial variable $x$,
\begin{equation}
\frac{d ^2 }{d x^2}\phi+(\omega ^2 -V)\phi=0.
\label{wavefreqintro}
\end{equation}
The form (\ref{fourierintro}) is not restrictive, since once we have a
solution for (\ref{wavefreqintro}), a general time dependent solution
can be given as a continuous Fourier transform of such solutions.  The
form (\ref{wavefreqintro}) is ideal to study QNMs in a
way that parallels a normal mode analysis. We shall now restrict
ourselves to asymptotically flat spacetimes, and defer a discussion of
asymptotically de Sitter (i.e., a spacetime with a positive cosmological constant) or anti-de Sitter (i.e., negative cosmological constant) 
for the next chapters. In
asymptotically flat spacetimes, the potential $V$ is positive and
satisfies
\begin{equation}
V \rightarrow 0\,\,, x \rightarrow -\infty
\label{Vh}
\end{equation}
\begin{equation}
V \rightarrow 0\,\,, x \rightarrow +\infty.
\label{Vi}
\end{equation}
Therefore, such potentials do not allow bound states, and this makes
it impossible to do a normal mode expansion. 
The idea
that the evolution of $\Psi$ will generally involve a superposition of
these QNMs can be shown to be correct by the use of
Laplace transforms. We refer the reader to \cite{kokkotas,nollert}.
Nevertheless, we saw that
the signal is somehow dominated by characteristic oscillations
(Figs.1-2), so we shall blindly continue with our analysis. 
Having in mind the form (\ref{Vh})-(\ref{Vi}) of the potential we have that
near the boundaries $-\infty$ and $+\infty$ the solution behaves as plane 
waves,
\begin{equation}
\phi \sim e^{\pm i\omega x}\,\,, x \rightarrow -\infty
\label{solhintro}
\end{equation}
\begin{equation}
\phi \sim e^{\pm i\omega x} \,\,, x \rightarrow +\infty.
\label{solintro}
\end{equation}
The boundary conditions defining QNMs are that toward
the boundaries the solutions should be purely outgoing at infinity
($x=+\infty$) and ingoing at the horizon ($x=-\infty$),
\begin{equation}
\phi \sim e^{- i\omega x}\,\,, x \rightarrow -\infty
\label{bound1intro}
\end{equation}
\begin{equation}
\phi \sim e^{+ i\omega x} \,\,, x \rightarrow +\infty.
\label{bound2intro}
\end{equation}
Here ingoing at the horizon means entering into the black hole,
therefore leaving the domain we are studying. These are physically
motivated boundary conditions. Only a discrete set of complex
frequencies $\omega_{\rm QN}$ satisfy these boundary conditions.
These are the QN frequencies, and the associated
wavefunctions $\phi$, solutions of (\ref{wavefreqintro}) are the QNMs.  It has been proved by
Vishveshwara \cite{vish2} that for the Schwarzschild geometry $\omega_{\rm QN}$
must have a negative imaginary part; this has also been found for
geometries other that Schwarzschild, for example
Schwarzschild-anti-de Sitter, Schwarzschild-de Sitter, Kerr (we shall go
through these geometries on the next chapters). 
This means on the one hand that QNMs decay exponentially
in time, and the physical significance of this is that the black hole
spacetime is loosing energy in the form of gravitational waves. On the 
other hand, this
also means that the spacetime is stable.  In addition, the imaginary part being negative
makes the numerical calculation of QN frequencies a
non-trivial task: according to the boundary conditions
(\ref{bound1intro})-(\ref{bound2intro}), and to the fact that
the QN frequencies have a negative imaginary part, one has 
that QNMs grow exponentially at the boundaries. Now, in
order to tell if a certain frequency is or not a QN frequency
one must check that, for example, there is only an out-going
$e^{i\omega x}$ piece at infinity, or in other words, one must check
that near infinity the $e^{-i\omega x}$ is absent. However, this last term
is exponentially suppressed in relation to the other, so one
must be able to distinguish numerically an exponentially small term
from an exponentially large one.
This has always been, and still is, a major obstacle when it comes
down to an actual computation of QN frequencies. Still, brute
numerical force sometimes works. Chandrasekhar and Detweiler
\cite{chandradet} have succeeded in finding some of the Schwarzschild
QN frequencies this way, in 1975.  Since then, numerous
techniques have been developed. Some of them are analytical tools 
like the WKJB technique of Schutz and Will
\cite{schutz}, later refined to 3rd order \cite{will}, and recently
extended to 6th order \cite{konoplyawkb}, or the ``potential fit''
\cite{ferrari} one, in which one tries to fit the Schwarzschild
potential to one which enables us to find exact results.  However, the
most successful attempt has been developed by Leaver \cite{L},
using a continued fraction form of the equations, which is rather easy
to implement numerically.  We refer the reader to
\cite{kokkotas,nollert} for a complete account of all these
techniques, and many others.  Exact results for QNMs of
certain black hole spacetimes only became available very recently, and
shall be later described in this thesis. They are however an
exception, the rule being that an exact solution to
(\ref{wavefreqintro}) satisfying the boundary conditions
(\ref{bound1intro})-(\ref{bound2intro}) is not available. So one has
to resort to any numerical or approximate method one can.
In Table 1.1 we present the first four QN frequencies for gravitational
perturbations of a Schwarzschild black hole, obtained using Leaver's \cite{L}
technique.
\vskip 1mm
\begin{table}
\caption{\label{tab:introqnm} The first four quasinormal frequencies
for a Schwarzschild black hole, measured in units of the black hole mass $M$.
To convert this to $Hz$ one must multiply the numbers in the Table
by $32310 \frac{M_{\odot}}{M}$. Thus, a one solar mass black hole has a typical 
ringing frequency of 10 $kHz$ in the quadrupole mode, and a damping timescale,
due to gravitational wave emission, of $\tau=3.74\times10^{-4} s$.
}
\begin{tabular}{llll}  \hline
\multicolumn{1}{c}{} &
\multicolumn{3}{c}{ $\omega_{\rm QN}M$}\\ \hline
$n$ & $l=2$:  &     $l=3$:   & $l=4$:\\ \hline
0   &  0.37367-0.08896$i$  &  0.59944-0.09270$i$ &  0.80918-0.09416$i$ \\ \hline 
1   &  0.34671-0.27391$i$  &  0.58264-0.28130$i$ &  0.79669-0.28449$i$ \\ \hline 
2   &  0.30105-0.47828$i$  &  0.55168-0.47909$i$ &  0.77271-0.47991$i$   \\ \hline 
3   &  0.25150-0.70514$i$  &  0.51196-0.69034$i$ &  0.73984-0.68392$i$   \\ \hline  
\end{tabular}
\end{table}
\vskip 1mm
In Table \ref{tab:introqnm} the angular quantum number $l$ gives the
angular dependence of the gravitational wave.  For example a wave with
$l=2$, the lowest gravitational radiatable multipole, has a quadrupole
distribution.  Some comments are in order. First,
QN frequencies come in pairs. That is, if $\omega=a+ib$ is a
QN frequency, then so is $\omega=-a+ib$. We only show in Table
\ref{tab:introqnm} frequencies with a positive real part.  There are a
countable infinity of QN frequencies. They are usually arranged by
their imaginary part. Thus, the frequency with lowest imaginary part
is called the fundamental frequency and is labeled with the integer
$n=0$; the one with second lowest imaginary part is the first overtone
($n=1$) and so on.  The real part can be shown to be, using a WKJB
type of reasoning \cite{schutz}, of the order of the height of the
potential barrier, i. e., $Re[\omega] \sim V_{\rm max}$, where $V_{\rm
max}$ is the maximum value attained by the potential $V$.  The real
part is, in the Schwarzschild geometry, bounded from above, whereas
the imaginary part seems to grow without bound for overtone number $n
\rightarrow \infty$. This is shown in Fig.\ref{fig:asymptintro},
where we show the first two thousand QNMs, using Leaver's
technique, plus an improvement devised by Nollert \cite{nollert2}.
\vskip 1mm

Notice that as the imaginary part of $\omega$ goes to infinity the real part seems
to go to a constant (in this case $0.0437123$) , and this has given rise to a lot of discussion
recently, in connection with the quantization of black hole area. This
issue will be addressed later on.  

\section{What is the importance of QNMs in fundamental physics}
Once one realizes the importance of black holes in fundamental
physics, one can grasp the meaning and significance of their
characteristic vibrations. Black holes are {\it the} general
relativistic object by excellence. They are pure objects, in the sense that
only a few parameters, like mass, charge and angular momentum are
enough to describe them. In this respect they come very close to our
notion of elementary particles.  They have been called the hydrogen
atom of general relativity and this is indeed a very good
comparison. Like in the hydrogen atom $\leftrightarrow$ quantum
mechanics dichotomy, a black hole being a solution to Einstein's
equations has all of the general relativistic properties embodied in
it, but still is simple enough to be a model for starting a complete
understanding of all the physics that go with Einstein's equations.
Moreover, probing black hole physics is probing the strong field
regime of general relativity, something that has still not been done
(here, gravitational waves may also play an important role. We defer
this discussion for Part II of this thesis.)

There are mainly three major motivations leading us to study in detail
the QNMs of black holes, namely estimating black hole parameters by
using the fact that QN frequencies depend only on intrinsic parameters,
estimating thermalization timescales in connection with the AdS/CFT
conjecture and semi-classical attempts to quantize the black hole
area, that are now making heavy use of highly damped QNMs.  These are
described below.

\subsection{QNMs and black hole parameter estimation}
Historically, one of the main reasons for studying QNMs
is that they may provide valuable help for identifying black hole
parameters.  Gravitational waves have not yet been detected, but their
indirect influence has been seen and measured with remarkable accuracy
by Hulse and Taylor for the binary pulsar PSR 1913+16.  One expects to
finally detect gravitational waves directly in the forthcoming years,
with gravitational wave detectors already taking data
\cite{geo,ligo,virgo}.  However, gravitational waves are so weak that
in order to ever hope for their detection one needs to have a very
strong source of gravitational waves, for instance stellar collapse or
black hole collisions. Numerical simulations have shown that in the
final stage of such processes (ringdown) QNMs dominate
the black hole response.  Of course, since the QNMs are
exponentially damped in time, only the mode corresponding to the
fundamental frequency (lowest imaginary part) should show up in the
signal. This is indeed the case.  Now, as we said earlier, the
QN frequencies depend only on the black hole fundamental
parameters. For example, if we are dealing with a Kerr black hole, the
QN frequencies will depend only on the black hole mass $M$
and angular momentum per unit mass $a$.
As such it should be possible to infer the black hole parameters solely
from the knowledge of the ringing frequency and the damping time. 
Consider a single interferometric gravitational wave detector and a perturbed
black hole. Focus attention on a single oscillation mode of the black hole,
for example the $l=m=2$ mode. Then the strain measured by the detector has the
time dependence of an exponentially damped sinusoid characterized by four parameters
$Q$, $f$, $V$ and $T$ \cite{echeverria}
\begin{equation}
h(t)=V^{-1/3}e^{-\pi f \frac{t-T}{Q}}\sin{2 \pi f(t-T)}.
\label{parestimation}
\end{equation}
Here $Q$ is the quality factor measured in units of the oscillation frequency $f$,
\begin{equation}
Q=\pi f \tau \,,
\label{Qdef}
\end{equation}
where $\tau$ is the damping time.
$V^{-1/3}$ is the amplitude of the waveform, which depends on the distance 
to the source, the size of the perturbation, and the relative orientation of
the detector and the source.
Finally, $T$ is the starting time of the perturbation.
Note the $Q$ and $f$ are related to the imaginary and real part of the 
fundamental QN frequency. In particular one finds that a good fit to the 
numerical results for the fundamental QN frequencies of a rotating hole is
\cite{echeverria}
\begin{equation}
f \sim \frac{1}{2\pi M} [1-\frac{63}{100}(1-a)^{3/10}]\,,
\label{wrestimate}
\end{equation}
\begin{equation}
Q=\pi f \tau =2(1-a)^{-9/20}\,.
\label{wiestimate}
\end{equation}
Now, by measuring the ringing frequency and decay time one could in principle
derive the parameters of the black hole, by inverting (\ref{wrestimate}) and (\ref{wiestimate}).
For a very recent work on how to effectively search for the gravitational ringing of black holes
see \cite{nakanoringing}.

\subsection{QNMs and thermalization in the AdS/CFT}
Quantum field theories are currently the very best we have for a
microscopic description of nature, and these are verified
experimentally to a high degree of accuracy. They treat particles as
point-like, capable of interacting locally with other
particles. Successful as quantum theories are, we have not yet been
able to incorporate gravity in such a formalism. Nevertheless, if one
gives up the notion that particles are point-like, and assumes that
the fundamental objects are strings, it is possible to quantize
gravity in a consistent scheme \cite{green}.  All string theories
include a particle with zero mass and spin two, and since gravity is
the only consistent interaction for massless spin two particles, this
means any string theory will contain gravity.  
Now, another theory that describes matter, namely quarks and gluons, is QCD.
QCD is a gauge theory based on the group
SU(3). This is sometimes stated by saying that quarks have three
colors.  QCD is asymptotically free, meaning that the effective
coupling constant decreases as the energy increases. At low energies
QCD becomes strongly coupled and it is not easy to perform
calculations. It was suggested by 't Hooft that the theory might
simplify when the number of colors $N$ is very large \cite{hooft}.
The plan was to try and solve exactly for $N= \infty$ and then do an
expansion in $1/N=1/3$. Furthermore there are strong indications that
the large $N$ theory is a free string theory with string coupling
constant $1/N$.  
The correspondence between QCD when $N\rightarrow \infty$ and string theory
is the AdS/CFT
correspondence conjecture
\cite{maldacena}.
It states that the bulk physics in AdS$_{n}$ is dual to the physics of a 
certain CFT$_{n-1}$, which may be thought of as living in the conformal boundary of
AdS$_{n}$.
For example, it conjectures that $U(N)$ Yang-Mills theory, with number
of spinor supercharges $\cal N =$ 4, which is a conformal
theory, is the same as a ten-dimensional superstring theory on AdS$_5$
$\times$ $S^5$.  The only particular feature of these strings is that
they are moving on an anti-de Sitter space. The
radius of curvature of this anti-de Sitter space depends on $N$ and
large $N$ corresponds to a large radius of curvature. Thus, by taking
$N$ to be large we can make the curvature as small as we want. The
theory in AdS includes gravity since any string theory includes
gravity. So it is claimed that there is an equivalence
between a gravitational theory and a field theory!  However the
mapping between the gravitational and the field theory is quite
non-trivial since the field theory lives in a lower dimension.  For
more details on this see the review \cite{maldacenareview}.  The
QN frequencies of AdS black holes have a direct
interpretation in terms of the dual conformal field theory. According
to the AdS/CFT correspondence, a large static black hole in AdS
corresponds to an approximately thermal state in the CFT, and
perturbing the black hole corresponds to perturbing this thermal
state.  The decay of the perturbation describes therefore the return
to equilibrium.  Now, we know that the fundamental QN
frequency should control the decay of any perturbation, and therefore
by computing QN frequencies in AdS spacetime (containing a
black hole), one obtains a prediction for the thermalization timescale
in the strongly coupled CFT.  It seems extremely difficult to compute
this timescale directly in the CFT.
\subsection{QNMs and black hole area quantization}
There are other attempts to quantize gravity, besides string theory.
As we remarked earlier, one has
the strong conviction that black holes may play a major role in our
attempts to shed some light on a quantum theory of gravity. 
The quest for a consistent quantization of black holes was initiated
many years ago by Bekenstein \cite{bek0,bek} who, based solely
on semi-classical arguments, conjectured that the horizon area of a
(non extremal) black hole should have a discrete eigenvalue spectrum.
By using a semi-classical version of Christodoulo's processes along
with Heisenberg's quantum uncertainty principle, Bekenstein deduced the
quantization condition of the black hole area,
\begin{equation}
A_n=\gamma l{_p}^2 n\,\,, n=1,2,...\,,
\label{quantconintro}
\end{equation}
where $\gamma$ is a dimensionless constant and
$l{_p}=\sqrt{\frac{G}{c^3}}\sqrt{\frac{h}{2\pi}}$ is the Planck length.
This formula has been arrived at only by using Sommerfeld's quantization procedure.
The quantization of horizon area in equal steps brings to mind an horizon formed
by patches of equal area $\gamma l{_p}^2$ which get added one at a time.
There is no need to think of a specific shape or localization of these patches.
It is their standard size which is important, and which makes them all equivalent.
This patchwork horizon can be regarded as having many degrees of freedom, one for
each patch. In quantum theory degrees of freedom independently manifest
distinct states. Since the patches are all equivalent, each will have the same number
of quantum states, say $k$. Therefore the total number of quantum states of the horizon is
\begin{equation}
N=k^{\frac{A}{\gamma l_p^2}}\,,
\label{numstatesintro}
\end{equation}
where $k$ is a positive integer. The $N$ states may not all be equally probable.
But if the $k$ states of each patch are all likely, then all $N$ states
are equally probable. In that case the statistical (Boltzmann) entropy associated
with the horizon is $\log N$ or
\begin{equation}
S_{BH}=\frac{\log k}{\gamma} \frac{A}{l_p^2}.
\label{entrintro}
\end{equation}
Comparing with Hawking's formula for the entropy, $S_{BH}=\frac{A}{4}$ one gets
\begin{equation}
\gamma=4\log k .
\label{gammaintro}
\end{equation}
Thus, statistical physics arguments force the dimensionless constant $\gamma$
to be of the form (\ref{gammaintro}). Still, a specific value of $k$ requires
further input, which was not available up until recently.
However, a sudden turning of events happened a few years ago, when Hod \cite{hod}
made a proposal to determine $\gamma$. His proposal made the spacing
of area levels consistent both with the area-entropy thermodynamic relation for
black holes, with the Boltzmann-Einstein formula in statistical physics, and
with Bohr's correspondence principle.
Hod builds his conjecture on Bohr's correspondence principle,
``transition frequencies at large quantum numbers should equal classical
 oscillation frequencies'', and then goes on to associate these classical oscillation 
frequencies with the highly damped QN frequencies, since 
``quantum transition do not take time'' \cite{bek2}.
As previously noted, the highly damped QN frequencies of a Schwarzschild
black hole were numerically found to be given by
\begin{equation}
M \omega _n =0.0437123-\frac{i}{4}(n+1/2)+{\mathcal O}(\frac{1}{\sqrt n})\,,
\label{qnfasymptnoll}
\end{equation}
and they are independent of $l$, the angular quantum number characterizing
the angular distribution. Hod observed that the numerical limit for the real part,
$0.0437123$ agreed (to the available data) with $\frac{\log3}{8\pi}$.
He therefore conjectured that asymptotically 
\begin{equation}
M \omega _n =\frac{\log3}{8\pi}-\frac{i}{4}(n+1/2)\,.
\label{Hofconjasymptnoll}
\end{equation}
Using $A=16\pi M^2$ and $dM=E=\frac{h}{2\pi}\omega$ he then concluded that
\begin{equation}
\gamma=4\log3\,,
\label{Hofconk}
\end{equation}
and therefore that
\begin{equation}
A_n=4l_p^2\log(3)  \times n\,.
\label{HofconA}
\end{equation}
Following Hod's suggestion, Dreyer \cite{dreyer}
recently used a similar argument to fix a free parameter (the
so-called Barbero-Immirzi parameter) appearing in Loop Quantum
Gravity, an alternative approach to a quantum gravity theory.
Supposing that transitions of a quantum black hole are characterized
by the appearance or disappearance of a puncture with lowest possible
spin $j_{\rm min}$. Dreyer found that Loop Quantum Gravity gives a
correct prediction for the Bekenstein-Hawking entropy if $j_{\rm
min}=1$, thereby fixing the Barbero-Immirzi parameter.  When Hod made
his original proposal, formula (\ref{Hofconjasymptnoll}) and therefore
also (\ref{Hofconk}) and (\ref{HofconA}) was just based on a curious
numerical coincidence. It has however been recently established by
Motl \cite{motl1} and Motl and Neitzke \cite{motl2} with analytical
techniques.
In this quantum gravity context, in which the correspondence principle is valid, 
relevant modes
are those which damp infinitely fast, do not significantly contribute to the
gravitational wave signal, and are therefore typically ignored in studies of 
gravitational radiation.
It is probably to early to assess the significance of these proposals.
An obvious route to follow is to generalize these results for the Kerr
black hole. However, as we shall describe, the Kerr black hole has
bravely resisted all attempts to do so. First, the most complete
numerical studies so far (see Chapter \ref{chap:qnmkerr}) do not seem to be able to
completely probe the asymptotic behaviour of the QN
frequencies.  The available results point to some strange behaviour.
\section{Outline of Part I}
In the first part of this thesis we shall make an extensive study of
QNMs and associated QN frequencies of black holes, both in
asymptotically flat and in asymptotically de Sitter or anti-de Sitter
backgrounds.  

In Chapter \ref{chap:qnmbtz} we shall begin by studying the QNMs and QN
frequencies of the so called BTZ black hole
\cite{banados}. This is a black hole solution to Einstein's equations
with a cosmological constant in (2+1) spacetime dimensions, and lives
in an asymptotically anti-de Sitter background.  We shall find that it
is possible to solve {\it analytically} the wave equation in this
background, and also that it is possible to find simple analytical,
exact expressions for the QN frequencies. This is the
first instance of an analytical solution for the QN frequencies of a
black hole, and was found by Cardoso and Lemos \cite{cardosoqnmbtz}.
Note that we shall study only scalar, electromagnetic and Weyl
perturbations, and not gravitational perturbations, since Einstein's
equations in (2+1) dimensions have no dynamical degrees of freedom,
and so there are no gravitational wave sin this spacetime.  The fact
that there is a simple exact expression for the quasinormal
frequencies makes it possible to make a number of studies which would
otherwise have to rely on numerical calculations. We shall see
examples of this in Part II.  Besides, it was possible, very recently,
to use these results to test for the validity of the QNM/black hole
area quantization relation, so the BTZ black hole is in fact a very
useful guide.

In Chapter \ref{chap:qnmsads} we go on to study the QNMs and QN
frequencies of the four dimensional Schwarzschild-anti-de Sitter black
hole. In order to analyze gravitational quasinormal modes we shall
decompose Einstein's equations in tensorial spherical harmonics and
repeat the procedure of Regge and Wheeler \cite{regge} and Zerilli
\cite{zerilli2} to arrive at the gravitational wave equation.  Notice
that both in Chapters
\ref{chap:qnmbtz} and \ref{chap:qnmsads} the regime of very high overtones, 
i.e., highly damped modes, is studied, and therefore it is possible to
make close contact with the AdS/CFT because these black holes live in
an asymptotically anti-de Sitter background, but also with the recent
conjectures relating highly damped QNMs with black hole area
quantization and Loop Quantum Gravity.

There are, however, families of black holes in general relativity with
a negative cosmological constant with horizons having topology
different from spherical. In Chapter \ref{chap:qnmtoro} we want to
focus on the family of black holes whose horizon has toroidal,
cylindrical or planar topology \cite{lemos1}, and make a complete
analysis of their QNMs. 
The analysis of gravitational wave is a non-trivial task, which we shall
carry by using Chandrasekhar's \cite{chandra} formalism.

In Chapter \ref{chap:qnmds} we will turn to
spacetimes which are asymptotically de Sitter. We will consider a near
extremal Schwarzschild-de Sitter black hole.  In this case it is also
possible to find an exact solution to the wave equation leading again
to exact analytical results, as was shown by Cardoso and Lemos
\cite{cardosoqnmtoro}. However, this procedure only yields the correct
QN frequencies up to some overtone number, i.e., it fails to give the correct
asymptotic behaviour. 
In Chapter \ref{chap:qnms} we shall make an exhaustive numerical and analytical
investigation of the QNMs of Schwarzschild black holes in four and higher dimensions.
We shall first start by reviewing results that are known for quite some time, and then
we shall take a look at new data. 
Finally, in Chapter \ref{chap:qnmkerr}, the last Chapter of Part I,
we will revisit the quasinormal frequencies of the Kerr black hole,
with the purpose to investigate the asymptotic behaviour of QN frequencies
with a large imaginary part. This will hopefully help us on deciding whther or not
there is some deep connection between QNMs and Loop Quantum Gravity,
or then if QNMs are really at the heart of black hole area quantization.
Here we will follow Berti et al \cite{cardosoqnmkerr}.

\thispagestyle{empty} \setcounter{minitocdepth}{1}
\chapter[Quasinormal modes of the three dimensional BTZ (AdS) black hole]{Quasinormal modes of the three dimensional BTZ (AdS) black hole} \label{chap:qnmbtz}
\lhead[]{\fancyplain{}{\bfseries Chapter \thechapter. \leftmark}}
\rhead[\fancyplain{}{\bfseries \rightmark}]{}
\minitoc \thispagestyle{empty}
\renewcommand{\thepage}{\arabic{page}}
\section{Introduction}

Up until very recently, all these works in QNMs dealt
with asymptotically flat spacetimes. In the past few years there has
been a growing interest in asymptotically AdS (anti-de Sitter)
spacetimes.  Indeed, the Ba\~nados-Teitelboim-Zanelli (BTZ) black hole
in 2+1-dimensions
\cite{banados}, as well as black holes in 
3+1 dimensional AdS spacetimes with nontrivial topology (see,
e.g. \cite{lemos1}), share with asymptotically flat spacetimes the
common property of both having well defined charges at infinity, such
as mass, angular momentum and electromagnetic charges, which makes
them a good testing ground when one wants to go beyond asymptotic
flatness.  Another very interesting aspect of these black hole
solutions is related to the AdS/CFT (Conformal Field Theory)
conjecture \cite{maldacena}. For instance, due to this AdS/CFT
duality, quasi-normal frequencies in the BTZ black hole spacetime
yield a prediction for the thermalization timescale in the dual
two-dimensional CFT, which otherwise would be very difficult to
compute directly. If one has, e.g., a 10-dimensional type IIB
supergravity, compactified into a ${\rm BTZ} \times S^3 \times T^4$
spacetime, the scalar field used to perturb the BTZ black hole, can be
seen as a type IIB dilaton which couples to a CFT field operator
$\cal{O}$. Now, the BTZ in the bulk corresponds to to a thermal sate
in the boundary CFT, and thus the bulk scalar perturbation corresponds
to a thermal perturbation with nonzero $<\cal O>$ in the CFT.

The study of QNMs in AdS spacetime was initiated, with
this motivation in mind, by Horowitz and Hubeny \cite{horowitz}, but
it focused on black holes living in AdS spacetime with dimension
greater or equal to four.  The QNMs of the BTZ black hole
were first computed by Cardoso and Lemos \cite{cardosoqnmbtz} who
investigate the non-rotating black hole solution, and turned out to
reveal a pleasant surprise: one can find an exact, analytical solution
for them.  In fact, it was the first example of a black hole spacetime
for which one could solve exactly the wave equation, and this makes it
an extremely important spacetime, where one can start to prove or
disprove conjectures relating QNMs and critical phenomena
or black hole area quantization.  This work was later extended to
rotating black holes by Birmingham
\cite{birmingham}, who could also establish a relation between
QN frequencies and the Choptuik parameter \cite{chop}, a
conjecture put forward in \cite{horowitz}.  Recently, and still in 2+1
dimensions, the QNMs of the more general dilatonic black
holes were studied in \cite{fernando}, and QNMs in three
dimensional time dependent anti-de Sitter spacetime were studied in
\cite{shen}
That the QN frequencies of the BTZ black hole lead indeed to
relaxation times in the CFT was proved by Birmingham, Sachs and
Solodukhin \cite{birmingham2}, who found agreement between
quasi-normal frequencies and the location of the poles of the retarded
correlation function of the corresponding perturbations in the dual
conformal field theory. This then provides a new quantitative test of
the AdS/CFT correspondence.  For previous work on BTZ black holes such
as entropy of scalar fields, see \cite{satoh} and references therein.

We shall now describe how to compute exactly the QNMs of the 3D
non-rotating BTZ black hole \cite{banados}.  This will follow closely
the work in \cite{cardosoqnmbtz}.  The non-rotating BTZ black hole
metric for a spacetime with negative cosmological constant, $\Lambda =
-\frac{1}{l^2}$, is given by
\begin{equation}
ds^{2}= (-M+\frac{r^2}{l^2})dt^{2}-
 (-M+\frac{r^2}{l^2})^{-1}dr^{2} - r^2d{\phi}^2                    
\,,
\label{2.1}
\end{equation}
where $M$ is the black hole mass. The horizon radius is given by $r_+
=M^{1/2}l$.  
The original article \cite{banados} considers
the static and rotating uncharged BTZ black hole and the static
electrically charged BTZ black hole.  The extension to consider the
rotating electrically charged BTZ black hole (mass, angular momentum
and electric charge different from zero) has been done by
Cl\'ement \cite{CL1} and by Mart\'{\i}nez, Teitelboim and
Zanelli \cite{BTZ_Q}. The static magnetic counterpart of the BTZ
solution has been
considered by  Cl\'ement \cite{CL1}, Hirschmann and Welch \cite{HW} and
Cataldo and
Salgado \cite{Cat_Sal}. The rotating magnetic solution has been found
by Dias and Lemos \cite{OscLemBTZ} and a physical interpretation for the
magnetic
field source has been given.

We shall in what follows suppose that the scalar,
electromagnetic and Weyl (neutrino) fields are a perturbation, i.e., they
propagate in a spacetime with a BTZ metric.  We will find that all
these fields obey a wave equation and the associated QNM 
are exactly
soluble yielding certain hypergeometric functions. As for the frequencies
one has exact and explicit results for scalar and electromagnetic
perturbations and numerical results for Weyl perturbations.
To our knowledge, this is the first exact solution of QNMs for 
a specific model (see \cite{beyer}).

In section 2 we give the wave equation for scalar and electromagnetic 
perturbations, and find the QNMs themselves and their frequencies. 
In section 3 we find the wave equation for Dirac and Weyl (neutrino) 
perturbations and analyze their QNMs.
We should note that Einstein's equations in $2+1$ dimensions
have no dynamical degrees of freedom, and therefore there is no
such thing as gravitational waves in $2+1$ dimensions. All the
gravitational degrees of freedom can be gauged away.


\noindent
\section{Perturbing a non-rotating BTZ black hole with scalar and electromagnetic fields}


\subsection{The wave equation}
In this subsection we shall analyze the scalar and electromagnetic  
perturbations, which as we shall see yield the same effective potential, 
and thus the same wave equation.

First, for scalar perturbations, we are interested in solutions to 
the minimally coupled scalar wave equation 
\begin{equation}
{\Phi^\mu}_{;\,\mu}=0 \,, 
\label{minimalscalareq1}
\end{equation}
where, a comma stands for ordinary derivative and a semi-colon stands
for covariant derivative.
We make the following ansatz for the field $\Phi$
\begin{equation}
\Phi=\frac{1}{r^{1/2}}\Psi(r)e^{-i\omega t}e^{im\phi}\,,
\label{ansatzforscalar}
\end{equation}
where $m$ is the angular quantum number. It is useful to use the tortoise coordinate  $r_*$
defined by the equation $dr_*=\frac{dr}{-M+\frac{r^2}{l^2}}$, and given implicitly by 
\begin{equation}
r=-M^{1/2}\coth(M^{1/2}r_*) \,,
\label{radiusastortoise}
\end{equation}
with $r_*\; \epsilon\;]-\infty,0]$, ($r_*=-\infty$ corresponds to $r=r_+$, and 
$r_*=0$ corresponds to $r=\infty$). 

With the ansatz (\ref{ansatzforscalar}) and the tortoise coordinate $r_*$, equation 
(\ref{minimalscalareq1}) is given by, 
\begin{equation}
\frac{d^2 \Psi(r)}{d {r_*}^2} + (\omega - V(r))\Psi(r)=0\,,
\label{minimalscalareq2}
\end{equation}
where,
\begin{equation}
V(r)=\frac{3r^2}{4 l^4} - \frac{M}{2 l^2}-\frac{M^2}{4 r^2}+\frac{m^2}{l^2}
 - \frac{Mm^2}{r^2}\,,
\label{potentialscalar}
\end{equation}
and it is implicit that $r=r(r_*)$. The rescaling to the the radial coordinate 
$\hat{r}=\frac{r}{l}$ and to the frequency $\hat{\omega}=\omega l$ is equivalent 
to take $l=1$ in (\ref{minimalscalareq2}) and (\ref{potentialscalar}), i.e., 
through this rescaling one measures the frequency and other quantities in 
terms of the AdS lengthscale $l$.

Now, the electromagnetic perturbations are governed by Maxwell's equations
\begin{equation}
{F^{\mu\nu}}_{\,;\nu}=0 \,\,, {\rm with} \,\, F_{\mu\nu}=A_{\nu,\mu}-
A_{\mu,\nu}\,,
\label{maxwellequation} 
\end{equation}
where $F_{\mu\nu}$ is the Maxwell tensor and $A_\mu$ is the electromagnetic 
potential. 
As the background is circularly symmetric, it would be advisable to expand $A_{\mu}$
in 3-dimensional vector spherical harmonics (see \cite{edmonds} 
and \cite{ruffini}):
\begin{eqnarray}
A_{\mu}(t,r,\phi)=
\left[ \begin{array}{c}g^{m}(t,r)\\h^{m}(t,r)
\\ k^{m}(t,r) \end{array}\right]e^{im\phi} \,, 
\label{empotentialdecomposition}
\end{eqnarray}
where $m$ is again our angular quantum number, and 
this decomposition is similar to the one in eigenfunctions
of the total angular momentum in flat space \cite{edmonds}. 

However, going through the same steps one finds that the equation 
for electromagnetic perturbations is the same as the one for scalars, 
equation (\ref{minimalscalareq2}). The reason is that in three dimensions 
the 2-form Maxwell field $F=F_{\mu\nu}dx^\mu\wedge dx^\nu$ is dual to 
a 1-form $d\Phi$.

\subsection{QNMs for scalar and electromagnetic perturbations}

Although a precise mathematical definition for a QNM can be given, as a pole
in the Green's function \cite{kokkotas}, we shall follow a more
phenomenological point of view. A QNM  describes the
decay of the field in question.  For the equation
(\ref{minimalscalareq2})  
it is defined as a corresponding solution 
which (i) near the horizon is purely ingoing, $\sim
e^{i\omega r_*}$, corresponding to the existence of a black hole,
 and (ii) 
near infinity is 
purely outgoing, $\sim e^{-i\omega r_*}$, (no
initial incoming wave from infinity is allowed). One can see that the potential $V(r)$ 
diverges at infinity, so we require that the perturbation
vanishes there (note that $r=\infty$ corresponds to a finite value of $r_*$, namely
$r_*=0$).

 \subsubsection{Exact calculation}

Putting $l=1$  and 
using the coordinate $r_*$, the wave equation (\ref{minimalscalareq2}) takes the form
\begin{eqnarray}
&\frac{\partial^{2} \Psi(r)}{\partial r_*^{2}} +
&\nonumber\\&
\left[\omega^2
 -\frac{3M}{4\sinh(M^{1/2}r_*)^2}+\frac{M}{4\cosh(M^{1/2}r_*)^2}+
\frac{m^2}{\cosh(M^{1/2}r_*)^2}\right]\Psi(r)=0 \,.&
\label{scalarmaxwellequation2}
\end{eqnarray}
On going to a new variable $x=\frac{1}{\cosh(M^{1/2}r_*)^2}$,
$x\; \epsilon\;[0,1]$ equation (\ref{scalarmaxwellequation2}) 
can also be written as 
\begin{equation}
4x(1-x)\partial_x^2 \Psi(r) +(4-6x)\partial_x\psi +\bar{V}(x)\Psi(r)=0 \,,
\label{scalarmaxwellequation3}
\end{equation}
where 
\begin{equation}
\bar{V}(x)=\frac{1}{4x(1-x)}\left[\frac{4\omega^2(1-x)}{M}-3x
-x(1-x)-\frac{4m^2x(1-x)}{M}\right]\,.
\end{equation}
By changing to a new wavefunction y (see \cite{uvarov} for details), through
\begin{equation}
\Psi \rightarrow \frac{(x-1)^{3/4}}{x^{\frac{i\omega}{2M^{1/2}}}} y \,,
\label{step}
\end{equation}
equation (\ref{scalarmaxwellequation3})
can be put in the canonical form \cite{uvarov,stegun}
\begin{equation}
x(1-x)y'' + [c-(a+b+1)x]y' -aby =0 \,,
\label{scalarmaxwellequation4}
\end{equation}
with 
$a=1+\frac{im}{2M^{1/2}}-\frac{i\omega}{2M^{1/2}} $, 
$b=1-\frac{im}{2M^{1/2}}-\frac{i\omega}{2M^{1/2}} $, 
and $c=1-\frac{i\omega}{M^{1/2}}$,
which is a
standard hypergeometric equation. The hypergeometric equation has
three regular singular points at $x=0,x=1,x=\infty$, and has two
independent solutions in the neighbourhood of each singular point. We
are interested in solutions of (\ref{scalarmaxwellequation4}) in the range [0,1],
satisfying the boundary conditions of ingoing waves near $x=0$, and
zero at $x=1$. One solution may be taken to be
\begin{equation}
y=(1-x)^{c-a-b}F(c-a,c-b,c;x) \,,
\label{hypersolution}
\end{equation}
where $F$ is a standard hypergeometric function of the second kind.
Imposing $y=0$ at $x=1$, and recalling that 
$F(a,b,c,1)=\frac{\Gamma(c)\Gamma(c-a-b)}{\Gamma(c-a)\Gamma(c-b)}$, we get
\begin{equation}
a=-n \,, {\rm or}\;  b=-n \,,
\label{parameters}
\end{equation}
with $n=0,1,2,...\,$, so that the quasi normal frequencies are given by
\begin{equation}
\omega=\pm m -2iM^{1/2}(n+1).
\label{frequency1}
\end{equation}
The lowest frequencies, namely those with $n=0$ and $m=0$ had already been
obtained by \cite{govinda} and agree with our results. 

\subsubsection{Numerical calculation of the frequencies}

In order to check our results, we have also computed numerically the
frequencies. By going to a
new variable $ z=\frac{1}{r} $, $h=\frac{1}{r_+}$ one can put the wave
equation (\ref{minimalscalareq2}) in the form (see
\cite{horowitz} for further details)
\begin{equation}
s(z)\frac{d^2}{dz^2}\Theta +t(z)\frac{d}{dz}\Theta+u(z)\Theta=0 \,, 
\label{scalarmaxwellequation5}
\end{equation}
where $\Theta=e^{i\omega r_*} \Psi(r)$, $s(z)=z^2-Mz^4$, $t(z)=2Mz^3-2i \omega z^2$ 
and $u(z)=\frac{V}{-M+\frac{1}{z^2}}$, with $V$ given by 
(\ref{potentialscalar}). 
Now, $z$ $\epsilon$ $[0,h]$ and one sees that in this range, the
differential equation has only regular singularities at $z=0$ and
$z=h$, so it has by Fuchs theorem a polynomial solution. We can now
use Fr\"{o}benius method (see for example \cite{arfken}) and look for
a solution of the form $\Theta(z)= \sum_{n=0}^{\infty}
\theta_{n(\omega)} (z-h)^n (z-h)^{\alpha} $, where $\alpha$ is to be
taken from the boundary conditions. Using the boundary condition of
only ingoing waves at the horizon, one sees \cite{horowitz} that
$\alpha=0$. So the final outcome is that $\Theta\,$ can be expanded as
\begin{eqnarray}
\Theta(z)= \sum_{n=0}^{\infty} \theta_{n(\omega)} (z-h)^n \,.
\label{expansion1} 
\end{eqnarray}
Imposing  now the second boundary condition, $\Theta=0$ at infinity ($z=0$) one gets
\begin{equation}
\sum_{n=0}^{\infty} \theta_{n(\omega)}(-h)^n=0 \,.
\label{expansion2} 
\end{equation}
\begin{table}
\caption{\label{btz1} Lowest ($n=0$) QNM frequencies for symmetric ($m=0$) scalar
and electromagnetic perturbations of a non-rotating BTZ black hole.}
\begin{tabular}{lllll}  \hline 
\multicolumn{5}{c}{$m=0$} \\  \hline 
\multicolumn{1}{l}{} &
\multicolumn{2}{l}{ Numerical} &
\multicolumn{2}{l}{ Exact} \\ \hline
$M^{1/2}$   &  $\omega_r$ &  $-\omega_i$ &  $\omega_r$  & $-\omega_i$ \\ \hline
$\frac{1}{2}$   & 0.000 & 1.000 & 0 & 1  \\ \hline
 1   & 0.000 & 2.000 & 0 & 2  \\ \hline
 5    & 0.000 & 10.000 & 0 & 10  \\ \hline
 10   &  0.000 & 20.000  & 0 & 20  \\ \hline
 50   & 0.000 & 100.000 & 0 & 100  \\ \hline
 100   & 0.000  & 200.000 & 0 & 200   \\ \hline
 1000   & 0.000  & 2000.000 & 0& 2000   \\ \hline
\end{tabular}
\end{table}
The problem is reduced to that of finding a numerical solution of the
polynomial equation (\ref{expansion2}).  The numerical roots for
$\omega$ of equation (\ref{expansion2}) can be evaluated resorting to
numerical computation. Obviously, one cannot determine the full sum in
expression (\ref{expansion2}), so we have to determine a partial sum
from 0 to N, say, and find the roots $\omega$ of the resulting
polynomial expression.  We then move onto the next term N+1 and
determine the roots.  If the method is reliable, the roots should
converge. We have stopped our search when a 3 decimal digit precision 
was achieved.
We have computed the lowest frequencies for some parameters of the
angular quantum number $m$ and horizon radius $r_+$.  The frequency is written as
$\omega = \omega_r + i\omega_i$, where $\omega_r$ is the real part of
the frequency and $\omega_i$ is its imaginary part.

In tables \ref{btz1} and \ref{btz2} we list the numerical values of the lowest
QNM frequencies, for $m=0$ and $m=1$, respectively, and 
for selected values of the black hole mass.
\begin{table}
\caption{\label{btz2} Lowest ($n=0$) QNM frequencies for $m=1$ scalar
and electromagnetic perturbations of a non-rotating BTZ black hole.}
\begin{tabular}{lllll}  \hline 
\multicolumn{5}{c}{$m=1$} \\  \hline 
\multicolumn{1}{l}{} &
\multicolumn{2}{l}{ Numerical} &
\multicolumn{2}{l}{ Exact} \\ \hline
$M^{1/2}$   &  $\omega_r$ &  $-\omega_i$ &  $\omega_r$  & $-\omega_i$ \\ \hline
$ \frac{1}{2} $  & 1.000 & 1.000 & 1 & 1  \\ \hline
 1   & 1.000 & 2.000 & 1 & 2  \\ \hline
 5    & 1.000 & 10.000 & 1 & 10  \\ \hline
 10   &  1.000 & $20.000$  & 1 & 20  \\ \hline
 50   & 1.000 & 100.000 & 1 & 100  \\ \hline
 100   & 1.000  & 200.000 & 1 & 200   \\ \hline
 1000   & 1.000  & 2000.000 & 1& 2000   \\ \hline
\end{tabular}
\end{table}
The numerical results agree perfectly with (\ref{frequency1}), and
one sees that the imaginary part of the frequency scales with the 
horizon whereas the real part depends only on the angular index $m$.

\section{Perturbing a black hole with Dirac and Weyl spinor fields}

\subsection{The wave equation}


We shall develop Dirac's equation for a massive spinor, and then
specialize to the massless case.
The two component massive spinor field $\Psi$, with mass
$\mu_s$ obeys the covariant
Dirac equation
\begin{equation}
i\gamma^\mu \nabla_{\mu}\Psi -\mu_s \Psi=0\, ,
\label{diracequation}
\end{equation}
where $\nabla_{\mu}$ is the spinor covariant derivative defined by
$\nabla_{\mu} = \partial_{\mu}+\frac{1}{4} \omega^{ab}_{\mu} \gamma _{[a} 
\gamma_{b]}$, and $\omega^{ab}_{\mu}$ is the spin connection, which 
may be given in terms of the tryad $e_a^\mu$. 

As is well known there are two inequivalent two dimensional irreducible
representations of the $\gamma$ matrices in three spacetime dimensions.
The first may be taken to be $\gamma^0=i\sigma^2\,,\gamma^1=\sigma^1$, and 
$\gamma^2=\sigma^3$, where the matrices $\sigma^k$ are the Pauli matrices.
The second representation is given in terms of the first by a minus 
sign in front of the Pauli matrices. From equation (\ref{diracequation}), 
one sees that a Dirac particle with mass $\mu_s$ in the first representation 
is equivalent to a Dirac particle with mass $-\mu_s$ in the second 
representation. To be definitive, we will use the first representation, but the 
results can be interchanged to the second one, by substituting $\mu_s\rightarrow-\mu_s$.
For Weyl particles, $\mu_s=0$, both representations yield the same results.

Again, one can separate variables by setting
\begin{eqnarray}
\Psi(t,r,\phi)=
\left[ \begin{array}{c}\Psi_1(t,r)\\ \Psi_2(t,r)
\end{array}\right]e^{-i\omega t}e^{im\phi} \,.
\label{diracdecomposition}
\end{eqnarray}
On substituting this decomposition into Dirac's equation (\ref{diracequation}) 
we obtain
\begin{eqnarray}
-\frac{i(M-2r^2)}{2\Delta^{1/2}}r\Psi_2 +i\Delta^{1/2}\partial _r \Psi_2
+\frac{r^2\omega}{\Delta^{1/2}}\Psi_2=(m+\mu_s)\Psi_1 \,,
\label{diracequation2a}
\\
-\frac{i(M-2r^2)}{2\Delta^{1/2}}r\Psi_1 +i\Delta^{1/2}\partial _r \Psi_1
+\frac{r^2\omega}{\Delta^{1/2}}\Psi_1=(m+\mu_s)\Psi_2 \,,
\label{diracequation2b}
\end{eqnarray}
where we have put $\Delta=-Mr^2+\frac{r^4}{l^2}$, we have restored the 
AdS lengthscale $l$, and in general we follow
Chandrasekhar's notation \cite{chandra}. Defining $R_1$, $R_2$, and $\hat{m}$
through the relations
\begin{eqnarray}
\Psi_1=i\Delta^{-1/4}R_1\,, 
\label{transform1}\\
\Psi_2= \Delta^{-1/4}R_2 \,,
\label{transform2}\\  
m=i\hat{m}\,,
\label{transform3}
\end{eqnarray} 
we obtain,
\begin{eqnarray}
(\partial_{r_*}-i\omega)R_2=\frac{i\Delta^{1/2}}{r^2}(\hat{m}-i\mu_sr)R_1 \,,
\label{diracequation3a}\\
(\partial_{r_*}+i\omega)R_1=\frac{i\Delta^{1/2}}{r^2}(\hat{m}+i\mu_sr)R_2 \,.
\label{diracequation3b}
\end{eqnarray}
Defining now $\nu$, $\Upsilon_1$ $\Upsilon_2$, and $\hat{r}_*$ through the 
relations
\begin{eqnarray}
\nu=\arctan(\frac{\mu_sr}{\hat{m}})\,,
\label{definition1a}\\
R_1=e^{\frac{i\nu}{2}}\Upsilon_1 \,,
\label{definition1b}\\
R_2=e^{\frac{-i\nu}{2}}\Upsilon_2 \,,
\label{definition1c}\\
\hat{r}_*=r_* +\frac{1}{2\omega}\arctan(\frac{\mu_sr}{\hat{m}})\,, 
\label{definition1d}
\end{eqnarray}
we get
\begin{eqnarray}
(\partial_{\hat{r}_*}-i\omega)\Upsilon_2=W\Upsilon_1 \,,
\label{diracequation4a}\\
(\partial_{\hat{r}_*}-i\omega)\Upsilon_2=W\Upsilon_2 \,,
\label{diracequation4b}
\end{eqnarray}
where,
\begin{eqnarray}
W=\frac{i\Delta^{1/2}(\hat{m}^2+\mu_s^2r^2)^{3/2}}{r^2(\hat{m}^2+\mu_s^2r^2)
+\frac{\hat{m}\mu_s\Delta}{2\omega}} \,. 
\label{potentialdirac1}
\end{eqnarray}
Finally, putting $Z_{\pm}=\Upsilon_1 \pm \Upsilon_2$ we have
\begin{eqnarray}
(\partial_{\hat{r}_*}^2+\omega^2)Z_{\pm}=V_{\pm}Z_{\pm} \,,
\label{diracequation5}
\end{eqnarray}
with
\begin{equation}
V_{\pm}=W^2 \pm \frac{dW}{d\hat{r}_*} \,.
\label{potentialdirac2}
\end{equation}
We shall be concerned
with massless spinors ($\mu_s=0$) for which $ \hat{r}_*=r_*$, and 
$W=\frac{i\Delta^{1/2}\hat{m}}{r^2}$. Thus, 
\begin{eqnarray}
V_{\pm}=\frac{m^2}{r^2}(\frac{r^2}{l^2}-M) \pm \frac{Mm}{r^2}
(\frac{r^2}{l^2}-M)^{1/2} \,.
\label{potentialdirac3}
\end{eqnarray}
In the form (\ref{potentialdirac2}) one immediatly recognizes that the two 
potentials $V_+$ and $V_-$ should yield the same spectrum. In fact
they are, in SUSY language, superpartner potentials derived
from a superpotential W (see \cite{cooper}).
Once again, we can rescale $r$ and take $l=1$, by measuring
everything in terms of l.

\subsection{ QNMs for Weyl perturbations}
Similarly, the wave equation (\ref{diracequation5}) for Weyl 
(until recently also called neutrino) perturbations
may be put in the form
\begin{equation}
\partial_{r_*}^2 Z_{\pm}+\left[\omega^2-
m\left(\frac{m}{\cosh(M^{1/2}r_*)^2} \pm M^{1/2}\frac{\sinh(M^{1/2}r_*)^2}
{\cosh(M^{1/2}r_*)^2}\right)\right]Z_{\pm}=0 \,.
\label{neutrinoequation1}
\end{equation} 
Going to a new independent variable, $x=-\sinh(M^{1/2}r_*)$,
$x$ $\epsilon$ $[\infty,0]$, we can write
\begin{equation}
(1+x^2)Z'' +xZ' +[\frac{\frac{\omega^2(1+x^2)}{M}-\frac{m^2}{M}
\pm\frac{mx}{M^{1/2}}}{1+x^2}]Z=0\,. 
\label{neutrinoequation2}
\end{equation}
By changing the wavefunction Z to $\chi$
\begin{equation}
\chi= e^{(\frac{M^{1/2}x-2m}{2M^{1/2}}-\frac{x}{2})\arctan(x)}\,,
\label{change}
\end{equation}
we have 
\begin{equation}
(1+x^2)\chi'' +(\frac{2m}{M^{1/2}}+x)\chi' +(\frac{\omega^2}{M})\chi=0 \,.
\label{neutrinoequation3}
\end{equation}
On putting $s=\frac{1+iz}{2}$, $s$ $\epsilon$ $[\frac{1}{2},i\infty]$, 
we have again the hypergeometric equation (\ref{scalarmaxwellequation4}), 
with $a=\frac{i\omega}{M^{1/2}}$, $b=-\frac{i\omega}{M^{1/2}} $, 
and $c=\frac{1}{2}\pm\frac{im}{M^{1/2}}$,
so that
the solution to the wave equation is again specified around each
singular point, and is given by the analytic continuation of the
standard hypergeometric function to the complex plane \cite{uvarov,stegun}.

Since infinity is located at $s=\frac{1}{2}$, there is no easy
way to determine the QNM frequencies, so we have to resort to
numerical calculations. If we put (\ref{potentialdirac1}) in the form
(\ref{scalarmaxwellequation5}) one again sees that it has no essential singularities,
so the numerical method just outlined in the previous section may be
applied. Moreover, since $V_+$ and $V_-$ have the same spectrum
\cite{cooper} and the same QNM frequencies \cite{chandra} we
need only to workout the frequencies for one of them. In table \ref{btz3}
we present the numerical results for the QNM frequencies for 
neutrino perturbations and for selected values of the black hole mass.

\begin{table}
\caption{\label{btz3}  Lowest QNM frequencies for $m=1$ Weyl 
perturbations of a non-rotating BTZ black hole.}
\begin{tabular}{lllll}  \hline 
\multicolumn{3}{c}{$m=1$} \\  \hline 
\multicolumn{1}{l}{} &
\multicolumn{2}{l}{ Numerical} \\ \hline
$M^{1/2}$   &  $\omega_r$ &  $-\omega_i$ \\ \hline
 2   & 0.378 & 2.174   \\ \hline
 5   & 0.316 & 5.027    \\ \hline
 10    & 0.224 & 10.006   \\ \hline
 50   &  0.099 & 50.001    \\ \hline
 100   & 0.071 & 100.000  \\ \hline
 500   & 0.0316  & 500.000    \\ \hline
\end{tabular}
\end{table}
\vskip 1mm

For large black holes one can see that the imaginary part of the 
frequencies scale with the horizon ($r_+=M^{1/2}$),
just as in the scalar and electromagnetic case. We have also computed
some higher modes, and the real part of the frequency $\omega_r$,  does not seem
to depend on which mode we are dealing with, just as in the scalar and
electromagnetic case.

\section{Discussion}
 We have computed the scalar,  electromagnetic and
neutrino QNM of BTZ black holes.
These modes dictate the late time behaviour of the fields.
In all cases, these modes scale with the horizon radius,
at least for large black holes and, since the decay of the perturbation
has a timescale $\tau = \frac{1}{\omega_i}$, this means
that the greater the mass, the less time it takes to approach
equilibrium. We have also found that for large black holes, 
the QNM frequencies are proportional to the black hole 
radius. Since the temperature of a BTZ black is proportional 
to the black hole radius, the QNM frequencies scale with 
the temperature, as a simple argument indicates \cite{horowitz}.
As we said earlier, it was proved in \cite{birmingham2} that the location
of the poles of the retarded
correlation function of the corresponding perturbations in the dual
conformal field theory match the QN frequencies.
This leads to a new quantitative realization of the AdS/CFT correspondence.

Can one use this exact expression for the QN frequencies
to test the correspondence made by Hod \cite{hod} and alluded to
earlier, that there should be a relation between the highly damped
QN frequencies and the quantum of area of a black hole?
Birmingham {\it et al.} \cite{birmingham3} have recently given intriguing
hints that one can indeed, and that the answer is positive. They considered 
the 2+1 dimensional BTZ black hole; They showed that the identification
of the fundamental quanta of black hole mass and angular momentum with
the real part of the QNM frequencies leads to the correct quantum
behaviour of the asymptotic symmetry algebra, and thus of the dual
conformal field theory.

\thispagestyle{empty} \setcounter{minitocdepth}{1}
\chapter[Quasinormal modes of the four dimensional Schwarzschild-AdS black hole]{Quasinormal modes of the four dimensional Schwarzschild-AdS black hole} \label{chap:qnmsads}
\lhead[]{\fancyplain{}{\bfseries Chapter \thechapter. \leftmark}}
\rhead[\fancyplain{}{\bfseries \rightmark}]{}
\minitoc \thispagestyle{empty}
\renewcommand{\thepage}{\arabic{page}}
\section{Introduction}
A great deal of effort has been spent to calculate the QNMs and their
associated frequencies. New powerful methods, both analytical and
numerical have been developed. The main interest in these studies  
is in the application to the analysis of the data from the gravitational 
waves to be detected by the forthcoming  gravitational wave
detectors. We refer the reader to \cite{kokkotas} for 
reviews.  In a different context, York \cite{york} tried to explain the
thermal quantum radiance of a Schwarzschild black hole in terms of 
quantum zero-point fluctuations of zero mean in the QNMs.
QNMs in
asymptotically flat spacetimes have recently acquired a further
importance since, as mentioned before, it has been proposed that 
the Barbero-Immirzi parameter, a factor introduced by hand in order that 
Loop Quantum Gravity reproduces correctly the black hole entropy, 
is equal to the real part of the QN frequencies 
with a large imaginary part \cite{hod, dreyer} (see \cite{baez} for a
short review).  For further developments in calculating QN frequencies
in Kerr spacetimes and in asymptotically flat
black holes spacetimes see \cite{motl1,motl2,dreyerall,bertikerr,cardosoqnmkerr,
brinkasympt,birminghamd}.

All these previous works deal with asymptotically flat spacetimes, but
the recent AdS/CFT correspondence conjecture \cite{maldacena} makes
the investigation of QNMs in anti-de Sitter spacetimes more appealing.
Horowitz and Hubeny
\cite{horowitz} began the study of QNMs 
in AdS, by thoroughly investigating scalar perturbations in 4, 5 and 7
spacetime dimensions.  Their work was completed by Konoplya
\cite{konoplyasmall}, who computed the QN frequencies of
small black holes in AdS space, something Horowitz and Hubeny did not
consider.  By now, the body of work in this field is immense
\cite{cardosoqnmbtz,fernando,qnmads,moss,cardosoqnmsads,cardosoqnmsads2,berti,cardosoqnmtoro}.

WE shall now consider the 4-dimensional Schwarzschild-AdS spacetime.
The lowest lying modes (i.e., the less damped ones parametrized by the 
overtone number $n=0$) for this spacetime were found by Horowitz and
Hubeny \cite{horowitz}, and completed by Konoplya \cite{konoplyasmall} for
the scalar case, and by Cardoso and Lemos \cite{cardosoqnmsads} and Cardoso, Lemos and Konoplya \cite{cardosoqnmsads2} for the
electromagnetic and gravitational case. 
Recently, Berti and Kokkotas
\cite{berti} have confirmed all these results and extended them to
Reissner-Nordstr\"om-AdS black holes. Here, we shall take a step
further in carrying on this program by computing numerically, through
an extensive search, the high overtone QN frequencies for scalar,
electromagnetic and gravitational perturbations in the
Schwarzschild-AdS black hole. We shall do an extensive search for the
high overtone QN frequencies, ($n\geq1$). We find that the modes are 
evenly spaced
for frequencies with a large imaginary part.  Moreover the scalar,
electromagnetic and gravitational perturbations all possess,
asymptotically for high overtones $n$, QN frequencies with the same
spacing, and this spacing is $l$-independent.  While we can
numerically prove this with great accuracy for large black holes, it
remains just a conjecture for small and intermediate black holes.
We shall also see that the QN frequencies of the toroidal black hole
with non-trivial topology \cite{lemos1} are identical to the QN
frequencies of a large Schwarzschild-AdS black hole
\cite{cardosoqnmtoro}. 

Electromagnetic
perturbations are of interest due to the AdS/CFT conjecture since they can be seen
as perturbations for some generic supergravity gauge field.
In addition, the Maxwell field is an important field with different features
from scalar or gravitational fields, which makes it worth studying. 
On the other hand,  gravitational perturbations
have the additional interest of arising from any other type of
perturbation, be it scalar, electromagnetic, Weyl, etc., which in turn
disturb the background geometry. Therefore, questions like the
stability of spacetime for scalar or other perturbations, have a
direct dependence on the stability to gravitational perturbations.

\noindent
\section{Scalar, Electromagnetic and Gravitational perturbations in a 
Schwarzschild AdS background}


\subsection{Maxwell perturbations}
The scalar wave equation has already been derived by Horowitz and Hubeny \cite{horowitz}
so we shall now focus on the electromagnetic and gravitational case.
We consider the evolution of a Maxwell field in a
Schwarzschild-anti-de Sitter spacetime with metric given by 
\begin{equation}
ds^{2}= f(r) dt^{2}- \frac{dr^{2}}{f(r)}-
r^{2}(d\theta^{2}+\sin^2\theta d\phi^{2})\,,
\label{lineelementc6}
\end{equation}
where, $f(r)=(\frac{ r^{2}}{R^2}+1-\frac{2M}{r})$,
$R$ is the AdS radius and $M$ the black hole mass.  
The evolution is governed by Maxwell's equations:
\begin{equation}
{F^{\mu\nu}}_{;\nu}=0 \quad, F_{\mu\nu}=A_{\nu,\mu}-A_{\mu,\nu}\,,
\label{maxwell} 
\end{equation}
where a comma stands for ordinary derivative and a semi-colon 
for covariant derivative.  As the background is spherically symmetric,
we can expand $A_{\mu}$ in 4-dimensional vector sphericall
harmonics (see \cite{ruffini}):

{\small
\begin{eqnarray}
A_{\mu}(t,r,\theta,\phi)=\sum_{l,m}\left( \begin{array}{cc}\left[
 \begin{array}{c} 0 \\ 0 \\
 \frac{a^{lm}(t,r)}{\sin\theta}\partial_\phi Y_{lm}\\
 -a^{lm}(t,r)\sin\theta\partial_\theta Y_{lm}\end{array}\right] &
 +\left[ \begin{array}{c}f^{lm}(t,r)Y_{lm}\\h^{lm}(t,r)Y_{lm} \\
 k^{lm}(t,r) \partial_\theta Y_{lm}\\ k^{lm}(t,r) \partial_\phi
 Y_{lm}\end{array}\right] \end{array}\right)\,,
\label{expansion}
\end{eqnarray}}

\noindent where the first term in the right-hand side has parity $(-1)^{l+1}$
and the second term has parity $(-1)^{l}$, $m$ is the azimuthal number
and $l$ the angular quantum number.  If we put this expansion into Maxwell's
equations (\ref{maxwell}) we get a second order differential
equation for the perturbation:
\begin{equation}
\frac{\partial^{2} \Psi(r)}{\partial r_*^{2}} +
\left\lbrack\omega^2-V(r)\right\rbrack\Psi(r)=0 \,,
\label{wavemaxwell}
\end{equation}
where the wavefunction $\Psi(r)$ is a linear combination of the functions
$f^{lm}$, $h^{lm}$, $k^{lm}$ and $a^{lm}$ as appearing in
(\ref{expansion}). $\Psi$ has a different functional dependence
according to the parity: for odd parity, i.e, $(-1)^{l+1}$, $\Psi$
is explicitly given by $\Psi=a^{lm}$ whereas for even parity $(-1)^l$
it is given by $\Psi=\frac{r^2}{l(l+1)}\left(-i\omega
h^{lm}-\frac{df^{lm}}{dr}\right)$, see \cite{ruffini} for further details.
It is assumed that the time dependence is $\Psi(t,r)=e^{-i\omega t}\Psi(r)$.
The potential $V$ appearing in equation (\ref{wavemaxwell}) is given by
\begin{equation}
V(r)=f(r)\left\lbrack\frac{l(l+1)}{r^2}\right\rbrack \,,
\label{potentialmaxwellc2}
\end{equation}
and the tortoise coordinate $r_*$ is defined as
\begin{equation}
\frac{\partial r}{\partial r_*}= f(r)\,.
\end{equation}
We can of course rescale $r$, $r\rightarrow\frac{r}{R}$ and if we do
this, the wave equation again takes the form (\ref{wavemaxwell}) with
rescaled constants i.e., $r_+ \rightarrow \frac{r_+}{R}$, $\omega
\rightarrow \omega R $, where $r_+$ is the horizon radius.  So, we can
take $R=1$ and measure everything in terms of $R$.
\subsection{ Gravitational perturbations}
When dealing with first order gravitational perturbations one supposes
that, at least in some restricted region of spacetime, the metric
functions can be written as
\begin{equation}
g_{ab}(x^\nu)= g^{(0)}_{ab}(x^\nu)+h_{ab}(x^\nu)\,,
\label{4.1}
\end{equation}
where the metric $g^{(0)}_{ab}(x^\nu)$ is the background metric,
given by some known solution of Einstein's equations, and $ h_{ab}(x^\nu)$ is
a small perturbation.  Our background metric is a
Schwarszchild-anti-de Sitter metric (\ref{lineelementc6}) and the metric
$g_{ab}(x^\nu)$ will follow Einstein's equations in vacuum with a
cosmological constant:
\begin{equation}
G_{ab}-\Lambda g_{ab}=0\,.
\label{einstein}
\end{equation}
Upon substituting (\ref{4.1}) in (\ref{einstein}) we will obtain some
differential equations for the perturbations. We use the same
perturbations as originally given by Regge and Wheeler \cite{regge},
retaining their notation. After a decomposition in tensorial spherical
harmonics (see Zerilli \cite{zerillimath} and Mathews \cite{mathews}),
these fall into two distinct classes - odd and even - with parities
$(-1)^{l+1}$ and $(-1)^l$ respectively, where $l$ is the angular
momentum of the particular mode.  While working in general relativity
one has some gauge freedom in choosing the elements $ h_{ab}(x^\nu)$
and one should take advantage of that freedom in order to simplify
the rather lengthy calculations involved in computing
(\ref{einstein}). We shall therefore work with the classical
Regge-Wheeler gauge in which the canonical form for the perturbations
is (see also \cite{vish1}):

\bigskip
\noindent { odd \, parity:}
\begin{eqnarray}
h_{\mu \nu}= \left[
 \begin{array}{cccc} 
 0 & 0 &0 & h_0(r) 
\\ 0 & 0 &0 & h_1(r)
\\ 0 & 0 &0 & 0
\\ h_0(r) & h_1(r) &0 &h_0(r)
\end{array}\right] e^{-i \omega t}
\left(\sin\theta\frac{\partial}{\partial\theta}\right)
P_l(\cos\theta)\,;
\label{odd}
\end{eqnarray}
\bigskip
{ even \, parity:}
\begin{eqnarray}
h_{\mu \nu}= \left[
 \begin{array}{cccc} 
 H_0(r) f(r) & H_1(r) &0 & 0 
\\ H_1(r) & H_2(r)/f(r)  &0 & 0
\\ 0 & 0 &r^2K(r) & 0
\\ 0 & 0 &0 & r^2K(r)\sin^2\theta
\end{array}\right] e^{-i \omega t}
P_l(\cos\theta).
\label{even}
\end{eqnarray}
Here $P_l(\cos\theta)$ is the Legendre polynomial with angular
momentum $l$. 
If we put this decomposition into Einstein's equations we get ten
coupled second order differential equations that fully describe the
perturbations: three equations for odd perturbations and seven for
even perturbations. It is however possible to circumvent the
task of solving these coupled equations. Regge and Wheeler
\cite{regge} and Zerilli \cite{zerilli2} showed how to combine these
ten equations into two second order differential equations, one for
each parity. So following Regge and Wheeler \cite{regge} we define, 
for odd parity the wave 
function $Q(r)$ given by
perturbations,
\begin{equation}
Q(r)=  \frac{f(r)}{r} h_1(r) \,.
\label{qodd}
\end{equation}
After some work, Einstein's equations yield
\begin{equation}
\frac{\partial^{2} Q}{\partial r_*^{2}} +
\left\lbrack\omega^2 -V_{\rm odd}(r)\right\rbrack Q=0 \,,
\label{waveodd}
\end{equation}
where
\begin{equation}
V_{\rm odd}=  f(r)    
\left\lbrack\frac{l(l+1)}{r^2}-\frac{6m}{r^3}\right\rbrack\,.
\label{vodd}
\end{equation}
Likewise, following Zerilli \cite{zerilli2} one can define for even modes the 
wavefunction $T(r)$ implicitly in terms of $H_0$, $H_1$ and $K$, through the 
equations
\begin{eqnarray}
K=
\frac{6m^2+c\left(1+c\right)r^2+m\left(3cr-3\frac{r^3}{R^2}\right)}
{r^2\left(3m+cr\right)} T
+\frac{dT}{dr_*} \,, \\
H_1= -\frac{i\omega\left(-3m^2-3cmr+cr^2-3m\frac{r^3}{R^2}\right)}
{r \left(3m+cr\right) f(r)} T -i \omega
\frac{r}{f(r)}\frac{dT}{dr_*}\,, 
\label{4.10}
\end{eqnarray}
where $c=\frac{1}{2}\left\lbrack l(l+1)-2\right\rbrack$.
Then Einstein's equations for even parity perturbations can be written as
\begin{equation}
\frac{\partial^{2} T}{\partial r_*^{2}} +
\left\lbrack \omega^2 -V_{\rm even}(r)\right\rbrack T=0 \,,
\label{waveeven}
\end{equation}
with
\begin{equation}
V_{\rm even}= \frac{2f(r)}{r^3}
\frac{9m^3+3c^2mr^2+c^2\left(1+c\right)r^3+3m^2\left(3cr+3\frac{r^3}{R^2}\right)}
{\left(3m+cr\right)^2} \,.
\label{veven}
\end{equation}
Now, by defining
\begin{equation} 
W=
\frac{2m}{r^2}+\frac{-3-2c}{3r}+\frac{3c^2+2c^2+27\frac{m^2}{R^2}}
{3c\left(3m+cr\right)}+j\,,
\label{Wc6}
\end{equation}
where $j=-\frac{1}{3}\left(\frac{c}{m}+\frac{c^2}{m}+\frac{9m}{cR^2}\right)$, we obtain
\begin{equation}
V_{\rm odd}=W^2 {+} \frac{dW}{dr_*} +\beta \,, 
\quad 
V_{\rm even}=W^2 {-} \frac{dW}{dr_*} +\beta \,,
\label{V}
\end{equation}
where 
$\beta=-\frac{c^2+2c^3+c^4}{9m^2}$.  It is interesting to note that the two
potentials, odd and even, can be written in such a simple form, a fact
which seems to have been discovered by Chandrasekhar
\cite{chandradet}. Potentials related in this manner are sometimes
called super-partner potentials \cite{cooper}).  We note that similar
equations were obtained by Mellor and Moss \cite{mellor} for
Schwarzschild-de Sitter spacetime, using a different approach.


\noindent
\section{QNMs and some of its properties}


\subsection{Analytical properties}
To solve equation (\ref{wavemaxwell}) for Maxwell fields 
and equations (\ref{waveodd}-\ref{waveeven}) for gravitational 
fields, one must
specify boundary conditions. 
Consider first the case of a Schwarzschild black hole in an
asymptotically flat spacetime (see, e.g., \cite{kokkotas}). Since 
in this case the
potential  vanishes at both infinity and horizon, the two
solutions near these points are plane waves of the type 
$\Psi \sim e^{\pm i\omega r_*}$, where the $r_*$ coordinate
in this case ranges from $-\infty $ to $\infty$.  QNMs 
are defined by the condition that at the horizon there are only ingoing waves,
i.e., $\Psi_{\rm hor}\sim e^{-i\omega r_*} $ . Furthermore, one does not want to
have fields coming in from infinity (where the potential in this case 
vanishes). So, there is only a purely outgoing wave at infinity, i.e., 
$\Psi_{\rm \infty}\sim e^{i\omega r_*} $. Only a discrete set of complex 
frequencies $\omega$ meet these requirements.

Consider now a Schwarzschild black hole in an 
asymptotically AdS spacetime.  The boundary condition at the horizon 
is the same, we want that near the horizon             
$\Psi_{\rm hor}\sim e^{-i\omega r_*} $. However, $r_*$ has a finite range,
so the second boundary condition needs to be modified. There have been
several papers discussing which boundary conditions one should impose at infinity
in AdS spacetimes (\cite{avis,breit,burgess}).  We
shall require energy conservation and thus adopt the reflective boundary
conditions at infinity \cite{avis}.
This means that the wavefunction is zero at infinity. For 
a different boundary condition see \cite{dasgupta}.

We now show that the imaginary part of the frequency $\omega$ is negative, for 
waves satisfying these boundary conditions, provided the potential $V$ is positive.
The proof proceeds as for the scalar field perturbation case  \cite{horowitz}, 
although there are some steps we think are useful to display explicitly here. 
Writing $\phi$ for a generic wavefunction as 
\begin{equation}
\phi=e^{i\omega r_*} Z\,, 
\label{generic}
\end{equation}
where, $Z$ can be $\Psi$, $Q$ or $T$, we find 
\begin{equation}
f(r)\frac{\partial^{2} \phi}{\partial r^{2}}+
\left\lbrack f'-2i\omega \right\rbrack
\frac{\partial \phi}{\partial r} -\frac{V}{f}\phi=0\,,
\label{waveeq}
\end{equation}
where $f=( r^{2}+1-\frac{2M}{r})$.
In the proof, we are going to need the asymptotic behavior of
the solutions of equation (\ref{waveeq}).
For $r \rightarrow r_+$ we have $f \sim (3r_+ +\frac{1}{r_+})(r-r_+)$
and $\frac{V}{f}\sim C$, where C is a constant which takes different values 
depending on the case, electromagnetic, odd or even gravitational perturbations.  
So equation (\ref{waveeq}) becomes, in this limit,
\begin{equation}
Ay\frac{\partial^{2} \phi}{\partial y^{2}}+ [A-2i\omega]\frac{\partial
\phi}{\partial y} -C\phi=0\,, 
\label{waveeq1}
\end{equation}
where $y=r-r_+$, and $A=3r_+ +\frac{1}{r_+}$.  This equation has an exact
solution in terms of the modified Bessel functions $ I_{\nu}(z)$ \cite{stegun}, 
\begin{equation}
\phi=C_1y^{i\frac{\omega}{A}}I_{-\frac{i\omega}{A}}
\left(2(\frac{C}{A}y)^{\frac{1}{2}}\right)
+C_2y^{i\frac{\omega}{A}}I_{\frac{i\omega}{A}}
\left(2(\frac{C}{A}y)^{\frac{1}{2}}\right).
\label{wavesol1}
\end{equation}
We want the asymptotic behavior of these functions when $y
\rightarrow 0$ which is given by  $I_\nu(z)
\rightarrow \frac{(\frac{z}{2})^\nu}{\Gamma(\nu +1)}\,,z \rightarrow
0$.  So, near the horizon the wavefunction $\phi$ behaves
as
\begin{equation}
\phi_{r_+}=C_1\frac{
\left(\frac{C}{A}\right)^{-\frac{i\omega}{A}}}
{\Gamma\left(1-\frac{2i\omega}{A}\right)}
+C_2\frac{y^{\frac{2i\omega}{A}}
\left(\frac{C}{A}\right)^{\frac{i\omega}{A}}}
{\Gamma\left(1+\frac{2i\omega}{A}\right)}.
\label{waveeqhor}
\end{equation}
We can see that if one wants to rule out outgoing modes at the
horizon, we must have $C_2=0$, so that $\phi$ in equation
(\ref{generic}) does not depend on $y$.  Let's now investigate the asymptotic
behavior at infinity.  For $r \rightarrow \infty$ we have
$\frac{V}{f} \rightarrow \frac{l(l+1)}{r^2}$. Therefore near infinity
equation (\ref{waveeq}) becomes
\begin{equation}
r^2\frac{\partial^{2} \phi}{\partial r^{2}}+
[2r-2i\omega]\frac{\partial \phi}{\partial r}
-\frac{l(l+1)}{r^2}\phi=0 \,.
\label{waveeq2c6}
\end{equation}
Putting $x=\frac{1}{r}$ we have
\begin{equation}
\frac{\partial^{2} \phi}{\partial x^{2}}+ 2i\omega\frac{\partial
\phi}{\partial x} -l(l+1)\phi=0 \,,
\label{waveeq3c6}
\end{equation} 
with solution $\phi=\phi_{\rm \infty}$ given by 
\begin{equation}
\phi_{\rm \infty}(x)=A \,
{\rm e}^{\left\lbrack -i\omega+i\left(\omega^2-l(l+1)\right)^{1/2}\right\rbrack x}
+ B\,
{\rm e}^{\left\lbrack-i\omega-i\left(\omega^2-l(l+1)\right)^{1/2}\right\rbrack x}
\label{wavesol2}
\end{equation}
Now, $\phi_{\rm \infty}(x=0)=0$, therefore $A=-B$, and thus,
\begin{equation}
\phi_{\infty}(x)=A\,{\rm e}^{-i\omega x}
\sin\left\lbrack \left(\omega^2-l(l+1)\right)^{\frac{1}{2}}x\right\rbrack
\label{wavesol3}
\end{equation}
We can now proceed in the proof. Multiplying equation (\ref{waveeq}) by $\bar{\phi}$ 
(the complex conjugate of $\phi$), and
integrating from $ r_+ $ to $\infty$ we obtain 
\begin{equation}
\int_{r_+}^{\infty}dr\left\lbrack\bar{\phi}
\frac{d\left(f\frac{d\phi}{dr}\right)}{dr}-
2i\omega\bar{\phi}\frac{d\phi}{dr}-\frac{V}{f}\bar{\phi}\phi \right\rbrack=0\,.
\label{1}
\end{equation}
Integrating by parts yields 
\begin{equation}
\int_{r_+}^{\infty}dr \left\lbrack
\frac{d[\bar{\phi}f\frac{d\phi}{dr}]}{dr}-
f|\frac{d\phi}{dr}|^2 - 2i\omega\bar{\phi}\frac{d\phi}{dr}-\frac{V}{f}|\phi|^2 
\right\rbrack=0\,.
\label{2}
\end{equation}
Now, one can show that $[\bar{\phi}f\frac{d\phi}{dr}]_{r_+}=0$, 
in order to satisfy the boundary conditions. Indeed,  at $r_+$, 
$\phi(r_+)={\rm constant}$ and $f(r_+)=0$.  Now, at infinity, even though
$\bar{\phi}(\infty)=0$, we have also $f(\infty)=\infty$, so we have to
show that $[\bar{\phi}f\frac{d\phi}{dr}]_{\infty}=0 $. From equation 
(\ref{wavesol3}) we can check that this is indeed true.  
Thus, equation (\ref{2}) gives 
\begin{equation}
\int_{r_+}^{\infty}dr \left\lbrack
f|\frac{d\phi}{dr}|^2 + 2i\omega\bar{\phi}\frac{d\phi}{dr}+\frac{V}{f}|\phi|^2 
\right\rbrack=0.
\label{3}
\end{equation}
Taking the imaginary part of (\ref{3}) we have
\begin{equation}
\int_{r_+}^{\infty}dr \left\lbrack
\omega\bar{\phi}\frac{d\phi}{dr}+ \bar{\omega}\phi\frac{d\bar{\phi}}{dr}
\right\rbrack=0\,,
\label{4}
\end{equation}
wich, after an integration by parts reduces to 
\begin{equation}
(\omega-\bar{\omega})\int_{r+}^{\infty}dr \left\lbrack
\bar{\phi}\frac{d\phi}{dr}
\right\rbrack
=\bar{\omega}|\phi(r_+)|^2.
\label{5}
\end{equation}
Finally, inserting this back into (\ref{3}) yields 
\begin{equation}
\int_{r_+}^{\infty}dr  \left\lbrack
f|\frac{d\phi}{dr}|^2 +\frac{V}{f}|\phi|^2 \right\rbrack
=-\frac{|\omega|^2|\phi(r_+)|^2}{{\rm Im}\, 
\omega}.
\label{6}
\end{equation}
 From this relation, one can infer that, if V is positive definite then
${\rm Im}\,\omega\,<0$ necessarily.
So, since electromagnetic and even gravitational perturbations have $V>0$ 
one always has ${\rm Im}\, \omega <0$. As for odd gravitational perturbations 
there are instances where $V<0$, making this theorem unreliable for these cases. 
However, for 
$r_+<\left\lbrack\frac{l(l+1)}{3}-1\right\rbrack^{\frac{1}{2}}$, 
i.e., small enough masses, 
$V>0$ (see equation (\ref{vodd})), and the theorem applies. 
 
Another important point concerns the late time behavior of these fields,
and the existence or not of power-law tails. As shown by Ching et al \cite{ching2}, 
for potentials that vanish exponentially near the horizon, there are no power-law tails,
so there  will be no such tails in our case.  


\section{Equations and numerical Method}
To summarize, the evolution of massless (scalar, electromagnetic
and gravitational) fields can be reduced to the following form
\begin{equation}
\frac{\partial^{2} \Psi(r)}{\partial r_*^{2}} +
\left\lbrack\omega^2-V(r)\right\rbrack\Psi(r)=0 \,,
\label{waveeqc2}
\end{equation}
where the tortoise coordinate $r_*$ is defined as
\begin{equation}
\frac{\partial r}{\partial r_*}= f(r)\,,
\end{equation}
and the potential $V$ appearing in (\ref{waveeqc2}) depends on the
specific field under consideration.
Explicitly, for scalar perturbations \cite{horowitz},
\begin{equation}
V_{\rm s}=f(r)\left\lbrack\frac{l(l+1)}{r^2}+
\frac{2M}{r^3}+\frac{2}{R^2}\right\rbrack \,,
\label{Vscalarc2}
\end{equation}
while for electromagnetic perturbations the potential
is given by (\ref{potentialmaxwellc2}).
The gravitational perturbations decompose into two sets 
the odd and the even parity one.
For odd perturbations the potential $V(r)$ in (\ref{waveeqc2}) is
given by (\ref{vodd}) while for even perturbations it is given by 
expression (\ref{veven}).

In all
cases, we denote by $l$ the angular quantum number, that gives the
multipolarity of the field.  We can of course rescale $r$,
$r\rightarrow\frac{r}{R}$. If we do this, the wave equation 
takes again the form (\ref{waveeq}) with rescaled constants i.e., $r_+
\rightarrow \frac{r_+}{R}$, $\omega \rightarrow \omega R $, where
$r_+$ is the horizon radius.  So, we can take $R=1$ and measure
everything in terms of $R$, the AdS radius. 

Eq. (\ref{waveeq}) should be solved under
appropriate boundary conditions, i.e., incoming waves near the horizon,
\begin{equation}
\Psi \sim e^{-i\omega r_*}\,,\, r \rightarrow r_+\,,
\label{b1}
\end{equation}
and no waves at infinity, 
\begin{equation}
\Psi=0\,,\,r \rightarrow \infty\,.
\label{b2}
\end{equation}
We note
that there are other reasonable boundary conditions at infinity, in
particular for the gravitational perturbations. For instance, 
one can define
Robin boundary conditions in such a way as to preserve certain
dualities between the odd and the even gravitational perturbations. 
However, it was verified numerically by Moss and Norman
\cite{moss} that Dirichlet or Robin boundary conditions yield
approximately the same result, so we shall keep $\Psi=0\,,\,r
\rightarrow \infty$. Moreover, Cardoso and Lemos \cite{cardosoqnmsads}
proved (we shall revisit thie proof in this chapter) 
that for high overtone QN frequencies the duality
is preserved, so in this regime the distinction is irrelevant.
Thus, to compute the
QN frequencies $\omega$ such that the boundary conditions
(\ref{b1}) and (\ref{b2}) are preserved, we follow the 
Horowitz-Hubeny approach \cite{horowitz}. Within this approach we need to
expand the solution to the wave equation around
$x_{+}=\frac{1}{r_{+}}$ ($x=1/r$), 
\begin{equation}\label{my1}
\Psi (x)=\sum_{k=0}^{\infty} a_{k}(\omega)(x-x_{+})^{k}\,,
\end{equation}
and to find the roots of the equation $\Psi(x=0)=0$. First, one
should substitute  (\ref{my1}) into the wave equation (\ref{waveeq})
in order to obtain a recursion relation for $a_{k}$ \cite{horowitz}.
Then, one has to truncate the
sum (\ref{my1}) at some large $k=N$ and check that for greater $k$ the
roots converge to some true root which is the sought QN frequency.
The higher the overtone number, and the smaller the
black hole size, the larger the number $N$ at which the
roots of the equation $\Psi(x=0)=0$ converge. Yet, since in the
series (\ref{my1}) each next term depends on all the preceding terms
through the recursion relations, when $N$ is too large, the tiny
numerical errors in the first terms, start growing as $N \sim
10^2-10^3$ or greater. As a result the roots suffer a sharp change for a
small change on any of the input parameters, displaying a ``noisy''
dependence. To avoid this we have to increase the precision of all the
input data and the recursion relation we are dealing with from the
standard 20-digit precision up to a  precision such that further
increasing of it will not influence the result for the QN
frequency. For small black holes the roots start converging 
at very large $N$ only,
for instance, when $r_{+}=1/20$ we can truncate the series at $N
\sim 3\cdot10^4$, but not before. 
Since for finding roots of (\ref{b2}) we have to
resort to trial and error method, the above procedure consumes much
time, but nevertheless allows to compute QNMs of small
AdS black holes \cite{konoplyasmall}, and to find the higher overtones we are
seeking.

\section{Numerical results}
In this section we will present the numerical results obtained using
the numerical procedure just outlined in the previous section.  The
results will be organized into three subsections: scalar,
electromagnetic and gravitational perturbations. 
For each field, we shall also
divide the results into three different regimes:  large,
intermediate and small black holes, since  the results depend
crucially on the regime one is dealing with.  Here a large black
hole stands for a black hole with $r_+ \gg 1$, an intermediate black
hole is one with $r_+ \sim 1$, and a small black hole has a horizon
radius $r_+ \ll 1$.  We shall then try to unify these results.  For
each horizon radius $r_+$ and angular quantum number $l$ there is an
infinity of QN frequencies (or overtones). 
We shall order them according to
the standard procedure, by increasing imaginary part. Accordingly,
the fundamental QN frequency is defined as the one having the lowest
imaginary part (in absolute value) 
and will be labeled with the integer $n=0$.  The first
overtone has the second lowest imaginary part and is labeled with
$n=1$, and so on. The QN frequencies also have a real part, which 
in general display an increase along with the imaginary part. To the lowest 
value of the imaginary part corresponds the lowest value of 
the real part, to the second lowest 
value of the imaginary part corresponds the second lowest value of 
the real part, and so on. Thus $n$, the overtone number,  
is also a  number that in general 
increases with the real part of the frequency 
(or energy) of the mode. This seems to be a characteristic of
AdS space only, due to the special boundary conditions
associated with this spacetime. 
This, in a sense, is to be expected since the 
wave equation to be studied is a Schr\"odinger type equation, where 
for quantum non-dissipative bound systems, such as the hydrogen atom 
or a particle in an infinite well potential, the 
principal quantum number $n$  (which here has been called the 
overtone number) appears due to the boundary conditions of 
the radial equation, a typical eigenvalue problem, 
and is related directly with the frequency 
of vibration of the orbital. The similarity is not full, though, 
since the boundary condition at the black hole is of a different 
kind. However, for pure AdS spacetimes, when there is no black 
hole and the boundary conditions are of infinite well type, 
the overtone number $n$ is is indeed a principal quantum number 
(see Appendix A).

\subsection{Scalar QN frequencies}
The fundamental scalar QN frequencies were first computed by Horowitz
and Hubeny \cite{horowitz} for intermediate and large black
holes. Konoplya \cite{konoplyasmall} extended these calculations to the
case of small black holes.  Recently Berti and Kokkotas \cite{berti}
rederived all these results.  Here we do for the first time an
extensive search for higher overtones of scalar perturbations. 
Some of the lowest lying modes we find are shown in Tables \ref{tab:1}, \ref{tab:2} and \ref{tab:3}
for large, intermediate and small black holes, respectively.

\vskip 2mm
\medskip
\noindent {\bf (i) Large black holes -} 
As proven by Horowitz and Hubeny \cite{horowitz} in the large black
hole regime the frequencies must scale as the horizon radius (this can
also be proven easily and directly from the differential equation
(\ref{waveeq})).  
We show in Table \ref{tab:1} the results for a spherically
symmetric mode ($l=0$) for a black hole with $r_+=100$ which is therefore
sufficient to infer the behaviour of all large black holes.  The
fundamental frequency agrees with previous results \cite{horowitz}.
Perhaps the most interesting result in this large black hole regime is
that asymptotically for high overtone number $n$ the frequencies
become evenly spaced and behave like, for $l=0$, 
\begin{equation}
\frac{\omega_{\rm s}}{r_+}=(1.299-2.25 i)n +1.856-2.673i
\,\,,\,\,\,\,(n\,,r_+)\rightarrow \infty \,.  \label{asymptoticscalar}
\end{equation}
Thus the spacing between frequencies is 
\begin{equation}
\frac{ {\omega_{{\rm s}_{\,n+1}}}-{\omega_{{\rm s}_{\,n}}}}{r_+}
=(1.299-2.25 i)
\,\,,\,\,\,\,(n\,,r_+)\rightarrow \infty \,.  \label{asymptoticscalarspacing}
\end{equation}
Moreover, although the offset $1.856-2.673i$ in
(\ref{asymptoticscalar}) is $l$-dependent (this number is different
for $l=1$ scalar perturbations for example), this asymptotic behaviour
for the spacing (\ref{asymptoticscalarspacing}) holds for any value of
$l$.  In fact our search of the QN frequencies for higher values of
$l$ reveal that the results are very very similar to those in Table \ref{tab:1}.
We have gone up to $l=4$ for scalar perturbations and the results were
quite insensitive to $l$.
The asymptotic behaviour sets in very quickly as one increases the
mode number $n$.  Typically for $n=10$ equation
(\ref{asymptoticscalar}) already gives a very good approximation.  Indeed,
for $n=10$ we find numerically (see Table 1) 
$\omega_{\rm s}=1486.23753 -2516.90740i$
for a $r_+=100$ black hole, while the asymptotic expression gives
$\omega_{\rm s}=1484.6 - 2517.3i$.

\medskip
\noindent {\bf (ii) Intermediate black holes -}
In Table \ref{tab:2} we show some of the lowest lying scalar QN frequencies
for an intermediate black hole with $r_+=1$. For a black with this size,
one finds again that the spacing does not depend on the angular number $l$
for very high overtone number $n$.
With an error of about 2\% the limiting value for the frequency is, for $l=0$,
\begin{equation}
\omega_{\rm s} \sim (1.97-2.35i)n+2.76-2.7i\,\,,\,\,\,\,n \rightarrow \infty \,.
\label{intscalar}
\end{equation}
For QN frequencies belonging to different $l$'s the offset in (\ref{intscalar})
is different, but as far as we can tell numerically, not the asymptotic spacing implied
by (\ref{intscalar}).
Expression (\ref{intscalar}) for the asymptotic behaviour works well again for $n>10$.

\begin{table}
\caption{\label{tab:1} QN frequencies corresponding to $l=0$ scalar
perturbations of a large Schwarzschild-AdS BH ($r_{+}=100$). It can be seen that
for large $n$ the modes become evenly spaced. Although not shown here,
our numerical data indicates that this happens for all values of $l$
and also that the spacing is the same, regardless of the value of $l$.
For $l=0$ and for high $n$ the QN frequencies go like 
$\frac{\omega_{\rm s}}{r_+}=(1.299-2.25 i)n +1.856-2.673i$. The corresponding
spacing between consecutive modes seems to be $l$-independent.}
\begin{tabular}{lll|lll}  \hline
$n$  &${\rm Re}[\omega_{QN}]$:&${\rm Im}[\omega_{QN}]:$ &
$n$  &${\rm Re}[\omega_{QN}]$:&${\rm Im}[\omega_{QN}]:$  \\ \hline
0  & 184.95344  & -266.38560 & 7   & 1096.44876 & -1841.88813     \\ \hline
1  & 316.14466  & -491.64354 & 8   & 1226.38317 & -2066.89596      \\ \hline
2  & 446.46153  & -716.75722 & 9   & 1356.31222 & -2291.90222      \\ \hline
3  & 576.55983  & -941.81253 & 10  & 1486.23753 & -2516.90740      \\ \hline
4  & 706.57518  & -1166.8440 & 50  & 6682.78814 & -11516.9823 \\  \hline
5  & 836.55136  & -1391.8641 & 299 & 39030.810  & -67542.308  \\ \hline
6  & 966.50635  & -1616.8779 & 300 & 39160.7272 & -67767.3091 \\ \hline   
\end{tabular}
\end{table}
\begin{table}
\caption{\label{tab:2} QN frequencies corresponding to $l=0$ scalar
perturbations of an intermediate Schwarzschild-AdS BH ($r_+=1$). Asymptotically for
large $n$ one finds approximately $\omega_{\rm s} \sim (1.97-2.35i)n+2.76-2.7i$.} 
\begin{tabular}{lll|lll}  \hline
$n$  &${\rm Re}[\omega_{QN}]$:&${\rm Im}[\omega_{QN}]:$ &
$n$  &${\rm Re}[\omega_{QN}]$:&${\rm Im}[\omega_{QN}]:$   \\ \hline
0  & 2.7982   & -2.6712   & 10 & 22.44671 & -26.20913 \\ \hline
1  & 4.75849  & -5.03757  & 11 & 24.41443   & - 28.55989  \\ \hline
2  & 6.71927  & -7.39449  & 12 & 26.38230   &  -30.91059 \\ \hline
3  & 8.46153  & -9.74852  & 13 & 28.35029   &  -33.26123 \\ \hline
4  & 10.6467 & -12.1012  & 14 & 30.31839   &  -35.61183 \\ \hline
5  & 12.6121 & -14.4533  & 15 & 32.28658   &  -37.96238 \\  \hline
6  & 14.5782 & -16.8049  & 16 & 34.25485   &  -40.31290 \\ \hline
7  & 16.5449 & -19.1562  & 17 & 36.22318   &  -42.66340 \\ \hline
8  & 18.5119 & -21.5073  & 18 & 38.19157   &  -45.01387 \\ \hline
9  & 20.4792 & -23.8583  & 19 & 40.16002   &  -47.36431 \\ \hline
\end{tabular}
\end{table}
\begin{table}
\caption{\label{tab:3} QN frequencies corresponding to $l=0$ scalar
perturbations of a small Schwarzschild-AdS BH ($r_{+}=0.2$). Asymptotically for
large $n$ one finds approximately $\omega_{\rm s} \sim (1.69-0.57i)n+2.29-0.46i$.}
\begin{tabular}{lll|lll}  \hline
$n$  &${\rm Re}[\omega_{QN}]$:&${\rm Im}[\omega_{QN}]:$ &
$n$  &${\rm Re}[\omega_{QN}]$:&${\rm Im}[\omega_{QN}]:$   \\ \hline
0  &  2.47511  & -0.38990 & 6  & 12.45222  & -3.89179 \\ \hline
1  &  4.07086  & -0.98966 & 7  & 14.14065  & -4.46714 \\ \hline
2  &  5.72783  & -1.57600 & 8  & 15.83026  & -5.04186 \\ \hline
3  &  7.40091  & -2.15869 & 9  & 17.52070  & -5.61610  \\  \hline
4  &  9.08118  & -2.73809 & 10 & 19.21191  & -6.18997 \\ \hline
5  &  10.7655  & -3.31557 & 11 & 20.90359  & -6.76355 \\ \hline
\end{tabular}
\end{table}
\vskip 5mm

\medskip 
\noindent {\bf (iii) Small black holes -} 
Our search for
the QN frequencies of small black holes, i.e, black holes with $r_+
\ll 1$ revealed what was expected on physical grounds, and was
uncovered numerically for the first time in \cite{konoplyasmall} for the
fundamental mode: for small black holes, the QN frequencies approach
the frequencies of pure AdS spacetime \cite{burgess} (see
also Appendix A).  In fact we find
\begin{equation} \omega_{\rm s} =2n +l+3 \,\,,r_+\rightarrow 0 \,.
\label{asymptoticscalarsmall} 
\end{equation} 
In Table \ref{tab:3} we show some results for a small black hole with $r_+=0.2$.
We stress that the values presented in Table \ref{tab:3} for the
asymptotic spacing between modes may have an error of about $2 \%$. In
fact it is extremely difficult to find very high overtones of small
black holes, and so it is hard to give a precise extimate of the value
they asymptote to.

\medskip
In summary, we can say that the QN
frequencies tend to be evenly spaced asymptotically as $n$ gets very
large, no matter if the black hole is large, intermediate or small.
Moreover the spacing between consecutive modes is, as far as we can
tell, independent of the angular quantum number $l$.

\subsection{Electromagnetic QN frequencies}
The fundamental electromagnetic QN frequencies were computed for the
first time by Cardoso and Lemos \cite{cardosoqnmsads}. Recently Berti
and Kokkotas \cite{berti} have redone the calculation showing 
excellent agreement. Here we extend the results to higher overtones.
Some of the lowest lying
electromagnetic frequencies are shown in Tables \ref{tab:4}-\ref{tab:8}.

\medskip
\noindent {\bf (i) Large black holes -} 
As found in \cite{cardosoqnmsads,cardosoqnmsads2} large black holes show a
somewhat peculiar behaviour: some of the lowest lying modes have pure
imaginary frequencies, and these are well described by an analytical
formula \cite{cardosoqnmsads}, due to Liu \cite{liu}.  
A surprising aspect unveiled for the
first time by the present search is that the number of such modes decreases
as the horizon radius becomes smaller, as can be seen from Tables \ref{tab:4} and \ref{tab:5}. 
In other words, for very large black holes the number of imaginary 
modes grows. 
For example, for $r_+=1000$ (Table \ref{tab:4}) there are eight pure imaginary modes, 
for $r_+=100$ there are four such modes (see Table \ref{tab:5}), and  
for $r_+=10$ there are only two.
If one wants to go for $r_{+}$  larger than $1000$, 
the computation is very time consuming since we use a trial
and error method for finding new modes.   

Again, we find that for large black holes and $l=1$, the 
frequencies are evenly spaced with 
\begin{equation}
\frac{\omega_{\rm em}}{r_+}=(1.299-2.25 i)n -11.501+12i
\,\,,\,\,\,\,(n\,,r_+)\rightarrow \infty \,.  \label{asymptoticel}
\end{equation}
And a spacing given by 
\begin{equation}
\frac{\omega_{{\rm em}_{\,n+1}}-\omega_{{\rm em}_{\,n}}}{r_+}=(1.299-2.25 i)
\,\,,\,\,\,\,(n\,,r_+)\rightarrow \infty \,.  \label{asymptoticelspacing}
\end{equation}
For different values of the angular quantum number $l$, we find the
same spacing (\ref{asymptoticelspacing}) between consecutive modes,
although the offset in (\ref{asymptoticel}) depends on $l$.
So, asymptotically for large $n$ and large horizon radius the spacing
is the same as for the scalar case!  This is surprising, specially since
the behaviour of the scalar and electromagnetic potentials are
radically different. It is even more surprising the fact that this
asymptotic behaviour does not depend on $l$, as the
electromagnetic potential is strongly $l$-dependent.  Furthermore from
the first electromagnetic overtones one could surely not anticipate
this behaviour.

\begin{table}
\caption{\label{tab:4} QNMs corresponding to $l=1$ electromagnetic
perturbations of a large Schwarzschild-AdS BH ($r_{+}=1000$). 
Notice that now there
are eight pure imaginary modes, still well described by Liu's formula.}
\begin{tabular}{lll|lll}  \hline
$n$  &${\rm Re}[\omega_{QN}]$:&${\rm Im}[\omega_{QN}]:$ &
$n$  &${\rm Re}[\omega_{QN}]$:&${\rm Im}[\omega_{QN}]:$     \\ \hline
0  & 0     & -1500.004789   & 5 &    0      &  -8985.232 \\ \hline
1  & 0     & -2999.982599   & 6 &    0      &  -10596.03 \\ \hline
2  & 0     & -4500.093600   & 7 &    0      &  -11644.76 \\ \hline
3  & 0     & -5999.513176   & 8 & 1219.7    &  -13566.42 \\  \hline
4  & 0     & -7502.69385    & 9 & 2494.6    &  -15847.06 \\ \hline
\end{tabular}
\end{table}
\begin{table}
\caption{\label{tab:5} QNMs corresponding to $l=1$ electromagnetic
perturbations of a large Schwarzschild-AdS BH ($r_{+}=100$). The first 
four modes
are pure imaginary and are well described by Liu's approximation
\cite{cardosoqnmsads,cardosoqnmsads2}. For high $n$ the QN freuencies obey, for $l=1$,
$\frac{\omega_{\rm em}}{r_+}=(1.299-2.25 i)n -11.501+12i$. The corresponding
spacing between consecutive modes seems to be $l$-independent.}
\begin{tabular}{lll|lll}  \hline
$n$  &${\rm Re}[\omega_{QN}]$:&${\rm Im}[\omega_{QN}]:$ &
$n$  &${\rm Re}[\omega_{QN}]$:&${\rm Im}[\omega_{QN}]:$     \\ \hline
0  & 0        & -150.0479  & 10 & 799.6171   &  -2171.826 \\ \hline
1  & 0        & -299.8263  & 11 & 927.812    &  -2398.208   \\ \hline
2  & 0        & -450.9458  & 12 & 1056.153   &  -2624.438  \\ \hline
3  & 0        & -595.3691  & 13 & 1184.620   &  -2850.543  \\ \hline
4  & 22.504   & -799.194    & 14 & 1313.192   &  -3076.546  \\  \hline
5  & 162.256  & -1035.098   & 15 & 1441.856   &  -3302.464  \\ \hline
6  & 289.028  & -1263.537   & 16 & 1570.601   &  -3528.310  \\ \hline
7  & 416.247  & -1491.223   & 17 & 1699.416   &  -3754.094  \\ \hline
8  & 543.792  & -1718.409   & 18 & 1828.295   &  -3979.824 \\ \hline
9  & 671.598  & -1945.246   &  19 & 1957.229   &  -4205.508  \\ \hline
\end{tabular}
\end{table}
\begin{table}
\caption{\label{tab:6} QNMs corresponding to $l=1$ electromagnetic
perturbations of an intermediate Schwarzschild-AdS BH ($r_{+}=1$). 
Asymptotically for 
large $n$ the modes become evenly spaced in mode number and behave as 
$\omega_{\rm em} \sim (1.96-2.36i)n+1.45-2.1i$.}
\begin{tabular}{lll|lll}  \hline
$n$  &${\rm Re}[\omega_{QN}]$:&${\rm Im}[\omega_{QN}]:$ &
$n$  &${\rm Re}[\omega_{QN}]$:&${\rm Im}[\omega_{QN}]:$     \\ \hline
0  & 2.163023  & -1.699093  & 10 & 21.067466 & -25.61714  \\ \hline
1  & 3.843819  & -4.151936  & 11 & 23.015470 & -27.98278 \\ \hline
2  & 5.673473  & -6.576456  & 12 & 24.965381 & -30.34713 \\ \hline
3  & 7.553724  & -8.980538  & 13 & 26.916889 & -32.71037 \\ \hline
4  & 9.458385  & -11.37238  & 14 & 28.869756 & -35.07265 \\  \hline
5  & 11.37722  & -13.75633  & 15 & 30.823790 & -37.43413 \\ \hline
6  & 13.30526  & -16.13482  & 16 & 32.778838 & -39.79488\\ \hline
7  & 15.23974  & -18.50933  & 17 & 34.734776 & -42.15499 \\ \hline
8  & 17.17894  & -20.88081  & 18 & 36.691500 & -44.51455 \\ \hline
9  & 19.12177  & -23.24993  & 19 & 38.648922 & -46.87360 \\ \hline
\end{tabular}
\end{table}
\begin{table}
\caption{\label{tab:7} QNMs corresponding to $l=1$ electromagnetic
perturbations of a small Schwarzschild-AdS BH ($r_{+}=0.2$). 
Asymptotically for large $n$ one finds approximately
$\omega_{\rm em} \sim (1.68-0.59i)n+1.87-0.04i$.}
\begin{tabular}{lll|lll}  \hline
$n$  &${\rm Re}[\omega_{QN}]$:&${\rm Im}[\omega_{QN}]:$ &
$n$  &${\rm Re}[\omega_{QN}]$:&${\rm Im}[\omega_{QN}]:$   \\ \hline
0  &  2.63842  & -0.05795 & 6  &  12.00066 & -3.53148  \\ \hline
1  &  3.99070  & -0.47770 & 7  & 13.66436  & -4.12974 \\ \hline
2  &  5.49193  & -1.08951 & 8  & 15.33370  & -4.72479 \\ \hline
3  &  7.07835  & -1.70859 & 9  & 17.00715  & -5.31725 \\  \hline
4  &  8.70165  & -2.32191 & 10 & 18.68370  & -5.90758 \\ \hline
5  &  10.3450  & -2.92920 & 11 & 20.36268  & -6.49615 \\ \hline
\end{tabular}
\end{table}
\begin{table}
\caption{\label{tab:8} The fundamental  ($n=0$) QNMs corresponding to $l=1$
electromagnetic perturbations of a small Schwarzschild-AdS BH 
for several values of $r_{+}$.}
\begin{tabular}{lll|lll}  \hline
$r_{+}$  &${\rm Re}[\omega_{QN}]$:&${\rm Im}[\omega_{QN}]:$ &
$r_{+}$  &${\rm Re}[\omega_{QN}]$:&${\rm Im}[\omega_{QN}]:$     \\ \hline
1/2  & 2.25913  & -0.65731  & 1/10 & 2.85188   &  -0.00064 \\ \hline
1/3  & 2.40171  & -0.29814  & 1/12 & 2.88058   &  -0.00030 \\ \hline
1/4  & 2.53362  & -0.13364  & 1/16 & 2.91363   &  -0.00016 \\ \hline
1/5  & 2.63842  & -0.05795  & 1/18 & 2.92406   &  -0.00009 \\  \hline
1/8  & 2.80442  & -0.00565  & 1/20 & 2.93200   &  -0.00002 \\ \hline
\end{tabular}
\end{table}

\medskip
\noindent {\bf (ii) Intermediate black holes -} 
In Table \ref{tab:6} we show some of the lowest lying electromagnetic QN frequencies
for an intermediate black hole with $r_+=1$. 
For a black with this size,
one finds again that the spacing does not depend on the angular number $l$
for very high overtone number $n$.
With an error of about 2\% the limiting value for the frequency is, for $l=1$,
\begin{equation}
\omega_{\rm em} \sim (1.96-2.36i)n+1.45-2.1i \,\,,\,\,\,\,n \rightarrow \infty \,.
\label{intel}
\end{equation}
We note that here too the offset in (\ref{intel}) does depend on $l$, but not
the asymptotic spacing.

\medskip
\noindent {\bf (iii) Small black holes -} 
For small black holes, see Tables \ref{tab:7} and \ref{tab:8}, 
the spacing seems also to be equal as for the scalar case, but
since it is very difficult to go very high in mode number $n$ in this
regime, the error associated in estimating the asymptotic behaviour is
higher, and one cannot be completely sure.
Again, the
electromagnetic QN frequencies of very small black holes asymptote to the
pure AdS electromagnetic modes (see Appendix A, where we sketch their
computation).  Indeed we find that \begin{equation} 
\omega_{{\rm em}_{\rm AdS}}
=2n +l+2 \,\,,r_+\rightarrow 0 \,.  \label{asymptoticelsmall}
\end{equation}
This can be clearly seen from Table \ref{tab:8}, where we show the fundamental
mode for small black holes of decreasing radius.  As the horizon
radius gets smaller and smaller, the fundamental frequency approaches
the value of $3+\,0\,i$, which is indeed the correct pure AdS mode for $l=1$,
$n=0$, electromagnetic perturbations.  It was conjectured by Horowitz
and Hubeny \cite{horowitz} that for very small black holes in 
AdS space, the imaginary part of the QN frequency for spherically
symmetric perturbations should scale with the horizon area, i.e., with
$r_+^2$.  Their argument was based on a previous result \cite{gibbons}
for the absorption cross section for the $l=0$ component. This conjecture 
was later verified numerically to be correct by Konoplya \cite{konoplyasmall}
for the $l=0$ case.
From Table \ref{tab:8} it is however apparent that this scaling is no longer
valid for $l=1$ perturbation, and indeed we find it is 
not valid for
$l\neq 0$ perturbations, be it scalar, electromagnetic or gravitational
perturbations.
The reason why the imaginary part no longer scales with the horizon
area for $l\neq 0$ perturbations is due to the fact that the partial
absorption cross section only scales with the horizon area for $l=0$
perturbations. For other $l$'s the behaviour is more complex, and it
could be that there is no simple scaling, or even that the behaviour
is oscillatory with the mass $M$ of the black hole. We refer the
reader to \cite{futterman} for details on the absorption cross section
of black holes.

\vskip 1cm

\subsection{Gravitational QN frequencies}
The fundamental gravitational QN frequencies were computed for the
first time by Cardoso and Lemos \cite{cardosoqnmsads}.
We remind that there are two sets of gravitational wave equations,
the odd and even ones. Although it was found \cite{cardosoqnmsads,cardosoqnmsads2}
that there is a family of the odd modes which is very slowly damped
and pure imaginary, it was possible to prove that for high frequencies
both odd and even perturbations must yield the same QN frequencies.
We present the results for higher overtones of odd perturbations 
in Tables \ref{tab:9}-\ref{tab:12}, and even perturbations in Tables \ref{tab:13}-\ref{tab:16}.

\subsubsection{Odd perturbations}

\medskip
\noindent {\bf (i) Large black holes -} As discussed for the first time
in \cite{cardosoqnmsads} these exhibit a pure imaginary fundamental mode 
(see Table \ref{tab:9}.).
For large black holes, this mode is slowly damped and scales as the inverse
of the horizon radius. Our analysis for higher $l$'s indicates that in the 
large black hole regime an excellent fit to this fundamental pure imaginary mode
is 
\begin{equation}
\omega_{{\rm odd}_{\,n=0}}=-\frac{(l-1)(l+2)}{3r_+}i \,\,,\,\,\,\,r_+\rightarrow \infty. 
\label{pureimaginaryodd}
\end{equation}
This generalizes a previous result by Berti and Kokkotas \cite{berti}
for the $l=2$ case.  
The simplicity of this formula (which is just a fit to our numerical data), 
leads us to believe it is possible to find an analytical explanation for it, 
but such expalnation is still lacking.
In the large black hole regime, asymptotically for high 
overtones one finds,
for $l=2$ for example,
\begin{equation}
\frac{\omega_{\rm odd}}{r_+}=(1.299-2.25 i)n +0.58-0.42i.
\,\,,\,\,\,\,(n\,,r_+)\rightarrow \infty \,.  \label{asymptoticodd}
\end{equation}
This leads to the spacing
\begin{equation}
\frac{\omega_{{\rm odd}_{\,n+1}}-\omega_{{\rm odd}_{\,n}}}{r_+}=(1.299-2.25 i)
\,\,,\,\,\,\,(n\,,r_+)\rightarrow \infty \,,  \label{asymptoticspacingodd}
\end{equation}
which, as our results indicate is again $l$-independent.
Again, the offset in (\ref{asymptoticodd}) depends on $l$.

\medskip
\noindent {\bf (ii) Intermediate black holes -}
Results for the odd QN frequencies of an intermediate ($r_+=1$) 
black hole are shown in Table \ref{tab:10}. 
With an error of about 5\% the limiting value for the frequency is, for $l=2$,
\begin{equation}
\omega_{\rm odd} \sim (1.97-2.35i)n+0.93-0.32i \,\,,\,\,\,\,n \rightarrow \infty \,.
\label{intodd}
\end{equation}
We note that here too the offset in (\ref{intodd}) does depend on $l$, but not
the asymptotic spacing, with a numerical error of about 5\%.

\medskip
\noindent {\bf (iii) Small black holes -} 
The behavior for small black holes is shown in Tables \ref{tab:11} and \ref{tab:12}.
As the black hole gets smaller, the pure imaginary mode gets more
damped: the imaginary part increases, as can be seen from Table \ref{tab:12},
where we show the two lowest QN frequencies for small black holes with
decreasing radius.  As mentioned by Berti and Kokkotas \cite{berti}
the ordering of the modes here should be different.  However, since
one can clearly distinguish this pure imaginary mode as belonging to a
special family, we shall continue to label it with $n=0$.  We have not
been able to follow this mode for black holes with $r_+ <0.5$, and so
Table \ref{tab:12} does not show any pure imaginary modes for horizon radius
smaller than $0.5$.  We note that, as for the scalar and
electromagnetic cases, here too the modes are evenly spaced, with a
spacing which seems to be independent of $l$ no matter if the black
hole is large or small.  For very small black holes, the frequencies
reduce to their pure AdS values, computed in Appendix A, to
wit
\begin{equation} 
\omega_{\rm odd}
=2n +l+2 \,,\;\;\;r_+\rightarrow 0 \,. 
\label{asymptoticoddsmall}
\end{equation}
One can see this more clearly from Table \ref{tab:12}, where in fact for very 
small black holes
the frequency rapidly approaches (\ref{asymptoticoddsmall}).
Again, for small black holes, the imaginary part does not scale with 
the horizon area, 
by the reasons explained before.

\begin{table}
\caption{\label{tab:9} QNMs corresponding to $l=2$ odd gravitational 
perturbations of a large Schwarzschild-AdS BH ($r_{+}=100$). The 
fundamental QN frequency
is pure imaginary and seems to be well described by the formula 
$\omega_{n=0}=\frac{-(l-1)(l+2)}{3r_+}i$ valid only in the large black hole regime.
In the large $n$ limit one finds 
$\frac{\omega_{\rm odd}}{r_+}=(1.299-2.25 i)n +0.58-0.42i$. The corresponding
spacing between consecutive modes seems to be $l$-independent.}
\begin{tabular}{lll|lll}  \hline
$n$  &${\rm Re}[\omega_{QN}]$:&${\rm Im}[\omega_{QN}]:$ &
$n$  &${\rm Re}[\omega_{QN}]$:&${\rm Im}[\omega_{QN}]:$ \\ \hline
0  & 0          & -  0.013255  & 6  & 836.55392    & -1391.86345 \\ \hline
1  & 184.95898  & - 266.38403  & 7   &  966.50872  & -1616.87735 \\ \hline
2  & 316.14887  & - 491.64242  & 8   &  1096.45098 & -1841.88755 \\ \hline
3  & 446.46505  & - 716.75629  & 9   &  1226.38527 & -2066.89540 \\ \hline
4  & 576.56293  & - 941.81172  & 10  &  1356.31422 & -2291.90170 \\  \hline
5  & 706.57797  & - 1166.8433  & 50  &  6552.87704 & -11291.9807  \\ \hline
\end{tabular}
\end{table}
\begin{table}
\caption{\label{tab:10} QNMs corresponding to $l=2$ odd gravitational
perturbations of an intermediate Schwarzschild-AdS BH ($r_{+}=1$). Asymptotically
for large $n$ one finds approximately
$\omega_{\rm odd} \sim (1.97-2.35i)n+0.93-0.32i$.}
\begin{tabular}{lll|lll}  \hline
$n$  &${\rm Re}[\omega_{QN}]$:&${\rm Im}[\omega_{QN}]:$ &
$n$  &${\rm Re}[\omega_{QN}]$:&${\rm Im}[\omega_{QN}]:$   \\ \hline
0  & 0         & -2         & 10 & 20.604949 & -23.803860 \\ \hline
1  & 3.033114  & -2.404234  & 11 & 22.567854   &  -26.157246 \\ \hline
2  & 4.960729  & -4.898194  & 12 & 24.531429   &  -28.510214 \\ \hline
3  & 6.905358  & -7.289727  & 13 & 26.495564   &  -30.862849  \\ \hline
4  & 8.854700  & -9.660424  & 14 & 28.460169   &  -33.215214 \\  \hline
5  & 10.80784  & -12.02344  & 15 & 30.425175   &  -35.567355 \\ \hline
6  & 12.76384  & -14.38266  & 16 & 32.390524   &  -37.919308 \\ \hline
7  & 14.72199  & -16.73969  & 17 & 34.356173   &  -40.271103  \\ \hline
8  & 16.68179  & -19.09530  & 18 & 36.322082   &  -42.622761 \\ \hline
9  & 18.64286  & -21.44994  & 19 & 38.288221   &  -44.974301 \\ \hline
\end{tabular}
\end{table}
\begin{table}
\caption{\label{tab:11} QNMs corresponding to $l=2$ odd gravitational
perturbations of a small Schwarzschild-AdS BH ($r_{+}=0.2$). Asymptotically for
large $n$ one finds approximately $\omega_{\rm odd} \sim (1.69-0.59i)n+2.49+0.06i$.}
\begin{tabular}{lll|lll}  \hline
$n$  &${\rm Re}[\omega_{QN}]$:&${\rm Im}[\omega_{QN}]:$ &
$n$  &${\rm Re}[\omega_{QN}]$:&${\rm Im}[\omega_{QN}]:$   \\ \hline
0  &  2.404    & -3.033   & 6  &  12.67161  & -3.43609 \\ \hline
1  &  4.91594  & -0.30408 & 7  &  14.33020  & -4.05366 \\ \hline
2  &  6.30329  & -0.89773 & 8  &  15.99881  & -4.66448 \\ \hline
3  &  7.82330  & -1.53726 & 9  &  17.67433  & -5.26955 \\  \hline
4  &  9.40720  & -2.17744 & 10 &  19.35465  & -5.86978\\ \hline
5  &  11.0279  & -2.81083 & 11 &  21.03839  & -6.46596  \\ \hline
\end{tabular}
\end{table}
\begin{table}
\caption{\label{tab:12} The fundamental ($n=0$) QNMs corresponding to $l=2$
odd gravitational perturbations of a small Schwarzschild-AdS BH 
for several values of $r_{+}$.}
\begin{tabular}{lll|lll}  \hline
$r_{+}$  &${\rm Re}[\omega_{QN}]$:&${\rm Im}[\omega_{QN}]:$ &
$r_{+}$  &${\rm Re}[\omega_{QN}]$:&${\rm Im}[\omega_{QN}]:$     \\ \hline
0.8 $(n=0)$ &  0              & -3.045373 & 0.5 $(n=1)$ & 3.03759  &  -0.71818\\ \hline
0.8 $(n=1)$ &  2.89739        & -1.69556  & 0.4         & 3.16209   & -0.43092\\\hline
0.7 $(n=0)$ & 0               & -3.83538  & 0.3         & 3.35487   & -0.17320\\ \hline
0.7 $(n=1)$ &  2.90665        & -1.34656  & 0.2         & 3.62697   & -0.01792\\ \hline
0.6 $(n=0)$ & 0               & -4.901973 & 0.1         & 3.84839   & -0.00005\\ \hline
0.6 $(n=1)$ & 2.95550         & -1.02196  & 1/15        & 3.90328   & -0.00001\\  \hline
0.5 $(n=0)$ & 0               & -6.40000  & 1/20        & 3.92882   & -0.000002\\ \hline
\end{tabular}
\end{table}

\medskip
In conclusion,
the higher overtones of odd perturbations follow a pattern very
similar to the scalar case. We note that the asymptotic behaviour
sets in very quickly, much like what happened for scalar and electromagnetic
perturbations. Typically the formulas yielding the asymptotic behaviour
work quite well for $n>10$.
We are now
able to prove that for sufficiently high
frequencies the scalar and gravitational perturbations are
isospectral, a mystery that remained in 
\cite{cardosoqnmsads},  This is done in section below.

\subsubsection{Even perturbations}

Let us now briefly discuss the even modes.  As found previously
\cite{cardosoqnmsads,cardosoqnmsads2} these modes behave very similar to the scalar
ones.  Yet, the even gravitational modes are stipulated by a more
complicated potential, and we have to truncate the series in power of
$x-x_{+}$ at larger $N$, which makes the whole procedure be more time
consuming. That is why when considering small black holes 
we were restricted only by first seven modes in that case. It is,
however, sufficient to see that even gravitational QNMs,
similar to other kind of perturbations, tend to arrange into
equidistant spectrum under the increasing of $n$.
We show in Tables \ref{tab:13}-\ref{tab:16} the numerical results for the QN frequencies
of even gravitational perturbations.

\medskip
\noindent {\bf (i) Large black holes -}
Results for the QN frequencies of large black holes are shown in
Table \ref{tab:13}.
In this regime one finds for $l=2$ even perturbations
\begin{equation}
\frac{\omega_{\rm even}}{r_+}=(1.299-2.25 i)n +1.88-2.66i.
\,\,,\,\,\,\,(n\,,r_+)\rightarrow \infty \,,  \label{asymptoticeven}
\end{equation}
leading to the spacing
\begin{equation}
\frac{\omega_{{\rm even}_{\,n+1}}-
\omega_{{\rm even}_{\,n}}}{r_+}=(1.299-2.25 i)
\,\,,\,\,\,\,(n\,,r_+)\rightarrow \infty \,,  \label{asymptoticspacingeven}
\end{equation}
which once more turns out to be $l$-independent!  All the results
concerning the spacing of frequencies for large black holes have a
very good precision, since in this regime it is possible to go very
far out in overtone number (typically $n=300$ is enough to achieve a
0.1\% accuracy for the spacing).

\medskip
\noindent {\bf (ii) Intermediate black holes -}
In Table \ref{tab:14} we show some of the lowest lying even gravitational QN frequencies
for an intermediate black hole with $r_+=1$. 
For a black with this size,
one finds again that the spacing does not seem to depend on the angular number $l$
for very high overtone number $n$.
With an error of about 5\% the limiting value for the frequency is, for $l=2$,
\begin{equation}
\omega_{\rm even} \sim (1.96-2.35i)n+2.01-1.5i \,\,,\,\,\,\,n \rightarrow \infty \,.
\label{inteven}
\end{equation}
We note that here too the offset in (\ref{intel}) does depend on $l$, but not
the asymptotic spacing.

\begin{table}
\caption{\label{tab:13} QNMs corresponding to $l=2$ even gravitational 
perturbations of a large Schwarzschild-AdS BH ($r_{+}=100$). For large $n$, one finds
$\frac{\omega_{\rm even}}{r_+}=(1.299-2.25 i)n +0.58-0.42i$. The corresponding
spacing between consecutive modes seems to be $l$-independent.}
\begin{tabular}{lll|lll}  \hline
$n$  &${\rm Re}[\omega_{QN}]$:&${\rm Im}[\omega_{QN}]:$ &
$n$  &${\rm Re}[\omega_{QN}]$:&${\rm Im}[\omega_{QN}]:$ \\ \hline
0  & 184.97400 & -266.351393  & 6   & 966.609780 & -1616.695872  \\ \hline
1  & 316.17838 & -491.584999  & 7   & 1096.56635 & -1841.681256 \\ \hline
2  & 446.50884 & -716.674054  & 8   & 1226.51495 & -2066.664293 \\ \hline
3  & 576.62103 & -941.70468   & 9   & 1356.45821 & -2291.645761 \\ \hline
4  & 706.65039 & -1166.71147  & 10  & 1486.39776 & -2516.626168 \\  \hline
5  & 836.64066 & -1391.70679  & 50  & 6683.51993 & -11515.70869 \\ \hline
\end{tabular}
\end{table}
\begin{table}
\caption{\label{tab:14} QNMs corresponding to $l=2$ even gravitational
perturbations of an intermediate Schwarzschild-AdS BH ($r_{+}=1$). Asymptotically
for large $n$ one finds approximately
$\omega_{\rm even} \sim (1.96-2.35i)n+2.01-1.5i$.}
\begin{tabular}{lll|lll}  \hline
$n$  &${\rm Re}[\omega_{QN}]$:&${\rm Im}[\omega_{QN}]:$ &
$n$  &${\rm Re}[\omega_{QN}]$:&${\rm Im}[\omega_{QN}]:$   \\ \hline
0  & 3.017795  & -1.583879  & 10 & 21.68949   &  -24.98271  \\ \hline
1  & 4.559333  & -3.810220  & 11 & 23.64402   &  -27.33549 \\ \hline
2  & 6.318337  & -6.146587  & 12 & 25.60052   &  -29.68799 \\ \hline
3  & 8.168524  & -8.500194  & 13 & 27.55860   &  -32.04026   \\ \hline
4  & 10.061220 & -10.85631  & 14 & 29.51796   &  -34.39234  \\  \hline
5  & 11.976813 & -13.21224  & 15 & 31.47838   &  -36.74424  \\ \hline
6  & 13.906140 & -15.56749  & 16 & 33.43969   &  -39.09600 \\ \hline
7  & 15.844371 & -17.92208  & 17 & 35.40174   &  -41.44762  \\ \hline
8  & 17.788721 & -20.27609  & 18 & 37.36444   &  -43.79914  \\ \hline
9  & 19.737469 & -22.62960  & 19 & 39.32769   &  -46.15057\\ \hline
\end{tabular}
\end{table}
\begin{table}
\caption{\label{tab:15} QNMs corresponding to $l=2$ even gravitational
perturbations of a small Schwarzschild-AdS BH ($r_{+}=0.2$). Asymptotically for
large $n$ one finds approximately $\omega_{\rm even} \sim (1.61-0.6i)n+2.7+0.37i$.}
\begin{tabular}{lll|lll}  \hline
$n$  &${\rm Re}[\omega_{QN}]$:&${\rm Im}[\omega_{QN}]:$ &
$n$  &${\rm Re}[\omega_{QN}]$:&${\rm Im}[\omega_{QN}]:$   \\ \hline
0  & 3.56571  & -0.01432 & 3  & 7.65872  & -1.42994 \\ \hline
1  & 4.83170  & -0.26470 & 4  &  9.20424 & -2.04345 \\ \hline
2  & 6.17832  & -0.82063 & 5  & 10.78800 & -2.65360 \\ \hline
\end{tabular}
\end{table}
\begin{table}
\caption{\label{tab:16} The fundamental ($n=0$) QNMs corresponding to $l=2$
even gravitational perturbations of a small Schwarzschild-AdS BH 
for several values of $r_{+}$.}
\begin{tabular}{lll|lll}  \hline
$r_{+}$  &${\rm Re}[\omega_{QN}]$:&${\rm Im}[\omega_{QN}]:$ &
$r_{+}$  &${\rm Re}[\omega_{QN}]$:&${\rm Im}[\omega_{QN}]:$       \\ \hline
0.8  & 2.91541  & -1.18894  & 0.3 & 3.29299   &  -0.14103       \\ \hline
0.7  & 2.90591  & -0.98953  & 0.2 & 3.56571   &  -0.01432       \\ \hline
0.6  & 2.92854  & -0.78438  & 0.1 & 3.80611   &  -0.00005       \\ \hline
0.5  & 2.98985  & -0.57089  &1/15 & 3.8735    &  -0.00001      \\  \hline
0.4  & 3.10317  & -0.35043  &1/20 & 3.90852   &  -0.000002       \\ \hline
\end{tabular}
\end{table}
\medskip
\noindent {\bf (iii) Small black holes -}
The behavior for small black holes is shown in Tables \ref{tab:15} and \ref{tab:16}.
Our search for
the QN frequencies of small black holes, i.e, black holes with $r_+
\ll 1$ revealed again what was expected on physical grounds: 
for small black holes, the QN frequencies approach
the frequencies of pure AdS spacetime (see Appendix A).  
In fact we find
\begin{equation} \omega_{{\rm even}_{\rm AdS}} =2n +l+2 \,\,,r_+\rightarrow 0 \,.
\label{asymptoticevensmall} 
\end{equation} 
In Table \ref{tab:15} we show the lowest lying QN frequencies for a small black
hole ($r_+=0.2$).  We stress that the values presented in Table \ref{tab:15} (as
a matter of fact, all the Tables containing data for small black
holes) for the asymptotic spacing between modes may have an error of
about $2 \%$. In fact it is extremely difficult to find very high
overtones of small black holes, and so it is hard to give a precise
extimate of the value they asymptote to.
In Table \ref{tab:16} we show some the fundamental even QN frequencies for  
small black holes of decreasing radius, and one can clearly see how
the fundamental frequency approaches the pure AdS value given
in Appendix A.

\section{Discussion of the results}
\noindent
\subsection{On the isospectrality breaking between odd and even perturbations }
As is well known \cite{chandra,chandradet} in the case of a
Schwarzschild black hole in an asymptotically flat space the two
potentials $V_{\rm even}$ and $ V_{\rm odd}$ give rise to the same QN
frequencies (in fact to the same absolute value of the reflection
and transmission coefficients).  This remarkable
property followed from a special relation (the equivalent for 
asymptotically flat spacetimes of our equation (\ref{V})) between the
potentials and the behavior of W at the boundaries.  However, as one
can see in the previous tables there is a isospectrality breaking between 
odd and
even perturbations in Schwarzschild anti-de Sitter spacetime.

We shall now treat this problem. 
The breaking of the isospectrality is intimately
related to the behavior of $W$ at infinity.  On taking
advantage of the machinery developed by Chandrasekhar, we seek a
relation between odd and even perturbations of the form
\begin{eqnarray}
Q= p_1T+q_1\frac{dT}{dr_*} \,, \\ T= p_2Q+q_2\frac{dQ}{dr_*}\,,
\label{relation1}
\end{eqnarray}
yielding (see \cite{chandra} for details), 
$q_1^2=\frac{1}{\beta-\omega^2}\,$, $p_1=qW\,$, $p_2=-p_1\,$ and 
$q_2=q_1=q$. Thus, we obtain
\begin{eqnarray}
Q= qWT+q\frac{dT}{dr_*} \,, \\ T= -qWQ+q\frac{dQ}{dr_*}\,.
\label{relation2}
\end{eqnarray}
Suppose now that $\omega$ is a QNM frequency of $T$, i.e., one for which 
\begin{eqnarray}
T \rightarrow A_{\rm even} e^{-i\omega r_*} \,,\,\, r\rightarrow r_+\,, \\
T \rightarrow 0 \,,\,\, r\rightarrow \infty\,.
\label{Tasymptotic}
\end{eqnarray}
Substituting this into equation (\ref{relation2}) we see that
\begin{eqnarray}
Q \rightarrow A_{\rm even}
q\left\lbrack W(r+)-i\omega\right\rbrack 
e^{-i\omega r_*} \,,\,\,r\rightarrow r_+\,, \\
Q \rightarrow q\left(\frac{dT}{dr_*}
\right)_{r=\infty} \,,\,\, r\rightarrow \infty\,.
\label{Qasymptotic}
\end{eqnarray}
However, from equation (\ref{wavesol3}), 
$\left(\frac{dT}{dr_*}\right)_{r=\infty}$ is in general
not zero so that $\omega$ will in general fail to be a QNM frequency
for Q.  Should $Q$ and $T$ be smooth functions of $\omega$, 
one expects that if $q$
is ``almost zero'' then $\omega$ should ``almost'' be a QNM frequency
for $Q$. Now, the condition that $q$ is almost zero is that
$\beta-\omega^2$ be very large, and one expects this to be true either
when $\omega$ is very large or else when $\beta$ is very large. And in
fact, as one can see in tables 3, 4 and 5 for very large $\omega$ the
frequencies are indeed almost identical. On the other hand, for very
small black holes ($\beta$ very large) one expects the frequencies to
be exactly the same, since both potentials have the same asymptotic
behavior in this regime, as we shall see in the next section.  One
would be tempted to account for the remarkable resemblances between QNM
frequencies of scalar and gravitational perturbations by a similar
approach, but the proof is still eluding us.  Should such an approach
work, it could be of great importance not only to this specific
problem, but also to the more general problem of finding the
asymptotic distribution of eigenvalues, by studying a different
potential with (asymptotically) the same eigenvalues, but more easy to
handle.

\subsection{Why are the scalar and gravitational perturbations
isospectral in the large black hole regime?}
In the previous subsection, we have
showed why the odd and even gravitational perturbations yield the same
QN frequencies for large frequencies.  The whole approach was
based on the fact that the odd and even gravitational potentials are
superpartner potentials \cite{cooper}, i.e., they are related to one
another via
$
V_{\rm odd}=W^2 {+} \frac{dW}{dr_*} +\beta \,,
\quad
V_{\rm even}=W^2 {-} \frac{dW}{dr_*} +\beta \,,
$
where $\beta=-\frac{\alpha^2+2\alpha^3+\alpha^4}{9M^2}$. The function $W$ is
$
W= \frac{2M}{r^2}+\frac{-3-2\alpha}{3r}+\frac{3\alpha^2+2\alpha^2+27M^2}
{3\alpha\left(3M+\alpha \,r\right)}-
\frac{1}{3}\left(\frac{\alpha}{M}+\frac{\alpha^2}{M}+\frac{9M}{\alpha}\right)$.
For more details we refer the reader to \cite{cardosoqnmsads}.
We shall now see that a similar method can be applied to show that in the
large black hole regime, scalar and gravitational perturbations
are isospectral for large QN frequencies.
To begin with, we note that the potentials $V1$ and $V2$ defined by 
\begin{equation}
V1={\tilde f}  \left( \frac{2}{R^2}+\frac{2M}{r^3} \right) \,,
\label{V1}
\end{equation}
and
\begin{equation}
V2={\tilde f} \left(\frac{a}{r^2}-\frac{6M}{r^3} \right)\,,
\label{V2}
\end{equation}
with $\tilde{f}=\frac{r^2}{R^2}+\frac{a}{2}-\frac{2M}{r}$, 
and $a$ any constant, 
are superpartner potentials.
The superpotential $\tilde {W}$ is in this case is given by 
\begin{equation}
\tilde {W}=\frac{r}{R^2}+\frac{a}{2r}-\frac{2M}{r^2}\,.
\label{Wtilde}
\end{equation}
Thus, the two superpartner potentials $V1$ and $V2$ can be expressed 
in terms of $\tilde{W}$ as
\begin{equation}
V1=\tilde{W}^2 {+} \frac{d\tilde{W}}{dr_*}  \,,
\quad
V2=\tilde{W}^2 {-} \frac{d\tilde{W}}{dr_*} \,.
\label{V12}
\end{equation}
Why are these two potentials of any interest?  Because in the
large $r_+$ limit, which we shall take to be $r_+ \gg a$, 
we have $\tilde {f} \sim
r^2-\frac{2M}{r}$. Notice now that in this large $r_+$ limit the scalar
potential (\ref{Vscalarc2}) is $V_{\rm s}\sim f(2+\frac{2M}{r^3})$, with
$f\sim r^2-\frac{2M}{r}$, since in this limit and with $r_+ \gg l$,
one has $\frac{l(l+1)}{r^2}\ll 2$. Thus, $V1$ reduces to the
scalar potential and $V2$ to the gravitational odd potential, provided
we take $a=l(l+1)$.  It then follows from the analysis in
\cite{cardosoqnmsads}  (section IIIC, which is repeated in the previous subsection) that for large black holes these
two potentials should yield the same frequencies.

\subsection{Future directions}
The preceding sections have shown that the QNMs of
Schwarzschild-AdS black holes have a universal behavior in
the asymptotic regime of high overtones.  This was verified explicitly
and with great accuracy for the large black hole regime, where we
showed numerically that the spacing does not depend on the
perturbation in question and is equal to 
\begin{equation}
\frac{\omega_{n+1}-\omega_{n}}{r_+}=(1.299-2.25 i)
\,\,,\,\,\,\,(n\,,r_+)\rightarrow \infty \,.
\label{asymptoticelspacingUniv} \end{equation} 
We conjecture that the asymptotic behavior is the same for all kinds
of perturbations irrespectively of the black hole size, i.e., a fixed
horizon radius $r_+$ Schwarzschild-AdS black hole will have
an asymptotic spacing between consecutive QN frequencies which is the
same for scalar, electromagnetic and gravitational perturbations. The
difficulty in extracting very high overtones for small black holes
however, prevents us from having an irrefutable numerical proof of
this.  It would be extremely valuable to have some kind of analytical
scheme for extracting the asymptotic behavior, much as has been done
for the asymptotically flat space by Motl and Neitzke
\cite{motl1,motl2}.  However it looks quite
difficult to make any analytical approximation in asymptotically
AdS spaces, although there have been some atempts at this
recently (see for example Musiri and Siopsis \cite{qnmads}).  
We also note that the spacing
(\ref{asymptoticelspacingUniv}) was already found to be true by Berti
and Kokkotas \cite{berti} for the scalar and gravitational
cases for the lowest radiatable multipole, i.e, $l=0$ and $l=2$
scalar and gravitational perturbations respectively.  We have
concluded that, surprisingly, the spacing (\ref{asymptoticelspacingUniv})
also works for electromagnetic case and for any value of $l$.
It was observed that,
despite having such different potentials the scalar, the
electromagnetic and gravitational QN frequencies have the same
asymptotic behavior.  Can one formulate some very general conditions
the potentials should obey in order to have the same asymptotic
solutions? This is still an open question.

There has been recently an exciting development trying to relate the
asymptotic QN frequencies with the Barbero-Immirzi parameter
\cite{dreyer,baez}.  In fact it was
observed, in the Schwarzschild case, that asymptotically for high
overtones, the real part of the QN frequencies was a constant,
$l$-independent, and using some (not very clear yet) correspondence
between classical and quantum states, was just the right constant to
make Loop Quantum Gravity give the correct result for the black hole
entropy.  Of course it is only natural to ask whether such kind of
numerical coincidence holds for other spacetimes.  We have seen that
apparently we are facing, in AdS space, a universal behavior, i.e.,
the asymptotic QN frequencies do not depend on the kind of
perturbations, and also don't depend on $l$.  However, and in contrast
with asymptotically flat space, the real part of the asymptotic QN
frequency is not a constant, but rather increases linearly with the
mode number $n$.  This is no reason to throw off the initial
motivation of seeking some kind of relation between Loop Quantum
Gravity and QNMs, after all, there are no predictions for AdS space.  

Finally we point out that the asymptotic behavior studied here for the 
Schwarzschild-AdS black hole will hold also for other black holes in 
asymptotically AdS. One example of these is the 
black hole
with non-trivial topology \cite{lemos1}. 
The general line element for this spacetime is
\cite{lemos1}: 
\begin{equation}
ds^{2}= f(r)\,dt^{2}- f(r)^{-1}dr^{2}-r^2\left( d\theta^{2}
+d\phi^{2}\right)\,
\label{lineelementlemos}
\end{equation}
where 
\begin{equation}
f(r)=\frac{
r^2}{R^2}-\frac{4MR}{r}\, ,
\label{f(r)}
\end{equation}
where $M$ is the ADM mass of the black hole, and $R$ is the AdS
radius. There is a horizon at $r_+=(4M)^{1/3}R$.  The range of the
coordinates $\theta$ and $\phi$ dictates the topology of the black hole
spacetime.  For a black hole with toroidal topology, a toroidal black
hole, the coordinate $\theta$  ranges from
$0$ to $2\pi$, and $\phi$ ranges from $0$ to $2\pi$ as well.  For the
cylindrical black hole, or black string, the coordinate $\theta$ has range
$-\infty<R\,\theta<\infty$, and $0\leq \phi <2\pi$. For the planar black hole,
or black membrane, the coordinate $\phi$ is further decompactified
$-\infty<R\,\phi<\infty$ \cite{lemos1}.  The fundamental QN frequencies
for these black holes were computed in \cite{cardosoqnmtoro}, where it
was verified that they follow the same pattern as for
Schwarzschild-AdS black holes. Indeed one easily sees that in the
large black hole regime they both should yield the same results as the
potentials are equal in this regime (compare the potentials in
\cite{cardosoqnmtoro} with the ones in the present work).  In
particular the asymptotic behavior will be the same.

\noindent
\section{The limit m $\rightarrow$ 0}
Although it is not possible to solve exactly for the QNM frequencies,
it is possible to gain some analytical insight in the special case of
very small black holes. There has been some discussion
about this regime (see \cite{horowitz,hubeny} 
and references therein). Here
we shall exploit the behavior of QN frequencies in this regime a
little further.
 For very small black holes one can easily see that both
potentials (electromagnetic and gravitational) look like,
 in the $r_*$ coordinate, a barrier with
unequal heights:
\begin{figure}
\includegraphics{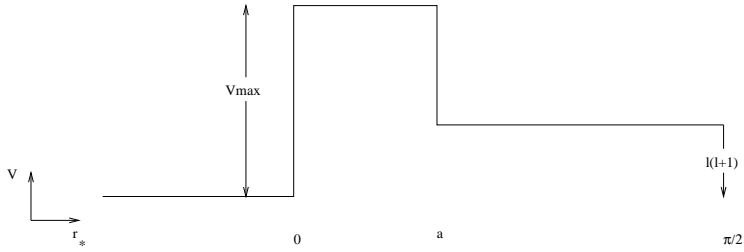}
\caption{\label{barreira} The potential for small black holes.
$V_{\rm max}=\frac{4l(l+1)}{27r_+^2}$ and $a \sim 3r_+$.}
\end{figure}
\vskip 3mm
It is trivial to obtain equations for the QN
frequencies in this limit. If $\Psi$ is a general wavefunction then
\begin{eqnarray}
\frac{d\Psi/dr_*}{\Psi}=-i\omega \;,\quad r_*<0\,, \\
\frac{d\Psi/dr_*}{\Psi}=\frac{ik_1Be^{ik_1r_*}-ik_1Ce^{-ik_1r_*}}
{Be^{ik_1r_*}+Ce^{-ik_1r_*}}\;, \quad 0<r_*<a\,, \\
\frac{d\Psi/dr_*}{\Psi}=\frac{ikDe^{ikr_*}-ik_1Ee^{-ikr_*}}
{De^{ikr_*}+Ee^{-ikr_*}}\;, \quad r_*>a \;,
\label{5.1}
\end{eqnarray}
where $k_1=(\omega^2-V_{max})^{1/2}$, and
$k=\left\lbrack\omega^2-l(l+1)\right\rbrack^{1/2}$. Imposing the continuity
of the logarithmic derivative and furthermore that $\Psi=0$ at
infinity ($r_*=\frac{\pi}{2}$), we get
\begin{eqnarray}
k_1\left\lbrack\frac{\frac{k_1-\omega}{k_1+\omega}e^{2ik_1a}-1}
{\frac{k_1-\omega}{k_1+\omega}e^{2ik_1a}+1}\right\rbrack
=
k\left\lbrack
\frac{e^{2ik(\frac{\pi}{2}-a)}+1}
{1-e^{2ik(\frac{\pi}{2}-a)}}\right\rbrack\;.
\label{5.2}
\end{eqnarray}
In the limit $a \rightarrow 0$, $k_1 \rightarrow \infty$
($m \rightarrow 0$) we have, supposing that $\omega$ stays small,
the condition $e^{2ik\frac{\pi}{2}}=1$ which means that
\begin{equation}
\omega_0^2=4n^2+l(l+1)\;, \quad n=1,2,...\;,
\label{5.3}
\end{equation}
corresponding to a bound state. This gives for the lowest 
QNM frequencies ($n=1$):
$\omega_0=2.45 \,$ for $l=1$ and $\omega_0=3.16 \,$ for $l=2$.
The above are to be compared with those in Tables 1-5.
The agreement seems excellent, and we can now go a step further
: If we linearize (\ref{5.2}) around the solution (\ref{5.3}), i.e,
if we write $\omega= \omega_0+i\delta$ and substitute back in (\ref{5.2}) we 
obtain, 
to third order in $\delta$, the values listed in Table \ref{tab:20} . 
\begin{table}
\caption{\label{tab:20} The linearized frequency $\delta\,$ for selected values of the 
angular quantum number $l$ and the potential width $a$.
}
\begin{tabular}{|l|l|l|l|l|}  \hline 
\multicolumn{1}{|c|}{} &
\multicolumn{1}{c|}{$ a=3/h$} &
\multicolumn{1}{c|}{$ a=6/h$} \\ \hline
\multicolumn{1}{|c|}{l}    & \multicolumn{1}{|c|}{ $\delta$} &\multicolumn{1}{|c|}{  $\delta$}
  \\ \hline
1   &$-2.1/h-i(1.42/h^2)$  &$ -1.92/h-i(0.05/h^2)$  
\\ \hline
2  &$ -0.859/h-i(0.04/h^2) $ &
$ -0.85/h-i(1.4{\rm x}10^{-4}/h^2)$   
\\ \hline
\end{tabular}
\end{table}

We have chosen a typical value of $a\sim \frac{3}{h}$, but we can see
that, although the real part does not depend very much on $a$, the
imaginary part is strongly sensitive to $a$. Nevertheless, one can be
sure that whatever value of $a$, the imaginary part goes as
$\frac{1}{h^2}$ and is always negative.  This question is by now
settled, with the numerical work of Konoplya \cite{konoplyasmall}.
The imaginary part does indeed go as $r_+^2$ for scalar perturbations,
and for scalar perturbations only.

\section{Conclusion}
We have done an extensive search for higher overtones $n$ of the
QNMs of Schwarzschild-AdS BH corresponding to scalar,
electromagnetic, and gravitational perturbations.
We have shown that:
(i) No matter what the size of the black hole is, the QN frequencies
are evenly spaced, both in the real and in the imaginary component,
for high overtone number $n$; 
(ii) The spacing between consecutive modes is independent of the perturbation.
This means that scalar, electromagnetic and gravitational perturbations all
have, asymptotically, the same spacing between modes. This is one of the major
findings in this work, together with the fact that this spacing seems to
be also independent of the angular quantum number $l$;
(iii) We were able to prove that the scalar and gravitational QN frequencies
must asymptotically be the same;
(iv) The electromagnetic QN frequencies of large black
holes have a number of first overtones with pure imaginary parts, and
the higher the black hole radius $r_{+}$, the higher the number of these first
pure damped, non-oscillating modes;
(v) Finally, we have computed analytically the electromagnetic and
gravitational pure AdS modes, and we have shown numerically that the
QN frequencies of very small black holes asymptote to these pure AdS
modes;


\vskip 2cm
\section{APPENDIX: Pure AdS normal modes for electromagnetic and 
gravitational perturbations}
\label{apendice}
In this appendix we shall briefly outline how to compute the pure
modes of AdS space (no black hole, $M=0$) for
electromagnetic and gravitational perturbations. The scalar case was
dealt with by Burgess and Lutken \cite{burgess}.  
In pure AdS space the electromagnetic
and gravitational potentials (both odd and even) are
\begin{equation}
V=\left(\frac{r^2}{R^2}+1\right)\frac{l(l+1)}{r^2}\,,
\label{potpureAdS}
\end{equation}
as can be seen by substituting $M=0$ in (\ref{potentialmaxwellc2})-(\ref{veven}).
Also in this case the relation $r(r_*)$ takes the simple form
\begin{equation}
r=R\tan{\frac{r_*}{R}}\,,
\label{rrtort}
\end{equation}
and therefore the potential (\ref{potpureAdS}) takes a simple form in the
$r_*$ coordinate, namely
\begin{equation}
V=\frac{l(l+1)}{R^2\sin\left({\frac{r_*}{R}}\right)^2}\,.
\label{potpureAdSrtort}
\end{equation}
To proceed, we note that the change of variable 
$x=\sin\left({\frac{r_*}{R}}\right)^2$
leads the wave equation to a hypergeometric equation,
\begin{equation}
\frac{\partial^{2} \Psi(x)}{\partial x^{2}} +
\frac{\tilde{\tau}}{\sigma}\frac{\partial \Psi(x)}{\partial x} +
\frac{\tilde{\sigma}}{\sigma ^2}\Psi(x)=0 \,,
\label{waveeqhyperg}
\end{equation}
with
\begin{eqnarray}
\tilde{\sigma}= 4(\omega R)^2 x(1-x) -4l(l+1)(1-x)\,,\\
\label{sigma}
\sigma=4x(1-x)\,,\\
\label{ttau}
\tilde{\tau}=2(1-2x)\,.
\end{eqnarray}
To put this in a more standard form, one changes wavefunction
by defining
\begin{equation}
\Psi(x)=\sqrt{x-1}\;x^{\frac{l+1}{2}}\;Z(x)\,,
\label{chgwavefunction}
\end{equation}
and one gets the following standard hypergeometric
differential equation for $Z$:
\begin{equation}
\sigma \frac{\partial^{2} Z(x)}{\partial x^{2}} +
\tau \frac{\partial Z(x)}{\partial x} +
\lambda Z(x)=0 \,,
\label{waveeqhyperg2}
\end{equation}
with $\sigma$ defined in (\ref{sigma}) and
\begin{eqnarray}
\tau=6-4l(x-1)-12x\,, \\
\lambda=-4-4l-l^2+\omega ^2 \,.
\label{hypergparameters2}
\end{eqnarray}
By requiring well behaved fields everywhere a simple analysis
\cite{uvarov} then shows that the following constraint needs
to be satisfied,
\begin{equation}
\omega \,R=2n+l+2 \,.
\label{frequenciespureadselectrgrav}
\end{equation}
These are the pure AdS frequencies for electromagnetic and gravitational
perturbations, corresponding to pure AdS
normal modes of the corresponding fields.
One can compare the frequencies in (\ref{frequenciespureadselectrgrav}) 
with the scalar frequencies corresponding to pure
AdS modes \cite{burgess},
$
\omega_{\rm s}\, R=2n+l+3 \,.
$

\vskip 2cm


\thispagestyle{empty} \setcounter{minitocdepth}{1}
\chapter[Quasinormal modes of the four dimensional toroidal (AdS) black hole]{Quasinormal modes of the four dimensional toroidal (AdS) black hole} \label{chap:qnmtoro}
\lhead[]{\fancyplain{}{\bfseries Chapter \thechapter. \leftmark}}
\rhead[\fancyplain{}{\bfseries \rightmark}]{}
\minitoc \thispagestyle{empty}
\renewcommand{\thepage}{\arabic{page}}
\section{Introduction}
Black holes in anti-de Sitter (AdS) spacetimes in several dimensions
have recently been the object of an intense study. 
One of the reasons for this intense study is
the AdS/CFT conjecture which states that there is a correspondence
between string theory in AdS spacetime and a conformal field theory
(CFT) on the boundary of that space. For instance, M-theory on ${\rm
AdS}_4\times S^7$ is dual to a non-abelian superconformal field theory
in three dimensions, and type IIB superstring theory on ${\rm
AdS}_5\times S^5$ seems to be equivalent to a super Yang-Mills theory
in four dimensions \cite{maldacena} (for a review see 
\cite{maldacenareview}).

All dimensions up to eleven are of interest in superstring theory, but
experiment singles out four dimensions as the most important. In
four-dimensional (4D) general relativity, an effective gravity theory 
in an appropriate  
string theory limit, the Kerr-Newman family of four-dimensional 
black holes can be extended to include a negative cosmological
constant \cite{bcarter}.  The horizon in this family has the topology
of a sphere. There are, however, other families of black holes in general
relativity with a negative cosmological constant with horizons having
topology different from spherical. Here we want to focus on the family
of black holes whose horizon has toroidal, cylindrical or planar
topology \cite{lemos1}.  
We are going to perturb these black holes with 
scalar, electromagnetic and gravitational fields.

Perturbations of known solutions are very important to perform in
order to study their intrinsic properties, such as the natural
frequencies of the perturbations, and to test for the stability of the
solutions themselves. For gravitational objects, such as a black hole,
the vibrational pattern set by the perturbation obliges the system to
emit gravitational waves. Thus, for black holes, the study of
perturbations is closely linked to the gravitational wave emission.
Due to the dissipative character of the emission of the gravitational
waves the vibrational modes do not for a normal set, indeed in the
spectrum each frequency is complex whose imaginary part gives the
damping timescale. These modes are called quasi-normal modes (QNMs).
The QNMs of a black hole appear naturally when one deals with the
evolution of some field in the black hole spacetime, and serve as a
probe to the dynamics outside its event horizon.

In this chapter we shall study scalar, electromagnetic and gravitational
perturbations of the toroidal, cylindrical or planar black holes in an
AdS spacetime found in \cite{lemos1}.  The motivation to perturb with
a scalar field can be seen as follows.  If one has, e.g.,
11-dimensional M-theory, compactified into a $({\rm toroidal\;
BH})_{4} \times ({\rm compact\; space})$ the scalar field used to
perturb the black hole, can be seen as a type IIB dilaton which
couples to a CFT field operator $\cal{O}$. Now, the black hole in the
bulk corresponds to a thermal sate in the boundary CFT, and thus the
bulk scalar perturbation corresponds to a thermal perturbation with
nonzero $<\cal O>$ in the CFT.  Similar arguments hold for the
electromagnetic perturbations since they can be seen as perturbations
for some generic gauge field in the low energy limit of 11-dimensional
M-theory.  On the other hand, gravitational perturbations are always
of importance since they belong to the essence of the spacetime
itself.

We will find that the QN frequencies for scalar perturbations scale
with the horizon radius, at least for large black holes.  In the case
of electromagnetic perturbations of large black holes, the
characteristic QN frequencies have only an imaginary part, and scale
with the horizon radius.  As for gravitational perturbations, there
are two important features. First, contrary to the asymptotically flat
spacetime case, odd and even perturbations no longer have the same
spectra, although in certain limits one can still prove that the
frequencies are almost the same.  The second  result is
that, for odd perturbations, there is a mode with a totally different
behavior from that found in the scalar and electromagnetic case: in
this mode the frequency scales with $\frac{1}{r_+}$, just as in
asymptotically flat Schwarzschild spacetime. 
These features were also found in our study of Schwarzschild-AdS 
black holes in the previous chapters. One could have predicted 
that these features would also appear here, at least for large 
black holes, because in the large horizon limit 
the spherical-AdS black holes have the geometry of the 
cylindrical (planar or toroidal) ones.

\vskip 0.5cm

\noindent
\section{Scalar, electromagnetic and gravitational perturbations in a 
toroidal, cylindrical or planar black hole in an  AdS background}

\vskip 3mm
Throughout this paper, we shall deal with the evolution of some 
perturbation in a spacetime geometry in general relativity 
with a background metric given by 
\cite{lemos1}: 
\begin{equation}
ds^{2}= f(r)\,dt^{2}- f(r)^{-1}dr^{2}-r^2\,dz^{2}
-r^{2}d\phi^{2}\,
\label{lineelementc9}
\end{equation}
where 
\begin{equation}
f(r)=\frac{
r^2}{R^2}-\frac{4MR}{r}\, ,
\label{fqnmtoro}
\end{equation}
$M$ is the ADM mass of the black hole, 
and $R$ is the AdS lengthscale $R^2=-\frac{3}{\Lambda}$, $\Lambda$ being
the cosmological constant. There is a horizon at $r_+=(4M)^{1/3}R$.
The range of the coordinates $z$ and $\phi$ dictates the topology of
the black hole spacetime.  For a black hole with toroidal topology, a
toroidal black hole, the coordinate $z$ is compactified such that
$z/R$ ranges from $0$ to $2\pi$, and $\phi$ ranges from $0$ to $2\pi$
as well.  For the cylindrical black hole, or black string, the
coordinate $z$ has range $-\infty<z<\infty$, and $0\leq \phi
<2\pi$. For the planar black hole, or black membrane, the coordinate
$\phi$ is further decompactified $-\infty<R\,\phi<\infty$
\cite{lemos1}.  We will work with the cylindrical topology but
the results are not altered for the other two topologies. 

According to the AdS/CFT correspondence solution 
(\ref{lineelementc9}) ($\times S^7$) is dual to a superconformal field theory 
in three dimensions with ${\cal N}=8$. 
These toroidal black holes with Ricci flat 
horizon we are considering in (\ref{lineelementc9}) can be seen as 
the large horizon radius limit, $r_+/R>>1$, of the spherical AdS black holes.
The large black holes are the ones that matter most to the AdS/CFT 
correspondence 
\cite{maldacena}.
Perturbations of spherical black holes were studied in the previous chapter. 
Therefore the results we obtain here 
should be similar to the results we have obtained in \cite{cardosoqnmsads} for 
large black holes. This will be confirmed below. For small black holes 
we will show that the spherical and toroidal yield different results.

\subsection{Scalar field perturbations}


For scalar perturbations, we are interested in solutions to 
the minimally coupled scalar wave equation 
\begin{equation}
{\Phi^{,\,\mu}}_{;\,\mu}=0 \,, 
\label{minimalscalareq1c9}
\end{equation}
where, a comma stands for ordinary derivative and a semi-colon stands
for covariant derivative.
We make the following ansatz for the field $\Phi$ 
\begin{equation}
\Phi=\frac{1}{r}P(r)e^{-i\omega t}e^{ikz}e^{il\phi}\,,
\label{ansatzforscalarc9}
\end{equation}
where $\omega$, $k$, and $l$, are the frequency, the wavenumber and
the angular quantum numbers of the perturbation. If one is dealing 
with the toroidal  topology then $k$ should be changed into  
an angular quantum number $\bar l$,   
$ e^{ikz} \rightarrow e^{i{\bar l}\frac{z}{R}}$. For the planar 
topology $e^{il\phi}\rightarrow e^{i{\bar k}R\,\phi}$, where 
$\bar k$ is now a continuous wave number. 

It is useful to use the tortoise 
coordinate  $r_*$
defined by the equation $dr_*=dr/\left({r^2}/{R^2}-{4MR}/{r}\right)$. 
With the ansatz (\ref{ansatzforscalarc9}) and the tortoise coordinate $r_*$, 
equation 
(\ref{minimalscalareq1c9}) is given by, 
\begin{equation}
\frac{d^2 P(r)}{d {r_*}^2} + \left\lbrack\omega^2 - 
V_{\rm scalar}(r)\right\rbrack
P(r)=0\,,
\label{minimalscalareq2c9}
\end{equation}
where,
\begin{equation}
V_{\rm scalar}(r)=f\left(\frac{l^2}{r^2}+
\frac{R^2k^2}{r^2}+\frac{f'}{r}\right)\,,
\label{potentialscalarc9}
\end{equation}
with $r=r(r_*)$ given implicitly and $f'\equiv df/dr$.
The rescaling to the radial coordinate 
$\hat{r}=\frac{r}{R}$ and to the frequency $\hat{\omega}=\omega R$ is 
equivalent 
to take $R=1$ in (\ref{minimalscalareq2c9}) and (\ref{potentialscalarc9}), i.e., 
through this rescaling one measures the frequency and other quantities in 
terms of the AdS lengthscale $R$.

\subsection{Maxwell field perturbations}
We consider the evolution of a Maxwell field in a
cylindrical-AdS black hole spacetime with a metric
given by (\ref{lineelementc9}).
The evolution
is governed by Maxwell's equations:
\begin{equation}
{F^{\mu\nu}}_{;\nu}=0\,,\quad F_{\mu\nu}=A_{\nu,\mu}-A_{\mu,\nu}\,.
\label{maxwellc9} 
\end{equation}
One can again separate variables by the ansatz:
\begin{eqnarray}
A_{\mu}(t,r,\phi,z)=
\left[ \begin{array}{c}g^{k\,l}(r)\\h^{k\,l}(r)
\\k^{k\,l}(r) \\ j^{k\,l}(r)\end{array}\right]
e^{-i\omega t}
e^{ikz}
e^{il\phi} \,, 
\label{empotentialdecompositionc9}
\end{eqnarray}
When we put this expansion into Maxwell's
equations (\ref{maxwellc9}) we get the $2^{\rm nd}$ order differential
equation for the perturbation:
\begin{equation}
\frac{\partial^{2} \Psi(r)}{\partial r_*^{2}} +\left\lbrack\omega^2
-V_{\rm maxwell}(r)\right\rbrack\Psi(r)=0 \,,
\label{wavemaxwellc9}
\end{equation}
where the wavefunction $\Psi$ is a  combination of the functions
$g^{lm}$, $h^{lm}$, $k^{lm}$ and $j^{lm}$ as appearing in
(\ref{empotentialdecompositionc9}), and $\Psi=-i\omega h^{lm}-
\frac{dg^{lm}}{dr}$ (see \cite{ruffini} for further details).
The potential $V_{\rm maxwell}(r)$ appearing in equation (\ref{wavemaxwellc9}) 
is given by
\begin{equation}
V_{\rm maxwell}(r)=f(r)\left(\frac{l^2}{r^2}
+\frac{R^2k^2}{r^2}\right) \,.
\label{potentialmaxwellc9}
\end{equation}
Again we can
take $R=1$ and measure everything in terms of $R$.
\subsection{ Gravitational perturbations}
In our analysis of gravitational perturbations, we shall adopt a
 procedure analogous to that of Chandrasekhar \cite{chandra}, generalizing
the calculation by the introduction of the cosmological constant 
$\Lambda=-\frac{3}{R^2}$.
The perturbed metric will be taken to be
\begin{equation}
ds^{2}= e^{2\nu}dt^{2}- e^{2\psi}(d\phi-wdt-q_2dr-q_3dz)^2-
e^{2\mu_2}dr^2-e^{2\mu_3}dz^2\,.
\label{lineelementchandra}
\end{equation}
where the unperturbed quantities are $e^{2\nu}=\frac{ r^{2}}{R^2}-
\frac{4MR}{r}$,
$e^{2\psi}=r^{2}$, $e^{2\mu_2}=(\frac{
r^2}{R^2}-\frac{4MR}{r})^{-1}$, $e^{2\mu_3}=\frac{r^2}{R^2}\,$ and all the 
other unperturbed quantities are zero.
By observing the effect of performing $\phi\rightarrow-\phi$, and
maintaining the nomenclature of the spherical symmetric case
\cite{chandra}, it can be seen that the perturbations fall into two
distinct classes: the odd perturbations (also called axial in the
spherical case) which are the quantities $w$, $q_2$, and $q_3$, and
the even perturbations (also called polar in the spherical case) which
are small increments $\delta\nu$, $\delta\mu_2$, $\delta\mu_3$ and
$\delta\psi$ of the functions $\nu$, $\mu_2$, $\mu_3$ and $\psi$,
respectively.  Note that since $z$ is also an ignorable coordinate,
one could, in principle, interchange $\phi$ with $z$ in the above argument. 

In what follows, we shall limit ourselves to  axially symmetric
perturbations, i.e., to the case in which the quantities listed above
do not depend on $\phi$. In this case odd and even perturbations
decouple, and it is possible to simplify them considerably.

\subsubsection{Odd perturbations}
We will deal with odd perturbations first. As we have stated, these
are characterized by the non vanishing of $w\,$, $q_2\,$ and
$q_3$. The equations governing these quantities are the following Einstein's
equations with a cosmological constant
\begin{equation}
G_{12}+\frac{3}{R^2}g_{12}=0\,,
\label{g01}
\end{equation}
\begin{equation}
G_{23}+\frac{3}{R^2}g_{23}=0\,.
\label{g12}
\end{equation}
The reduction of these two equations to a one dimensional second order
differential equation is well known (see \cite{chandra}), and we only
state the results. By defining
\begin{equation}
Z^-(r)=r\left(\frac{ r^{2}}{R^2}-\frac{4MR}{r}\right)
\left(\frac{dq_2}{dz}-\frac{dq_3}{dr}\right)\,,
\label{Z-}
\end{equation}
One can easily check that $Z^-(r)$ satisfies
\begin{equation}
\frac{\partial^{2} Z^-(r)}{\partial r_*^{2}} +\left\lbrack\omega^2
-V_{\rm odd}(r)\right\rbrack Z^-(r)=0 \,,
\label{waveaxial}
\end{equation}
where $V_{\rm odd}(r)$ appearing in equation (\ref{waveaxial}) is given by
\begin{equation}
V_{\rm odd}(r)=f(r)\left(\frac{R^2k^2}{r^2}-\frac{12MR}{r^3}\right) \,.
\label{potentialaxial}
\end{equation}

\subsubsection{Even perturbations}
Even perturbations are characterized by non-vanishing increments 
in the metric functions $\nu\,$, $\mu_2\,$, $\mu_3\,$ and $\psi\,$.
The equations which they obey are obtained by linearizing 
$G_{01}+\frac{3}{R^2}g_{01}$,
$G_{03}+\frac{3}{R^2}g_{03}$, 
$G_{13}+\frac{3}{R^2}g_{13}$, 
$G_{11}+\frac{3}{R^2}g_{11}$ and
$G_{22}+\frac{3}{R^2}g_{22}$ about their unperturbed values.
Making the ansatz
\begin{eqnarray}
\delta \nu=N(r)e^{ikz}\,,\\
\delta \mu_2=L(r)e^{ikz}\,,\\
\delta \mu_3=T(r)e^{ikz}\,, \\ 
\delta \psi=V(r)e^{ikz}  \,, 
\label{ansatzpolar}
\end{eqnarray}
we have from $\delta(G_{03}+\frac{3}{R^2}g_{03})=0$ that
\begin{equation}
V(r)=-L(r)\,.
\label{relpol1}
\end{equation}
Inserting this relation and using (\ref{ansatzpolar}) in 
$\delta(G_{01}+\frac{3}{R^2}g_{01})=0\,$,
we have  
\begin{equation}
\left(\frac{3}{r}-\frac{f'}{2f}\right)L(r)
+\left(\frac{f'}{2f}-\frac{1}{r}\right)T(r)+L(r)_{,r}-
T(r)_{,r}=0\,.
\label{Ac9}
\end{equation}
 From $\delta(G_{13}+\frac{3}{R^2}g_{13})=0\,$ we have
\begin{equation}
\left(\frac{1}{r}+\frac{f'}{2f}\right)L(r)+
\left(\frac{1}{r}-\frac{f'}{2f}\right)N(r)-N(r)_{,r}+
L(r)_{,r}=0\,,
\label{C}
\end{equation}
and from $\delta(G_{11}+\frac{3}{R^2}g_{11})=0\,$ we obtain
\begin{eqnarray}
&-\frac{k^2R^2}{fr^2}N(r)+
\left(\frac{k^2R^2}{fr^2}-\frac{\omega^2}{f^2}-\frac{6}{f}\right)L(r)+
\frac{\omega^2}{f^2}T(r)+
&\nonumber\\&
\left(\frac{1}{r}+\frac{1}{Rr}\right)N(r)_{,r}+
\left(-\frac{f'}{2f}-\frac{1}{Rr}\right)L(r)_{,r}
+ \left(\frac{f'}{2f}+\frac{1}{r}\right)T(r)_{,r}=0\,.
\label{D}
\end{eqnarray}
Multiplying equation (\ref{C}) by $\frac{2}{r}$ and adding (\ref{D}) we 
can obtain $N(r)$
and $N(r)_{,r}$ in terms of $L(r)$, $L(r)_{,r}$, $T(r)$ and $T(r)_{,r}$. 
Using (\ref{Ac9}) and (\ref{C}) we can express $L(r)\,$, $N(r)\,$, and up to
their second derivatives in terms of $T(r)$, $T(r)_{,r}$ and $T(r)_{,rr}$.
Finally, we can look for a function
\begin{equation}
Z^+=a(r)T(r)+b(r)L(r)\,.
\label{defZ}
\end{equation}
which satisfies the second order differential equation
\begin{equation}
\frac{\partial^{2} Z^+(r)}{\partial r_*^{2}} +\left\lbrack\omega^2
-V_{\rm even}(r)\right\rbrack Z^+(r)=0 \,,
\label{wavepolar}
\end{equation}
Substituting (\ref{defZ}) into (\ref{wavepolar}) and expressing $L(r)$ and
its derivatives in terms of $T(r)$ and its derivatives, we obtain an equation
in $T(r)$, $T(r)_{,r}$ and $T(r)_{,rr}$ whose coefficients must vanish 
identically.
If we now demand that $a(r)$ and $b(r)$ do not depend on the frequency 
$\omega$ we find 
\begin{eqnarray}
a(r)=\frac{r}{12Mr+k^2r^2}\,,\\
b(r)=\frac{6M+k^2r}{72M^2+6k^2Mr}\,,
\label{relation}
\end{eqnarray}
and the potential $V_{\rm even}(r)$ in (\ref{wavepolar}) is
\begin{equation}
V_{\rm even}(r)=f(r)\left\lbrack\frac{576M^3+12k^4Mr^2+k^6r^3+
144M^2r(k^2+2r^2)}{r^3(12M+k^2r)^2}\right\rbrack \,.
\label{potentialpolar}
\end{equation}

As a final remark concerning the wave equations obeyed by odd and even
gravitational perturbations, we note that it can easily be checked that the 
two 
potentials can be expressed in the form
\begin{eqnarray}
V_{\stackrel{\rm odd}{\rm even}}= 
W^2 \pm \frac{dW}{dr_*} +\beta \,, \\
W=\frac{96M^2(k^2+3r^2)}{2k^2r^2(12M+k^2r)}+j\,,
\label{Wc9}
\end{eqnarray}
where $j=-\frac{k^6+288M^2}{24k^2M}$, and
$\beta=-\frac{k^8}{576M^2}$.  It is worth of notice that the two
potentials can be written in such a simple form (potentials related in
this manner are sometimes called superpartner potentials
\cite{cooper}), a fact which seems to have been discovered by
Chandrasekhar \cite{chandradet}.

\vskip 0.5cm

\noindent
\section{Quasi-normal modes and some of its properties}

\vskip 3mm


\subsubsection{Boundary conditions}
To solve (\ref{minimalscalareq2c9}), (\ref{wavemaxwellc9}),
(\ref{waveaxial}) and (\ref{wavepolar}) one must specify boundary
conditions, a non-trivial task in AdS spacetimes.  Consider first the
case of a Schwarzschild black hole in an asymptotically flat spacetime
(see \cite{kokkotas}). Since the potential now vanishes at both
infinity and the horizon, two independent solutions near these points
are $ \Psi_1 \sim e^{-i\omega r_*}$ and $\Psi_2 \sim e^{i\omega r_*}$,
where the $r_*$ coordinate now ranges from $-\infty $ to $\infty$.
QNMs are defined by the condition that at the horizon
there are only ingoing waves, $\Psi_{\rm hor}\sim e^{-i\omega r_*} $.
Furthermore, one wishes to have nothing coming in from infinity (where
the potential now vanishes), so one wants a purely outgoing wave at
infinity, $\Psi_{\rm infinity}\sim e^{i\omega r_*} $. Clearly, only a
discrete set of frequencies $\omega$ meet these requirements.

Consider now our asymptotically AdS spacetime.  The first boundary
condition stands as it is, so we want that near the horizon
$\Psi_{\rm hor}\sim e^{-i\omega r_*} $. However $r_*$ has a finite range,
so the second boundary condition needs to be changed. There have been
a number of papers on which boundary conditions to impose at infinity
in AdS spacetimes (\cite{avis}-\cite{burgess}).  We
shall require energy conservation and adopt reflective boundary
conditions at infinity \cite{avis}
which means that the wavefunction is zero at infinity
(see however \cite{dasgupta}).

\subsection{Numerical calculation of the QN frequencies}
To find the frequencies $\omega$ that satisfy the previously stated
boundary conditions we follow exactly the same method as outlined
in the previous chapter. We shall dwell on this no further.

We present the results in tables \ref{toro1}-\ref{toro4}.

\vskip 0.5cm
\subsubsection{Scalar:}
\vskip 0.5cm
\begin{table}
\caption{\label{toro1} Lowest QN frequencies of scalar perturbations for $l=0\,$ and $k=0$ of the topologically
non-trivial black hole \cite{lemos1}.}
\begin{tabular}{lllll}  \hline 
\multicolumn{1}{c}{} &
\multicolumn{2}{c}{ Numerical} \\ \hline
$r_+$    &  $-\omega_i$ &  $\omega_r$ \\ \hline
0.1  & 0.266 & 0.185   \\ \hline
 1   &  2.664  &  1.849   \\ \hline
5  & 13.319 & 9.247  \\ \hline
10  & 26.638  & 18.494    \\ \hline
50  &133.192 & 92.471   \\ \hline
100 &  266.373 &184.942  \\ \hline
\end{tabular}
\end{table}
\vskip 1mm

In table \ref{toro1} we list the numerical values of the lowest ($n=1$) QN
frequencies for the $l=0$ scalar field and for selected values of
$r_+$. As discussed in Horowitz and Hubeny (HH) \cite{horowitz} the
frequency should be a function of the scales of the problem, $R$ and
$r_+$. However, they showed and argued that due to additional
symmetries in the scalar field case and for large Schwarzschild-AdS
black holes the frequency scales as $\omega\propto T$, with the
temperature of the large black hole given by $T\propto r_+/R^2$, i.e.,
$\propto r_+$ in our units. This behavior is a totally different
behavior from that of asymptotically flat space, in which the
frequency scales with $\frac{1}{r_+}$.  Now, for cylindrical (planar
or toroidal) black holes and scalar fields this symmetry is present
for any horizon radius, so $\omega\propto r_+$ always.  For these
black holes the temperature is also proportional to $r_+$, no matter
how small the black hole is \cite{lemos1,peca}. Thus, the scalar field
QN frequencies are proportional to $T$, as can be directly seen from
table \ref{toro1}, $\omega\propto r_+\propto T$.  The imaginary part of the
frequency determines how damped the mode is, and according to the
AdS/CFT conjecture is a measure of the characteristic time
$\tau=\frac{1}{\omega_i}$ of approach to thermal equilibrium in the
dual CFT (moreover, the frequencies do not seem to depend on the
angular quantum number $l$, we have performed calculations for higher
values of $l$).  In the dual CFT the approach to thermal equilibrium
is therefore faster for higher temperatures, i.e., larger black holes.
This scaling for all horizon radii with temperature only happens in
the scalar field case.  For the electromagnetic and some of the
gravitational perturbations the frequency scales with the temperature
only in the large black hole regime, as we will show.

In table \ref{toro2} we list the numerical values of the lowest ($n=1$)
QN frequencies for $l=1$ and for selected values of
$r_+$.  For frequencies with no real part, we list the values
obtained in the ``highly damped approximation'' \cite{liu,Liu2}. 
We can note from table \ref{toro2} that $\omega$ is proportional to $r_+$ and thus to 
the temperature for large black holes, $r_+ \buildrel>\over\sim  
5$, say.

\vskip 0.5cm
\subsubsection{Electromagnetic:}
\vskip 0.5cm
\begin{table}
\caption{\label{toro2} Lowest QN frequencies of electromagnetic 
perturbations for $l=1\,$ and $k=0$ of the topologically
non-trivial black hole \cite{lemos1}.}
\begin{tabular}{lllll}  \hline 
\multicolumn{1}{c}{} &
\multicolumn{2}{c}{ Numerical} &
\multicolumn{2}{c}{ Highly Damped} \\ \hline
$r_+$    &  $-\omega_i$ &  $\omega_r$ &  $-\omega_i$  & $\omega_r$ \\ \hline
0.1  & 0.104 &1.033  & $-$ & $-$  \\ \hline
 1   &  1.709  & 1.336  & $-$ & $-$   \\ \hline
5  &7.982  & $\sim0$ & 7.500 & $\sim0$  \\ \hline
10  &15.220   & $\sim0$  & 15.000 & $\sim0$  \\ \hline
50  &75.043 & $\sim0$ & 75.000 & $\sim0$  \\ \hline
100 & 150.021  & $\sim0$ & 150.000 & $\sim0$  \\ \hline
\end{tabular}
\end{table}
\vskip 1mm

In the large black hole regime, both scalar and electromagnetic QN
frequencies for this geometry are very similar to those of the
Schwarzschild-anti-de Sitter black hole 
\cite{horowitz,cardosoqnmsads,cardosoqnmsads2}. This
is a consequence of the fact that in this regime, the wave equation
for the fields become identical in both geometries.  Indeed we can
compare table \ref{toro1} for the scalar field with table 1 of 
HH  \cite{horowitz} for the same field. We see that for $r_+=1$
one has for the toroidal black hole $\omega=1.849-2.664i$, whereas HH
find $\omega=2.798-2.671 i$. Thus they differ, $r_+=1$ is not a large
black hole. For $r_+>>1$ one expects to find very similar QN
frequencies.  For instance, for $r_+=100$ we obtain
$\omega=184.942-266.373i$ for the toroidal black hole, while HH obtain
$\omega=184.953-266.385 i$ for the Schwarzschild-anti de Sitter black
hole.  
The same thing happens for electromagnetic perturbations.  We
can compare table \ref{toro2} for the electromagnetic field with table I of
Cardoso and Lemos \cite{cardosoqnmsads}. For large black holes one can see
that the frequencies for toroidal black holes are very similar to the
frequencies of the Schwarzschild-anti-de Sitter black hole. For
instance, for $r_+=100$ we find $\omega=-150.021i$ for the toroidal
black hole, while in \cite{cardosoqnmsads} we found $\omega=-150.048 i$ for
the Schwarzschild-anti-de Sitter black hole.  We can also see that the
QN frequencies in the electromagnetic case are in excellent agreement with
the analytical approximation for strongly damped modes.

\subsubsection{Gravitational:}
The numerical calculation of the QN frequencies for gravitational
perturbations proceeds as outlined previously (the associated
differential equation has only regular singularities, so it is
possible to use an expansion such as that given in the previous chapter. In tables \ref{toro3}
and \ref{toro4} we show the two lowest lying ($n=1\,,2$) QN frequencies for
$l=2$ and $l=3$ gravitational perturbations.

We first note that there is clearly a distinction between odd and even
perturbations: they no longer have the same spectra, contrary to the
asymptotically flat space case (see \cite{chandra}).  This problem
was studied in some detail by Cardoso and Lemos \cite{cardosoqnmsads} 
(the derivation was given in the previous chapter) who
showed that it is connected with the behavior of $W$ (see equation
(\ref{Wc9})) at infinity.
\noindent 
\begin{table}
\caption{\label{toro3} Lowest and second lowest 
QN frequencies of gravitational odd perturbations for 
$k=2$ of the topologically
non-trivial black hole \cite{lemos1}.}
\begin{tabular}{lllll}  \hline 
\multicolumn{1}{c}{} &
\multicolumn{2}{c}{ lowest QNM} &
\multicolumn{2}{c}{ second lowest QNM} \\ \hline
$r_+$    &  $-\omega_i$ &  $\omega_r$ &  $-\omega_i$  & $\omega_r$ \\ \hline
1  & 2.646  & $\sim0$  &2.047  & 2.216   \\ \hline
5  & 0.2703 & $\sim0$ &13.288  &9.355   \\ \hline
10   &  0.13378 & $\sim0$  & 26.623 & 18.549  \\ \hline
50   & 0.02667 & $\sim0$ & 133.189 & 92.482 \\ \hline
100  & 0.0134  & $\sim0$ & 266.384 & 184.948  \\ \hline
\end{tabular}
\end{table}
\noindent 
\begin{table}
\caption{\label{toro4} Lowest QN frequencies of gravitational even perturbations for 
$k=2$ of the topologically
non-trivial black hole \cite{lemos1}.}
\begin{tabular}{lllll}  \hline 
\multicolumn{1}{c}{} &
\multicolumn{2}{c}{ lowest QNM, $k=2$} \\ \hline
$r_+$    &  $-\omega_i$ &  $\omega_r$ \\ \hline
1   & 1.552 & 2.305   \\ \hline
5  &  12.633 & 9.624     \\ \hline
10  & 26.296  & 18.696     \\ \hline
50  & 133.124 & 92.512       \\ \hline
100  & 266.351  & 184.963     \\ \hline
\end{tabular}
\end{table}
For odd gravitational QNMs the lowest one scales with
$\frac{1}{r_+}\propto \frac{1}{T}$. This is odd, but one can see that
it is a reflection of the different behavior of the potential $V_{\rm
odd}$ for odd perturbations.  For the second lowest odd gravitational
QN frequency table \ref{toro3} shows that for large black holes it scales with $T$.  For
the lowest even gravitational QN frequency table \ref{toro4} shows that it also scales
with $T$ for large black holes.  Note further that the scalar, odd
second lowest and even gravitational QNMs are very similar in the
large black hole regime.  Indeed, tables \ref{toro1}, \ref{toro3} and \ref{toro4} show a remarkable
resemblance even though the potentials are so different.  Finally,
let us compare tables \ref{toro3} and \ref{toro4} with tables III-V of \cite{cardosoqnmsads}.  We
see that for large black holes the frequencies of toroidal black holes
are again very similar to those of the Schwarzschild-anti-de Sitter
black hole.  For instance, for odd perturbations and $r_+=100$ we find
from table 3 $\omega=-0.0134i$ for the toroidal black hole, while in
\cite{cardosoqnmsads} we found (table III) $\omega=-0.0132i$ for the
Schwarzschild-anti-de Sitter black hole. 

\vskip 0.5cm

\noindent
\section{Conclusions}
\vskip 3mm

We have computed the scalar, electromagnetic and gravitational QN
frequencies of the toroidal, cylindrical or planar black hole in four
dimensions. These modes dictate the late time behaviour of a minimally
coupled scalar, electromagnetic field and of small gravitational
perturbations, respectively.  The main conclusion to be drawn from
this work is that these black holes are stable with respect to small
perturbations.  In fact, as one can see, the frequencies all have a
negative imaginary part, which means that these perturbations will
decay exponentially with time.  For odd gravitational perturbations in
the large black hole regime, the imaginary part of the frequency goes
to zero scaling with $\frac{1}{r_+}$, just as in asymptotically flat
space and in the odd gravitational perturbations of Schwarzschild-Ads
black hole.  In terms of the AdS/CFT correspondence, this implies that
the greater the mass, the more time it takes to approach equilibrium,
a somewhat puzzling result.  Apart from this interesting result, the
frequencies all scale with the horizon radius, at least in the large
black hole regime, supporting the arguments given in \cite{horowitz}.
The QNM for toroidal, cylindrical or planar black holes (in anti-de
Sitter space) are quite similar to those of the Schwarzschild-anti-de
Sitter black hole \cite{horowitz,cardosoqnmsads}. 
As for the highly damped modes, and as was seen in the previous chapter,
we expect them to coincide with those of the Schwarzschild-anti-de Sitter
black hole. 

\thispagestyle{empty} \setcounter{minitocdepth}{1}
\chapter[Quasinormal modes of the near extremal four dimensional 
Schwarzschild-dS black hole]{Quasinormal modes of the near extremal four dimensional Schwarzschild-dS black hole} \label{chap:qnmds}
\lhead[]{\fancyplain{}{\bfseries Chapter \thechapter. \leftmark}}
\rhead[\fancyplain{}{\bfseries \rightmark}]{}
\minitoc \thispagestyle{empty}
\renewcommand{\thepage}{\arabic{page}}
\section{Introduction}
As we remarked earlier the QNMs have acquired a different
importance recently.  Following an observation by Hod \cite{hod}, it has been
proposed \cite{dreyer,dreyerall} that the Barbero-Immirzi
parameter \cite{immirzi}, a factor introduced by hand in order that
Loop Quantum Gravity reproduces correctly the black hole entropy, is
equal to the real part of the QN frequencies with a large
imaginary part.  The identification came from what first seemed to be
a numerical coincidence, but which has been proved to be exact for
Schwarzschild black holes by Motl \cite{motl1} and Motl and Neitzke \cite{motl2}, 
assuming the gauge group of the theory to be $SO(3)$.

It is now important to see
whether the agreement works only for Schwarzschild black holes, or if
it continues to be true in different spacetimes.
Anti-de Sitter spacetimes have been tested and it seems, although it
is still too early too claim something, that there is a chance of
getting a relation between QN frequencies and quantization
\cite{birmingham3,cardosoqnmbtz}.  It has been conjectured by
Kunstatter \cite{dreyerall}, using Dreyer's \cite{dreyer} arguments
that the real part of the QN frequency cooresponding to
highly damped modes of higher dimensional Schwarzschild black holes
should have a particular form \cite{dreyerall}.  This conjecture has
very recently been confirmed by Birmingham \cite{birminghamd},
following earlier work on qiasinormal modes for this spacetime
\cite{vitoroscarjose,konoplyawkb}.  One of the important black hole
spacetimes that has been resisting is the Kerr black hole, which we
shall study in the next section, and the Schwarzschild-de Sitter black
hole.

We shall take a step further on carrying on
this program by computing analytically the QN frequencies of the
near extremal Schwarzschild-de Sitter black hole, which is the
spacetime for which the black hole horizon and the cosmological
horizon are close to each other, in a manner to be defined latter.
Sometimes the extreme Schwarzschild-dS black hole is labelled as
Nariai black hole. However, we remark that the true Nariai solution
is significantly different from the Schwarzschild-dS black hole:
it is not a black hole solution, and has a rather different topology.
The Nariai solution solves exactly the Einstein equations with
$\Lambda>0$, without or with a
Maxwell field, and has been discovered by Nariai in 1951
\cite{Nariai}. It is the direct topological product of $dS_2
\times S^2$, i.e., of a (1+1)-dimensional dS spacetime with a
round 2-sphere of fixed radius.
Three decades after Nariai's paper, Ginsparg and Perry
\cite{GinsPerry} connected the Nariai solution with the
Schwarzschild-dS solution. They showed that the Nariai solution
can be generated from a near-extreme dS black hole, through an
appropriate non-trivial limiting procedure in which the black hole
horizon
approaches the cosmological horizon. One of the aims of Ginsparg
and Perry was to study the quantum stability of the Nariai and
the Schwarzschild-dS solutions \cite{GinsPerry}.  It was shown
that the Nariai solution is in general unstable and, once created,
decays through a quantum tunnelling process into a slightly
non-extreme Schwarzschild-dS black hole
(for a complete review and references on this subject see, e.g.,
Bousso \cite{Bousso60y} and Dias and Lemos \cite{OscLemNariai}.

For this spacetime, we find that it is possible to solve the field
equations exactly in terms of hypergeometric functions, and therefore
an exact analytical expression for the QN frequencies of
scalar, electromagnetic and gravitational perturbations is also
possible. In particular this will give us the QN frequencies
with very large imaginary part. We demonstrate why an approach by Moss
and Norman \cite{moss} based on fitting the potential to the
P\"oshl-Teller potential works well in the Schwarzschild-de Sitter
spacetime.
This chapter will follow the work in \cite{cardosoqnmds}.
This work has been generalized to higher dimensions in \cite{molina} and
improved to second order in \cite{brinkds}.
Very recently, Yoshida and Futamase \cite{shijunds} have taken a completely numerically
look at this problem, and solved numerically for the highly damped
QNMs. Their conclusion is that while the results we shall 
present are correct for the low lying QN frequencies, they do not
describe the highly damped modes. This means that our results are not correct to all
orders in perturbation theory, although they certainly are correct up to second order,
as was demonstrated in \cite{brinkds}.
For other works related to QNMs in the Schwarzschild-de Sitter
spacetime see \cite{suneeta,zhidenko}.

\section{Equations}
Our notation will follow that of \cite{brady} which we have found
convenient.  The metric of the Schwarzschild-de Sitter (SdS) spacetime
is given by
\begin{equation}
ds^2 = -f\, dt^2 + f^{-1}\, dr^2 + r^2 (d\theta^2 
+ \sin^2\theta\, d\phi^2), 
\label{2.1qnmds}
\end{equation}
where 
\begin{equation}
f = 1 - \frac{2M}{r} - \frac{r^2}{a^2},
\label{2.2}
\end{equation}
with $M$ denoting the black-hole mass, and $a^2$ is given in terms of the
cosmological constant $\Lambda$ by $a^2 = 3/\Lambda$.  The spacetime
possesses two horizons: the black-hole horizon is at $r=r_b$ and the
cosmological horizon is at $r = r_c$, where $r_c > r_b$. The function
$f$ has zeroes at $r_b$, $r_c$, and $r_0 = -(r_b + r_c)$. In terms of
these quantities, $f$ can be expressed as
\begin{equation}
f = \frac{1}{a^2 r}\, (r-r_b)(r_c-r)(r-r_0).
\label{2.3}
\end{equation}
It is useful to regard $r_b$ and $r_c$ as the two fundamental
parameters of the SdS spacetime, and to express $M$ and $a^2$ as
functions of these variables. The appropriate relations are
\begin{equation}
a^2 = {r_b}^2 + r_b r_c + {r_c}^2
\label{2.4}
\end{equation}
and
\begin{equation}
2M a^2 = r_b r_c (r_b + r_c).
\label{2.5}
\end{equation}
We also introduce the surface gravity $\kappa_b$ associated with the
black hole horizon $r = r_b$, as defined by the relation 
$\kappa_b = \frac{1}{2} 
df/dr_{r=r_b}$. Explicitly,  we have
\begin{equation}
\kappa_b = \frac{ (r_c-r_b)(r_b-r_0) }{ 2a^2 r_b }.
\label{surface}
\end{equation}
After a Fourier decomposition in frequencies and a multipole expansion, 
the scalar, electromagnetic and gravitational perturbations all
obey a wave equation of the form \cite{cardosoqnmbtz,cardosoqnmsads,cardosoqnmtoro}
\begin{equation}
\frac{\partial^{2} \phi(\omega,r)}{\partial r_*^{2}} +
\left\lbrack\omega^2-V(r)\right\rbrack
\phi(\omega,r)=0 \,,
\label{waveequation}
\end{equation}
where the tortoise coordinate is given by 
\begin{equation}
r_* \equiv \int f^{-1}\, dr\,, 
\label{tortoise}
\end{equation}
and the potential $V$
depends on the kind of field under consideration. 
Explicitly, for scalar
perturbations 
\begin{equation}
V_{\rm s}=f\left[\frac{l(l+1)}{r^2}+\frac{2M}{r^3}-\frac{2}{a^2}\right] \,,
\label{potentialscalarqnmds}
\end{equation}
while for electromagnetic perturbations
\begin{equation}
V_{\rm el}=f\left[\frac{l(l+1)}{r^2}\right] \,.
\label{potentialelectromagnetic}
\end{equation}
The gravitational perturbations decompose into two sets \cite{cardosoqnmsads}, 
the odd and the even parity one. We find however that for this spacetime,
they both yield the same QN frequencies, so it is enough
to consider one of them, the odd parity ones say, for which the potential is 
\cite{cardosoqnmsads}
\begin{equation}
V_{\rm grav}=f\left[\frac{l(l+1)}{r^2}-\frac{6M}{r^3}\right] \,.
\label{potentialgravitational}
\end{equation}
In all cases, we denote by $l$ the angular quantum number,
that gives the multipolarity of the field.
Let us now specialize to the near extremal SdS black hole, which is
defined as the spacetime for which the cosmological horizon $r_c$
is very close (in the $r$ coordinate) to the black hole horizon $r_b$, i.e.
$\frac{r_c-r_b}{r_b}<<1$.
For this spacetime one can make the following approximations
\begin{equation}
r_0 \sim -2r_{b}^2\,\,;\,a^2\sim 3r_{b}^2;\,\,
M \sim \frac{r_b}{3}\,\,;\,\kappa_b \sim \frac{r_c-r_b}{2r_{b}^2}\,.
\label{approximation1}
\end{equation}
Furthermore, and this is the key point, since $r$ is constrained to
vary between $r_b$ and $r_c$, we get $r-r_0 \sim r_b -r_0 \sim 3r_0$
and thus 
\begin{equation}
f \sim \frac{(r-r_b)(r_c-r)}{r_{b}^2}\,.
\label{approximation2}
\end{equation}
In this limit, one can invert the relation $r_*(r)$ 
of (\ref{tortoise}) to get
\begin{equation}
r= \frac{r_c e^{2\kappa_b r_*}+r_b}{1+e^{2\kappa_b r_*}}\,.
\label{rtortoise}
\end{equation}
Substituting this on the expression (\ref{approximation2})
for $f$ we find 
\begin{equation}
f = \frac{(r_c-r_b)^2}{4r_{b}^2\cosh{(\kappa_b r_*)}^2}\,.
\label{approximation3}
\end{equation}
As such, and taking into account the functional form of the potentials
(\ref{potentialscalarqnmds})-(\ref{potentialgravitational}) we see that for
the near extremal SdS black hole the wave equation (\ref{waveequation}) is 
of the form
\begin{equation}
\frac{\partial^{2} \phi(\omega,r)}{\partial r_*^{2}} +
\left\lbrack\omega^2-\frac{V_0}{\cosh{(\kappa_b r_*)}^2}\right\rbrack
\phi(\omega,r)=0 \,,
\label{waveequation2}
\end{equation}
with
\begin{equation}
V_0=\left\{ \begin{array}{ll}
            \kappa_{b}^2 l(l+1)\,,   & {\rm \,scalar\, and\, electromagnetic}\\
                                  &  {\rm perturbations}. \\ 
            \kappa_{b}^2 (l+2)(l-1)\,,   &{\rm \, gravitational} \\
                                     &  {\rm perturbations}\,.
\end{array}\right.
\label{V0}
\end{equation}
The potential in (\ref{waveequation2}) is the well known 
P\"oshl-Teller potential \cite{teller}. The solutions to 
(\ref{waveequation2})
were studied and they are of the hypergeometric type, 
(for details see Ferrari and Mashhoon \cite{ferrari}).
It should be solved under appropriate boundary conditions:
\begin{eqnarray}
\phi \sim e^{-i\omega r_*} \,,r_* \rightarrow -\infty \\
\phi \sim e^{i\omega r_*}  \,,r_* \rightarrow \infty. 
\label{behavior1}
\end{eqnarray}
These boundary conditions impose a non-trivial condition on
$\omega$ \cite{ferrari}, and those that satisfy both simultaneously
are called QN frequencies. For the P\"oshl-Teller potential
one can show \cite{ferrari} that they are given by
\begin{equation}
\omega=\kappa_{b} \left [ -(n+\frac{1}{2})i+
\sqrt{\frac{V_0}{\kappa_b^2}-\frac{1}{4}} \right ]\,, n=0,1,...\,.
\label{solution}
\end{equation}
Thus, with (\ref{V0}) one has 
\begin{equation}
\frac{\omega}{\kappa_b}=-(n+\frac{1}{2})i+
\sqrt{l(l+1)-\frac{1}{4}}\,,n=0,1,...\,.
\label{finalsclarelectr}
\end{equation}
for scalar and electromagnetic perturbations.
And
\begin{equation}
\frac{\omega}{\kappa_b}=-(n+\frac{1}{2})i+\sqrt{(l+2)(l-1)-
\frac{1}{4}}\,,n=0,1,...\,.
\label{finalgrav}
\end{equation}
for gravitational perturbations.
Our analysis shows that
Eqs. (\ref{finalsclarelectr})-(\ref{finalgrav}) are correct up to
terms of order $O(r_c-r_b)$ or higher.  Moss and Norman \cite{moss} have
studied the QN frequencies in the SdS geometry numerically
and also analytically, by fitting the potential to a P\"oshl-Teller
potential. Their analytical results (see their Figs 1 and 2) were in
excellent agreement with their numerical results, and this agreement
was even more remarkable for near extremal black holes and for high
values of the angular quantum number $l$.  We can now understand why:
for near extremal black holes the true potential is indeed given by
the P\"oshl-Teller potential!  Furthermore for near extremal SdS black
holes and for high $l$ our formula (\ref{finalgrav}) is approximately
equal to formula (19) of Moss and Norman \cite{moss}. With their
analytical method of fitting the potential one can never be sure if
the results obtained will continue to be good as one increases the
mode number $n$. But we have now proved that if one is in the near
extremal SdS black hole, the P\"oshl-Teller is the true potential, and
so Eq. (\ref{finalsclarelectr})-(\ref{finalgrav}) is exact.  For
example, Moss and Norman obtain numerically, and for gravitational
perturbations with $l=2$ of nearly extreme SdS black holes, the result
\begin{equation}
\frac{\omega_{\rm num}}{\kappa_b}=1.93648-i(n+\frac{1}{2})\,,
\label{mossnumel2}
\end{equation}
and we obtain, from (\ref{finalgrav})
\begin{equation}
\frac{\omega}{\kappa_b}=1.936492-i(n+\frac{1}{2})\,.
\label{meuexactl2}
\end{equation}
For $l=3$ Moss and Norman \cite{moss} obtain
\begin{equation}
\frac{\omega_{\rm num}}{\kappa_b}=3.12249-i(n+\frac{1}{2})\,,
\label{mossnumel3}
\end{equation}
and we obtain, from (\ref{finalgrav})
\begin{equation}
\frac{\omega}{\kappa_b}=3.122499-i(n+\frac{1}{2})\,.
\label{meuexactl3}
\end{equation}
So this remarkable agreement allows us to be sure that 
(\ref{finalsclarelectr})-(\ref{finalgrav}) are indeed correct.
\section{Conclusions}
We have found for the first time an analytical expression for the
QNMs and frequencies of a nearly extreme schwarzschild-de
Sitter black hole. This expression,
Eqs. (\ref{finalsclarelectr})-(\ref{finalgrav}) are correct up to
terms of order $O(r_c-r_b)$ or higher for all $n$.  We should stress
however that our results may not be correct to all orders in
$(r_c-r_b)$ (although they are indeed correct for terms up to second order
\cite{brinkds}), and as such it is not guaranteed that our results give
the correct behaviour of the highly damped QNMs.  As a
matter of fact it would be strange indeed if they described correctly
highly damped modes. In that case there would be an $l$-dependent
quantization scheme, which does not fit easily into the existing
framework.
Some recent work \cite{shijunds} seems indeed to confirm what we said:
for the lowest lying modes our expressions (\ref{finalsclarelectr})-(\ref{finalgrav})
are correct. However, the highly damped modes show a different behaviour.

\thispagestyle{empty} \setcounter{minitocdepth}{1}
\chapter[Quasinormal modes of Schwarzschild black holes in four and 
higher dimensions ]{Quasinormal modes of Schwarzschild black holes in four and higher dimensions} \label{chap:qnms}
\lhead[]{\fancyplain{}{\bfseries Chapter \thechapter. \leftmark}}
\rhead[\fancyplain{}{\bfseries \rightmark}]{}
\minitoc \thispagestyle{empty}
\renewcommand{\thepage}{\arabic{page}}
\section{Introduction}
The study of quasinormal modes of black holes began more than thirty
years ago, when Vishveshwara \cite{vish} noticed that the signal from
a perturbed black hole is, for most of the time, an exponentially
decaying ringing signal. It turns out that the ringing frequency and
damping timescale are characteristic of the black hole, depending only
on its parameters (like the mass and angular momentum).  We call these
characteristic oscillations the quasinormal modes (QNMs) and the
associated frequencies are termed quasinormal frequencies (QN
frequencies), because they are really not stationary perturbations.  
Not surprisingly, QNMs play an important role in the
dynamics of black holes, and consequently in gravitational wave
physics. In fact, it is possible \cite{echeverria,nakanoringing} to
extract the parameters of the black hole simply by observing these
QN frequencies, using for example gravitational wave
detectors.  The discovery that QNMs dominate the answer of a black
hole to almost any exterior perturbation was followed by a great
effort to find, numerically and analytically, the QN frequencies.
For excellent reviews on the status of QNMs, prior to 2000,
we refer the reader to Kokkotas and Schmidt
\cite{kokkotas} and Nollert \cite{nollert}.
It is important to note that on the astrophysical aspect, the most
important QN frequencies are the lowest ones, i.e., frequencies 
with smaller imaginary part, and the most important spacetimes are the 
asymptotically flat
and perhaps now the asymptotically de Sitter.  
However, three years ago Horowitz and Hubeny \cite{horowitz}
pointed out that QNMs of black holes in anti-de Sitter space have a
different importance.  According to the AdS/CFT correspondence
\cite{maldacena}, a black hole in anti-de Sitter space may be viewed
as a thermal state in the dual theory. Perturbing this black hole
corresponds to perturbing the thermal state, and therefore the typical
timescale of approach to thermal equilibrium (which is hard to compute
directly in the dual theory) should as well be governed by the  lowest
QN frequencies.  That this is indeed the case was proved by Birmigham,
Sachs and Solodukhin \cite{birmingham2} for the BTZ black hole,
taking advantage that this is one of the few spacetimes where one can
compute exactly its QN frequencies, as showed by Cardoso and Lemos
\cite{cardosoqnmbtz}.  A similar study was made by Kurita and Sakagami
\cite{kurita} for the D-3 brane.
This interpretation for the imaginary part of the QN frequencies 
in terms of timescales of approach
to thermal equilibrium in the dual conformal field theory has
motivated a generalized search for the quasinormal modes of different
black holes in anti-de Sitter spacetime, over the last three years 
\cite{horowitz,cardosoqnmbtz,qnmads}.

Recently, the motivation for studying QN modes of black holes has grown
enormously with the conjectures \cite{hod,dreyer,dreyerall,ling}
relating the highly damped QNMs (i.e., QN frequencies with large
imaginary parts) to black hole area quantization and to the Barbero
Immirzi parameter appearing in Loop Quantum Gravity. The seeds of that idea
were planted some
time ago by Bekenstein \cite{bek0}. A semi-classical reasoning of
the conjecture \cite{bek2} that the black hole area spectrum
is quantized leads to
\begin{equation}
A_{\,n}=\gamma\, l_{P}^2\, n\,,\quad n=1,2,...\quad.
\label{areaspectrum}
\end{equation}
Here $l_{P}$ is the Planck length and $\gamma$ is an undetermined constant.
However, statistical physics arguments impose a constraint on $\gamma$ 
\cite{bek2}:
\begin{equation}
\gamma=4 \log k\,,
\label{areaspectrum1}
\end{equation}
where $k$ is an integer.
The integer $k$ was left undetermined (although there were some suggestions
for it \cite{bek2}), until Hod \cite{hod}, supported
by some of Bekenstein's ideas, put forward the proposal to determine
$k$ via a version of Bohr's correspondence principle, in which one admits
that the real part of QN frequencies with a large imaginary
part plays a fundamental role. 
It was seen numerically by Nollert \cite{nollert,nollert2}
that QN frequencies with a large imaginary part behave,
in the Schwarzschild geometry as
\begin{equation}
\omega M= 0.0437123+\frac{i}{8}\,(2n+1)\,,
\label{asymptQNnollert}
\end{equation}
where $M$ is the black hole mass, and $n$ the mode number.
Hod first realized that $ 0.0437123 \sim \frac{\ln3}{8\pi}$, and 
went on to say that, if one supposes that the emission of a quantum
with frequency $\frac{\ln3}{8\pi}$ corresponds to the least possible energy 
a black hole can emit, then the change in surface area will be
(using $A=16\pi M^2$)
\begin{equation}
\Delta A= 32\pi MdM=32\pi M{\hbar} \omega= 4{\hbar}\ln 3 \,.
\label{surfacearea}
\end{equation}
Comparing with (\ref{areaspectrum}) we then get $k=3$ and therefore the area
spectrum is fixed to
\begin{equation}
A_{\rm n}=4\ln 3 l_{P}^2 n\,;\,\,n=1,2,...
\label{areaspectrumfinal}
\end{equation}
It was certainly a daring proposal to map $0.0437123$ to $\ln
3$, and even more to use the QN frequencies to quantize the black hole
area, by appealing to ``Bohr's correspondence principle''. The
risk paid off: recently, Dreyer \cite{dreyer} put forward the
hypothesis that if one uses such a correspondence it is possible to
fix a formerly free parameter, the Barbero-Immirzi parameter, appearing
in Loop Quantum Gravity.  
This fixing of the Barbero-Immirzi parameter was made at the expense 
of changing the gauge group from $SU(2)$, which is the most natural one, to
$SO(3)$. 
It may be possible however to keep $SU(2)$ as gauge group, as Ling and Zhang
\cite{ling} recently remarked, by considering the supersymmetric
extension of Loop Quantum Gravity.

All of these proposals and conjectures could be useless
if one could not prove that the real part of the QN frequencies
do approach $\frac{\ln3}{8\pi}$, i.e., that the number $0.0437123$ in
Nollert's paper was exactly $\frac{\ln3}{8\pi}$. This
was accomplished by Motl \cite{motl1} some months ago, using an
ingenious technique, working with the continued fraction
representation of the wave equation.  Subsequently, Motl and Neitzke
\cite{motl2} used a more flexible and powerful approach, called 
the monodromy method, and were not only able to rederive the four
dimensional Schwarzschild value $\frac{\ln3}{8\pi}$, but also to
compute the asymptotic value for generic $D$-dimensional Schwarzschild
black holes thereby confirming a previous conjecture by Kunstatter
\cite{dreyerall}.
They have also predicted the form of the asymptotic QN frequencies
for the Reissner-Nordstr\"om geometry, a prediction which was verified
by Berti and Kokkotas \cite{bertikerr}.

It is interesting to speculate why, in thirty years investigating QNMs
and QN frequencies, no one has ever been able to deduce analytically
the asymptotic value of the QN frequencies, nor even for the
Schwarzschild geometry, which is the simplest case.  In our opinion,
there are three important facts that may help explain this: first,
the concern was mostly with the lowest QN frequencies, the ones
with smaller imaginary part, since they are the most important in the
astrophysical context where QNMs were inserted until three years
ago. In fact, most of the techniques to find QN frequencies, and there
were many, have been devised to compute the lowest QN frequencies (see
\cite{kokkotas,nollert} for a review).  Secondly, there was no 
suggestion for the asymptotic $\ln{3}$ value. 
To know this value serves as
a strong stimulation. The third reason is obvious: this was a very difficult
technical problem.

It is thus satisfying that Motl and Neitzke have proved exactly
that the limit is $\ln{3}$. In addition, Van den Brink \cite{brinkasympt} 
used a
somewhat different approach to rederive the $\frac{\ln3}{8\pi}$
value, and at the same time computed the first order correction to
gravitational QN frequencies, confirming analytically Nollert's
\cite{nollert2} numerical results.  However,
Motl and Neitzke's monodromy method is more flexible and powerful: it
has been easily adapted to other spacetimes.  For example, Castello-Branco 
and Abdalla \cite{karlucio} have generalized the results to the
Schwarzschild-de Sitter geometry, while Schiappa, Nat\'ario and
Cardoso \cite{ricardo} have used it to compute the asymptotic values
in the $D$-dimensional Reissner-Nordstr\"om geometry.
Moreover the first order corrections are also easily obtained, as has
been showed by Musiri and Siopsis \cite{musiri2} for the four
dimensional Schwarzschild geometry. Here we shall also generalize
these first order corrections so that they can encompass a general
$D$-dimensional Schwarzschild black hole.
A body of work has been growing on the subject of computing highly
damped QNMs, both in asymptotically flat and in de Sitter or anti-de
Sitter spacetimes (see the references in \cite{cardosoqnms}.

As an aside, we note that one of the most intriguing
features of Motl and Neitzke's technique is that the region beyond the
event horizon, which never enters in the definition of quasinormal
modes, plays an extremely important role.  In particular, the
singularity at $r=0$ is decisive for the computation.  At the
singularity the effective potential for wave propagation blows
up. Somehow, the equation, or the singularity, knows what we are
seeking! It is also worth of note the following: for high frequencies, or
at least frequencies with a large imaginary part, the important region
is therefore $r=0$ where the potential blows up, whereas for low
frequencies, of interest for late-time tails \cite{cardosotails}, it is the
other limit, $r \rightarrow \infty$ which is important.

The purpose of this work is two-fold: first we want to enlarge and settle
the results for the four dimensional Schwarzschild black hole.  We shall
first numerically confirm Nollert's results for the gravitational
case, and Berti and Kokkotas'\cite{bertikerr} results for the scalar
case, and see that the match between these and the analytical
prediction is quite good. Then we shall fill-in a gap left behind in
these two studies: the electromagnetic asymptotic QN frequencies, for
which there are no numerical results, but a lot of analytical
predictions.  This will put the monodromy method on a firmer
ground.  One might say that in the present context (of black hole
quantization and relation with Loop Quantum Gravity) electromagnetic
perturbations are not relevant, only the gravitational ones are. This
is not true though, because when considering the Reissner-Nordstr\"om case
gravitational and electromagnetic perturbations are coupled
\cite{chandra} and it is therefore of great theoretical interest to
study each one separately, in the case they do decouple, i.e., the
Schwarzschild geometry.
We find that all the analytical predictions are correct: scalar and
gravitational QN frequencies asymptote to 
$\frac{\omega}{T_H}=\ln{3}+i(2n+1)\pi$ 
with $\frac{1}{\sqrt n}$ corrections, where 
$T_H=\frac{1}{8\pi M}$ is the Hawking temperature.  The
corrections are well described by the existing analytical formulas.
On the other hand, electromagnetic QN frequencies asymptote to
$\frac{\omega}{T_H}=0+i2n\pi$ with no $\frac{1}{\sqrt n}$ corrections, as
predicted by Musiri and Siopsis \cite{musiri2}.  Indeed we find that
the corrections seem to appear only at the $\frac{1}{n^{3/2}}$ level.

The second aim of this paper is to establish numerically the validity of the
monodromy method by extending the numerical calculation to higher
dimensional Schwarzschild black holes, in particular to the 
five dimensional one.  We obtain numerically in the five dimensional case
the asymptotic
value of the QN frequencies, which turns out to be $\ln 3+i(2n+1)\pi$ (in
units of the black hole temperature), for scalar, gravitational tensor
and gravitational vector perturbations.  Equally important, is that
the first order corrections appear as $\frac{1}{n^{2/3}}$ for these
cases.  We shall generalize Musiri and Siopsis' method to higher
dimensions and find that (i) the agreement with the numerical data is
very good (ii) In generic $D$ dimensions the corrections appear at the
$\frac{1}{n^{(D-3)/(D-2)}}$ level.  Electromagnetic perturbations in
five dimensions seem to be special: according to the analytical
prediction by Motl and Neitzke, they should not have any asymptotic QN
frequencies. Our numerics seem not to go higher enough in mode number
to prove or disprove this. They indicate that the real part approaches
zero, as the mode number increases, but we have not been able to
follow the QN frequencies for very high overtone number.  We also give
complete tables for the lowest lying QN frequencies of the five
dimensional black holes, and confirm previous results
\cite{vitoroscarjose,konoplyawkb,roman2,ida} for the fundamental QN
frequencies obtained using WKB-like methods.  We find no purely
imaginary QN frequencies for the gravitational QNMs, which may be
related to the fact that the potentials are no longer superpartners.
This chapter will follow \cite{cardosoqnms}.

\section{QN frequencies of the four-dimensional Schwarzschild black hole }

The first computation of QN frequencies of four-dimensional
Schwarzschild black holes was carried out by Chandrasekhar and
Detweiler \cite{chandradet}.  They were only able to compute the
lowest lying modes, as it was and still is quite hard to find
numerically QN frequencies with a very large imaginary part.  After
this pioneering work, a number of analytical, semi-analytical and
numerical techniques were devised with the purpose of computing higher
order QNMs and to establish the low-lying ones using different methods
\cite{kokkotas,nollert}. The most
successful numerical method to study QNMs of black holes was proposed
by Leaver \cite{L}. In a few words his method amounts to reducing the
problem to a continued fraction, rather easy to implement
numerically. Using this method Leaver was able to determine the first
60 modes of the Schwarzschild black hole, which at that time was about
an order of magnitude better than anyone had ever gone before.  The
values of the QN frequencies he determined, and we note he only
computed gravitational ones, suggested that there is an infinite
number of QN modes, that the real part of the QN frequencies approach
a finite value, while the imaginary part grows without bound.  
That the real part of the QN frequencies approach
a finite value was however opened to debate \cite{guinn}: it seemed one still 
had to
go higher in mode number in order to ascertain the true asymptotic
behaviour.  Nollert \cite{nollert2} showed a way out
of these problems: he was able to improve Leaver's method in a very
simple fashion so as to go very much higher in mode number: he was
able to compute more than 2000 modes of the gravitational QNMs.  His
results were clear: the real part of the gravitational QN frequencies
approaches a constant value $\sim 0.0437123/M $, and this value is
independent of $l$. The imaginary part grows without bound as
$(2n+1)/8$ where $n$ is any large integer.  These asymptotic
behaviours have both leading corrections, which were also determined
by Nollert. He found that to leading order the asymptotic behaviour
of the gravitational QN frequencies of a Schwarzschild black hole is
given by Eq. (\ref{asymptQNnollert}) plus a leading correction, 
namely,
\begin{equation}
\omega M=0.0437123+\frac{i}{8}\,(2n+1)+ \frac{a}{\sqrt{n}}\,,
\label{nono}
\end{equation}
where the correction term $a$ depends on $l$.  About ten years
later, Motl \cite{motl1} proved analytically that this is indeed the
correct asymptotic behaviour. To be precise he proved that
\begin{equation}
\frac{\omega}{T_H}=\ln{3}+i\,(2n+1)\pi+\,{\rm corrections}\,.
\label{nonolubos}
\end{equation}
Since the Hawking temperature $T_H$ is 
$T_H=\frac{1}{8\pi M}$, Nollert's
result follows.
He also proved that this same result also holds for scalar QN
frequencies, whereas electromagnetic QN frequencies asymptote to a
zero real part.  Motl and Neitzke \cite{motl2} using a completely
different approach, have rederived this result whereas Van den Brink
\cite{brinkasympt} was able to find the leading asymptotic correction
term $a$ in (\ref{nono}) for the gravitational case.  Very
recently, Musiri and Siopsis \cite{musiri2}, leaning on Motl and
Neitzke's technique, have found the leading correction term for any
field, scalar, electromagnetic and gravitational.
This correction term is presented in section \ref{apendiceqnms}, where
we also generalize to higher dimensional black holes. 
In particular, for electromagnetic perturbations they find that the 
first order
corrections are zero.

Here we use Nollert's method to compute scalar, electromagnetic and
gravitational QN frequencies of the four-dimensional Schwarzschild
black hole.  The details on the implementation of this method in four
dimensions are well known and we shall not dwell on it here any more.
We refer the reader to the original references \cite{nollert2,L}.
\begin{figure}
\centerline{\includegraphics[width=8 cm,height=8 cm]
{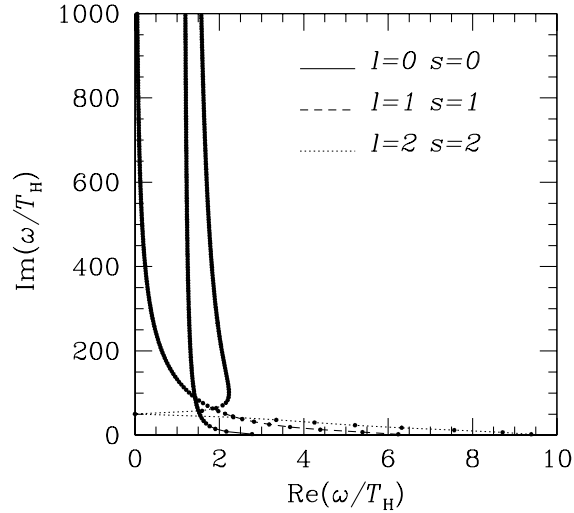}}
\caption{Scalar ($s=0$), electromagnetic ($s=1$) and gravitational ($s=2$)
QN frequencies of a four-dimensional Schwarzschild black hole. 
Note that the dots indicate the QN frequencies there, and the lines
connecting the dots only help to figure out
to which multipole ($l$, $s$) they belong to. 
We show
the lowest 500 modes for the lowest radiatable multipoles of each
field, i.e., $l=0$ for the scalar field, $l=1$ for the electromagnetic
and $l=2$ for gravitational.  However, the asymptotic behaviour is
$l$-independent. It is quite clear from this plot that highly damped
electromagnetic QN frequencies have a completely different behaviour
from that of scalar or gravitational highly damped QN frequencies. In
fact, the electromagnetic ones asymptote to $\frac{\omega}{T_H} \rightarrow
i(2n+1)\pi\,\,,n\rightarrow \infty$, as predicted by Motl
\cite{motl1} and Motl and Neitzke \cite{motl2}. The scalar and
gravitational ones asymptote to $\frac{\omega}{T_H}\rightarrow
\ln{3}+i\pi(2n+1)\,\,,n\rightarrow \infty$.}
\label{fig:qnall4d}
\end{figure}

\begin{figure}
\centerline{\includegraphics[width=8 cm,height=8 cm]
{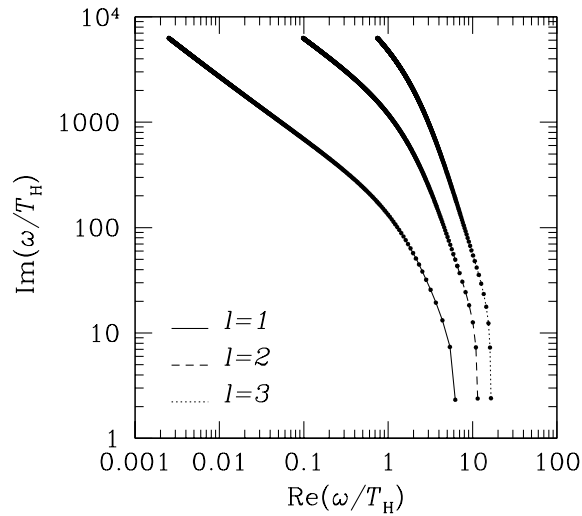}}
\caption{Electromagnetic QN frequencies of  a 
four-dimensional Schwarzschild black hole.}
\label{fig:qnelec4d}
\end{figure}

Our results are summarized in Figures
\ref{fig:qnall4d}-\ref{fig:qnelec4d}, and Tables \ref{tab:corrd4j0}-
\ref{tab:corrd4j2}.  
In Fig. \ref{fig:qnall4d} we
show the behaviour of the 500 lowest QN frequencies for the lowest
radiatable multipole of each field 
(we actually have computed the first 5000 QN frequencies, but for 
a better visualization we do not show them all): $l=0$ for scalar, $l=1$ for
electromagnetic and $l=2$ gravitational.  
Note that in Fig. \ref{fig:qnall4d}, as well as in all other plots in this 
work, 
the dots indicate the QN frequencies there, and the lines
connecting the dots only help to figure out
to which multipole ($l$, $s$) they belong to. 
In what concerns the asymptotic behaviour, our numerics show it is not
dependent on $l$, and that formula (\ref{nono}) holds. The results for
the scalar and gravitational QN frequencies are not new: the
gravitational ones have been obtained by Nollert \cite{nollert2}, as
we previously remarked, and the scalar ones have been recently arrived
at by Berti and Kokkotas \cite{bertikerr}.  Here we have verified both.
Both scalar and gravitational QN frequencies have a real part
asymptoting to $\ln{3}\times T_H$.  Electromagnetic QN frequencies,
on the other hand, have a real part asymptoting to $0$. This is
clearly seen in Fig. \ref{fig:qnall4d}.  For gravitational
perturbations Nollert found, and we confirm, that the correction term
$a$ in (\ref{nono}) is
\begin{eqnarray}
a= 0.4850\,,\,\,l=2
\nonumber \\
\!\!\!a= 1.067 \,,\,\,l=3
\label{corrtermsnum}
\end{eqnarray}
We have also computed the correction terms for the scalar case, and we have 
obtained
results in agreement with Berti and Kokkotas' ones \cite{bertikerr} .
Our results are summarized in Tables \ref{tab:corrd4j0}-\ref{tab:corrd4j2}, 
where
we also show the analytical prediction \cite{musiri2} (see section 
\ref{apendiceqnms}).
To fix the conventions adopted throughout the rest of the paper,
we write
\begin{equation}
\frac{\omega}{T_H}=\ln{3}+i(2n+1)\pi+\frac{{\rm Corr}}{\sqrt{n}}\,.
\label{conv}
\end{equation}
Thus, the correction term $a$ in (\ref{nono}) is related to {\rm Corr}
by $a={\rm Corr}\times T_H$. 

\vskip 1mm
\begin{table}
\caption{\label{tab:corrd4j0} The correction coefficients for the
four-dimensional Schwarzschild black hole, both numerical, here
labeled as ``${\rm Corr_{4}^{N}}$'' and analytical, labeled as ``${\rm
Corr_{4}^{A}}$''. These results refer to scalar
perturbations.
The analytical results are extracted from \cite{musiri2}
(see also formula (\ref{corrd4} below).  Notice the very good agreement
between the numerically extracted results and the
analytical prediction.These values were obtained using the first
5000 modes.}
\begin{tabular}{l|ll}  \hline
\multicolumn{1}{c}{} &
\multicolumn{2}{c}{ $s=0$}\\ \hline
$l$ &${\rm Corr_4^N}$:&${\rm Corr_4^A}$\\ \hline
0   &1.20-1.20i  &1.2190-1.2190i  \\ \hline 
1   &8.52- 8.52i  &8.5332-8.5332i   \\ \hline 
2   &23.15-23.15i  &23.1614-23.1614i   \\ \hline 
3   &44.99-44.99i  &45.1039-45.1039i \\ \hline 
4   &74.34-74.34i &74.3604-74.3605i  \\ \hline 
\end{tabular}
\end{table}
\vskip 1mm
\vskip 1mm
\begin{table}
\caption{\label{tab:corrd4j2} The correction coefficients for the
four-dimensional Schwarzschild black hole, both numerical, here
labeled as ``${\rm Corr_{4}^{N}}$'' and analytical, labeled as ``${\rm
Corr_{4}^{A}}$''. These results refer to gravitational
perturbations.
The analytical results are extracted from \cite{musiri2}
(see also formula (\ref{corrd4} below). Notice the very good agreement
between the numerically extracted results and the
analytical prediction. These values were obtained using the first
5000 modes.}
\begin{tabular}{l|ll}  \hline
\multicolumn{1}{c}{} &
\multicolumn{2}{c}{ $s=2$}\\ \hline
$l$ &${\rm Corr_4^N}$:&${\rm Corr_4^A}$\\ \hline
2   & 6.08-6.08i   & 6.0951-6.0951i  \\ \hline 
3   & 13.39-13.39i & 13.4092-13.4092i \\ \hline 
4   & 23.14-23.14i & 23.1614-23.1614i \\ \hline 
5   & 35.33-35.33i & 35.3517-35.3517i\\ \hline 
6   & 48.90-48.90i & 49.9799-49.9799i \\ \hline 
\end{tabular}
\end{table}
\vskip 1mm
As for the electromagnetic correction terms, we found it was not very easy 
to determine
them. In fact, it is hard, even using Nollert's method, to go very high in mode
number for these perturbations (we have made it to $n=5000$).
The reason is tied to the fact that these frequencies asymptote to zero,
and it is hard to determine QN frequencies with a vanishingly small real part.
Nevertheless, there are some features one can be sure of.
We have fitted the electromagnetic data to the following form
$\frac{\omega}{T_H}=\ln{3}+(2n+1)i+\frac{b}{\sqrt{n}}\,,$
and found that this gave very poor results. Thus one saw numerically that
the first order correction term is absent, as predicted in \cite{musiri2}
(see also section \ref{apendiceqnms} where we rederive these corrections, 
generalizing them
to arbitrary dimension).
Our data seems to indicate that the leading correction term is of the form
$\frac{b}{n^{3/2}}$. This is more clearly seen in Fig. \ref{fig:qnelec4d}
where we plot the first electromagnetic QN frequencies on a $\ln$ plot.
For frequencies with a large imaginary part, the slope is about $-2/3$ as
it should if the asymptotic behaviour is of the form
\begin{equation}
\frac{\omega}{T_H}=\ln{3}+(2n+1)i+\frac{b}{n^{3/2}}\,.
\label{true}
\end{equation}
Our numerics indicate that $b$ has an $l$-dependence going like
\begin{equation}
b \sim c_1 l(l+1)+c_2 \left[l(l+1)\right]^2+c_3 \left[l(l+1)\right]^3\,,
\label{bcoef}
\end{equation}
where $c_1$, $c_2$, and$c_3$ are constants. This is also the expected
behaviour, since a $\frac{1}{n^{3/2}}$ dependence means we have to go
to third order perturbation theory, where we get corrections of the
form (\ref{bcoef}) as one can easily convince oneself.
\section{QN frequencies of $D$-dimensional Schwarzschild black holes }

\subsection{Equations and conventions}
The metric of the $D$-dimensional
Schwarzschild black hole in ($t,r,\theta_1,\theta_2,..,\theta_{D-2}$)
coordinates is \cite{tangherlini,myersperry}
\begin{equation}
ds^2= -fdt^2+
f^{-1}dr^2
+r^2d\Omega_{D-2}^2\,,
\label{metrictangqnms} 
\end{equation}
with
\begin{equation}
f=1-\frac{m}{r^{D-3}}.
\label{fdefqnms}
\end{equation}
The mass of the black hole is given by
$M=\frac{(D-2)\Omega_{D-2} m}{16\pi{\cal G}}$, where
$\Omega_{D-2}=\frac{2\pi^{(D-1)/2}}{\Gamma[(D-1)/2]}$ is the area of a
unit $(D-2)$ sphere, $d\Omega_{D-2}^{2}$ is the line element on
the unit sphere $S^{D-2}$, and ${\cal G}$ is Newton's constant in 
$D$-dimensions.  
In the following it will prove more useful to rescale variables so that
the form of the metric is (\ref{metrictangqnms}) but with $f=1-\frac{1}{r^{D-3}}$,
i.e., we shall choose $m=1$, following the current fashion.
We will only consider the linearized
approximation, which means that we are considering wave fields outside
this geometry that are so weak they do not alter this
background. 
The evolution equation for a massless scalar field follows
directly from the Klein-Gordon equation (see
\cite{cardosoradadsddim} for details).  The gravitational evolution
equations have recently been derived by Kodama and Ishibashi
\cite{kodama}. There are three kinds of gravitational perturbations,
according to Kodama and Ishibashi's terminology: the scalar
gravitational, the vector gravitational and the tensor gravitational
perturbations.  The first two already have their counterparts in
$D=4$, which were first derived by Regge and Wheeler \cite{regge} and
by Zerilli \cite{zerilli}. The tensor type is a new kind appearing in
higher dimensions. However, it obeys exactly the same equation as
massless scalar fields, as can easily be seen.  Due to the complex
form of the gravitational scalar potential, we shall not deal with
it. Instead, we shall only consider the tensor and vector type of
gravitational perturbations. In any case, if the analytic results are correct,
then the gravitational scalar QN frequencies should have the same asymptotic
form as the gravitational vector and tensor QN frequencies.  
The evolution equation for the electromagnetic field in the higher
dimensional Schwarzschild geometry was arrived at for the first time
by Crispino, Higuchi and Matsas \cite{matsas}. It has recently
been rederived by Kodama and Ishibashi, in a wider context of charged
black hole perturbations.  We shall follow Kodama and Ishibashi's
terminology. According to them, there are two kinds of electromagnetic
perturbations: the vector and scalar type. If one makes the charge of
the black hole $Q=0$ in Kodama and Ishibashi's equations one recovers
the equations by Crispino, Higuchi and Matsas, although this seems to
have been overlooked in the literature. 
The evolution equation for all kinds of
fields (scalar, gravitational and electromagnetic) 
can be reduced to the second order differential equation
\begin{equation}
\frac{d^2\Psi}{dr_{*}^2}+(\omega^2-V)\Psi=0 \,,
\label{eveqqnms}
\end{equation}
where $r$ is a function of the tortoise coordinate $r_*$, defined through 
$\frac{\partial r}{\partial
r_*}=f(r)$, and the potential can be written in compact form as
\begin{eqnarray}
V &=& f(r)
{\biggl [}
\frac{l(l+D-3)}{r^2}+\frac{(D-2)(D-4)}{4r^2}+
\nonumber \\
& &
\frac{(1-j^2)(D-2)^2}{4r^{D-1}}
{\biggr ]}
\,.
\label{potentialj}
\end{eqnarray}
The constant $j$ depends on what kind of field one is studying:
\begin{equation}
j=\left\{ \begin{array}{llll}
            0\,,   & {\rm \,scalar\, and\, gravitational\, tensor}\\
                                  &  {\rm perturbations}. \\ 
            1\,,   &{\rm \, gravitational\, vector} \\
                                     &  {\rm perturbations}. \\
            \frac{2}{D-2}\,,   &{\rm \, electromagnetic \,vector} \\
                                     &  {\rm perturbations}. \\
            2- \frac{2}{D-2}\,,   &{\rm \, electromagnetic\, scalar} \\
                                     &  {\rm perturbations}\,.
\end{array}\right.
\label{jdef}
\end{equation}
Notice that in four dimensions, $j$ reduces to the usual values
given by Motl and Neitzke \cite{motl2}. 
According to our conventions, the Hawking temperature of a
$D$-dimensional Schwarzschild black hole is
\begin{equation}
T_H=\frac{D-3}{4\pi}
\label{hawktemp}
\end{equation}

Our purpose here is to investigate the QN frequencies of higher
dimensional black holes.  Of course one cannot study every $D$. We
shall therefore focus on one particular dimension, five, and make a
complete analysis of its QN frequencies. The results should be
representative. In particular, one feature that distinguishes 
an arbitrary $D$ from the four dimensional case is that the correction
terms come with a different power, i.e., whereas in four dimensions
the correction is of the form $\frac{1}{\sqrt{n}}$ we shall find that in 
generic 
$D$ it is of the form $\frac{1}{n^{(D-3)/(D-2)}}$. Thus, if one verifies 
this for
$D=5$ for example, one can ascertain it will hold for arbitrary $D$.
We shall now briefly sketch our numerical procedure for finding
QN frequencies of a five dimensional black hole. This is just Leaver's and
Nollert's technique with minor modifications.

The QNMs of the higher-dimensional Schwarzschild black hole are characterized 
by the boundary conditions of incoming waves at the black hole horizon and 
outgoing waves at spatial infinity, written as
\begin{equation}
\Psi(r)\rightarrow\left\{
\begin{array}{ll}
e^{-i\omega r_*}&{\rm as}\ r_*\rightarrow \infty\\
e^{ i\omega r_*}&{\rm as}\ r_*\rightarrow-\infty\,, 
\end{array}
\right.
\label{def-bc}
\end{equation}
where the time dependence of perturbations has been assumed to be 
$e^{i\omega t}$.
\subsection{Perturbative calculation of QNMs of $D$-dimensional 
Schwarzschild BHs}
\label{apendiceqnms}
In this section we shall outline the procedure for computing the
first order corrections to the asymptotic value of the QN frequencies
of a $D$-dimensional Schwarzschild black hole.  This will be a 
generalization of Musiri and Siopsis' method \cite{musiri2}, so we
adopt all of their notation, and we refer the reader to their paper
for further details.
The computations are however rather tedious and the final expressions are too
cumbersome, so we shall refrain from giving explicit
expressions for the final result.

We start with the expansion of the potential $V$ near the singularity
$r=0$. One can easily show that near this point the potential 
(\ref{potentialj}) 
may be approximated as
\begin{equation}
V \sim -\frac{\omega ^2}{4z^2}\left[1-j^2+\frac{A}{(-z 
\omega)^{(D-3)/(D-2)}} \right]\,,
\label{potr0}
\end{equation}
where we adopted the conventions in \cite{musiri2} and therefore 
$z=\omega r_*$.
The constant $A$ is given by
\begin{eqnarray}
\!\!\!\!A\!&=&\frac{4l(l+D-3)+(D-2)(D-4)-(D-2)^2(1-j^2)}{(D-2)^{(D-1)/(D-2)}}+
\nonumber \\
& &
\frac{2(1-j^2)(D-2)^{(2D-5)/(D-2)}}{2D-5} \,.
\label{constA}
\end{eqnarray}
In $D=4$ this reduces to the usual expression
\cite{musiri2,neitzke} for the potential near $r=0$.  Expression
(\ref{potr0}) is a formal expansion in
$\frac{1}{\omega^{(D-3)/(D-2)}}$, so we may anticipate that indeed the
first order corrections will appear in the form
$\frac{1}{n^{(D-3)/(D-2)}}$.  So now we may proceed in a direct
manner:
we expand the wavefunction to first order in $\frac{1}{\omega^{(D-3)/(D-2)}}$ 
as
\begin{equation}
\Psi= \Psi^{(0)}+\frac{1}{\omega^{(D-3)/(D-2)}}\Psi^{(1)}\,,
\label{expansionpsi}
\end{equation}
and find that the first-order correction obeys the equation
\begin{equation}
\frac{d^2\Psi^{(1)}}{dz^2}+(\frac{1-j^2}{4z^2}+1)\Psi^{(1)}=
\omega^{(D-3)/(D-2)}\delta V \Psi^{(0)}\,,
\label{firstordereq}
\end{equation}
with 
\begin{equation}
\delta V=-\frac{A}{\omega^{(D-3)/(D-2)}(-z)^{(D-1)/(D-2)}}\,.
\label{deltaV}
\end{equation}
All of Musiri and Siopsis' expressions follow directly to the $D$-dimensional
case, if one makes the replacement $\sqrt{-\omega r_0} \rightarrow 
\omega^{(D-3)/(D-2)}$
in all their expressions. 
For example, the generalsolution of (\ref{firstordereq}) is
\begin{equation}
\Psi^{(1)}_{\pm}={\cal C} \Psi^{(0)}_{+} \int_{0}^z \Psi^{(0)}_{-}\delta V 
\Psi^{(0)}_{\pm}-
{\cal C} \Psi^{(0)}_{-} \int_{0}^z \Psi^{(0)}_{+}\delta V \Psi^{(0)}_{\pm}
\,,
\label{solfirstorder}
\end{equation}
where
 \begin{equation}
{\cal C}=\frac{\omega^{(D-3)/(D-2)}}{\sin{j\pi/2}}\,,
\label{calC}
\end{equation}
and the wavefunctions $\Psi^{(0)}_{\pm}$ are 
\begin{equation}
\Psi^{(0)}_{\pm}=\sqrt{\frac{\pi z}{2}}J_{\pm j/2}(z)\,.
\label{psi0}
\end{equation}
The only minor modification is their formula (30).
In $D$-dimensions, it follows that
\begin{equation}
\Psi^{(1)}_{\pm}=z^{1 \pm j/2 +k} G_{\pm}(z)\,,
\label{psi1}
\end{equation} 
where $k=\frac{D-4}{2(D-2)}$, and $G_{\pm}$ are even analytic functions of $z$.
So we have all the ingredients to construct the first order corrections in the 
$D$-dimensional Schwarzschild geometry. Unfortunately the final expressions 
turn out to be quite cumbersome, and we have not managed to simplify them.
We have worked with the symbolic manipulator {\it Mathematica}. It is 
possible to 
obtain simple expression for any particular $D$, but apparently not for a 
generic
$D$. For generic $D$ we write 
\begin{equation}
\frac{\omega}{T_H}=\ln{(1+2\cos{\pi j})}+i(2n+1)\pi+
\frac{{\rm Corr_D}}{\omega^{(D-3)/(D-2)}}
\label{xxx}
\end{equation}
So the leading term is 
$\frac{\omega}{T_H}=\ln{3}+i(2n+1)\pi$, for scalar, gravitational tensor and 
gravitational tensor perturbations (and the same holds for gravitational 
scalar).
However, for electromagnetic perturbations (vector or scalar) in $D=5$, 
$1+2\cos{\pi j}=1+2\cos{\frac{2\pi}{3}}=0$. So there seem to be no QN 
frequencies
as argued by Motl and Neitzke \cite{motl2}.
We obtained for $D=4$ the following result for the correction coefficient
in (\ref{xxx}),
\begin{equation}
{\rm Corr_4}=- \frac {(1-i)\pi ^{3/2}(-1+j^2-3l(l+1))\frac{\cos{j\pi}}{2}
\Gamma[1/4]} 
{ 3(1+2\cos{j\pi})\Gamma[3/4]\Gamma[3/4-j/2]\Gamma[3/4+j/2] }\,,
\label{corrd4}
\end{equation}
which reduces to Musiri and Siopsis' expression. For $D=5$
it is possible also to find a simple expression for the
corrections:
\begin{eqnarray}
\!\!\!{\rm Corr_5}\! &=\!\!\!&\frac {i(9j^2-(24+20l(l+2)))\pi^{3/2}} 
{ 120\times 3^{1/3}\Gamma[2/3]\Gamma[2/3-j/2]\Gamma[2/3+j/2] }
\nonumber \\ 
& &
\times \Gamma[1/6] 
\frac{ \frac{e^{ij\pi}}{\sin{((1/3+j/2)\pi)}}-
\frac{1}{\sin{((2/3+j/2)\pi)}}}{-1+e^{ij\pi}}\,.
\label{corrd5}
\end{eqnarray}
The limits $j \rightarrow 0,2$ are well defined and yield
\begin{equation}
{\rm Corr_5}^{j=0}= \frac {(1-i\sqrt{3})(6+10l+5l^2)\pi^{3/2}\Gamma[1/6]} 
{ 45\times 3^{1/3}\Gamma[2/3]^3} \,,
\label{corrd5j0}
\end{equation}
\begin{equation}
{\rm Corr_5}^{j=2}= \frac {(-1+i\sqrt{3})(-3+10l+5l^2)\pi^{3/2}\Gamma[1/6]} 
{ 45\times 3^{1/3}\Gamma[-1/3]\Gamma[2/3]\Gamma[5/3]} \,,
\label{corrd5j2}
\end{equation}
We list in Tables \ref{tab:corrj0}-\ref{tab:corrj2} some values of this 
five dimensional
correction for some values of the parameter $j$ and $l$ and compare
them with the results extracted numerically.  The agreement is quite
good.  The code for extracting the generic $D$-dimensional corrections
is available from the authors upon request.
It should not come as a surprise that for $j=2/3$ the correction term blows up:
indeed already the zeroth order term is not well defined.
\subsection{Numerical procedure and results }
\label{numtechniqueresults}

\subsubsection{Numerical procedure }
\label{numtechnique}

 In order to numerically obtain the QN frequencies, 
in the present investigation, we make use of Nollert's method 
\cite{nollert}, since asymptotic behaviors of QNMs in the 
limit of large imaginary frequencies are prime concern in the present study.

In our numerical study, we only consider the QNMs of the Schwarzschild black 
hole in the five dimensional spacetime, namely the $D=5$ case. 
As we have remarked, this should be representative. 
The tortoise 
coordinate is then reduced to 
\begin{equation}
r_*=x^{-1}+{1\over 2x_1}\ln(x-x_1)-{1\over 2x_1}\ln(x+x_1)\,, 
\end{equation}
where $x=r^{-1}$ and $x_1=r_h^{-1}$. Here, $r_h$ stands for the horizon 
radius of the black hole. The perturbation function $\Psi$ may be expanded 
around the horizon as 
\begin{equation}
\Psi=e^{-i\omega x^{-1}}(x-x_1)^\rho(x+x_1)^\rho\sum_{k=0}^\infty 
a_k\left({x-x_1\over-x_1}\right)^k\,, 
\label{expansionqnms}
\end{equation}
where $\rho=i\omega/2x_1$ and $a_0$ is taken to be $a_0=1$. The 
expansion coefficients $a_k$ in equation (\ref{expansionqnms}) are determined 
via the four-term recurrence relation (it's just a matter of substituting
expression (\ref{expansionqnms}) in the wave equation (\ref{eveqqnms})), given by
\begin{eqnarray}
&&\alpha_0a_1+\beta_0a_0=0\,, \nonumber \\
&&\alpha_1a_2+\beta_1a_1+\gamma_1a_0=0\,, \\
&&\alpha_ka_{k+1}+\beta_ka_k+\gamma_ka_{k-1}+\delta_ka_{k-2}=0\,,
\ k=2,3,\cdots, \nonumber
\end{eqnarray}
where 
\begin{eqnarray}
\alpha_k&=& 2(2\rho+k+1)(k+1)\,, \nonumber \\
\beta_k &=&-5(2\rho+k)(2\rho+k+1)-l(l+2)-{3\over 4} \nonumber \\
&&-{9\over 4}\,(1-j^2) \,, \nonumber \\
\gamma_k&=& 4(2\rho+k-1)(2\rho+k+1)+{9\over 2}\,(1-j^2)\,, \nonumber \\
\delta_k&=&-(2\rho+k-2)(2\rho+k+1)-{9\over 4}\,(1-j^2)\,. \nonumber
\end{eqnarray}
It is seen that since the asymptotic form of the perturbations as 
$r_*\rightarrow\infty$ is written in terms of the variable $x$ as
\begin{equation}
e^{-i\omega r_*}=e^{-i\omega x^{-1}}(x-x_1)^{-\rho}(x+x_1)^{\rho}\,,
\label{expwr} 
\end{equation}
the expanded perturbation function $\Psi$ defined by equation 
(\ref{expansionqnms}) automatically satisfy the QNM boundary conditions 
(\ref{def-bc}) if the power series converges for $0\le x\le x_1$.
Making use of a Gaussian elimination \cite{Le90}, we can reduce the 
four-term recurrence relation to the three-term one, given by
\begin{eqnarray}
&&\alpha'_0a_1+\beta'_0a_0=0\,, \nonumber \\
&&\alpha'_ka_{k+1}+\beta'_ka_k+\gamma'_ka_{k-1}=0\,, 
\ k=1,2,\cdots, 
\end{eqnarray}
where $\alpha'_k$, $\beta'_k$, and $\gamma'_k$ are given in terms of 
$\alpha_k$, $\beta_k$, $\gamma_k$ and $\delta_k$ by 
\begin{eqnarray}
\alpha'_k=\alpha_k,\quad \beta'_k=\beta_k,\quad \gamma'_k=\gamma_k, 
\quad {\rm for\ } k=0,1, 
\end{eqnarray}
and
\begin{eqnarray}
\alpha'_k&=&\alpha_k, \nonumber \\
 \beta'_k&=&\beta_k-\alpha'_{k-1}\delta_k/\gamma'_{k-1}\,, \\
\gamma'_k&=&\gamma_k-\beta'_{k-1}\delta_k/\gamma'_{k-1}\,,\quad 
{\rm for\ }k\ge 2\,. \nonumber  
\end{eqnarray}
Now that we have the three-term recurrence relation for determining the 
expansion coefficients $a_k$, the 
convergence condition for the expansion (\ref{expansionqnms}), namely the 
QNM conditions, can be written in terms of the continued fraction 
as \cite{Gu67,L}
\begin{eqnarray}
\beta'_0-{\alpha'_0\gamma'_1\over\beta'_1-}
{\alpha'_1\gamma'_2\over\beta'_2-}{\alpha'_2\gamma'_3\over\beta'_3-}
...\equiv 
\beta'_0-\frac{\alpha'_0\gamma'_1}{\beta'_1-
\frac{\alpha'_1\gamma'_2}{\beta'_2-\frac{\alpha'_2\gamma'_3}{\beta'_3-...}}}
=0 \,,
\label{a-eq}
\end{eqnarray}
where the first equality is a notational definition commonly 
used in the literature for infinite continued 
fractions.
Here we shall adopt such a convention.
In order to use Nollert's method, with which relatively high-order 
QNM with large imaginary frequencies can be obtained, we have to know 
the asymptotic behaviors of $a_{k+1}/a_k$ in the limit of 
$k\rightarrow\infty$. According to a similar consideration as that by 
Leaver \cite{Le90}, it is found that 
\begin{eqnarray}
{a_{k+1}\over a_k}=1\pm2\sqrt{\rho}\,k^{-1/2}+
\left(2\rho-{3\over 4}\right)k^{-1}+\cdots\,, 
\label{exp-R}
\end{eqnarray}
where the sign for the second term in the right-hand side is chosen so as 
to be 
\begin{eqnarray}
{\rm Re}(\pm2\sqrt{\rho}) < 0\,. 
\end{eqnarray}
In actual numerical computations, it is convenient to solve the $k$-th 
inversion of the continue fraction equation (\ref{a-eq}), given by 
\begin{eqnarray}
&&\beta'_k-{\alpha'_{k-1}\gamma'_k\over\beta'_{k-1}-}
{\alpha'_{k-2}\gamma'_{k-1}\over\beta'_{k-2}-}\cdots
{\alpha'_0\gamma'_1\over\beta'_0} \nonumber \\
&=& {\alpha'_{k}\gamma'_{k+1}\over\beta'_{k+1}-}
{\alpha'_{k+1}\gamma'_{k+2}\over\beta'_{k+2}-}\cdots\,. \quad 
\label{a-eq2}
\end{eqnarray}
The asymptotic form (\ref{exp-R}) plays an important role in Nollert's 
method when the infinite continued fraction in the right-hand side of
equation (\ref{a-eq2}) is evaluated \cite{nollert}.

\subsubsection{Numerical results}
\label{numericalresults}
Using the numerical technique described in section \ref{numtechnique}
we have extracted the 5000 lowest lying QN frequencies for the five
dimensional Schwarzschild black hole, in the case of scalar, gravitational 
tensor
and gravitational vector perturbations.
For electromagnetic vector perturbations the situation is different:
the real part rapidly approaches zero, and we have not been able to compute
more than the first $30$ modes.

\bigskip\medskip
\centerline {\it (i) Low-lying modes}
\medskip

Low lying modes are important since they govern the intermediate-time
evolution of any black hole perturbation. As such they may play a role
in TeV-scale gravity scenarios \cite{hamed}, and higher dimensional
black hole formation \cite{vitoroscarjose,bhprod}.  For example, it is
known \cite{vitoroscarjose,cardosorads,cardosoradkerrs,cardosoradkerr,cardosoradel,bertihighd} 
that if one forms black holes
through the high energy collision of particles, then the fundamental
quasinormal frequencies serve effectively as a cuttof in the energy
spectra of the gravitational energy radiated away.
In Tables \ref{tab:qnf5dj0}-\ref{tab:qnf5dj23} we list the five lowest lying
QN frequencies for some values of the multipole index $l$.
The fundamental modes are in excellent agreement with the ones presented
by Konoplya \cite{konoplyawkb,roman2} using a high-order WKB approach.
Notice he uses a different convention so one has to be careful when 
comparing the results. For example, in our units Konoplya obtains for
scalar and tensorial perturbations ($j=0$) with $l=2$ a fundamental
QN frequency $\omega _{0}/T_H=9.49089+2.24721i$, whereas we get, from 
Table \ref{tab:qnf5dj0} the number $9.4914+2.2462i$ so the WKB approach
does indeed yield good results, at least for low-lying modes, since it
is known it fails for high-order ones.

\begin{table}
\caption{\label{tab:qnf5dj0} The first lowest QN frequencies
for scalar and gravitational tensor ($j=0$) perturbations 
of the five dimensional Schwarzschild black hole. The frequencies
are normalized in units of black hole temperature, so the Table
really shows $\frac{\omega}{T_H}$, 
where $T_H$ is the Hawking temperature of the black hole. }
\begin{tabular}{l|lll}  \hline
\multicolumn{1}{c}{} &
\multicolumn{3}{c}{ $j=0$}\\ \hline
$n$ &$l=0$:         &$l=1$:      &$l=2$\\ \hline
0&3.3539+2.4089i &6.3837+2.2764i &9.4914+2.2462i  \\ \hline 
1&2.3367+8.3101i &5.3809+7.2734i &8.7506+6.9404i  \\ \hline 
2&1.8868+14.786i &4.1683+13.252i &7.5009+12.225i \\ \hline 
3&1.6927+21.219i &3.4011+19.708i &6.2479+18.214i \\ \hline 
4&1.5839+27.607i &2.9544+26.215i &5.3149+24.597i \\ \hline 
\end{tabular}
\end{table}

\begin{table}
\caption{\label{tab:qnf5dj2} The first lowest QN frequencies
for gravitational vector ($j=2$) perturbations 
of the five dimensional Schwarzschild black hole. The frequencies
are normalized in units of black hole temperature, so the Table
really shows $\frac{\omega}{T_H}$, 
where $T_H$ is the Hawking temperature of the black hole. }
\begin{tabular}{l|lll}  \hline
\multicolumn{1}{c}{} &
\multicolumn{3}{c}{ $j=2$}\\ \hline
$n$ &$l=2$:       &$l=3$:           &$l=4$\\ \hline
0&7.1251+2.0579i  &10.8408+2.0976i  &14.3287+2.1364i  \\ \hline 
1&5.9528+6.4217i  &8.8819+1.0929i   &12.8506+1.0574i   \\ \hline 
2&3.4113+12.0929i &7.2160+16.3979i  &11.5069+16.0274i   \\ \hline 
3&2.7375+19.6094i &5.5278+22.3321i  &9.99894+21.5009i  \\ \hline 
4&2.5106+26.2625i &3.9426+28.6569i  &8.56386+27.4339i  \\ \hline 
\end{tabular}
\end{table}
\vskip 1mm

\vskip 1mm
\begin{table}
\caption{\label{tab:qnf5dj23} The first lowest QN frequencies
for electromagnetic vector ($j=2/3$) perturbations 
of the five dimensional Schwarzschild black hole. The frequencies
are normalized in units of black hole temperature, so the Table
really shows $\frac{\omega}{T_H}$, 
where $T_H$ is the Hawking temperature of the black hole. }
\begin{tabular}{l|lll}  \hline
\multicolumn{1}{c}{} &
\multicolumn{3}{c}{ $j=2$}\\ \hline
$n$ &$l=1$:        &$l=2$:           &$l=3$\\ \hline
0&5.9862+2.2038i &9.2271+2.2144i &12.4184+2.2177i\\ \hline 
1&4.9360+7.0676i &8.4728+6.8459i &11.8412+6.7660i \\ \hline 
2&3.6588+12.9581i&7.1966+12.0804i&10.7694+11.6626i\\ \hline 
3&2.8362+19.3422i&5.9190+18.0343i&9.4316+17.1019i \\ \hline 
4&2.3322+25.7861i&4.9682+24.3832i&8.1521+23.0769i  \\ \hline 
\end{tabular}
\end{table}

In the four dimensional case, and for gravitational vector 
perturbations, there is for is each $l$, a purely
imaginary QN frequency, which had been coined an ``algebraically
special frequency'' by Chandrasekhar \cite{chandra}.  For further
properties of this special frequencies we refer the reader to
\cite{MVDBdoublet,MVDBas}. 
The existence of these purely imaginary
frequencies translates, in the four dimensional case, a relation
between the two gravitational wavefunctions (i.e., between the
Regge-Wheeler and the Zerilli wavefunction, or between the
gravitational vector and gravitational scalar wavefunctions,
respectively).  It is
possible to show for example that the associated potentials are
related through supersymmetry.  Among other consequences, this
relation allows one to prove that the QN frequencies of both
potentials are exactly the same \cite{chandra}.

We have not spotted any purely imaginary QN frequency for this
five-dimensional black hole.  In fact the QN frequency with the
lowest real part is a gravitational vector QN frequency with $l=4$ and
overtone number $n=13$, $\omega=8.7560\times 10^{-2}+6.9934i$.  This
may indicate that in five dimensions, there is no relation between the
wavefunctions, or put another way, that the potentials are no longer
superpartner potentials.  This was in fact already observed by Kodama
and Ishibashi \cite{kodama} for any dimension greater than four. It
translates also in Tables \ref{tab:qnf5dj0}-\ref{tab:qnf5dj2},
which yield different values for the QN frequencies. Although we have not
worked out the gravitational scalar QN frequencies, a WKB approach
can be used for the low-lying QN frequencies, and also yields different values
\cite{konoplyawkb,roman2}.

\bigskip\medskip
\centerline {\it (ii) Highly damped modes}
\medskip

\begin{figure}
\centerline{\includegraphics[width=8 cm,height=8 cm]
{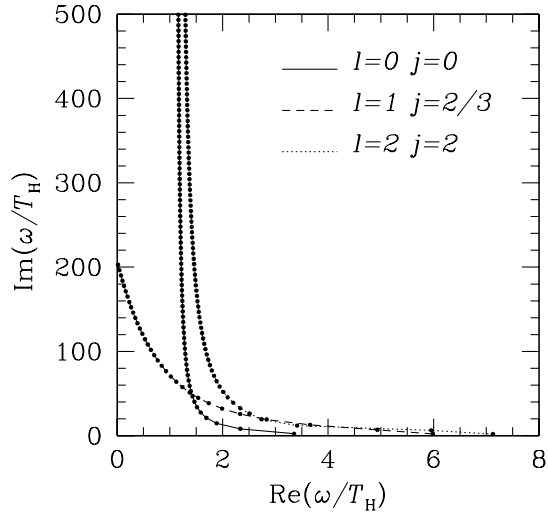}}
\caption{The QN frequencies of scalar ($j=0$), gravitational vector ($j=2$) 
and electromagnetic vector ($j=2/3$) perturbations
of a five-dimensional Schwarzschild black hole. We only show the 
lowest radiatable multipoles,
but the asymtptoic behaviour is $l$-independent.}
\label{fig:qnall5d}
\end{figure}

\begin{figure}
\centerline{\includegraphics[width=8 cm,height=8 cm]
{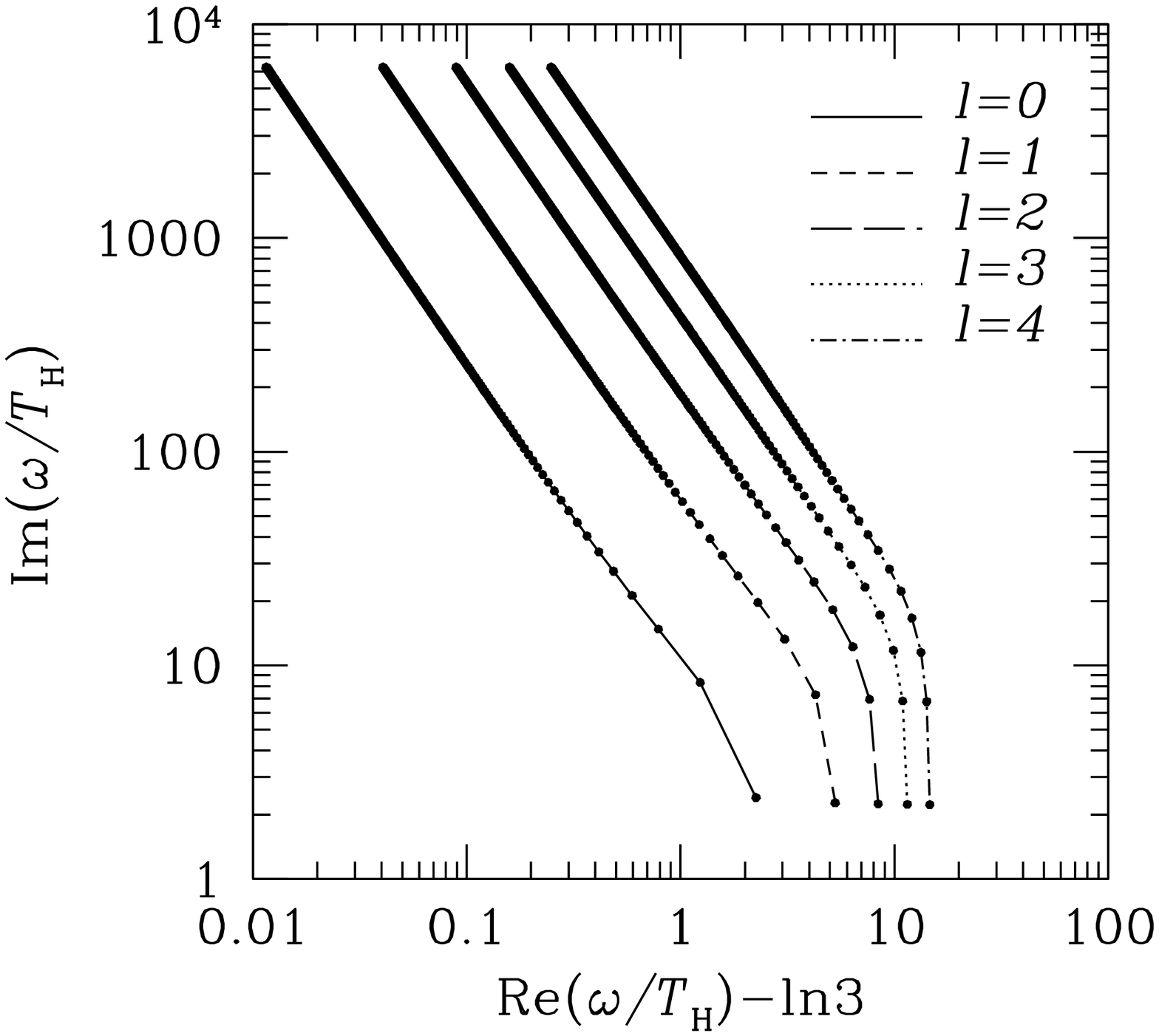}}
\caption{The QN frequencies of 
scalar and gravitational tensor perturbations ($j=0$) 
of a five-dimensional Schwarzschild black hole.
Notice that for frequencies with a large imaginary part
the slope is $l$-independent and equals $3/2$.}
\label{fig:qntensor5d}
\end{figure}

\begin{figure}
\centerline{\includegraphics[width=8 cm,height=8 cm]
{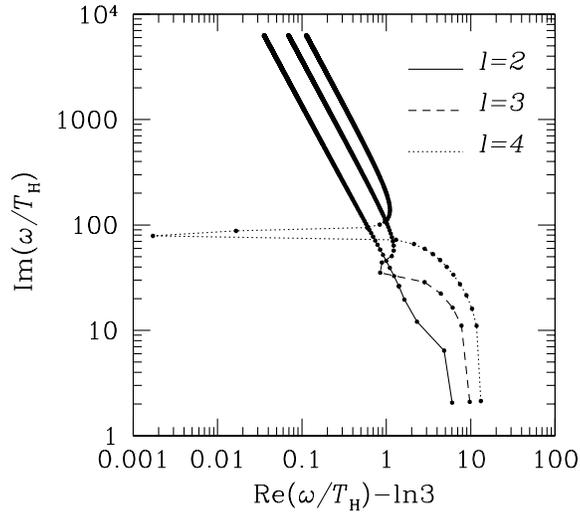}}
\caption{The QN frequencies of gravitational vector perturbations
($j=2$) of a five-dimensional Schwarzschild black hole.}
\label{fig:qnvector5d}
\end{figure}

Our numerical results for the highly damped modes, i.e., QN
frequencies with a very large imaginary part, are summarized in
Figures \ref{fig:qnall5d}-\ref{fig:qnvector5d} and in Tables
\ref{tab:corrj0}-\ref{tab:corrj2}.

\vskip 1mm

In Fig. \ref{fig:qnall5d} we show our results for scalar,
gravitational tensor ($j=0$), gravitational vector ($j=2$) and
electromagnetic vector ($j=2/3$) QN frequencies.  For $j=0\,,2$ the
real part approaches $\ln{3}$ in units of the black hole temperature, a
result consistent with the analytical result in \cite{motl2} (see also
section \ref{apendiceqnms}).  Electromagnetic vector ($j=2/3$) QN
frequencies behave differently: the real part rapidly approaches zero,
and this makes it very difficult to compute the higher modes.
In fact, the same kind of problem appears in the Kerr geometry \cite{cardosoqnmkerr},
and this a major obstacle to a definitive numerical characterization of the
highly damped QNMs in this geometry. 
We have
only been able to compute the first $30$ modes with accuracy.  The
prediction of \cite{motl2} for this case (see section \ref{apendiceqnms})
is that there are no asymptotic QN frequencies.
It is left as an open question whether this is true or not.
Our results indicate that the real part rapidly approaches zero, but we
cannot say whether the modes die there or not, or even if they perform
some kind of oscillation. The imaginary part behaves in the usal manner,
growing linearly with mode number.

Let us now take a more detailed look at our numerical data for
scalar and gravitational QN frequencies.
In Fig. \ref{fig:qntensor5d} we show in a $\ln$ plot
our results for scalar and gravitational tensor ($j=0$) QN frequencies.
\vskip 1mm
\begin{table}
\caption{\label{tab:corrj0} The correction coefficients for the
five-dimensional Schwarzschild black hole, both numerical, here
labeled as ``${\rm Corr_{5}^{N}}$'' and analytical, labeled as ``${\rm
Corr_{5}^{A}}$''. These results refer to scalar or gravitational tensor
perturbations ($j=0$).
The analytical results are extracted from the
analytical formula (\ref{corrd5j0}).  Notice the very good agreement, to
within 0.5\% or less, between the numerically extracted results and the
analytical prediction.}
\begin{tabular}{l|ll}  \hline
\multicolumn{1}{c}{} &
\multicolumn{2}{c}{ $j=0$}\\ \hline
$l$ &${\rm Corr_5^N}$:&${\rm Corr_5^A}$\\ \hline
0   &  1.155 - 1.993i &  1.15404-1.99886i \\ \hline 
1   &  4.058 - 6.956i &  4.03915-6.99601i \\ \hline 
2   &  8.894 - 15.26i &  8.84765-15.3246i \\ \hline 
3   &  15.64 - 26.79i &  15.5796-26.9847i \\ \hline 
4   &  24.33 - 41.67i &  24.2349-41.9761i \\ \hline 
\end{tabular}
\end{table}
\vskip 1mm

Our numerical results are very clear: asymptotically the QN
frequencies for scalar and gravitational tensor perturbations $(j=0)$ behave
as
\begin{equation}
\frac{\omega}{T_H}=\ln{3}+i\,(2n+1)\pi+\frac{{\rm Corr_5}}{n^{2/3}}\,\,,
\label{asympt5dScalar}
\end{equation}
So the leading terms $\ln{3}+(2n+1)i$ are indeed the ones predicted
in \cite{motl2}. Interestingly, the first corrections do not appear as
$\frac{1}{\sqrt{n}}$, but as $\frac{1}{n^{2/3}}$. This was shown to be
the expected analytical result in section \ref{apendiceqnms}, where we
generalized Musiri and Siopsis' \cite{musiri2} results to higher dimensions.
In table
\ref{tab:corrj0} we show the coefficient ${\rm Corr_5}$ extracted
numerically along with the predicted coefficient (see expression
(\ref{corrd5j0})).  The table is very clear: the numerical values
match the analytical ones.  Another confirmation that the corrections
appear as $\frac{1}{n^{2/3}}$ is provided by
Fig. \ref{fig:qntensor5d}. In this figure the QN frequencies are plotted
in a $\ln$ plot for an easier interpretation. One sees that for large
imaginary parts of the QN frequencies, the slope of the plot is
approximately $-3/2$, as it should be if the corrections are of the
order $\frac{1}{n^{2/3}}$.

In Fig. \ref{fig:qnvector5d} we show in a $\ln$ plot our results for
gravitational vector ($j=2$) QN frequencies.  Again, vector QN
frequencies have the asymptotic behaviour given by expression
(\ref{asympt5dScalar}), with a different correction term ${\rm
Corr_5}$. Again, the $\frac{1}{n^{2/3}}$ corrections show themselves
in the $\ln$ plot of Fig. \ref{fig:qnvector5d}: for very large imaginary
parts, the slope is $-3/2$, as it should.  The numerically extracted
coefficient ${\rm Corr_5}$ for $j=2$ perturbations is listed in Table
\ref{tab:corrj2}, along with the analytically predicted value (see
section \ref{apendiceqnms}).  the agreement is very good.
\vskip 1mm
\begin{table}
\caption{\label{tab:corrj2} The correction coefficients for the
five-dimensional Schwarzschild black hole, both numerical, here
labeled as ``${\rm Corr_{5}^{N}}$'' and analytical, labeled as ``${\rm
Corr_{5}^{A}}$'', for gravitational vector perturbations ($j=2$).  
The analytical results are extracted from the
analytical formula (\ref{corrd5j2}).  Notice the very good agreement, to
within 0.5\% or less, between the numerically extracted results and the
analytical prediction.}
\begin{tabular}{l|ll}  \hline
\multicolumn{1}{c}{} &
\multicolumn{2}{c}{ $j=2$}\\ \hline
$l$ &${\rm Corr_5^N}$:&${\rm Corr_5^A}$:\\ \hline
2   & 3.568 - 6.139i  & 3.5583 -6.16316i \\ \hline 
3   & 6.947 - 11.94i  & 6.92425-11.9932i \\ \hline 
4   & 11.29 - 19.38 i & 11.2519-19.4889i \\ \hline 
\end{tabular}
\end{table}
\vskip 1mm

\section{Discussion of results and future directions}
We have made an extensive survey of the QNMs of the four and five
dimensional Schwarzschild black hole.  The investigation presented
here makes a more complete characterization of the highly damped QNMs in the
Schwarzschild geometry.  In the four-dimensional case, we confirmed
previous numerical results regarding the scalar and gravitational
asymptotic QN frequencies. We found that both the leading behaviour
and the first order corrections for the scalar and gravitational
perturbations agree extremely well with existing analytical formulas.  We
have presented new numerical results concerning electromagnetic QN frequencies
of the four-dimensional Schwarzschild geometry. In particular, this is
the first work dealing with highly damped electromagnetic QNMs.
Again, we find that the leading behaviour and the first order
corrections agree with the analytical calculations. The first order
corrections appear at the $\frac{1}{n^{3/2}}$ level.
In the five-dimensional case this represents the first study on highly
damped QNMs.  We have seen numerically that the asymptotic behaviour
is very well described by $\frac{\omega}{T_H} = \ln{3}+\pi i(2n+1)$ for
scalar and gravitational perturbations, and agrees with the predicted
formula.  Moreover the first order corrections appear at the level
$\frac{1}{n^{2/3}}$, which is also very well described by the
analytical calculations, providing one more consistency check on the
theoretical framework.  
In generic $D$ dimensions the corrections appear at the
$\frac{1}{n^{(D-3)/(D-2)}}$ level.
We have not been able to prove or disprove the analytic result
for electromagnetic QN frequencies in five dimensions ($j=2/3$).
Other conclusions that can be taken from
our work are: the monodromy method by Motl and Neitzke is correct.
It is important to test it numerically, as we did, since there are some 
ambiguous 
assumptions in that method. Moreover, their method is highly flexible, 
since it allows an
easy computation of the correction terms, as shown by Musiri and
Siopsis, and generalized here for the higher dimensional Schwarzschild
black hole.
We have basically showed numerically that the monodromy method
by Motl and Neitzke \cite{motl2}, and its extension by Musiri and
Siopsis \cite{musiri2} to include for correction terms in overtone
number, are excellent techniques to investigate the highly damped QNMs
of black holes. 

Highly damped QNMs of black holes are the bed rocks that the recent 
conjectures
\cite{hod,dreyer,ling} are built on, relating these to black hole area
quantization.  However, to put on solid ground the conjecture that
QNMs can actually be of any use to black hole area quantization, one
has to do much better then one has been able to do up to now.  In
particular, it is crucial to have a deeper understanding of the 
Kerr and Reissner-Nordstr\"om QNMs. 
These are, next to the Schwarzschild, the simpler asymptotically
flat geometries. If these conjectures hold, they must to do also
for these spacetimes.
Despite the fact that there are convincing numerical results for the Kerr 
geometry
\cite{cardosoqnmkerr}, there seems to be for the moment no serious analytical
investigation of the highly damped QNMs for this spacetime.  Moreover,
even though we are in possession of an analytical formula for the
highly damped QNMs of the Reissner-Nordstr\"om geometry \cite{motl2},
and this has been numerically tested already \cite{bertikerr}, we have
no idea what it means!  In particular, can one use it to quantize the
black hole area? In this case the electromagnetic perturbations
are coupled to the gravitational ones. How do we proceed with the
analysis, that was so simple in the Schwarzschild geometry? 
Is the assumption of equally spaced eigenvalues a correct one, in this case?
Can these ideas on black hole area quantization be translated to 
non-asymptotically 
flat spacetimes, as the
de Sitter or anti-de Sitter spacetime?
These are the fundamental issues that remain to be solved in this
field. Should any of them be solved satisfactorily, then these
conjectures would gain a whole new meaning. As they stand, it may just
be a numerical coincidence that the real part of the QN frequency goes
to $\ln{3}$.


\thispagestyle{empty} \setcounter{minitocdepth}{1}
\chapter[A complete survey on the quasinormal modes of the four dimensional Kerr black hole]{Quasinormal modes of the four dimensional Kerr black hole} \label{chap:qnmkerr}
\lhead[]{\fancyplain{}{\bfseries Chapter \thechapter. \leftmark}}
\rhead[\fancyplain{}{\bfseries \rightmark}]{}
\minitoc \thispagestyle{empty}
\renewcommand{\thepage}{\arabic{page}}
\section{Introduction}
The study of linearized perturbations of black hole solutions in
general relativity has a long history \cite{chandra}. The development of
the relevant formalism, initially motivated by the need of a formal
proof of black hole stability, gave birth to a whole new research
field. A major role in this field has been played by the concept of
QNMs: oscillations having purely ingoing wave
conditions at the black hole horizon and purely outgoing wave
conditions at infinity. These modes determine the late-time evolution
of perturbing fields in the black hole exterior. Numerical simulations
of stellar collapse and black hole collisions in the ``full''
(non--linearized) theory have shown that in the final stage of such
processes (``ringdown'') QNM's dominate the black hole response to any
kind of perturbation. Since their frequencies are uniquely determined
by the black hole parameters (mass, charge and angular momentum),
QNM's are likely to play a major role in the nascent field of
gravitational wave astronomy, providing unique means to ``identify''
black holes \cite{kokkotas,nollert}.

An early attempt at relating classical black hole oscillations to the
quantum properties of black holes was carried out by York \cite{york}.
More recently Hod made an interesting proposal \cite{hod}, which has been
described in the previous chapter. 
In light of these exciting new results, Hod's conjecture seems to be a
very promising candidate to shed light on quantum properties of black
holes. However, it is natural to ask whether the conjecture applies to
more general (charged and/or rotating) black holes. If asymptotic
frequencies for ``generic'' black holes depend (as they do) on the
hole's charge, angular momentum, or on the presence of a cosmological
constant, should Hod's proposal be modified in some way? And what is
the correct modification? The hint for an answer necessarily comes
from analytical or numerical calculations of highly damped QNM's for
charged and rotating black holes, or for black holes in
non-asymptotically flat spacetimes. Some calculations in this
direction have now been performed, revealing unexpected and puzzling
features \cite{bertikerr,motl2,cardosoqnmds,brinkds,abdallads,cardosoqnmsads2}.

In particular, the technique originally developed by Nollert to study
highly damped modes of Schwarzschild black holes has recently been
extended to the RN case \cite{berti}, showing that highly damped RN QNM's
show a peculiar spiralling behaviour in the complex-$\omega$ plane as
the black hole charge is increased. Independently, Motl and Neitzke
obtained an analytic formula for the asymptotic frequencies of scalar
and electromagnetic-gravitational perturbations of a RN black hole
whose predictions show an excellent agreement (at least for large
values of the charge) with the numerical results \cite{motl2}. For
computational convenience, Motl and Neitzke fixed their units in a
somewhat unconventional way: they introduced a parameter $k$ related
to the black hole charge and mass by $Q/M=2\sqrt{k}/(1+k)$, so that
$\beta=4\pi/(1-k)=1/T_H$ is the inverse black hole Hawking temperature
and $\beta_I=-k^2\beta$ is the inverse Hawking temperature of the
inner horizon. Their result is an implicit formula for the QNM
frequencies,
\begin{equation}
e^{\beta \omega}+2+3e^{-\beta_I \omega}=0,
\label{MNf}
\end{equation}
and has now been confirmed by independent calculations
\cite{andersson2}. However, its interpretation in terms of the suggested
correspondence is still unclear. QN frequencies of a charged
black hole, according to formula (\ref{MNf}), depend not only on the
black hole's Hawking temperature, but also on the Hawking temperature
of the (causally disconnected) inner horizon.  Perhaps more worrying
is the fact that the asymptotic formula does not yield the correct
Schwarzschild limit as the black hole charge $Q$ tends to
zero. However this may well be due to higher-order corrections in
$\omega_I^{-1/2}$; as we shall see, the numerical study of Kerr modes
we present in this chapter (first presented in \cite{cardosoqnmkerr}, but see also the more recent developments in \cite{cardosokerr2}) 
seems to support such a conjecture.  Finally
and most importantly, it is not at all clear which are the
implications of the non--periodic behaviour of asymptotic RN modes for
the Hod conjecture. Maybe the complicated behaviour we observe is an
effect of the electromagnetic--gravitational coupling, and we should
only consider {\it pure gravitational perturbations} for a first
understanding of black hole quantization based on Hod's
conjecture. The latter suggestion may possibly be ruled out on the
basis of two simple observations: first of all, in the large damping
limit ``electromagnetic'' and ``gravitational'' perturbations seem to
be isospectral \cite{motl2}; secondly, Kerr modes having $m=0$ show a
very similar spiralling behaviour, which is clearly {\it not} due to any
form of electromagnetic--gravitational coupling.

The available numerical calculations for highly damped modes of black
holes in non--asymptotically flat spacetimes are as puzzling as those
for RN black holes in flat spacetime.  Cardoso and Lemos \cite{cardosoqnmds}
have studied the asymptotic spectrum of Schwarzschild black holes in a
de Sitter background. They found that, when the black hole radius is
comparable to the cosmological radius, the asymptotic spectrum depends
not only on the hole's parameters, but also on the angular separation
index $l$. Quite worryingly the universality of the formula ,
that does not depend on dimensionality and gives the same limit for
``scalar'' and ``gravitational'' modes (loosely using the standard
four dimensional terminology; see \cite{kodama,birminghamd} for a more precise
formulation in higher dimensions) seems to be lost when the
cosmological constant is non--zero. This study has recently been
generalized to higher dimensional Schwarzschild-de Sitter black holes
in \cite{molina}.  Higher--order corrections to the behaviour
predicted in \cite{cardosoqnmds} have been found by Van den Brink
\cite{brinkds}.  The issue is not settled yet however, and the
asymptotics may be different from what was predicted in
\cite{cardosoqnmds}. Indeed, recent numerical calculations \cite{shijunds} seem to
suggest that the analytical result in \cite{cardosoqnmds} is correct only when
the overtone index $n$ satisfies $n k \ll 1$, where $k$ is the surface
gravity at the Schwarzschild-de Sitter black hole horizon. For higher
overtones, the behaviour seems to be different. One still needs a
complete numerical verification, as it is seems difficult to extract
QNM frequencies for $nk>1$ \cite{shijunds}.  Calculations of QNM's for
Schwarzschild--anti--de Sitter black holes were performed in various
papers \cite{qnmads}, showing that the nature of the QNM spectrum in this
case is remarkably different (basically due to the ``potential
barrier'' arising because of the cosmological constant, and to the
changing QNM boundary conditions at infinity). Those calculations were
recently extended to encompass asymptotic modes \cite{cardosoqnmsads2}. The basic
result is that consecutive highly damped modes (whose real part goes
to infinity as the imaginary part increases) have a uniform {\it
spacing} in both the real and the imaginary part; this spacing is
apparently independent of the kind of perturbation considered and of
the angular separation index $l$.

The aim of this chapter is to study in depth the behaviour of highly
damped Kerr QNM's, complementing and clarifying results that were
presented in previous works \cite{bertikerr,onozawa}.  The plan of the chapter is as
follows, and is based on the work presented in \cite{cardosoqnmkerr}. 
In section \ref{sec1} we briefly introduce our numerical
method. In section \ref{sec2} we discuss some results presented in
\cite{bertikerr} and show a more comprehensive calculation of gravitational
QNM's, considering generic values of $m$ and higher multipoles
(namely, $l=3$). In section \ref{sec3} we display some results for
scalar and electromagnetic perturbations, showing that the asymptotic
behaviour we observe for $l=m=2$ gravitational perturbations seems to
be very special. In section \ref{sec4} we briefly summarize our
results and we discuss the asymptotic behaviour of the modes'
imaginary part. Finally, in section \ref{sec5} we turn our attention
to a different open problem concerning Kerr perturbations. Motivated
by some recent, surprising developments arising from the study of the
branch cut in the Schwarzschild problem \cite{MVDBdoublet} and by
older conjectures steaming from analytical calculations of the
properties of algebraically special modes \cite{MVDBas}, we turn our
attention to Kerr QNM's in the vicinity of the Schwarzschild
algebraically special frequencies. As the black hole is set into
rotation, we find for the first time that a QNM multiplet appears
close to the algebraically special Schwarzschild modes. A summary,
conclusions and an overlook on possible future research directions
follow.

\section{Numerical method}\label{sec1}

A first numerical study of Kerr QNM's was carried out many years ago
by Detweiler \cite{De}. Finding highly damped modes through a
straightforward integration of the perturbation equations is
particularly difficult even for non--rotating black holes
\cite{kokkotas,nollert}. In the Kerr case the situation is even worse, because, due
to the non--spherical symmetry of the background, the perturbation
problem does not reduce to a single ordinary differential equation for
the radial part of the perturbations, but rather to a couple of
differential equations (one for the angular part of the perturbations,
the other for the radial part).

A method to find the eigenfrequencies without resorting to
integrations of the differential equations was developed by Leaver,
and has been extensively discussed in the literature \cite{L,onozawa,bertikerr}. In
this chapter we will apply exactly the same method. Following Leaver, we
will choose units such that $2M=1$. Then the perturbation equations
depend on a parameter $s$ denoting the spin of the perturbing field
($s=0,-1,-2$ for scalar, electromagnetic and gravitational
perturbations respectively), on the Kerr rotation parameter $a$
($0<a<1/2$), and on an angular separation constant $A_{lm}$. In the
Schwarzschild limit the angular separation constant can be determined
analytically, and is given by the relation $A_{lm}=l(l+1)-s(s+1)$.
The basic idea in Leaver's method is the following. Boundary
conditions for the radial and angular equations translate into
convergence conditions for the series expansions of the corresponding
eigenfunctions. In turn, these convergence conditions can be expressed
as two equations involving continued fractions. Finding QNM
frequencies is a two-step procedure: for assigned values of
$a,~\ell,~m$ and $\omega$, first find the angular separation constant
$A_{lm}(\omega)$ looking for zeros of the {\it angular} continued
fraction equation; then replace the corresponding eigenvalue into the
{\it radial} continued fraction equation, and look for its zeros as a
function of $\omega$.  Leaver's method is relatively well convergent
and numerically stable for highly damped modes, when compared to other
techniques \cite{KerrQNM}. We mention that an alternative, approximate
method to find Kerr QN frequencies, that has the advantage of
highlighting some physical features of the problem, has recently been
presented \cite{GA}.

In the next sections we will use Leaver's technique to complement
numerical studies of Kerr QN overtones carried out by some of
us in the past \cite{onozawa,bertikerr}. As we just said, the method we use for our
analysis is the one described in those papers. However, since we are
pushing our numerics to their limits, we have systematically
cross--checked all the results we present here using two independent
codes. As we shall see, our study will uncover a plethora of
interesting new features.

\section{Gravitational perturbations}\label{sec2}

\subsection{Modes having $l=m=2$: a more extensive discussion}

Let us consider rotating black holes, having angular momentum per unit
mass $a=J/M$. The black hole's (event and inner) horizons are given in
terms of the black hole parameters by $r_\pm=M\pm\sqrt{M^2-a^2}$. The
hole's temperature $T_H=(r_+-r_-)/A$ where $A=8\pi Mr_+$ is the
hole's surface area, related to its entropy $S$ by the relation
$A=S/4$. Introducing the angular velocity of the horizon $\Omega=4\pi
a/A$, applying the first law of black hole thermodynamics,
\begin{equation}
\Delta M=T_H\Delta S+\Omega \Delta J,
\label{FLaw}
\end{equation}
and {\it assuming that the formula for the area spectrum derived for a
Schwarzschild black hole still holds in this case}, Hod conjectured
that the real parts of the asymptotic frequencies for rotating black
holes are given by:
\begin{equation}
\omega_R=\tilde \omega_R=T_H \ln 3+m\Omega,
\label{Hod}
\end{equation}
where $m$ is the azimuthal eigenvalue of the field \cite{hod}.  Hod
later used a systematic exploration of moderately damped Kerr black
hole QNM's carried out a few years ago by one of us \cite{onozawa} to lend
support to formula (\ref{Hod}), at least for modes having $l=m$
\cite{H2}.  His conclusions were shown to be in contrast with the
observed behaviour of modes having stronger damping in \cite{bertikerr}: the
deviations between the numerics and formula (\ref{Hod}) were indeed
shown to {\it grow} as the mode order grows (see figure 7 in
\cite{bertikerr}).  Hod even used (\ref{FLaw}), {\it without including the
term due to variations of the black hole charge $\Delta Q$}, to
conjecture that (\ref{Hod}) holds for Kerr--Newman black holes as well
\cite{hod}. This second step now definitely looks as a bold
extrapolation. Not only does formula (\ref{Hod}) disagree with the
observed numerical behaviour for perturbations of Kerr black holes
having $l=m=2$ \cite{bertikerr} (not to mention other values of $m$, as we
shall see in the following); by now, analytical and numerical
calculations have now shown that RN QNM's have a much more rich and
complicated behaviour \cite{bertikerr,motl2,neitzke}. Summarizing, there is now
compelling evidence that the conjectured formula (\ref{Hod}) must be
wrong. However it turns out \cite{bertikerr}, quite surprisingly, that an
extremely good fit to the numerical data for $l=m=2$ is provided by an
even simpler relation, not involving the black hole temperature:
\begin{equation}
\omega_R=m\Omega.
\label{mOm}
\end{equation}

\begin{figure}[htbp]
\centering
\includegraphics[angle=270,width=8cm,clip]{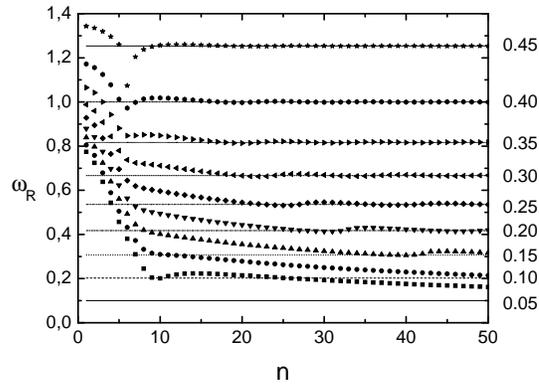}
\caption{
Each different symbol corresponds to the (numerically computed) value
of $\omega_R$ as a function of the mode index $n$, at different
selected values of the rotation parameter $a$. The selected values of
$a$ are indicated on the right of the plot. Horizontal lines
correspond to the predicted asymptotic frequencies $2\Omega$ at the
given values of $a$. Convergence to the asymptotic value is clearly
faster for larger $a$. In the range of $n$ allowed by our numerical
method ($n\lesssim 50$) convergence is not yet achieved for $a\lesssim
0.1$.
}\label{fig1}
\end{figure}

Is this just a coincidence? After all, this formula does not yield the
correct Schwarzschild limit; why should we trust it when it is only
based on numerical evidence?  A convincing argument in favour of
formula (\ref{mOm}) is given in figure \ref{fig1}. There we show the
real part of modes having $l=m=2$ as a function of $n$ {\it at fixed
$a$}, for some selected values of $a$ (namely,
$a=0.00,0.05,..0.45$). The convergence to the limiting value
$\omega_R=2\Omega$ (horizontal lines in the plot) is
evident. Furthermore, the convergence is much faster for holes
spinning closer to the extremal limit, and becomes slower for black
holes which are slowly rotating. In our opinion, this lends support to
the idea that the Schwarzschild limit may not be recovered
straightforwardly as $a\to 0$, and that some order-of-limits issues
may be at work, as recently claimed in \cite{neitzke} to justify the
incorrect behaviour of formula (\ref{MNf}) as the black hole charge
$Q\to 0$.

Is formula (\ref{mOm}) just an approximation to the ``true''
asymptotic behaviour, for example a lowest--order expansion in powers
of $\Omega$? To answer this questions we can try and replace
(\ref{mOm}) by some alternative relation. Since in the Schwarzschild
limit (\ref{mOm}) doesn't give the desired ``$\ln 3$'' behaviour, we
would like a higher--order correction that {\it does} reproduce the
non--rotating limit, while giving a good fit to the numerical data.
Therefore, in addition to formulas (\ref{Hod}) and (\ref{mOm}), we
tried out the following fitting relations:
\begin{equation}
\omega_R =4\pi T_H^2\ln 3+ m\Omega=T_H\ln 3(1-\Omega^2)+m\Omega, \\
\label{Om2}
\end{equation}
\begin{equation}
\omega_R =T_H\ln 3(1-m^2\Omega^2)+m\Omega.
\label{mOm2}
\end{equation}

\begin{figure}
\centering
\includegraphics[angle=270,width=8cm,clip]{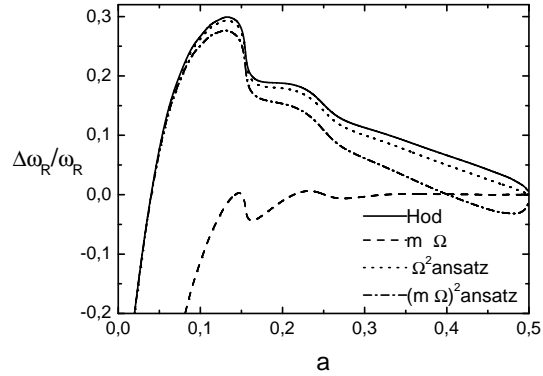}
\caption{
Relative difference between various fit functions and numerical
results for the mode having index $n=40$. From top to bottom in the
legend, the lines correspond to the relative errors for formulas
(\ref{Hod}), (\ref{mOm}), (\ref{Om2}) and (\ref{mOm2}).
}
\label{fig2}
\end{figure}

Formula (\ref{Om2}) enforces the correct asymptotic limit at $a=0$,
and can be seen as an $\Omega^2$--correction of Hod's relation. Since
numerical results suggest a dependence on $m\Omega$ we also used the
slight modification given by formula (\ref{mOm2}), hoping it could
turn out to fit better our numerical data. The relative errors of the
various fitting formulas with respect to the numerical computation for
a the $n=40$ QNM are given in figure \ref{fig2}. Once again, formula
(\ref{mOm}) is clearly the one which performs better. All relations
are seen to fail quite badly for small rotation rate, but this
apparent failure is only due to the onset of the asymptotic behaviour
occurring {\it later} (that is, when $n>40$) for small values of $a$.

We think that the excellent fitting properties and the convergence
plot, when combined together, are very good evidence in favour of
formula (\ref{mOm}). Therefore, let us assume (\ref{mOm}) {\it is} the
correct formula (at least for $l=m=2$, and maybe for large enough
$a$), and let's look at its consequences in computing the area
spectrum for Kerr black holes. Modes having $l=m$ may be the relevant
ones to make a connection with quantum gravity, as recently claimed in
\cite{H2}. The proportionality of these modes to the black hole's
angular velocity $\Omega$ seems to suggest that something deep is at
work in this particular case.

In the following, we will essentially repeat the calculation carried
out by Abdalla {\it et al.}  \cite{abdallads} for extremal ($a=M$) Kerr
black holes. We will argue that the conclusion of their calculation
may in fact be wrong, since those authors did not take into account
the (then unknown) functional behaviour of $\omega_R(a)$, but rather
assumed $\omega_R=m/2M$(=constant) in the vicinity of the extremal
limit.  In following the steps traced out in \cite{abdallads} we will
restore for clarity all factors of $M$. This means, for example, that
the asymptotic frequency for $m>0$ in the extremal limit is
$\omega=m/2M$.  Let us also define $x=a/M$. The black hole inner and
outer horizons are $r_\pm=M\left[1\pm\sqrt{1-x^2}\right]$. The black
hole temperature is
\begin{equation}
T=\frac{r_+-r_-}{A}=\frac{1}{4\pi M}\frac{\sqrt{1-x^2}}{1+\sqrt{1-x^2}},
\end{equation}
where 
\begin{equation}
A=8\pi Mr_+=8\pi M^2\left[1+\sqrt{1-x^2}\right]
\end{equation}
is the hole's surface area, related to its entropy $S$ by the relation
$A=S/4$. The hole's rotational frequency is 
\begin{equation}
\Omega=\frac{4\pi a}{ A}=\frac{a}{ 2Mr_+}=\frac{1}{ 2M}\frac{x}{1+\sqrt{1-x^2}}.
\end{equation}
Let us now apply the first law of black hole thermodynamics and the
area--entropy relation to find
\begin{equation}
\Delta A=\frac{4}{T}\left(\Delta M-\Omega \Delta J\right).
\end{equation}
The authors of \cite{abdallads} focused on the extremal limit. They used
$\Delta J=\hbar m$ and $\Delta M=\hbar \omega_R(x=1)=\hbar m/2M$ to deduce
that
\begin{equation}
\Delta A=4\hbar m\left[\frac{1/2M-\Omega}{ T}\right]=\hbar m {\cal A}.
\label{dA}
\end{equation}
where {\cal A} is the area quantum. Now, the square parenthesis is
undefined, since $\Omega\to 1/2M$ when $x\to 1$ . Taking the limit $x\to
1$ {\it and keeping $\Delta M=\hbar m/2M$ constant} leads to
\begin{equation}
{\cal A}=8\pi\left(1+\sqrt{\frac{1-x}{2}}\right)\simeq 8\pi,
\end{equation}
which is the final result in \cite{abdallads}. The fundamental assumption in
this argument is that the asymptotic frequency $\omega_R=m/2M$, which
is true only for $x=1$.  However one has to consider how the QNM
frequency changes with $x$, and our numerical QNM calculations show
that assuming it is constant ($\omega_R=m/2M$) is not justified.  What
is the effect of assuming $\omega_R=m\Omega$ on the area spectrum?
The calculation is exactly the same, but the equation $\Delta M=\hbar
m/2M$ is replaced by $\Delta M=\hbar m\Omega$, and we conclude
\begin{equation}
\Delta A=0.
\label{reversible}
\end{equation} 

The area variation is {\it zero} at any black hole rotation rate.  At
first sight, this result may look surprising. It is not, and it
follows from fundamental properties of black holes. Indeed, we are
looking at reversible black hole transformations. It is well known
that the gain in energy $\Delta E$ and the gain in angular momentum
$\Delta J$ resulting from a particle with negative energy $-E$ and
angular momentum $-L_z$ arriving at the event horizon of a Kerr black
hole is subject to the inequality
\begin{equation}
\Delta M\geq \Omega \Delta J;
\label{ineqc5}
\end{equation}
see, for example, formula (352) on page 373 in \cite{chandra} and the
related discussion. This inequality is equivalent to the statement
that the irreducible mass $M_{irr}\equiv (Mr_+/2)^{1/2}$ of the black
hole can only increase \cite{CR}, or that by no continuous
infinitesimal process involving a single Kerr black hole can the
surface area of the black hole be decreased (Hawking's area
theorem). Assuming the validity of Hod's conjecture,
and using the result (\ref{mOm}) for asymptotic QNM's, we are
saturating the inequality (\ref{ineqc5}); in other words, we are in
presence of a {\it reversible} process, in which the area (or the
irreducible mass, which is the same) is conserved. This is indeed the
content of equation (\ref{reversible}).

The result (\ref{reversible}) can either mean that using modes having
$l=m$ in Hod's conjecture is wrong, or that we cannot use Bohr's
correspondence principle to deduce the area spectrum for Kerr.  A
speculative suggestion may be to {\it modify Bohr's correspondence
principle as introduced by Hod}. For example, if we do not interpret
the asymptotic frequencies as a change in {\it mass} ($\Delta M=\hbar
\omega_R$) but rather impose $T\Delta S=\hbar \omega_R$ (this is of
course equivalent to Hod's original proposal when $a=0$), the
asymptotic formula would imply, using the first law of black hole
mechanics, that the minimum possible variation in mass is $\Delta
M=2m\hbar \Omega$.

\subsection{Modes having $l=2$, $l\neq m$}

From the discussion in the previous paragraph we feel quite confident
that the real part of modes having $l=m=2$ approaches the limit
$\omega_R=m\Omega$ as the mode damping tends to infinity. What about
modes having $l\neq m$? In \cite{bertikerr} it was shown that modes having
$m=0$ show a drastically different behaviour.  As the damping
increases, modes show more and more loops; pushing the calculation to
very high dampings is not easy, but the trend strongly suggests a
spiralling asymptotic behaviour, reminiscent of RN modes.  In this
section we present results for the cases not considered in \cite{bertikerr},
concentrating on the real parts of modes having $l=2$ and
$m=1,~-1,~-2$.

\begin{figure}[htbp]
\centering
\includegraphics[angle=270,width=8cm,clip]{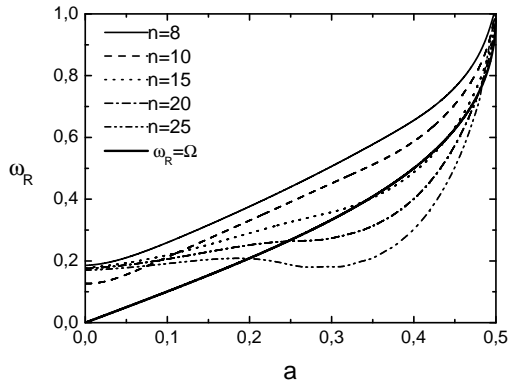}
\caption{
Real part of the frequency for modes with $l=2$, $m=1$.  The modes
``bend'' downwards as $n$ increases; the bold solid line is
$\omega_R=\Omega$. Mode frequencies tend to $\omega_R=m=1$ in the
extremal limit.
}\label{fig3}
\end{figure}

Modes having $l=2$, $m=1$ are displayed in figure \ref{fig3}. They do
not seem to approach the limit one could naively expect, that is,
$\omega_R=\Omega$. The real part of the frequency shows instead a
minimum as a function of $a$, and approaches the limit $\omega_R=m$ as
$a\to 1/2$. The fact that the real part of modes having $l=2$ and
$m=1$ approaches $\omega_R=m=1$ as $a\to 1/2$ has not been observed
before, to our knowledge. We will see later that this behaviour is
characteristic of QNM's due to perturbation fields having arbitrary
spin as long as $m>0$.

\begin{figure}[htbp]
\centering
\includegraphics[angle=270,width=8cm,clip]{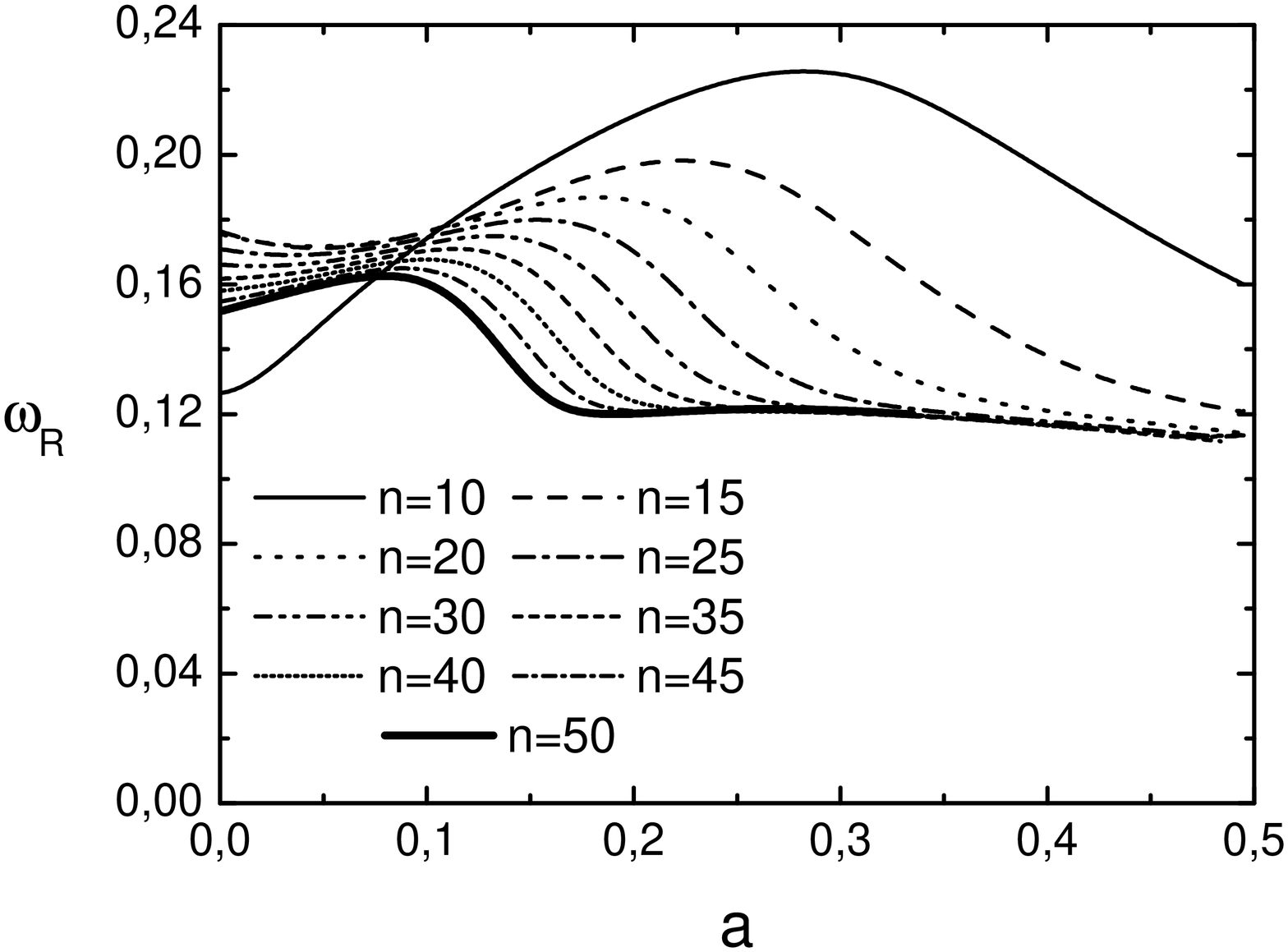}
\includegraphics[angle=270,width=8cm,clip]{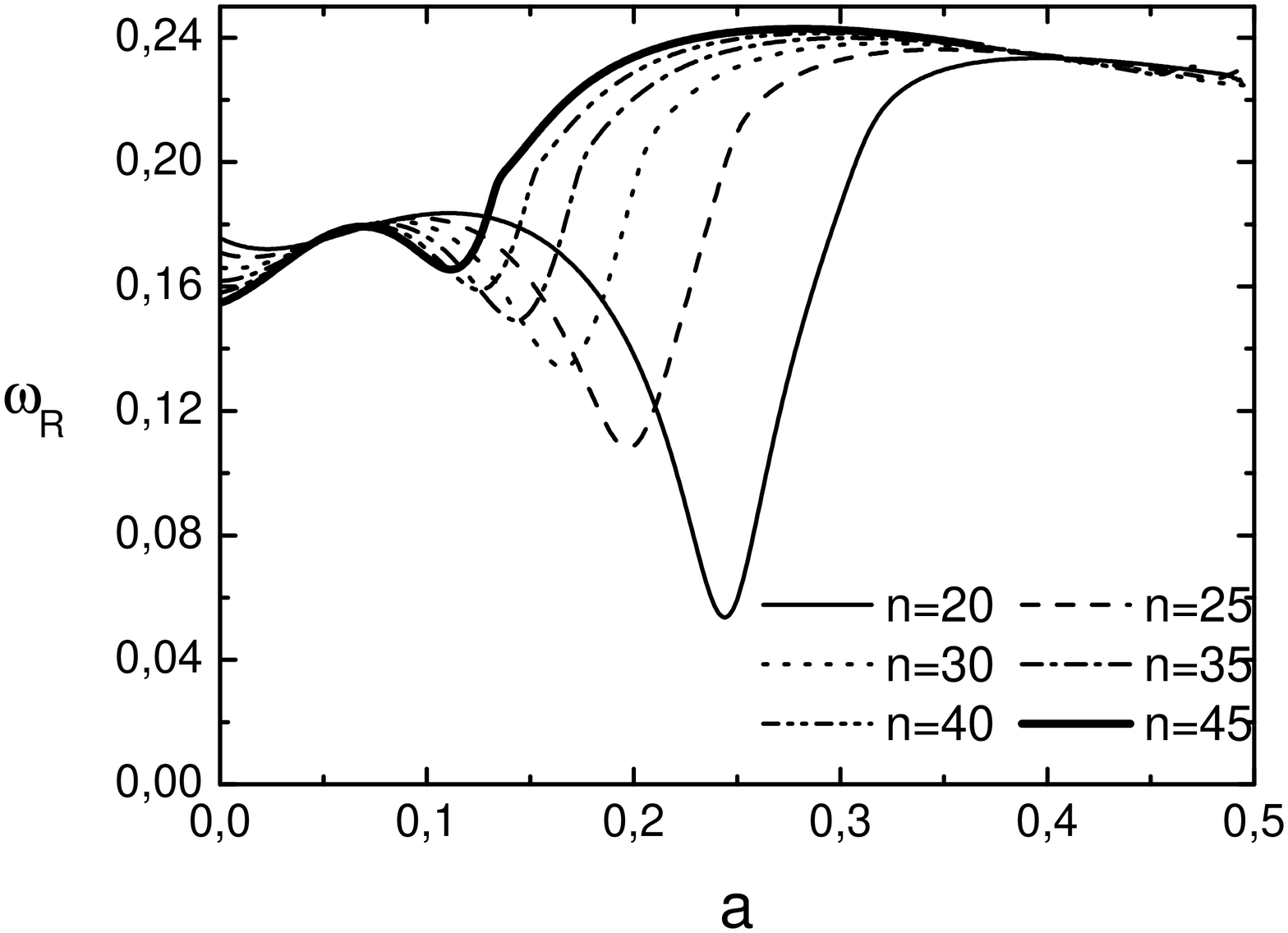}
\caption{
Real part of the first few modes with $l=2$ and $m<0$. Modes having
$m=-1$ are shown in the left panel, modes having $m=-2$ in the right
panel. As the mode order $n$ increases, $\omega_R$ seems to approach a
(roughly) constant value $\omega_R=-m \varpi$, where $\varpi\simeq
0.12$.  Convergence to this limiting value is faster for large values
of the rotation parameter $a$ (compare figure \ref{fig1}).
}\label{fig4}
\end{figure}

The real parts of modes having $l=2$, $m<0$ as functions of $a$ (for
some selected values of $n$) are displayed in figure \ref{fig4}.  From
the left panel, displaying the real part of modes having $m=-1$, we
infer an interesting conclusion: the frequencies tend to approach a
constant (presumably $a$--independent) limiting value, with a
convergence rate which is faster for large $a$, as in the $l=m=2$
case. The limiting value is approximately given by $0.12$. A similar
result holds for modes having $l=2$, $m=-2$ (right panel). Once again
the frequencies seem to asymptotically approach a (roughly) constant
value, with a convergence rate which is faster for large $a$.  The
limiting value is now approximately given by $\omega_R=0.24$, about
twice the value we got for $m=-1$.  Concluding, an extremely
interesting result emerges from our calculations: the real part of
modes having $m<0$ seems to asymptotically approach the limit
\begin{equation}
\omega_R=-m \varpi, 
\end{equation}
where $\varpi\simeq 0.12$ is to a good approximation independent of
$a$. We have not yet been able to obtain this result analytically, but
an analytical calculation is definitely needed. It may offer some
insight on the physical interpretation of the result, and help explain
the surprising qualitative difference in the asymptotic behaviour of
modes having different values of $m$.

\subsection{Modes having $l=3$}

Results for a few highly--damped modes having $l=3$, $m=0$ were shown
in \cite{bertikerr}. Those modes exhibit the usual ``spiralling'' behaviour
in the complex plane as the damping increases. In this paragraph we
present a more complete calculation of modes having $l=3$.
Unfortunately we did not manage to push the calculation to values of
$n$ larger than about 50. The algebraically special frequency
separating the lower QNM branch from the upper branch corresponds to
$n=41$ when $l=3$, so we cannot be completely sure that our
calculations are indicative of the asymptotic behaviour.

\begin{figure}[htbp]
\centering
\includegraphics[angle=270,width=8cm,clip]{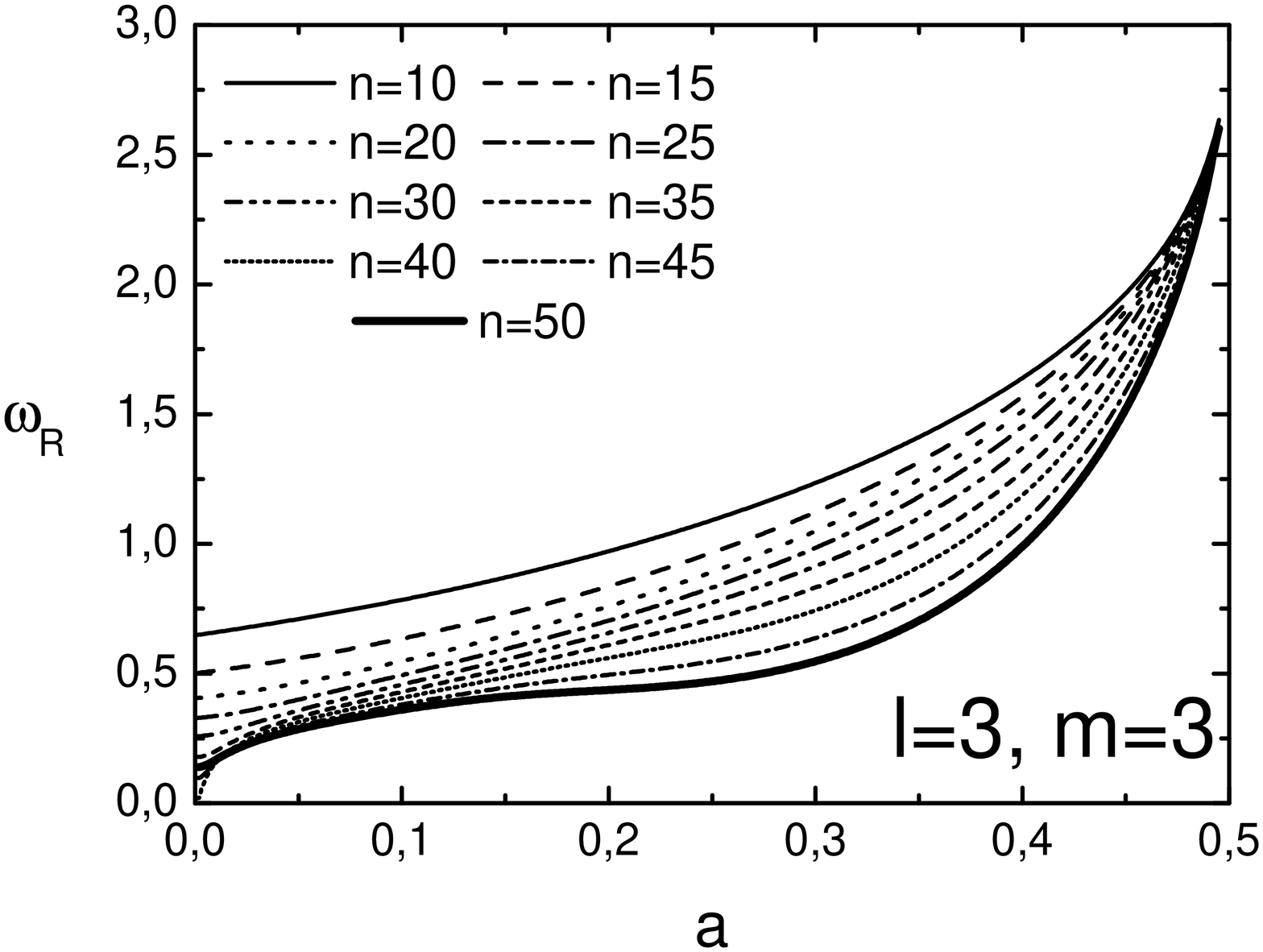}
\includegraphics[angle=270,width=8cm,clip]{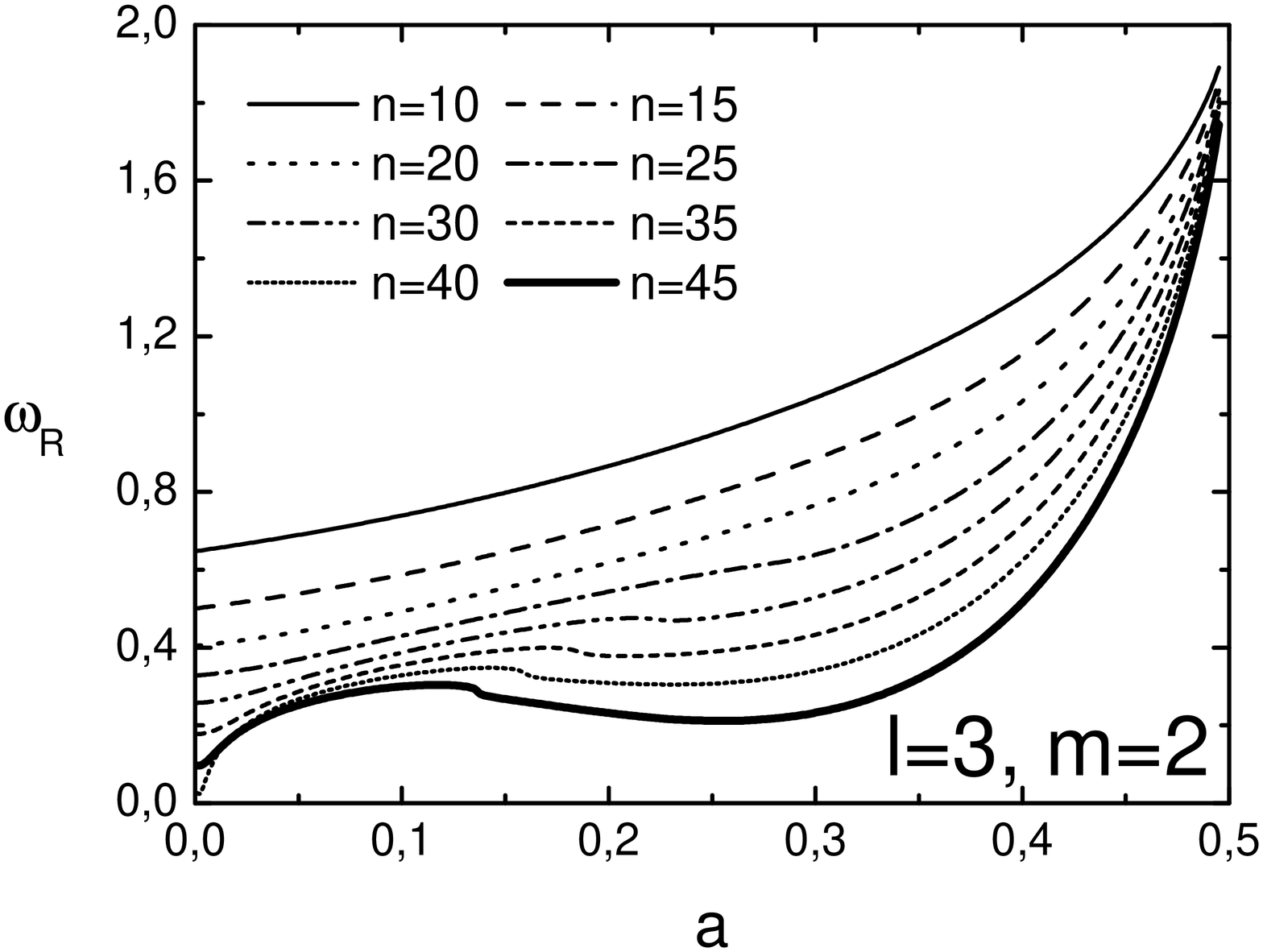}
\includegraphics[angle=270,width=8cm,clip]{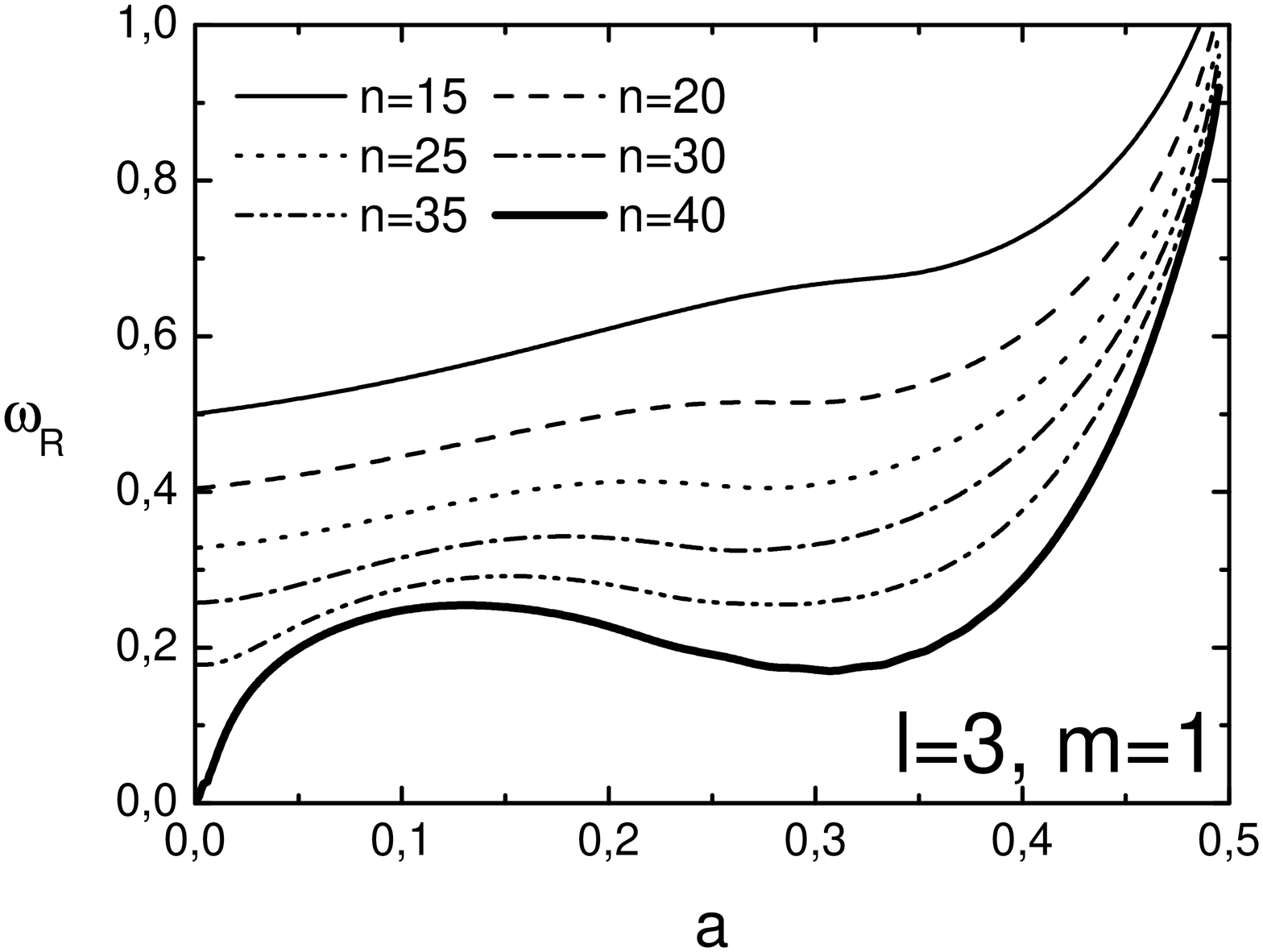}
\includegraphics[angle=270,width=8cm,clip]{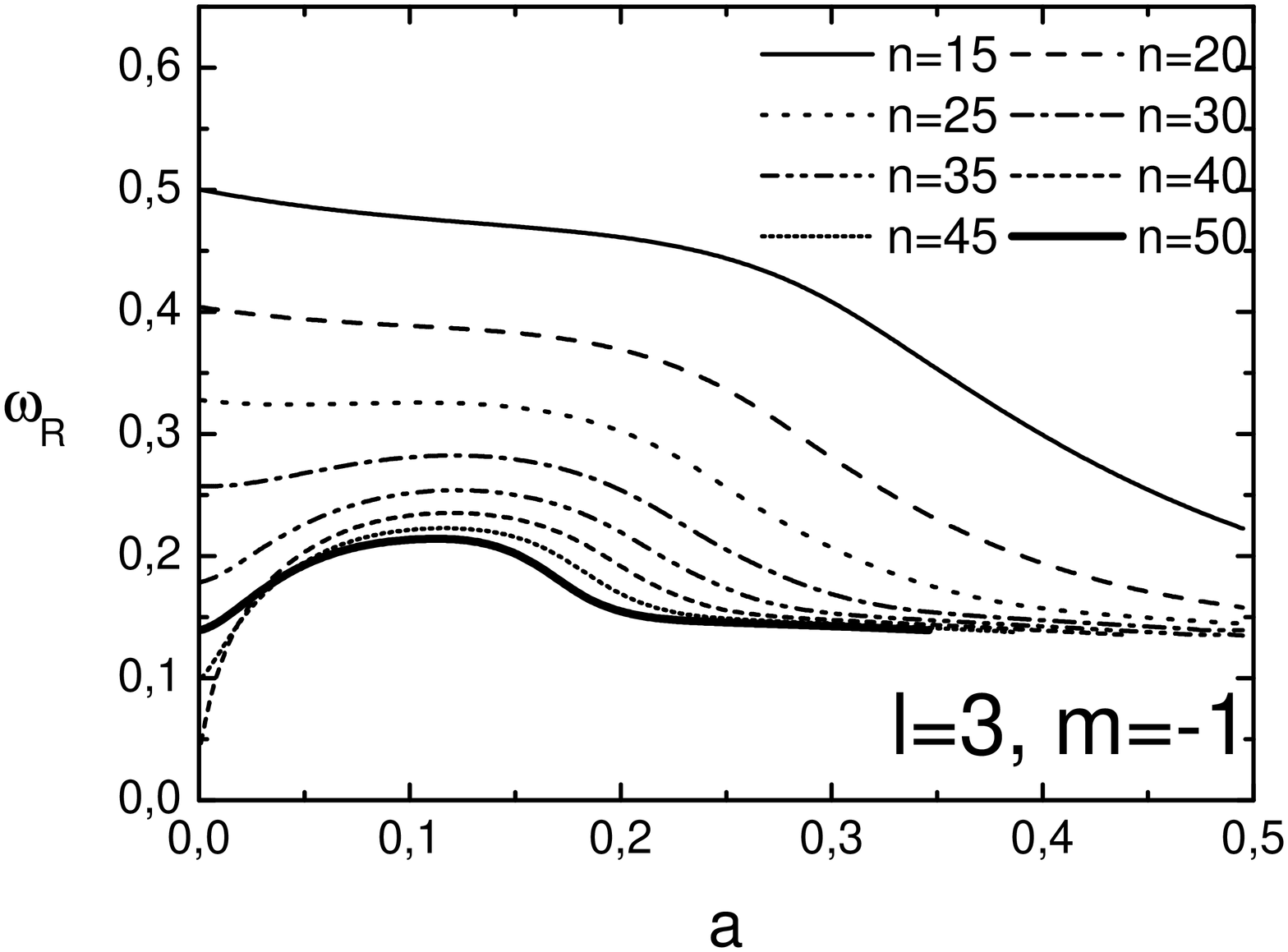}
\caption{
Real parts of some modes having $l=3$ and different values of $m$
(indicated in the plots). When $m>0$, the observed behaviour is
reminiscent of modes having $l=2$, $m=1$ (see figure \ref{fig3}).
Modes having $m<0$ approach a (roughly) constant value
$\omega_R=-m\varpi$ (we only show modes having $m=-1$), as they do for
$l=2$ (see figure \ref{fig4}).
}\label{fig5}
\end{figure}

However, some prominent features emerge from the general behaviour of
the real part of the modes, as displayed in the different panels of
figure \ref{fig5}. First of all, contrary to our expectations, neither
the branch of modes having $m=3$ nor the branch having $m=2$ seem to
approach the limit $m\Omega$. These modes show a behaviour which is
more closely reminiscent of modes having $l=2$, $m=1$: the modes' real
part ``bends'' towards the zero--frequency axis, shows a minimum as a
function of $a$, and tends to $\omega_R=m$ as $a\to 1/2$. We stress
that the observed behaviour should not be taken as the final word on
the problem, since we should probably look at modes having $n\gtrsim
100$ to observe the true asymptotic behaviour. However, should not the
qualitative features of the modes at larger $n$ drastically change, we
would face a puzzling situation. Indeed, gravitational modes with
$l=m=2$ would have a rather unique asymptotic behaviour, that would
require more physical understanding to be motivated.

Another prominent feature is that, whenever $m>0$, there seems to be
an infinity of modes approaching the limit $\omega_R=m$ as the hole
becomes extremal. This behaviour confirms the general trend we
observed for $l=2$, $m>0$.

Finally, our calculations of modes having $m<0$ show, once again, that
these modes tend to approach $\omega_R=-m\varpi$, where $\varpi\simeq
0.12$. We display, as an example, modes having $l=3$ and $m=-1$ in the
bottom right panel of figure \ref{fig5}.

\section{Scalar and electromagnetic perturbations}\label{sec3}

The calculations we have performed for $l=3$ hint at the possibility
that modes having $l=m=2$ are the only ones approaching the limit
$\omega_R=m\Omega$. However, as we have explained, reaching the
asymptotic regime for modes having $l>2$ is difficult. The reason is
quite simple to understand. The (pure imaginary) Schwarzschild
algebraically special mode, located at
\begin{equation}
\tilde \Omega_l=\frac{-i{(l-1)l(l+1)(l+2)}}{ 6}
\label{AlgSp}
\end{equation}
and marking the onset of the asymptotic regime moves quickly downwards
in the complex plane as $l$ increases. Therefore, calculations in the
asymptotic regime rapidly become impracticable as $l$ grows.

This technical difficulty is a hindrance if we want to test whether
the observed behaviour of gravitational modes having $l=m=2$ is in
some sense ``unique''. An alternative idea to check this
``uniqueness'' is to look, not at gravitational modes having different
$l$'s, but at perturbations due to fields having different spin and
$l\leq 2$. In particular, here we show some results we obtained
extending our calculation to scalar ($s=0$) and electromagnetic
($s=-1$) modes. Results for Kerr scalar modes, to our knowledge, have
only been published in \cite{GA}; some highly damped electromagnetic
modes were previously computed in \cite{onozawa}.

\subsection{Scalar modes}

\begin{figure}[htbp]
\centering
\includegraphics[angle=270,width=8cm,clip]{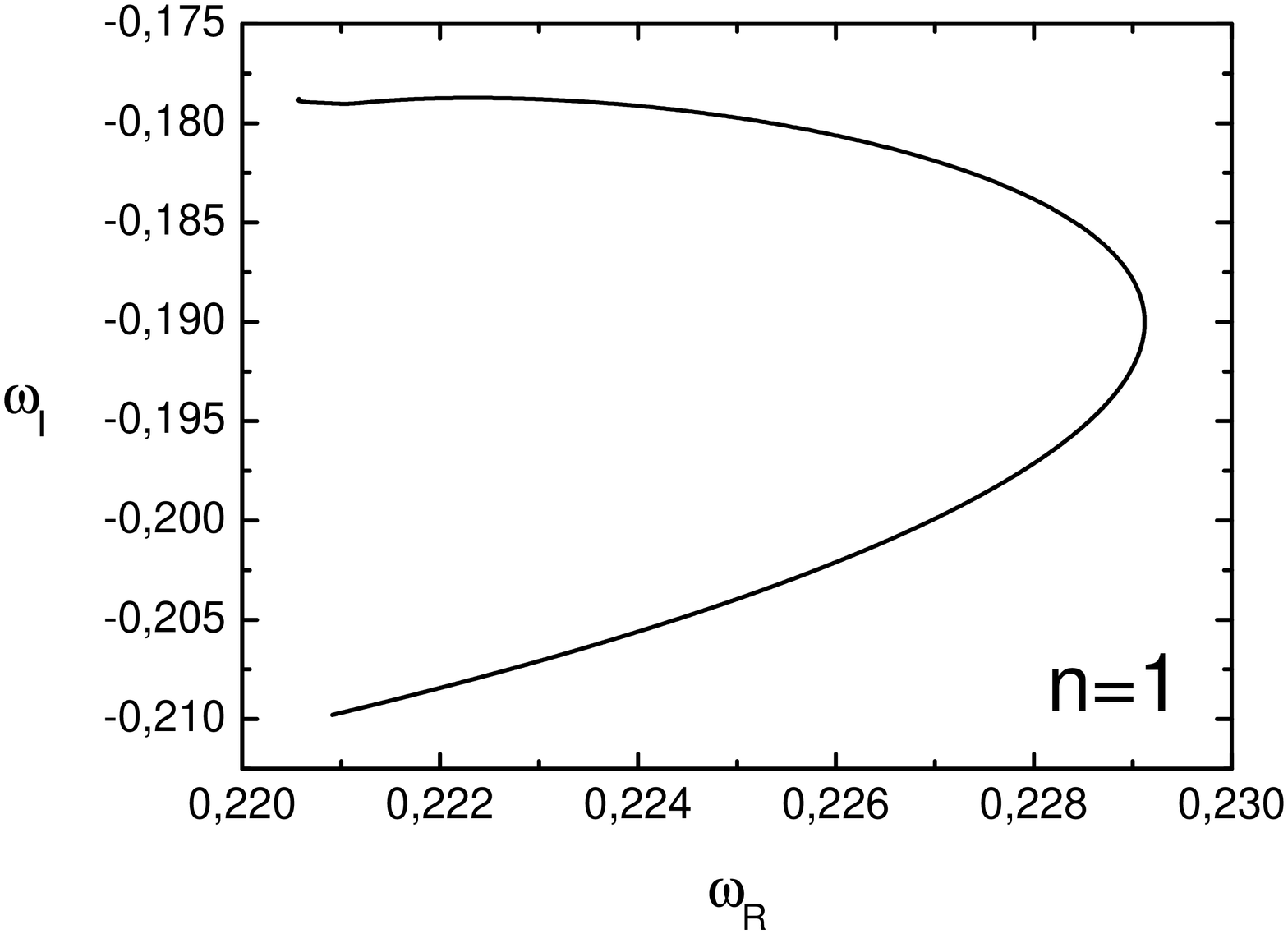}
\includegraphics[angle=270,width=8cm,clip]{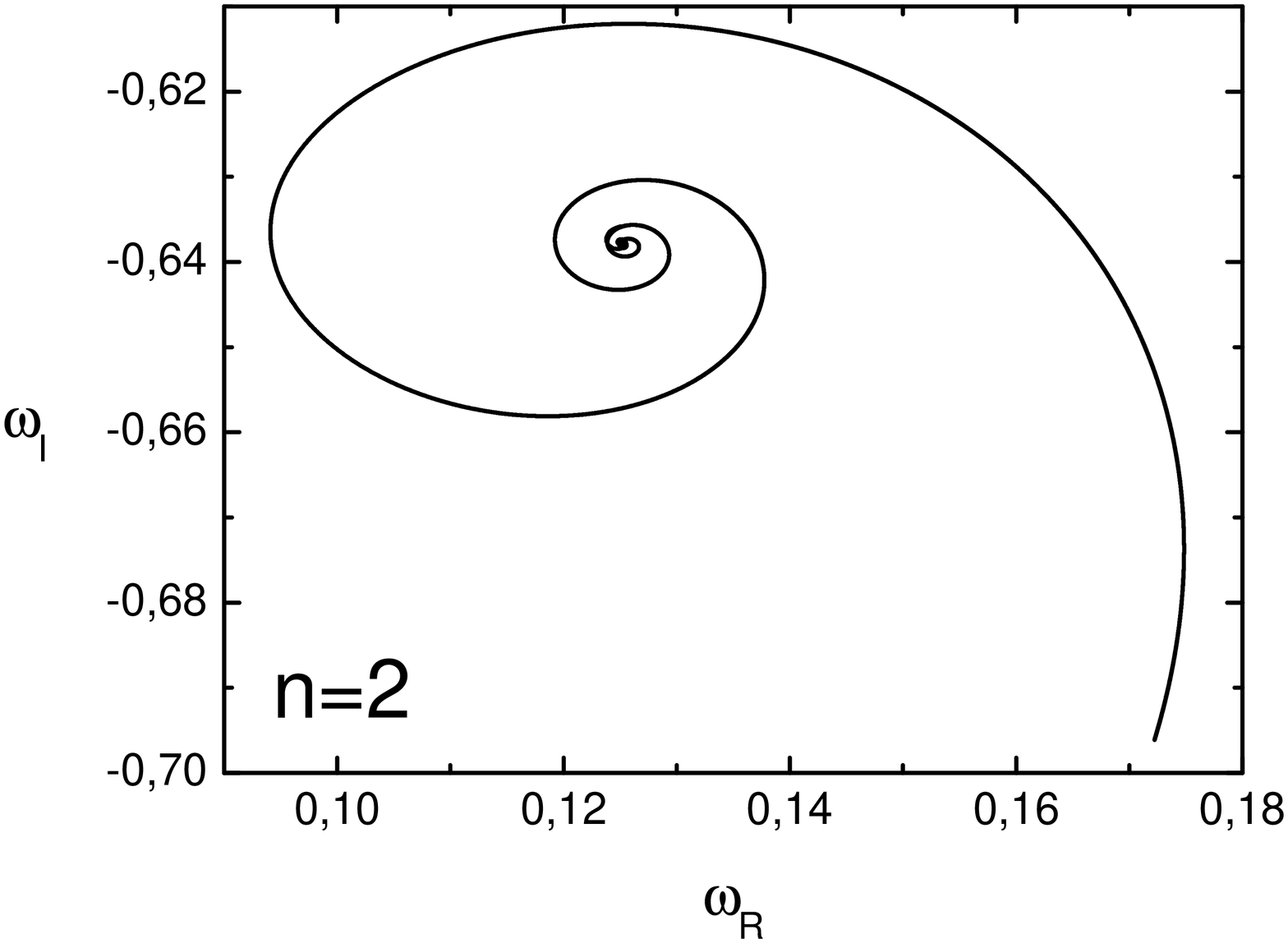}
\includegraphics[angle=270,width=8cm,clip]{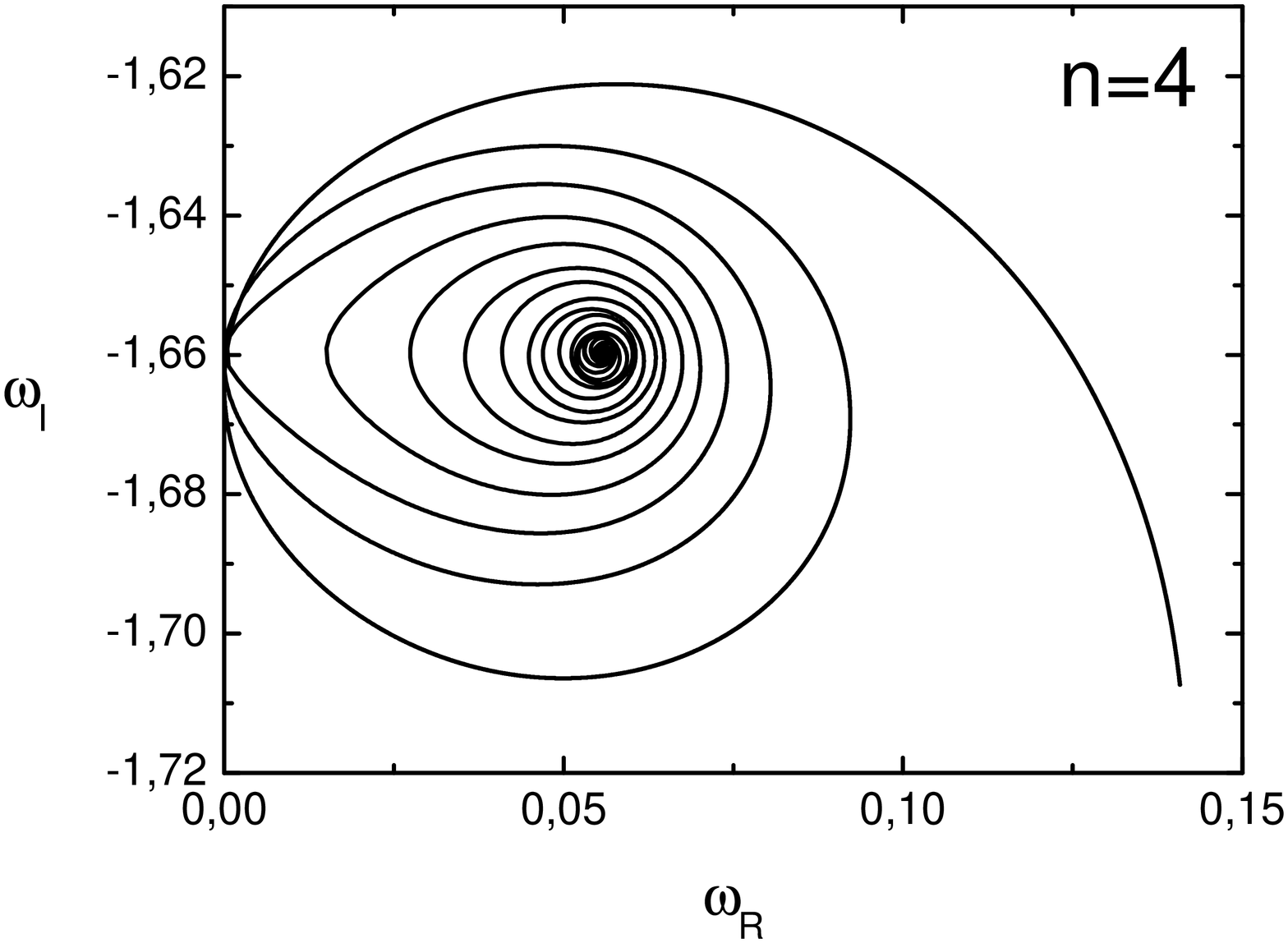}
\includegraphics[angle=270,width=8cm,clip]{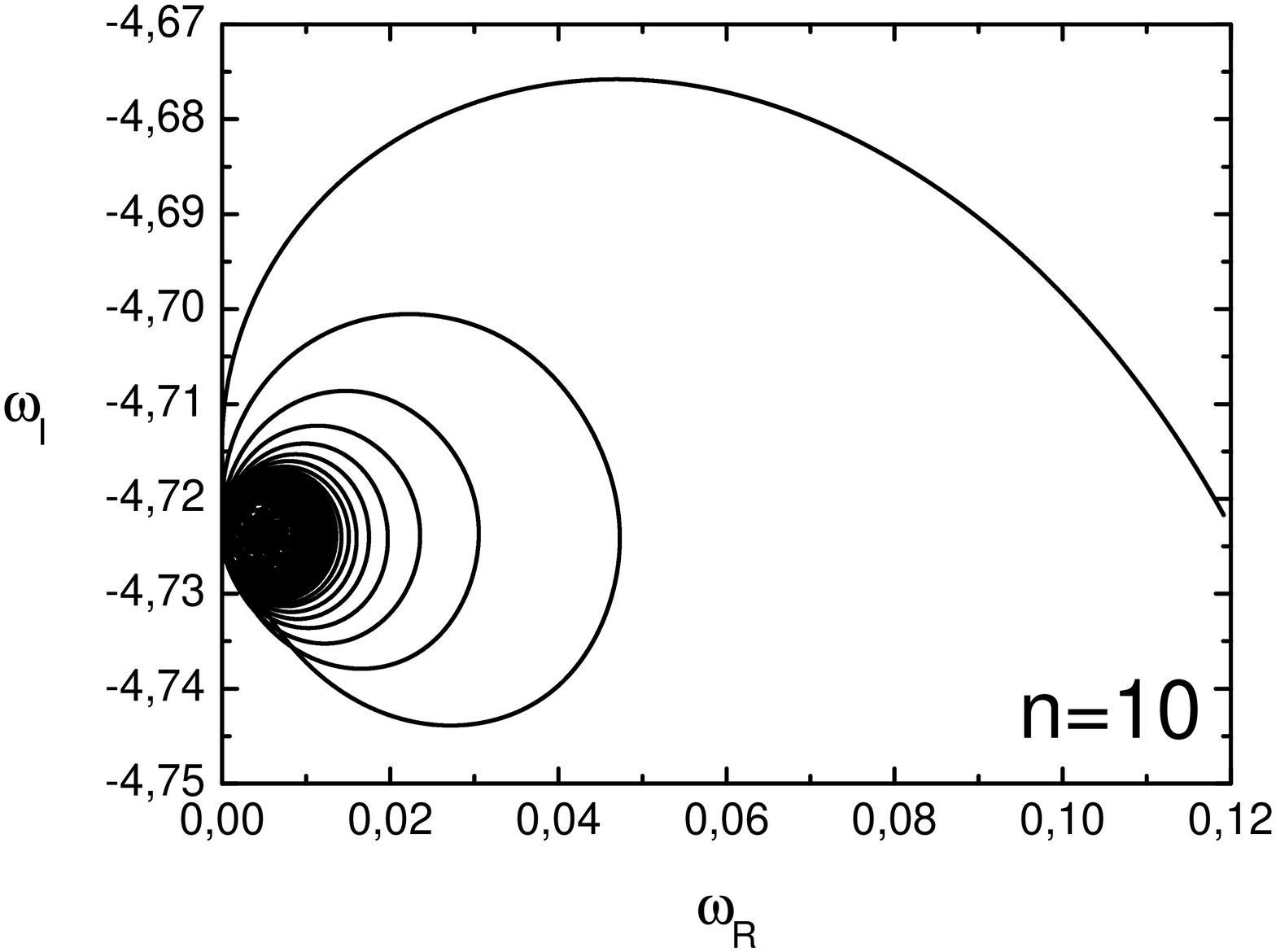}
\caption{
Trajectories of a few scalar modes having $l=m=0$. The different
panels correspond to the fundamental mode (top left), showing no
spiralling behaviour, and to modes having overtone indices
$n=2,~4,~10$.
}\label{fig6}
\end{figure}

In figure \ref{fig6} we show a few scalar modes having $l=m=0$. As
we could expect from existing calculations \cite{bertikerr,GA} the modes show
the typical spiralling behaviour; the surprise here is that this
spiralling behaviour sets in very quickly, and is particularly
pronounced even if we look at the first overtone ($n=2$). As the mode
order grows, the number of spirals grows, and the centre of the spiral
(corresponding to extremal Kerr holes) moves towards the pure
imaginary axis, at least for $n\lesssim 10$.

\begin{figure}[htbp]
\centering
\includegraphics[angle=270,width=8cm,clip]{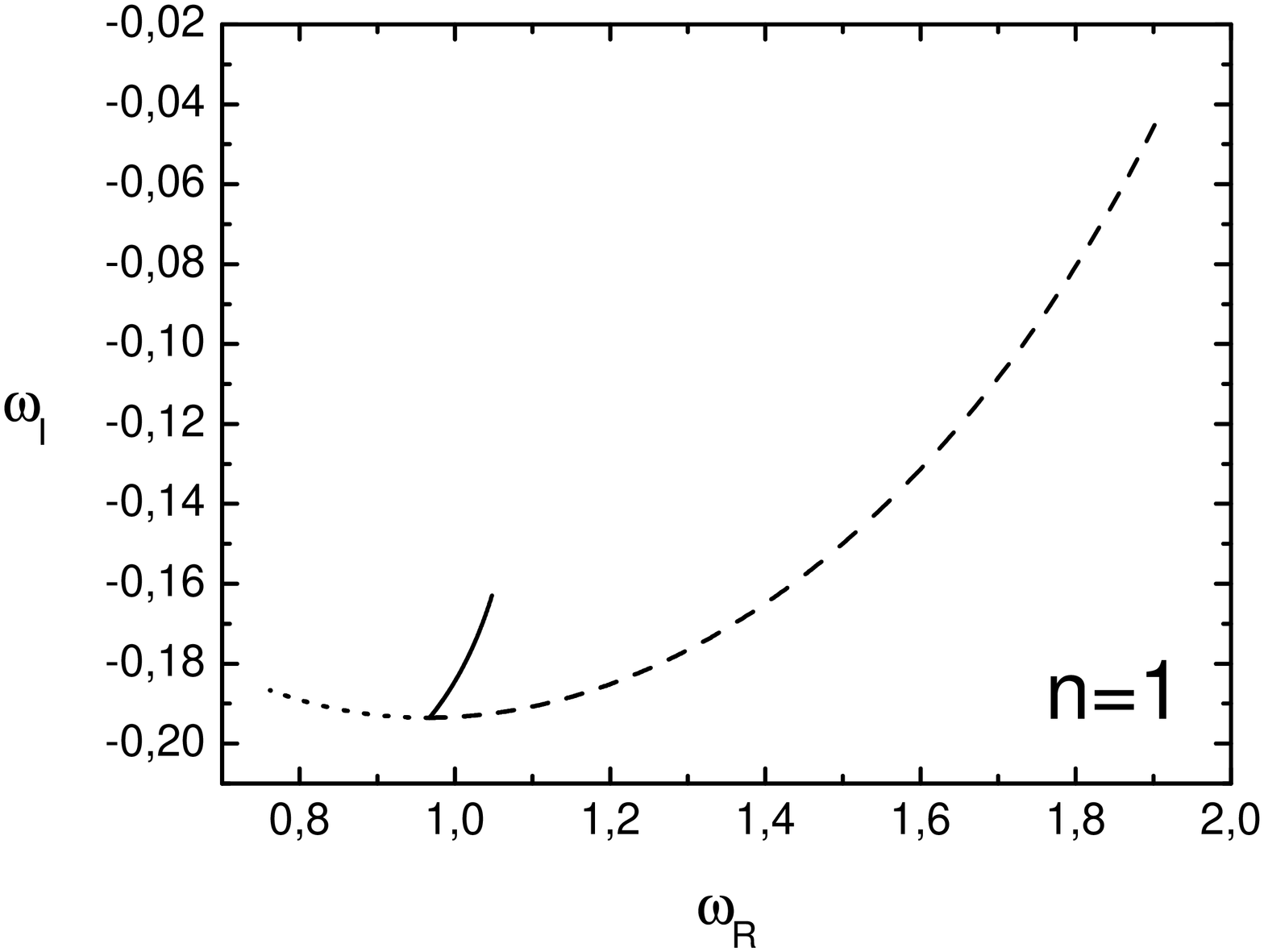}
\includegraphics[angle=270,width=8cm,clip]{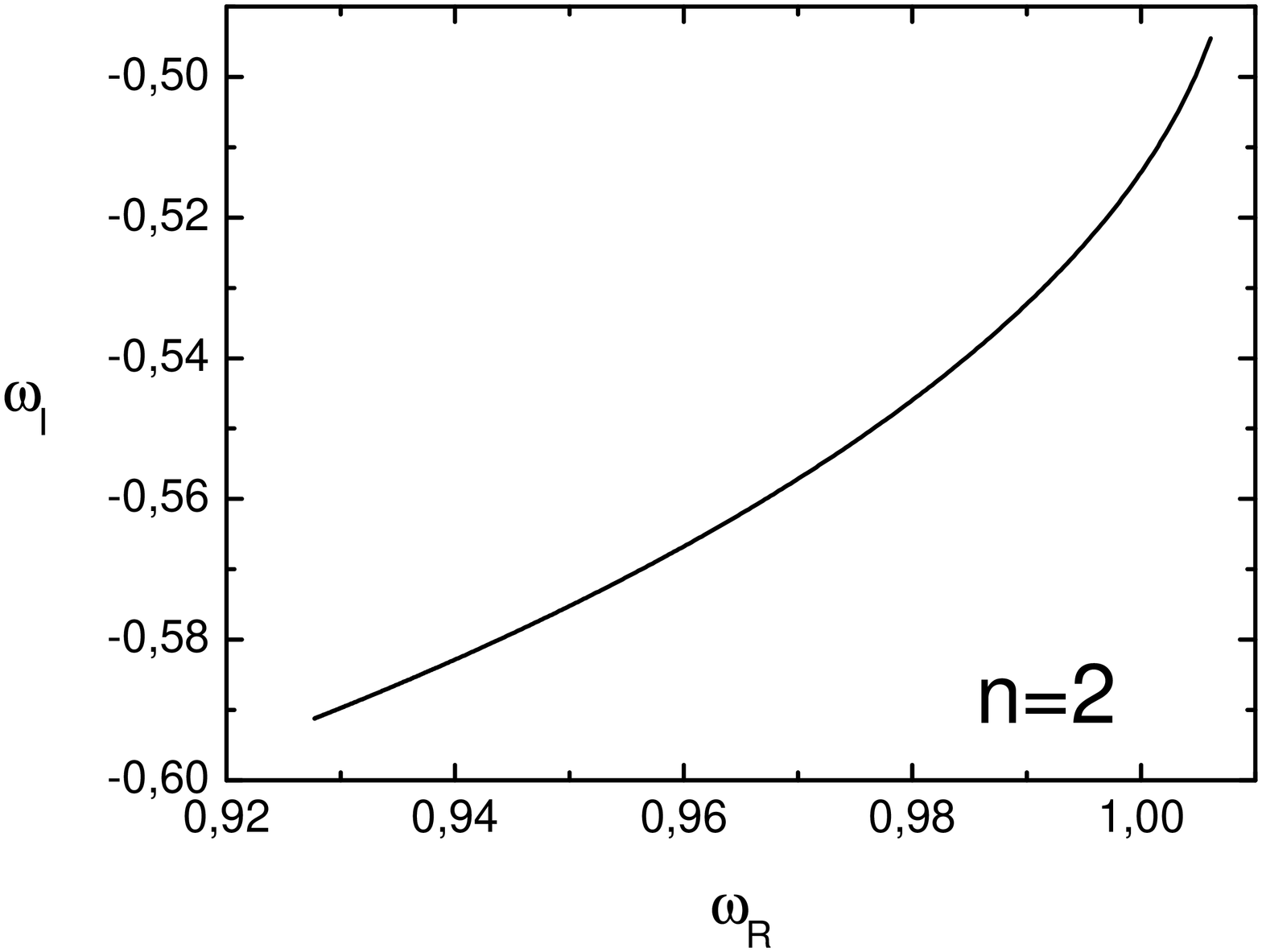}
\includegraphics[angle=270,width=8cm,clip]{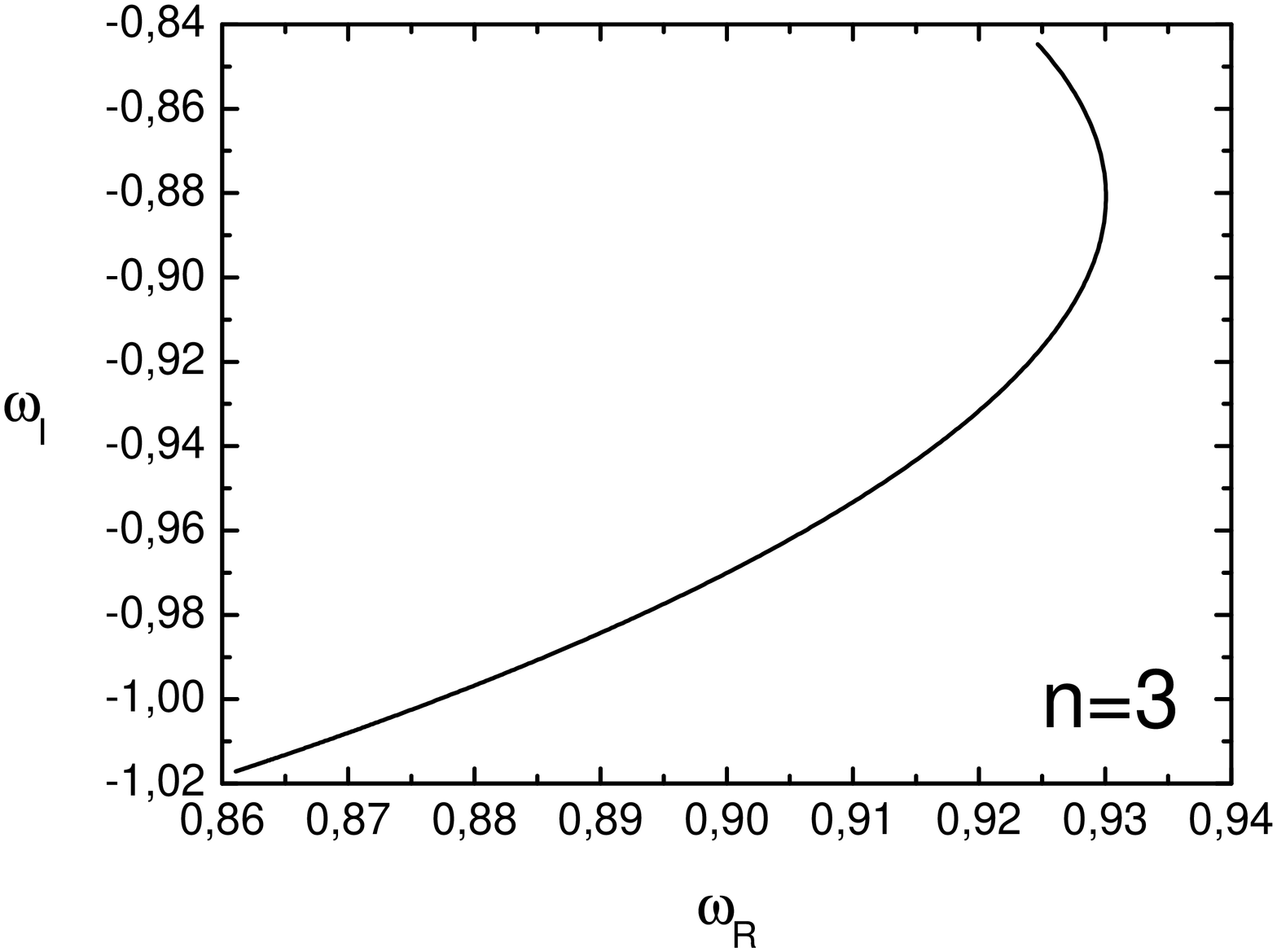}
\includegraphics[angle=270,width=8cm,clip]{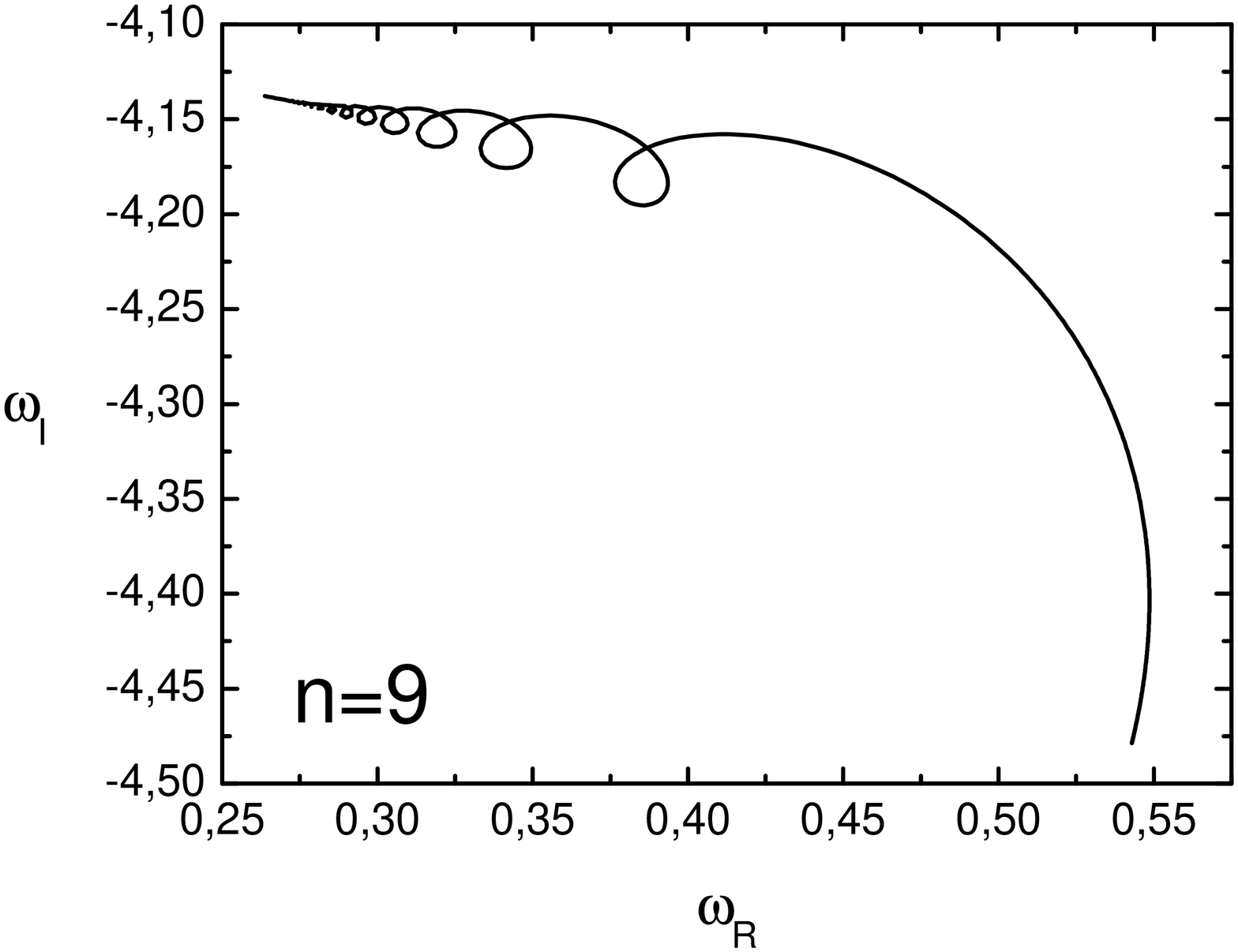}
\caption{
Trajectories of a few scalar modes for $l=2$. In the top left panel we
show how rotation removes the degeneracy of modes having different
$m$'s, displaying three branches (corresponding to $m=2,~0,~-2$)
``coming out of the Schwarzschild limit'' for the fundamental mode
($n=1$). In the top right and bottom left panel we show the
progressive ``bending'' of the trajectory of the $m=0$ branch for the
first two overtones ($n=2,~3$). Finally, in the bottom right panel we
show the typical spiralling behaviour for a mode with $m=0$ and
$n=9$. This plot can be compared to figure 6 in \cite{GA} (notice that
their scales have to be multiplied by two to switch to our units). The
continued fraction method allows us to compute modes for larger values
of $a$ (and is presumably more accurate) than the Pr\"ufer method.
}\label{fig7}
\end{figure}

In figure \ref{fig7} we show the trajectories of some scalar modes for
$l=2$. As can be seen in the top left panel, rotation removes the
degeneracy of modes having different values of $m$. If we follow modes
having $m=0$ we see the usual spiralling behaviour, essentially
confirming results obtained in \cite{GA} using the Pr\"ufer
method. However our numerical technique seems to be more accurate than
Pr\"ufer's method, which is intrinsically approximate, and we are able
to follow the modes up to larger values of the rotation parameter
(compare the bottom right panel in our figure 7 to figure 6 in
\cite{GA}, remembering that their numerical values must be multiplied
by a factor 2 due to the different choice of units).  On the basis of
our numerical results, it is quite likely that the asymptotic
behaviour of scalar modes having $l=m=0$ is described by a formula
similar to (\ref{MNf}). However, at present, no such formula has been
analytically derived.

\begin{figure}[htbp]
\centering
\includegraphics[angle=270,width=8cm,clip]{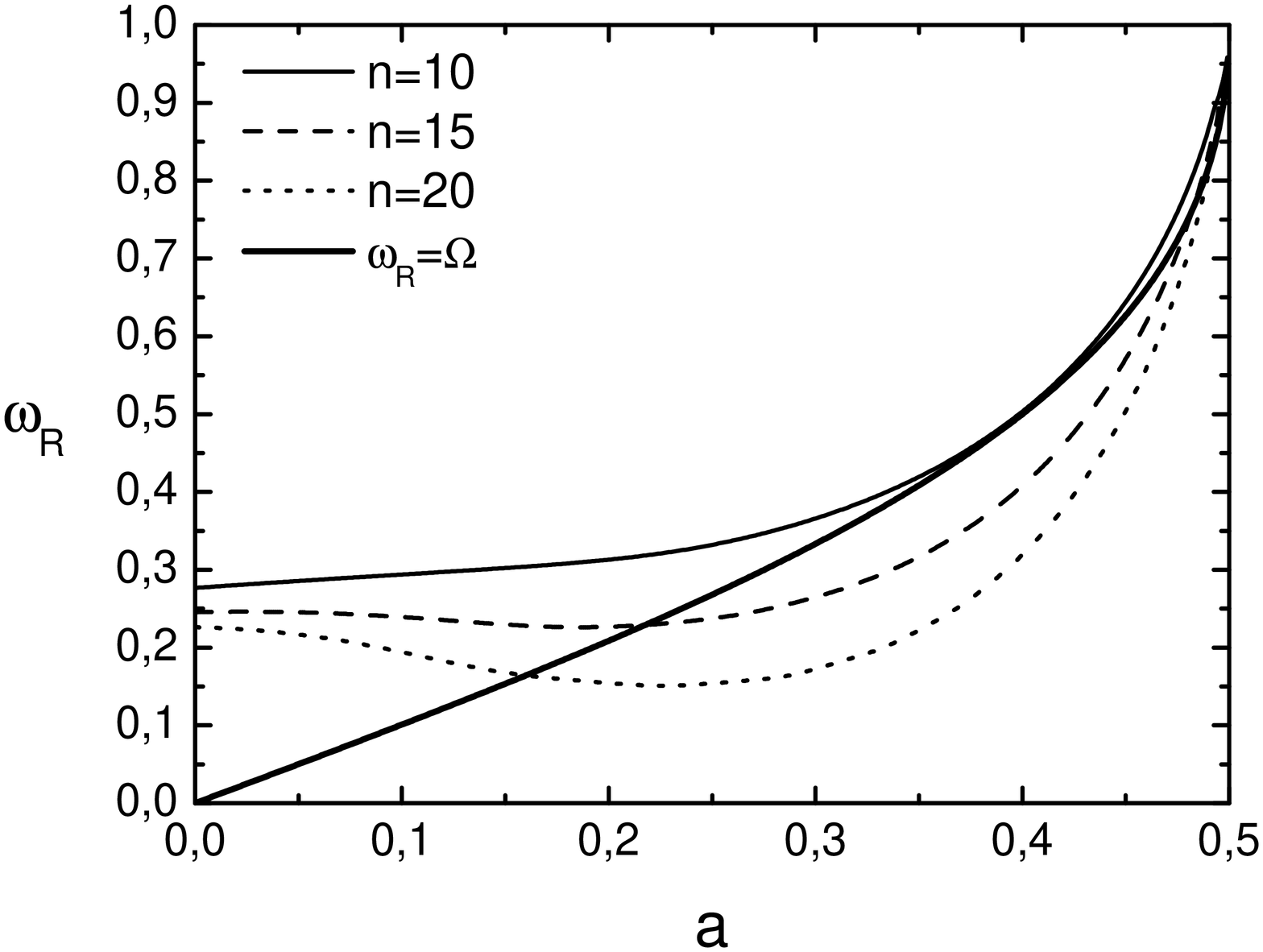}
\includegraphics[angle=270,width=8cm,clip]{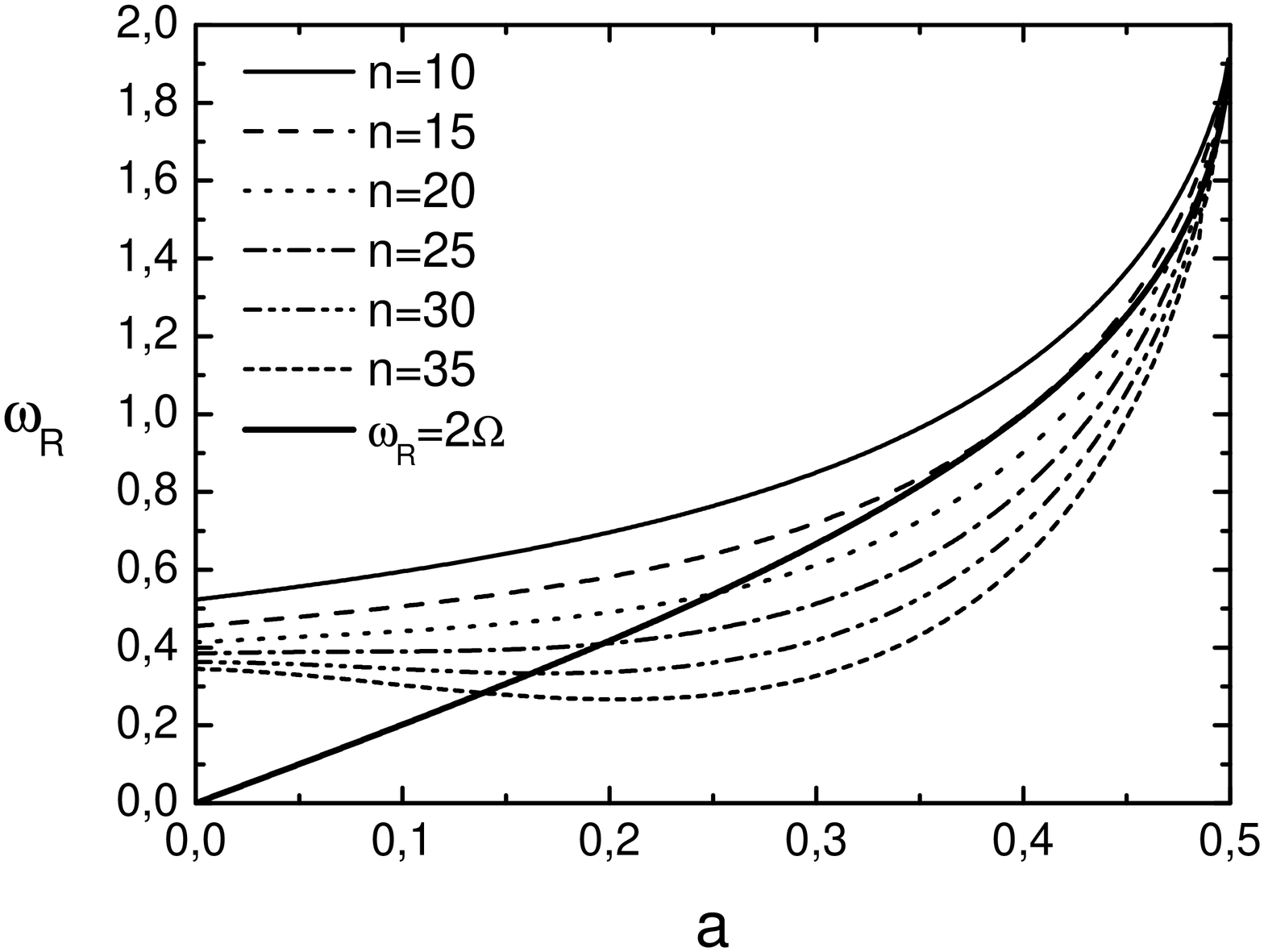}
\caption{
Real parts of the scalar modes having $l=m=1$, $l=m=2$. The observed behaviour is reminiscent of figures \ref{fig3} and \ref{fig5}.
}\label{fig8}
\end{figure}

In figure \ref{fig8} we show the real part of scalar modes having
$l=m=1$ and $l=m=2$ as a function of $a$ for increasing values of the
mode index.  In both cases modes do not show a tendency to approach
the $\omega_R=m\Omega$ limit suggested by gravitational modes with
$l=m=2$. Once again (as we observed for modes having $l=3$ and $m>0$)
the behaviour is more similar to that of gravitational modes with
$l=2$ and $m=1$. This may been considered further evidence that
gravitational perturbations having $l=m=2$ are indeed a very special
case.

\subsection{Electromagnetic modes}

\begin{figure}[htbp]
\centering
\includegraphics[angle=270,width=8cm,clip]{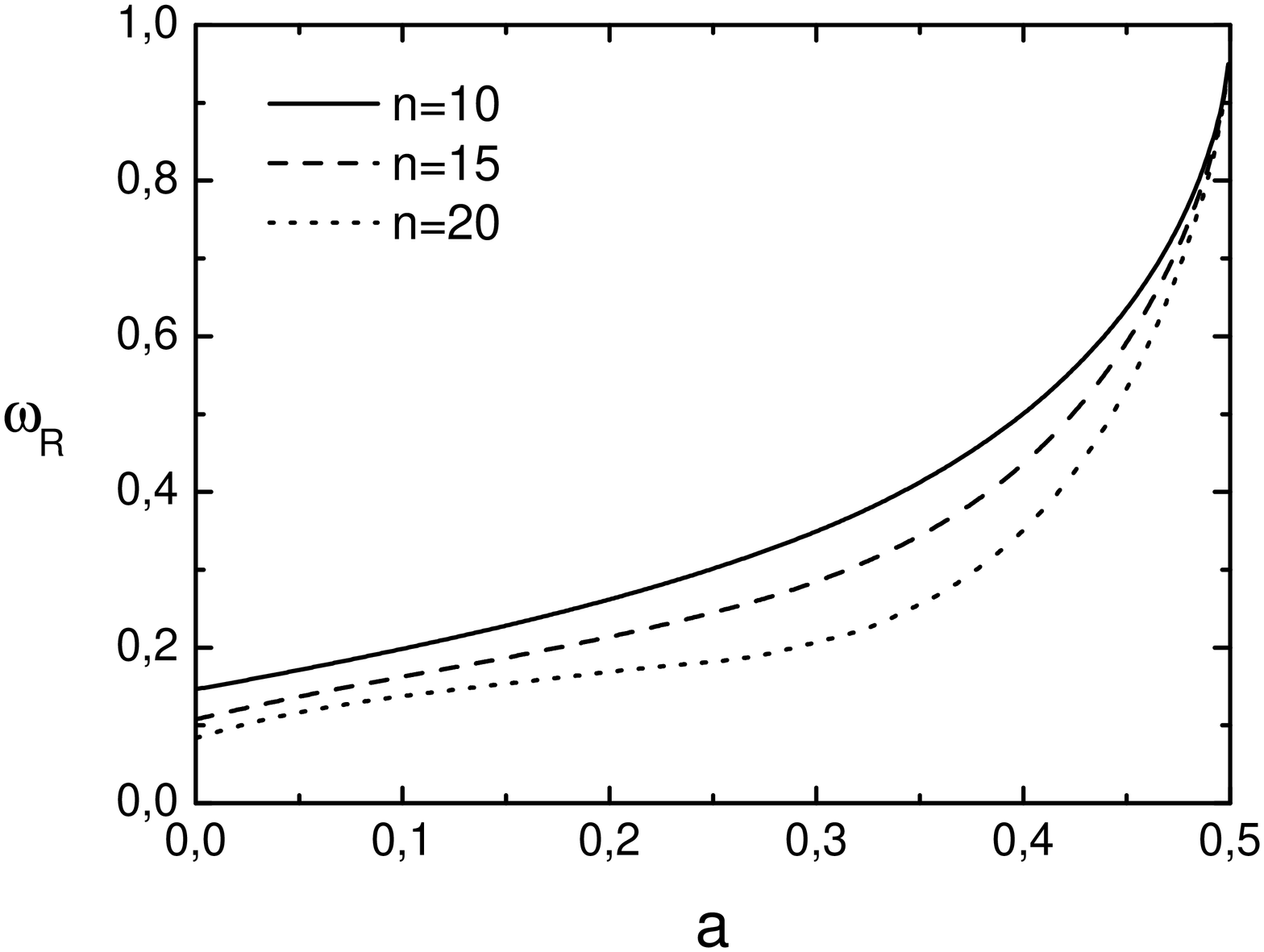}
\includegraphics[angle=270,width=8cm,clip]{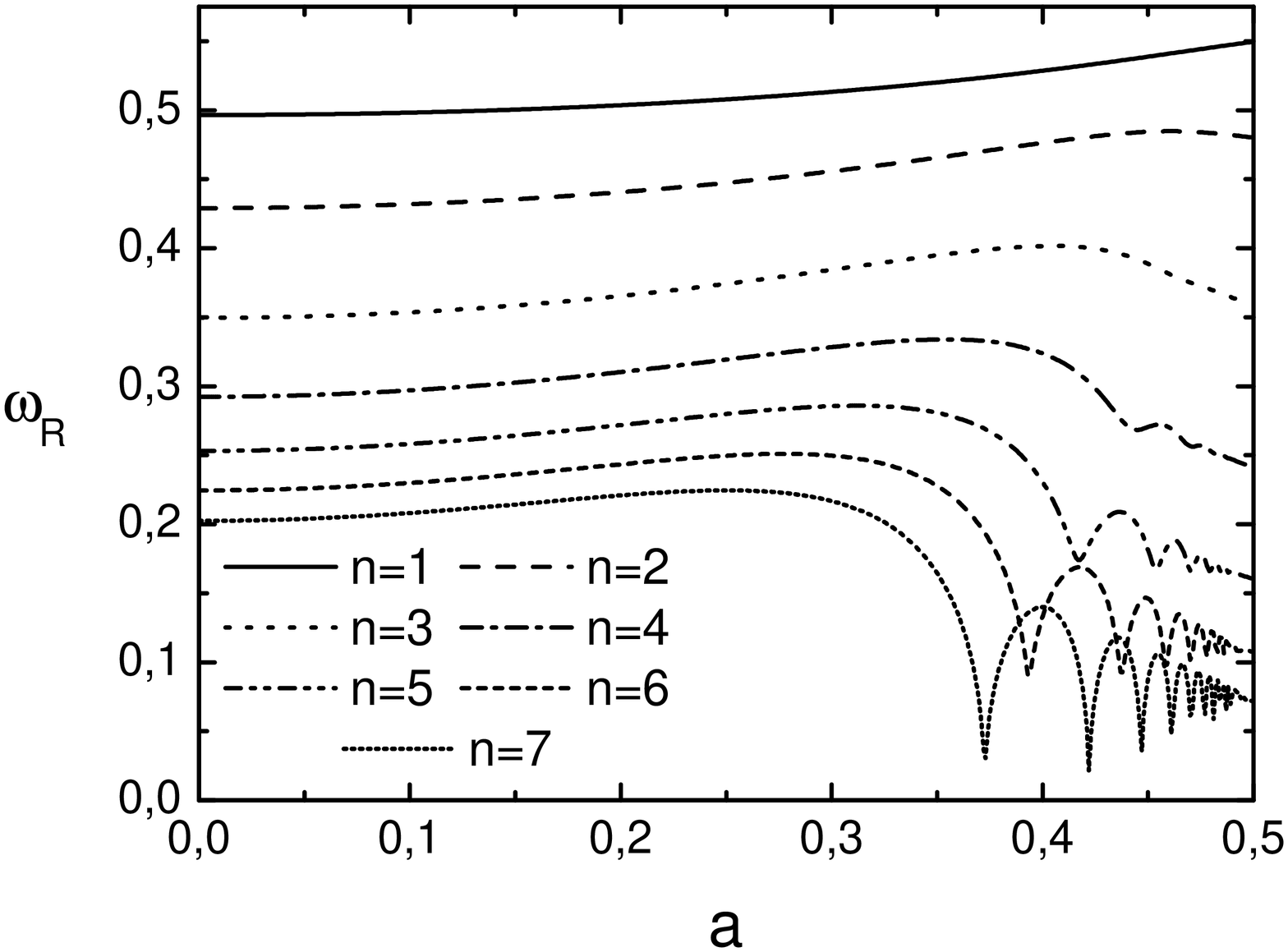}
\includegraphics[angle=270,width=8cm,clip]{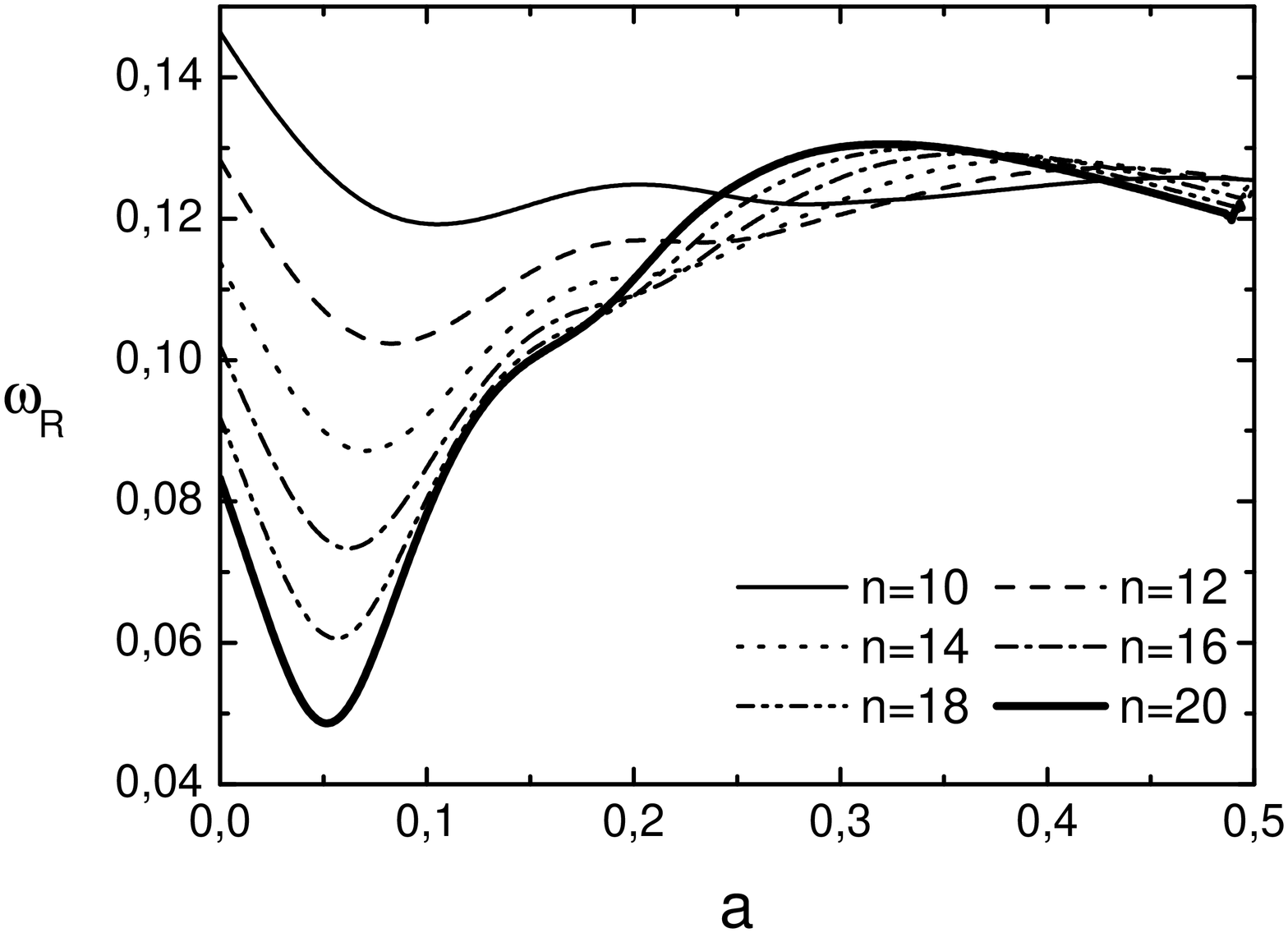}
\includegraphics[angle=270,width=8cm,clip]{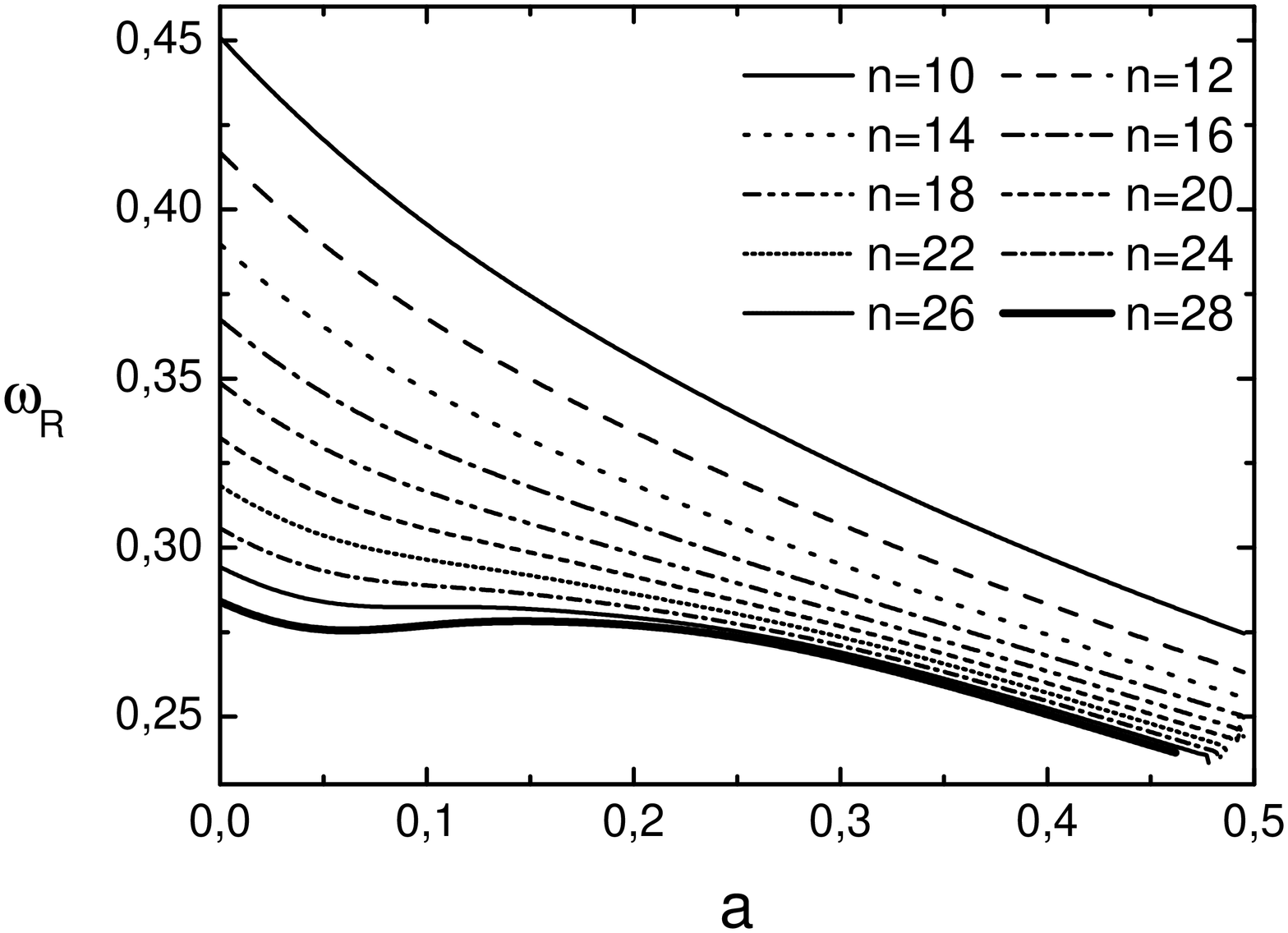}
\caption{
Real part of electromagnetic modes having $l=m=1$ (top left), $l=1$,
$m=0$ (top right), $l=1$, $m=-1$ (bottom left) and $l=2$, $m=-2$
(bottom right) as a function of the rotation parameter $a$, for
increasing values of the mode index.
}\label{fig9}
\end{figure}

The calculation of highly damped electromagnetic QNM's basically
confirms the picture we obtained from the computation of scalar QNM's
presented in the previous section. We show some selected results in
figure \ref{fig9}. The top left panel of figure \ref{fig9} shows that
the real part of electromagnetic QNM's having $m>0$ tend, for large
damping, to show a local minimum and approach the limit $\omega_R=m$
as $a\to 1/2$. The top right panel shows that the real parts of modes
having $l=1$ and $m=0$ quickly start oscillating (that is, QNM's
display spirals in the complex-$\omega$ plane). Finally, the bottom
plots show the behaviour of modes having $l=1$, $m=-1$ (left) and
$l=2$, $m=-2$ (right). Once again, modes seem to approach a roughly
constant value $\omega\simeq -m\varpi$. However, our computations
probably break down too early to draw any final conclusion on this.

\section{The asymptotic behaviour of the modes' imaginary part}\label{sec4}

The evidence for a universality of behaviour emerging from the
calculations we have presented in the previous sections is
suggestive. Calling for some caution, due to the fact that we may not
always be in the asymptotic regime when our numerical codes become
unreliable, we can still try and draw some conclusions. Our results
suggest that, whatever the kind of perturbation (scalar,
electromagnetic or gravitational) that we consider, asymptotic modes
belong to one of three classes:

1) Modes having $m>0$: their real part probably approaches the limit
$\omega_R=m \Omega$ only for gravitational modes having $l=m$ (our
calculation for $l=m=3$ is not conclusive in this respect, since we
are not yet in the asymptotic regime). For other kinds of
perturbations, or for $m\neq l$, $\omega_R$ apparently shows a minimum
as a function of $a$. This may be a real feature of asymptotic modes,
but it may as well be due to the asymptotic behaviour emerging only
for larger values of $n$. To choose between the two alternatives we
would either require better numerical methods or the development of
analytical techniques. A ``universal'' feature is that, whatever the
spin of the perturbing field, QNM frequencies approach the limiting
value $\omega_R=m$ as $a\to 1/2$.

2) Modes having $m=0$: these modes show a spiraling behaviour in the
complex plane, reminiscent of RN QNM's.

3) Modes having $m<0$: their real part seems to asymptotically
approach a constant (or weakly $a$--dependent) limit $\omega_R\simeq
-m\varpi$, where $\varpi\simeq 0.12$, whatever the value of $l$ and
the spin of the perturbing field. Maybe this limit is not exactly
independent of $a$, but on the basis of our numerical data we are
quite confident that highly damped modes with $m<0$ tend to a
universal limit $\omega_R\simeq -m \varpi_{ext}$, where $\varpi_{ext}$
has some value between $0.11$ and $0.12$, as $a\to 1/2$.

What about the modes' imaginary part? 
In \cite{bertikerr} we observed that the following formula holds
for gravitational modes having $l=m=2$:
\begin{equation}
\omega^{Kerr}_{l=m=2}=2\Omega+i2\pi T_H n.
\end{equation}

Our numerical data show that, in general, all modes having $m>0$, {\it
whatever the kind of perturbation} (scalar, electromagnetic or
gravitational) we consider, have an asymptotic separation equal to
$2\pi T_H$ if $m>0$. For $m=0$ the imaginary part oscillates, and this
beatiful, general result does not hold.  It turns out that it doesn't
hold as well for modes having $m<0$.  An analysis of our numerical
data did not lead us, up to now, to any conclusion on the asymptotic
separation of modes having $m<0$. This may hint at the fact that for
$m<0$ our calculations are not yet indicative of the asymptotic
regime. Therefore, some care is required in drawing conclusions on
asymptotic modes from our results for $m<0$.


\section{Algebraically special modes}\label{sec5}

\subsection{An introduction to the problem}

Algebraically special modes of Schwarzschild black holes have been
studied for a long time, but only recently their understanding has
reached a satisfactory level. Among the early studies rank those of
Wald \cite{W} and of Chandrasekhar \cite{Cas}, who gave the exact
solution of the Regge--Wheeler, Zerilli and Teukolsky equations at the
algebraically special frequency. The nature of the QNM boundary
conditions at the Schwarzschild algebraically special frequency is
extremely subtle, and has been studied in detail by Van den Brink
\cite{MVDBas}. Black hole oscillation modes belong to three
categories:

1) ``standard'' QNM's, which have outgoing wave boundary conditions at
both sides (that is, they are outgoing at infinity and ``outgoing into
the horizon'', using Van den Brink's ``observer-centered definition''
of the boundary conditions); 

2) total transmission modes from the left (TTM$_L$'s) are modes
incoming from the left and outgoing to the other side;

3) total transmission modes from the right (TTM$_R$'s) are modes
incoming from the right and outgoing to the other side.


In our units, the Schwarzschild ``algebraically special'' frequency is
given by formula (\ref{AlgSp}), and has been traditionally associated
with TTM's.  However, when Chandrasekhar found the exact solution of
the perturbation equations at the algebraically special frequency he
did not check that these solutions satisfy TTM boundary conditions. In
\cite{MVDBas} it was shown that, in general, they don't. An important
conclusion reached in \cite{MVDBas} is that the Regge--Wheeler
equation and the Zerilli equation (which are known to yield the same
QNM spectrum, being related by a supersymmetry transformation) have to
be treated on different footing at $\tilde \Omega_l$, since the
supersymmetry transformation leading to the proof of isospectrality is
singular there. In particular, the Regge-Wheeler equation has {\it no
modes at all} at $\tilde \Omega_l$, while the Zerilli equation has
{\it both a QNM and a TTM$_L$}.

Numerical calculations of algebraically special modes have yielded
some puzzling results. Leaver \cite{L} (studying the Regge-Wheeler
equation, that should have no QNM's at all according to Van den
Brink's analysis, and not the Zerilli equation) found a QNM which is
very close, but not exactly located {\it at}, the algebraically
special frequency. Namely, he found QNM's at frequencies $\tilde
\Omega'_l$ such that 
\begin{equation}
\tilde \Omega'_2=0.000000-3.998000i, \qquad
\tilde \Omega'_3=-0.000259-20.015653i.
\end{equation}
Notice that the ``special'' QNM's $\tilde \Omega'_l$ are such that
$\Re i\tilde \Omega'_2<|\tilde \Omega_2|$, $\Re i\tilde
\Omega'_3>|\tilde \Omega_3|$, and that the real part of $\tilde
\Omega'_3$ is not zero. Van den Brink \cite{MVDBas} speculated that
the numerical calculation may be inaccurate and the last three digits
may not be significant, so that no conclusion can be drawn on the
coincidence of $\tilde \Omega_l$ and $\tilde \Omega'_l$, ``if the
latter does exist at all''.

An independent calculation was carried out by Andersson
\cite{A}. Using a phase--integral method, he found that the
Regge--Wheeler equation has pure imaginary TTM$_R$'s which are very
close to $\tilde \Omega_l$ for $2\leq l\leq 6$. He therefore suggested
that the modes he found coincide with $\tilde \Omega_l$, which would
then be a TTM.  Van den Brink \cite{MVDBas} observed that, if all
figures in the computed modes are significant, the coincidence of
TTM's and QNM's is not confirmed by this calculation, since $\tilde
\Omega'_l$ and $\tilde \Omega_l$ are numerically (slightly) different.


Onozawa \cite{onozawa} showed that the Kerr mode having $n=9$ tends to
$\tilde \Omega_l$ as $a\to 0$, but suggested that modes approaching
$\tilde \Omega_l$ from the left and the right may cancel each other at $a=0$,
leaving only the special (TTM) mode. He also calculated this (TTM)
special mode for Kerr black holes, solving the relevant condition that
the Starobinsky constant should be zero and finding the angular
separation constant by a continued fraction method; his results
improved upon the accuracy of those previously obtained in \cite{Cas}.


The analytical approach adopted in \cite{MVDBas} clarified many
aspects of the problem for Schwarzschild black holes, but the
situation concerning Kerr modes branching from the algebraically
special Schwarzchild mode is still far from clear. In \cite{MVDBas}
Van den Brink, using slow--rotation expansions of the perturbation
equations, drew two basic conclusions on these modes. The first is
that, for $a>0$, the so--called Kerr special modes (that is, solutions
to the condition that the Starobinsky constant should be zero
\cite{Cas,onozawa}) are all TTM's (left or right, depending on the sign of
the spin). The TTM$_R$'s cannot survive as $a\to 0$, since they do not
exist in the Schwarzschild limit; this is related to the limit $a\to
0$ being a very tricky one. In particular, in this limit, the special
Kerr mode becomes a TTM$_L$ for $s=-2$; furthermore, the special mode
and the TTM$_R$ cancel each other for $s=2$. Studying the limit $a\to
0$ in detail, Van den Brink reaches a second important conclusion: the
Schwarzschild special frequency $\tilde \Omega_l$ is a limit point for
a multiplet of ``standard'' Kerr QNM's, which for small $a$ are well
approximated by
\begin{equation}
\omega=-4i-{\frac{33078176}{700009}}ma+{\frac{3492608}{ 41177}}ia^2
+{\cal O}(ma^2)
+{\cal O}(a^4)
\label{VDBsmalla}
\end{equation}
when $l=2$, and by a more complicated formula -- his equation (7.33)
-- when $l>2$. We will see in the next section that, unfortunately,
none of the QNM's we numerically found seems to approach this
limit when the rotation rate $a$ is small.

Van den Brink suggested (see note [46] in \cite{MVDBas}) that QNM's
corresponding to the algebraically special frequency and having $m>0$
may have one of the following three behaviours in the Schwarzschild
limit: they may merge with those having $m<0$ at a frequency $\tilde
\Omega'_l$ such that $|\tilde \Omega'_l|<|\tilde \Omega_l|$ (but
$|\tilde \Omega'_l|>|\tilde \Omega_l|$ for $l\geq 4$) and disappear,
as suggested by Onozawa \cite{onozawa}; they may terminate at some (finite)
small $a$; or, finally, they may disappear towards $\omega=-i\infty$.
Recently Van den Brink {\it et al.} \cite{MVDBdoublet} put forward
another possibility: studying the branch cut on the imaginary axis,
they found that in the Schwarzschild case a pair of ``unconventional
damped modes'' should exist. For $l=2$, these modes are identified by
a fitting procedure to be located at
\begin{equation}
\omega_\pm=\mp0.027+(0.0033-4)i.  
\label{unconv}
\end{equation}
An approximate analytical calculation confirms the presence of these
modes, yielding 
\begin{equation}
\omega_+\simeq-0.03248+(0.003436-4)i, 
\end{equation}
in reasonable agreement with (\ref{unconv}).  If their prediction is
true, an {\it additional} QNM multiplet should emerge, as $a$
increases, near $\tilde \Omega_l$; this multiplet {\it ``may well be
due to $\omega_\pm$ splitting (since spherical symmetry is broken) and
moving through the negative imaginary axis as $a$ is tuned''}
\cite{MVDBdoublet}.  In the following paragraph we will show that a
careful numerical search reveals, indeed, the emergence of such
multiplets, but these do not seem to behave exactly as predicted in
\cite{MVDBdoublet}.

\subsection{Numerical search and QNM multiplets}

As we have summarized in the previous paragraph, the situation for
Kerr modes branching from the algebraically special Schwarzschild mode
is still unclear. Is a multiplet of modes emerging from the
algebraically special frequency when $a>0$? Can QNM's be matched by
the analytical prediction (\ref{VDBsmalla}) at small values of $a$?
If a doublet does indeed appear, as recently suggested in
\cite{MVDBdoublet}, does it tend to the algebraically special
frequency $\tilde \Omega_2=-4i$ as $a\to 0$, does it tend to the values
predicted by formula (\ref{unconv}), or does it go to some other
limit?

\begin{figure}[htbp]
\centering
\includegraphics[angle=270,width=8cm,clip]{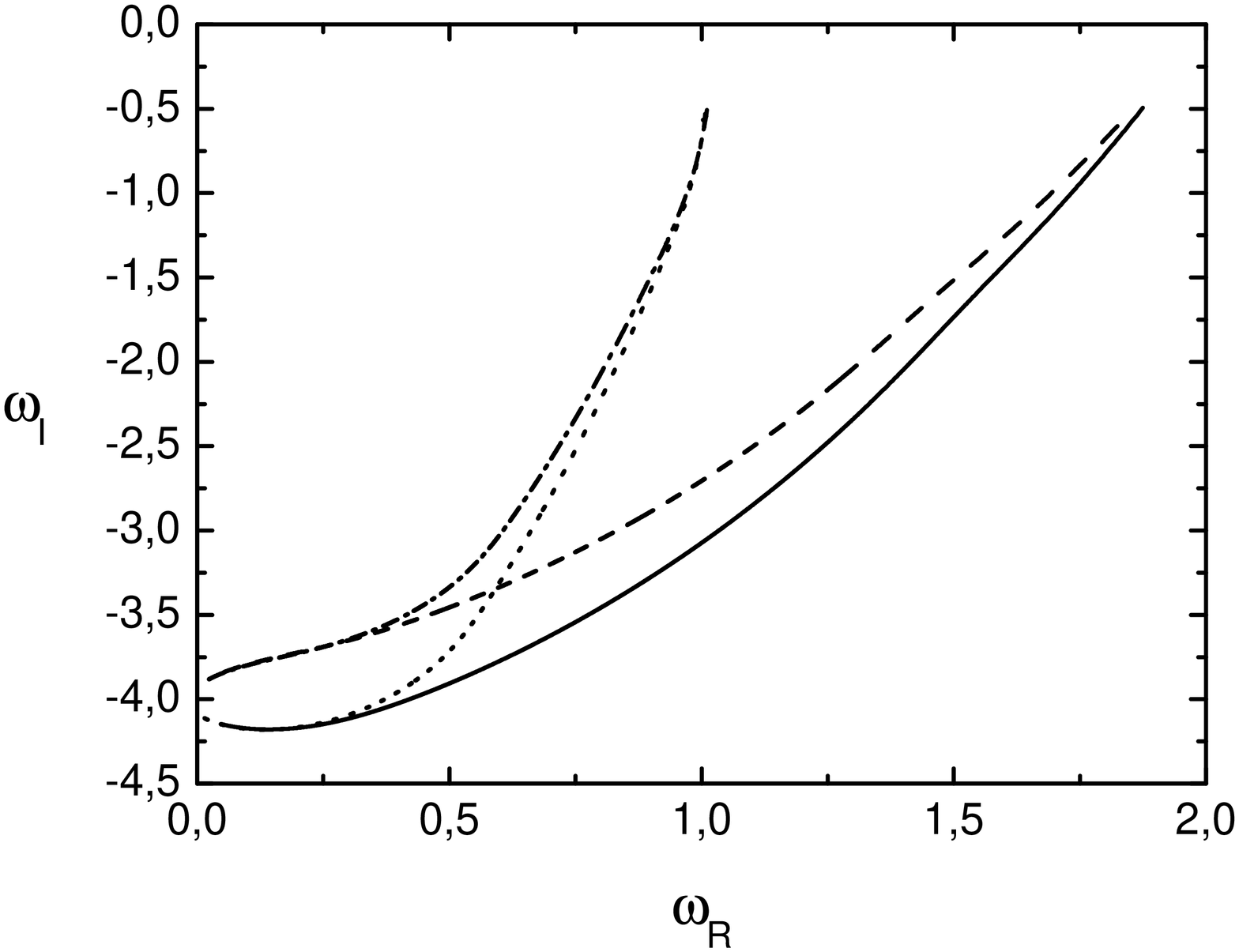}
\includegraphics[angle=270,width=8cm,clip]{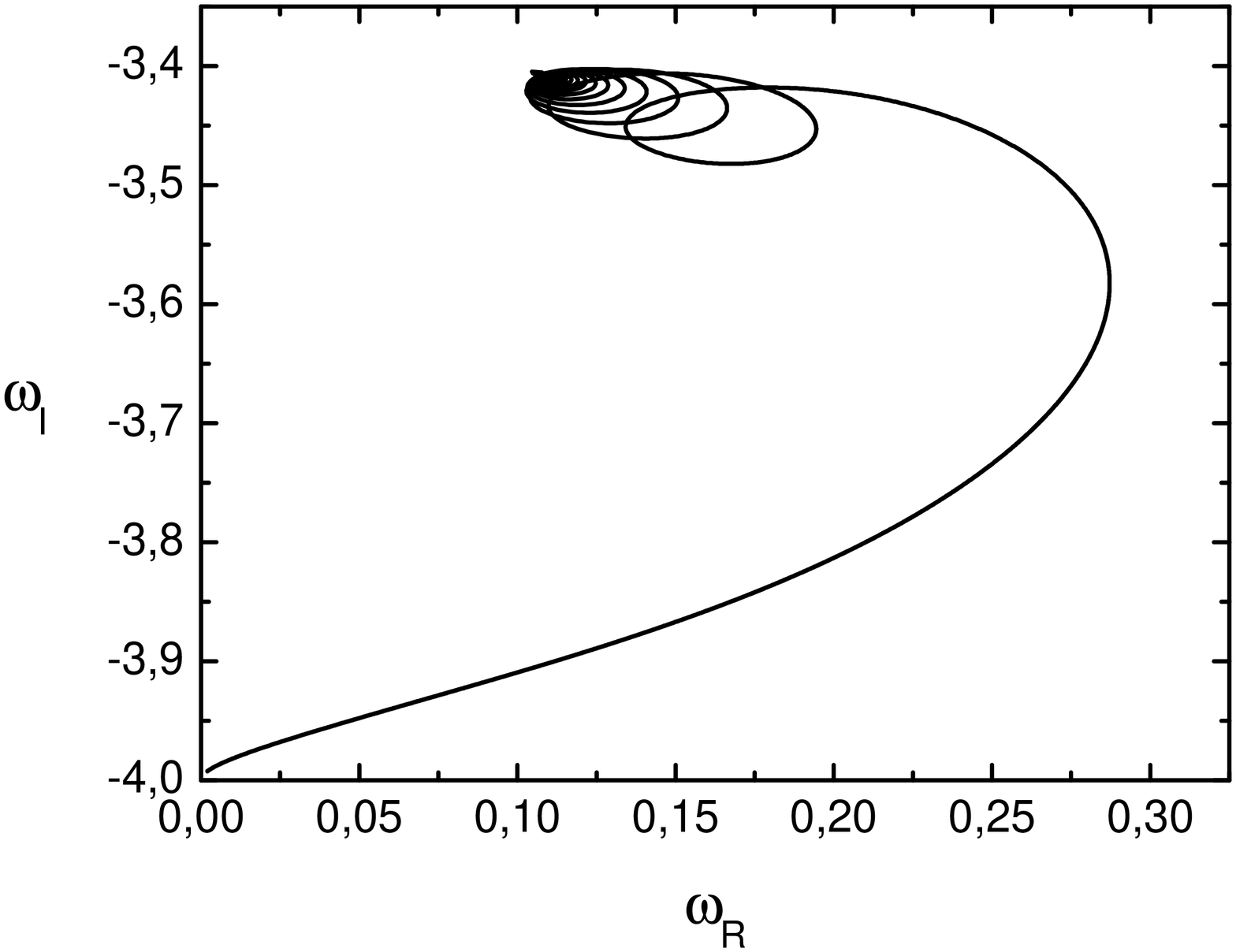}
\caption{
The left panel shows the trajectories described in the
complex-$\omega$ plane by the doublets emerging close to the
Schwarzschild algebraically special frequency ($\tilde \Omega_2=-4i$)
when $m>0$ and $l=2$. Notice that the real part of modes having $m>0$
tends to $\omega_R=m$ as $a\to 1/2$.  The right panel shows the
spiralling trajectory of the mode having $m=0$.
}\label{fig10}
\end{figure}

After carrying out an extensive numerical search with both of our
numerical codes, we have indeed found some surprises. Our main new
result is shown in the left panel of figure \ref{fig10}. There we
show the trajectories in the complex plane of QNM's having $l=2$ and
$m>0$: a {\it doublet} of modes does indeed appear close to the
algebraically special frequency. Both modes in the doublet tend to the
usual limit ($\tilde \Omega_2=m$) as $a\to 1/2$. We have tried to match
these ``twin'' modes with the predictions of the analytical formula
(\ref{VDBsmalla}). Unfortunately, none of the two branches we find
seems to agree with (\ref{VDBsmalla}) at small $a$.  Our searches
succeeded in finding a mode doublet only when $m>0$. For $m\leq 0$ the
behaviour of the modes is, in a way, more conventional. For example,
in the right panel of figure \ref{fig10} we see the $l=2$, $m=0$
mode coming out of the standard algebraically special frequency
$\tilde \Omega_2$ and finally describing the ``usual'' spirals as $a$
increases.

\begin{figure}[htbp]
\centering
\includegraphics[angle=270,width=8cm,clip]{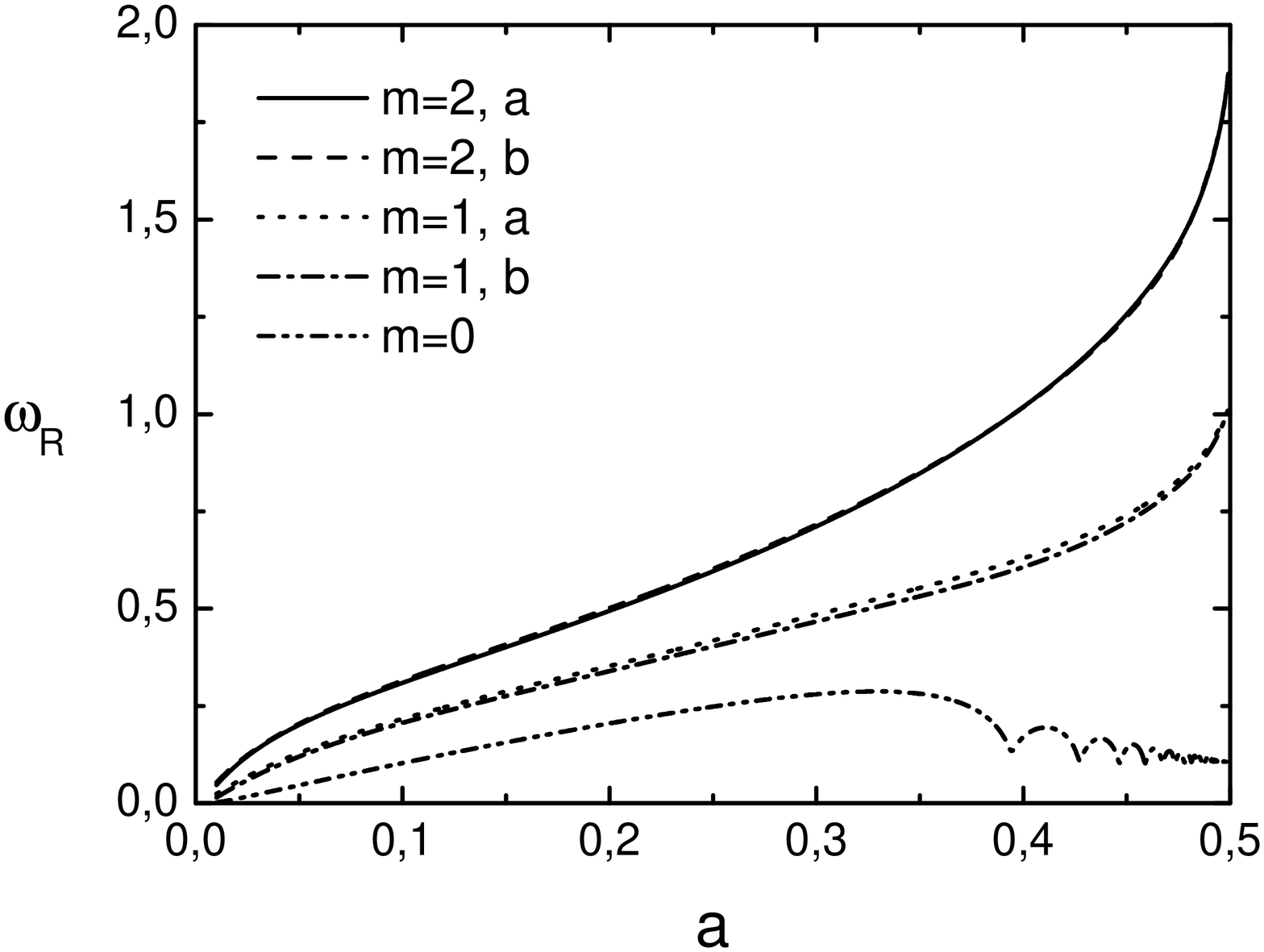}
\includegraphics[angle=270,width=8cm,clip]{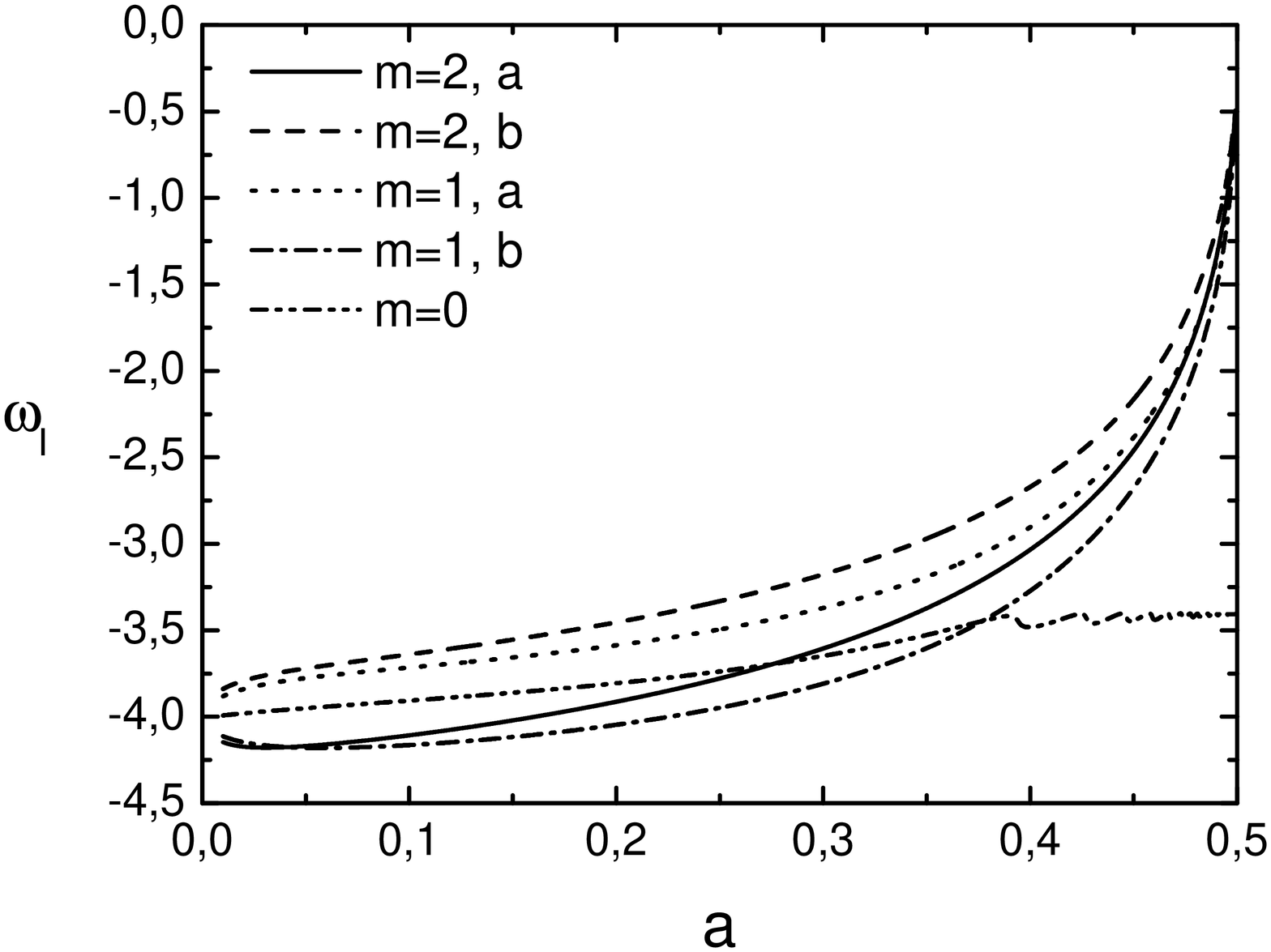}
\includegraphics[angle=270,width=8cm,clip]{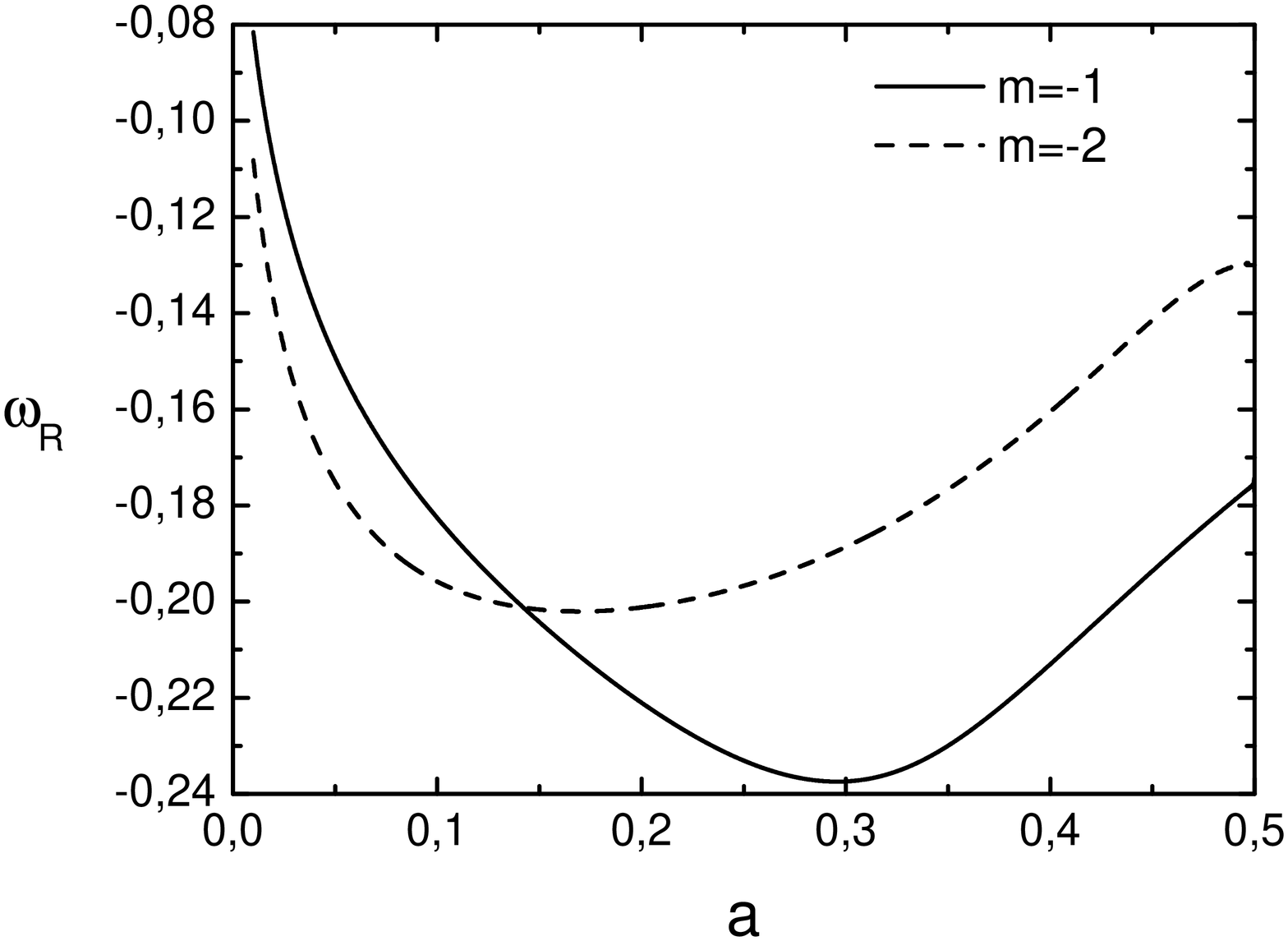}
\includegraphics[angle=270,width=8cm,clip]{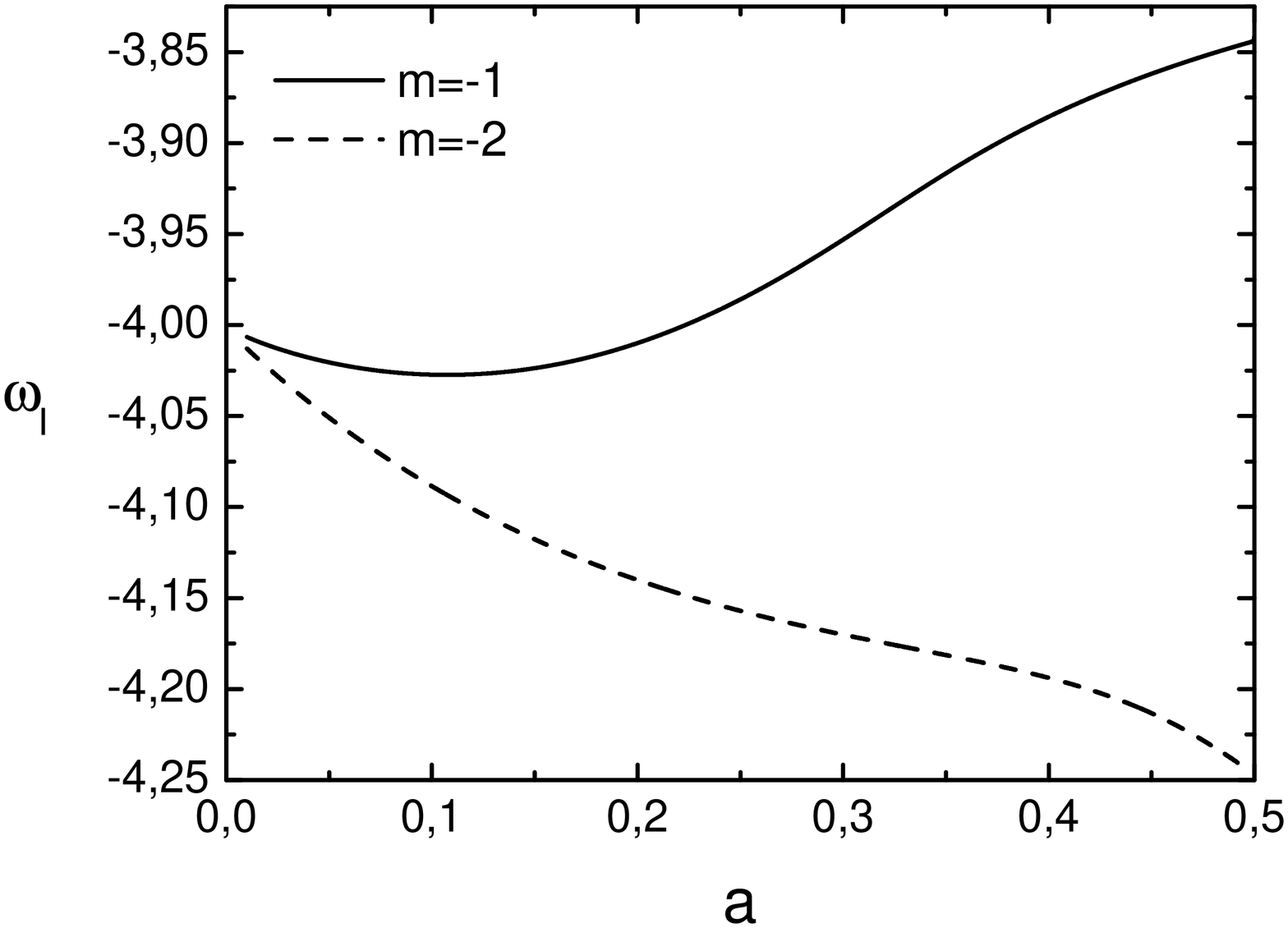}
\caption{
The top row shows the real and imaginary parts (left and right,
respectively) of the ``doublet'' of QNMs emerging from the
algebraically special frequency as functions of $a$.  The doublets
only appear when $m>0$. We also overplot the real and imaginary parts
of the mode having $l=2$, $m=0$ (showing the usual oscillatory
behaviour). The bottom row shows, for completeness, the real and
imaginary parts (left and right, respectively) of modes having
negative $m$ and branching from the algebraically special frequency.
}\label{fig11}
\end{figure}

In the top left panel of figure \ref{fig11} we see that the real part
of all modes having $m\geq 0$ does indeed go to zero as $a\to 0$, with
an $m$--dependent slope. However, the top right panel in the same
figure shows that the imaginary part of the modes does {\it not} tend
to $-4$ as $a\to 0$. This behaviour agrees quite well, at least
qualitatively, with that predicted by formula (\ref{unconv}).
Extrapolating our numerical data to the limit $a\to 0$ yields,
however, slightly different numerical values; our extrapolated values
for $l=2$ are $\omega=(-4-0.10)i$ and $\omega=(-4+0.09)i$.

At present, we have no explanation for the appearance of the doublet
only when $m>0$. A numerical confirmation of this behaviour comes from
numerical searches we have carried out close to the algebraically
special frequency $\tilde \Omega_3$. There we observed a similar
behaviour: a multiplet of modes appears only when $m>0$. In
particular, we see the appearance of a doublet that behaves quite
similarly to the modes shown in the left panel of figure
\ref{fig10}. Extrapolating the numerical data for the $l=3$ doublet
yields the values $\omega=(-20-0.19)i$ and $\omega=(-20+0.24)i$ as
$a\to 0$.

A more careful search around the algebraically special frequency
$\tilde \Omega_3$ surprisingly revealed the existence of other
QNM's. However, we don't trust our numerics enough to present our
findings in this chapter. The additional modes we find may well be
``spurious'' modes due to numerical inaccuracies, since we are pushing
our method to its limits of validity (very high dampings and very
small imaginary parts). Our numerical data will be made available on
request to those who may be interested.

\section{Conclusions}

In this chapter we have numerically investigated the behaviour of highly
damped QNM's for Kerr black holes, using two independent numerical
codes to check the reliability of our results. Our findings do not
agree with the simple behaviour conjectured by Hod's for the real part
of the frequency \cite{hod,H2} as given in formula (\ref{Hod}). We did
not limit our attention to gravitational modes, thus filling some gaps
in the existing literature. 

Our main results concerning highly damped modes can be summarized as
follows. Scalar, electromagnetic and gravitational modes show a
remarkable universality of behaviour in the high damping limit. The
asymptotic behaviour crucially depends, for any kind of perturbation,
on whether $m>0$, $m=0$ or $m<0$. As already observed in \cite{bertikerr},
the frequency of gravitational modes having $l=m=2$ tends to
$\omega_R=2 \Omega$, $\Omega$ being the angular velocity of the black
hole horizon. We showed that, if Hod's conjecture is valid, this
asymptotic behaviour is related to {\it reversible black hole
transformations}, that is, transformations for which the black hole
irreducible mass (and its surface area) does not change.  Other
(gravitational and non-gravitational) modes having $m>0$ do {\it not}
show a similar asymptotic behaviour in the range of $n$ allowed by our
numerical method. In particular, in the high--damping limit, the real
part of (gravitational and non--gravitational) modes having $m>0$
typically shows a minimum as a function of the rotation parameter $a$,
and then approaches the limit $\omega_R=m$ as the black hole becomes
extremal.  At present we cannot exclude the possibility that our
calculations actually break down {\it before} we reach the asymptotic
regime. Better numerical methods or analytical techniques are needed
to give a final answer concerning the asymptotic behaviour of modes
having $m>0$.  

An interesting new finding of this chapter is that for all values of
$m>0$, and for any kind of perturbing field, there seems to be an
infinity of modes tending to the critical frequency for superradiance,
$\omega_R=m$, in the extremal limit. This finding generalizes a
well--known analytical result by Detweiler for QNM's having $l=m$
\cite{De,GA}. It would be interesting to generalize Detweiler's proof,
which only holds for $l=m$, to confirm our conjecture that for {\it
any} $m>0$ there is an infinity of QNM's tending to $\omega_R=m$ as
$a\to 1/2$.  

The real part of modes having $l=2$ and {\it negative} $m$
asymptotically approaches a value $\omega_R\simeq -m\varpi$,
$\varpi\simeq 0.12$ being (almost) independent of $a$. Maybe this
limit is not exactly independent of $a$, but on the basis of our
numerical data we feel quite confident that highly damped modes with
$m<0$ do tend to a universal limit $\omega_R\simeq -m \varpi_{ext}$
(where $\varpi_{ext}$ has some value between $0.11$ and $0.12$) as
$a\to 1/2$.

This is an
interesting prediction, and it would again be extremely useful to
confirm it using analytical techniques. Up to now, we have not been
able to find any simple explanation for this limiting value. For
example, we have tentatively explored a possible connection between
$\varpi$ and the frequencies of marginally stable counterrotating
photon orbits, but we could not find any obvious correlation between
the two.

Both for gravitational and for non--gravitational perturbations, the
trajectories in the complex plane of modes having $m=0$ show a
spiralling behaviour, strongly reminiscent of the one observed for
Reissner-Nordstr\"om (RN) black holes, and probably well approximated
in the high damping limit by a formula similar to (\ref{MNf}).  

Last but not least, an important result concerning highly damped modes
is that, for any perturbing field, the asymptotic separation in the
imaginary part of consecutive modes having $m>0$ is given by $2\pi
T_H$ ($T_H$ being the black hole temperature). This is presumably
related to the fact that QNM's determine the position of the poles of
a Green's function on the black hole background, and that the
Euclidean black hole solution converges to a thermal circle at
infinity having temperature $T_H$; so it is not surprising that the
spacing in asymptotic QNM's coincides with the spacing, $2\pi i T_H$,
expected for a thermal Green's function \cite{motl2}.  This simple
relation concerning the mode spacing does not seem to hold when $m\leq
0$.


Finally, we studied numerically in some detail modes branching from
the so--called ``algebraically special frequency'' of Schwarzschild
black holes. We found numerically for the first time that QNM {\it
multiplets} emerge from the algebraically special modes as the black
hole rotation increases, confirming a recent speculation
\cite{MVDBdoublet}. However, we found some quantitative disagreement
with the analytical predictions presented in
\cite{MVDBas,MVDBdoublet}, that deserves further investigation.

Our numerical results will hopefully serve as an important guide in
the analytical search for asymptotic QNM's of Kerr black holes.
Although one can in principle apply Motl and Neitzke's \cite{motl2}
method in the present case, the Kerr geometry has some special
features that complicate the analysis. The Teukolsky equation
describing the field's evolution no longer has the
Regge-Wheeler-Zerilli (Schr\"odinger--like) form; however, it can be
reduced to that form by a suitable transformation of the radial
coordinate. The main technical difficulty concerns the fact that the
angular separation constant $A_{lm}$ is not given analytically in
terms of $l$, as it is in the Schwarzschild or RN geometry; even
worse, it depends on the frequency $\omega$ in a non linear
way. Therefore, an analytical understanding of the problem must
encompass also an understanding of the asymptotic properties of the
separation constant. The scalar case is well studied, both
analytically and numerically \cite{SepSca}, but a similar
investigation for the electromagnetic and gravitational perturbations
is still lacking. An idea we plan to exploit in the future is to use
a numerical analysis of the angular equation as a guideline to find
the asymptotic behaviour of $A_{lm}$. Once the asymptotic behaviour of
$A_{lm}$ is determined, the analysis of the radial equation may
proceed along the lines traced in \cite{motl2}.

\newpage

\begin{table}
\centering
\caption{
Caption of the table.
}
\vskip 12pt
\begin{tabular}{@{}ccccccc@{}}
\multicolumn{1}{c}{$r_+$} 
&\multicolumn{2}{c}{Scalar ($l=0$)} 
&\multicolumn{2}{c}{Axial ($l=2$)} 
&\multicolumn{2}{c}{Polar ($l=2$)} \\
\hline
$r_+$  &min &max 
       &min &max
       &min &max
\\
\hline
100  & 0.366   & 0.474  & 0.366 & 0.474 & 0.366 & 0.474\\
50   & 0.366   & 0.474  & 0.366 & 0.474 & 0.366 & 0.474\\
10   & 0.367   & 0.475  & 0.368 & 0.476 & 0.369 & 0.477\\
5    & 0.372   & 0.480  & 0.376 & 0.483 & 0.376 & 0.488\\
1    & 0.468   & 0.571  & -     & -     & 0.503 & -    \\
\end{tabular}
\label{minmax}
\end{table}

\part{Encounters of black holes with particles in four and higher dimensions:
quasinormal modes, gravitational and scalar radiation}

\thispagestyle{empty} \setcounter{minitocdepth}{1}
\chapter[Gravitational radiation: an introduction]{Gravitational radiation: an 
introduction} \label{chap:Intro2}
\lhead[]{\fancyplain{}{\bfseries Chapter \thechapter. \leftmark}}
\rhead[\fancyplain{}{\bfseries \rightmark}]{}
\minitoc \thispagestyle{empty}
\renewcommand{\thepage}{\arabic{page}}
\section{Introduction}
The existence of electromagnetic waves was predicted by Maxwell in the
1860's, and verified experimentally by Hertz less than two decades
later.  The usefulness of electromagnetic waves for our understanding
of the Universe is enormous, in fact, most of our knowledge is due to
their existence.  Each time one opened a new electromagnetic window
for our telescopes, for example the radio band or the X-ray band, some
revolution occurred.  Take for example the discovery of cosmic radio
waves in the 1930's and their detailed study in the 1940's, 50's and
60's.  Previously, the Universe, as seen by visible light, was very
serene and quiescent. It was, so it seemed, composed in the vast
majority by stars revolving smoothly in their orbits, and
evolving on timescales of millions or billions of years.  The radio
window showed us a completely new aspect of the Universe, a violent
one: galaxies were colliding, jets thrown away from galactic nuclei,
quasars with luminosities far greater than our galaxy evolving on
timescales of hours, pulsars with gigantic radio beams rotating many
times per second, and many others.  
The history of gravitational waves is not as
glamorous, but we are witnessing a dramatic turning of
events.  Gravitational waves were theoretically predicted
by Einstein in 1916, soon after the formulation of general relativity.
However, a series of failed attempts to analyze theoretically 
radiation reaction in
some systems in the late 1940's and 1950's shook physicist's faith in
the ability of the waves to carry off energy.  It required a clever
thought experiment by Bondi \cite{bondi} to restore faith in the
energy of the waves.  By now, we have a soundly based theory of
gravitational waves, an account of which may be found in the review
papers by Thorne
\cite{thorne} and Damour \cite{damour}.
Gravitational waves have not been experimentally detected
directly, however there has been an indirect verification, in the form
of the observed inspiral of a binary pulsar due to gravitational
radiation reaction
\cite{weisbergtaylor}).
We are still trying to detect them directly, an effort that began 
40 years ago with Weber's pioneering work \cite{weber}, and that continues
nowadays with more modern gravitational wave detectors, such as 
GEO \cite{geo}, LIGO \cite{ligo} or VIRGO \cite{virgo}.
The reason for the difficulty in detecting gravitational waves is that
gravity is about
$10^{40}$ times weaker than electromagnetism.  This number comes from
comparing the gravitational and electromagnetic forces between a
proton and an electron.  
This extreme
feebleness is the major obstacle to the technological manipulation of
gravity and almost certainly means that the study of gravitational
radiation will have to rely on powerful natural sources in the
Universe.  Unfortunately, the same feebleness that inhibits the
generation of gravitational waves also afflicts their detection, so
that even a flux of gravitational energy at the Earth's surface
comparable to the heat and light of the Sun is unlikely to be detected
by present terrestrial equipment.
So weak is the interaction between gravitational waves and matter that
only one graviton in $10^{23}$ registers on a typical detector. This
means that enormously powerful processes are necessary to render any
practical interest to the subject of gravitational waves.  But these
processes do exist, and the payoff may be extremely high.
I shall now describe briefly some of these processes, and the 
information they may give us regarding the nature of our Universe. 
It is likely this information may be collected in forthcoming years,
since our gravitational wave detectors are already quite sensible.

We have talked about astrophysical sources. There is yet another 
process of producing gravitational waves and indirectly detect them.
It has been proposed that the Universe may have large extra dimensions.
If this is so, one consequence may be the production of black holes
from collisions with center of mass energy of about $1TeV$. Thus, the Large Hadron
Collider (LHC) at CERN might be a black hole factory.
An enormous fraction (mabe about 25\% or more) of the center of mass energy
is expected to be radiated as gravitational radiation during the black hole 
production. This will appear as a missing energy, and thus one can indirectly observe gravitational
waves at accelerators.
In the following i will describe both sources of gravitational waves,
astronomical and microscopic, in more detail.
\section{Astronomical sources of gravitational waves}
Among the sources that can be detected using present day technology,
some of the most promising are (see the review papers by Schutz
\cite{schutz1} and by Hughes \cite{hughes}).

Supernovae and gravitational collapse.
Supernovae are triggered by the gravitational collapse of the interior
degenerate core of an evolved star. According to current theory the
final outcome should be a neutron star or a black hole.  The collapse
releases an enormous amount of energy, about $0.15 M_{\odot}c^2$, most
of which is carried away by neutrinos. An uncertain fraction is
carried by gravitational waves. It is ironic that, although detecting
supernovae was the initial goal of detector development when it
started 4 decades ago, little more is known today about what to expect
than scientists knew then.  The problem with gravitational collapse is
that perfectly spherical motions do not emit gravitational waves, and
it is still not possible to estimate is a reliable way the amount of
asymmetry in gravitational collapse. 

Binary systems.
These systems have provided us with the best proof, up to the moment,
of the existence of gravitational waves. This has been an indirect
proof, through the decrease of the orbital period due to the emission
of gravitational radiation. For example, for the famous binary pulsar
PSR 1916+16, discovered by Hulse and Taylor in 1974, the observed
value is $2.4\times 10^{-12} s/s$. The theoretical prediction is
$2.38\times 10^{-12} s/s$, which is in agreement within the
measurement errors.
Detection of gravitational waves from these systems seems however impossible
using the current detectors, since the frequency of the radiation is too low
to be detected on Earth.


Chirping binary systems.
When a binary looses enough energy so that its orbit shrinks by an
observable amount during an observation, it is said to {\it chirp}. As
the orbit shrinks, the frequency and amplitude go up.  Chirping binary
systems are more easily detected than gravitational collapse events
because one can model with great accuracy the gravitational waveform
during the inspiral phase.  There will be radiation, probably with
considerable energy, during the poorly understood plunge phase (when
the objects reach the last stable circular orbit and fall rapidly
toward one another) and during the merger merger event, but the
detectability of such systems rests on tracking their orbital
emissions.

Big Bang. Quantum fluctuations in the spacetime metric are
parametrically amplified during inflation to relatively high
amplitudes.

There are many other important astrophysical sources of gravitational waves,
the reader can find a more complete account of them in \cite{schutz1,hughes}.

A certain fraction of such systems could contain black holes instead
of neutron stars.  Astronomers now recognize that there is an
abundance of black holes in the Universe.  Observations of various
kinds have located black holes in X-ray binary systems in the galaxy
and in the centers of galaxies.  These two classes of black holes have
very different masses, stellar black holes typically have masses of
around $10 M_{\odot}$, and are thought to have been formed by the
gravitational collapse of the center of a large, evolved red giant
star, perhaps in a supernova explosion.  Massive black holes in
galactic centers seem to have masses between $10^6 M_{\odot}$ and
$10^{10} M_{\odot}$, but their history and method of formation are not
yet understood. 
It was generally believed these were the only two kinds of black holes
in the Universe.
In the last few years, however, evidence has accumulated for an
intermediate-mass class of black holes, with hundreds to thousands of
solar masses.  If such objects exist they have important implications
for the dynamics of stellar clusters, the formation of supermassive
black holes, and the production and detection of gravitational waves.
We refer the reader to \cite{miller} for a review on the observational
evidence of such black holes, and the implications their existence may have for
our knowledge of the Universe.

Radiation from stellar black holes is expected mainly from coalescing
binary systems, when one or both of the components is a black hole.
Gravitational radiation is expected from supermassive black holes in
two ways.  In one scenario, two massive black holes spiral together in
a powerful version of the coalescence we have discussed.  A second
scenario for the production of radiation by massive black holes is the
swallowing of a stellar mass black hole or a neutron star by the large
hole.  Such captures emit a gravitational wave signal that may be
approximated by studying the motion of a point mass near a black
hole. This process is well understood theoretically, and will be
revisited in this thesis. (As an aside, and as expected from Part I of
this thesis, this process is also dominated by a quasinormal ringing
at intermediate times).

What can we gain from gravitational wave observations? First,
the weakness with which gravitational waves interact with matter is a
great advantage for astronomy. It means that gravitational waves
arriving at Earth have not been corrupted by any matter they may
have encountered since their generation. As such they carry clean
information about their birth place.  

Gravitational waves are emitted by motions of their sources as a
whole, not by individual atoms or electrons, as is normally the case
for electromagnetic waves.  Astrophysical electromagnetic radiation
typically arises from the incoherent superposition of many
emitters. This radiation directly probes the thermodynamic state of a
system or an environment. On the other hand, gravitational waves are coherent
superpositions arising from the bulk dynamics of a dense source of
mass-energy. These waves directly probe the dynamical state of a
system.

The direct observable of gravitational radiation is the waveform $h$,
a quantity that falls off with distance as $\frac{1}{r}$. Most
electromagnetic observables are related to the energy flux, and so
fall off with a $\frac{1}{r^2}$ law. This means that a small
improvement in the gravitational wave detector's sensitivity can have
a large impact on their science: doubling the sensitivity of a
detector doubles the distance to which sources can be detected (and
this increases the volume of the Universe to which sources are
measurable by a factor of $8$).

Black holes can emit gravitational waves, and indeed gravitational
waves provide the only way to make direct observation of these
objects.  Since there is now strong indirect evidence that giant black
holes inhabit the centers of many, or most, galaxies and since smaller
ones are common in the galaxy (and not to mention intermediate ones,
which also seem to be abundant, as has recently been recognized
\cite{miller}) , there is great interest in making direct observations
of them.

Gravitational waves can come from extraordinarily early in the history of the
Universe. Gravitational waves, if they can be detected, would picture the Universe
when it was only perhaps $10^{-24}$ seconds old, just at the end of Inflation.

Gravitational radiation is the last fundamental prediction of
Einstein's general relativity that has not yet been directly
verified. If another theory of gravity is correct, then differences
could in principle show up in the properties of gravitational waves,
such as their polarization.

\section{Microscopic sources of gravitational waves}
Studying highly relativistic collisions between astrophysical
objects is not interesting: it is highly unlikely we will ever 
observe such event. However, a new scenario, described below,
as been recently proposed in which the gravitational radiation 
generated during 
highly relativstic is extremely interesting.

Another mechanism by which gravitational waves may be
indirectly detected is provided by the TeV-scale gravity scenario.  
This scenario
tries to solve some problems in high energy physics, namely the
hierarchy problem, by postulating the existence of large extra dimensions in
our Universe.

Traditionally, high energy physics seeks the nature of the strong and
weak subnuclear interactions, which can be understood through the 
Standard Model. However, the foundations of the Standard Model are still mysterious.  
Over the last two
decades, one of the greatest driving forces for the construction of
theories beyond the Standard Model has been trying to explain the huge
difference between the two fundamental energy scales in nature: the
electroweak scale $m_{\rm EW} \sim 300\, {\rm GeV}$ and the Planck scale
$M_{\rm 4Pl} \sim 10^{19} \,{\rm GeV} $. This is the so called hierarchy problem.  
There is an important difference in
these two scales, though \cite{hamed}: while electroweak interactions
have been probed at distances $\sim m_{\rm EW}^{-1}=10^{-16}\, {\rm cm}$, gravitational
forces have not been probed at distances $\sim M_{\rm Pl}^{-1}=10^{-33} \, {\rm cm}$.
For instance, from $e^+ e^- \rightarrow e^+ e^-$ at LEP200, we know
that the electromagnetic interactions are $\frac{1}{r}$ down to $5
\times 10^{-17} \,{\rm cm}$.  However, gravity has only been accurately
measured in the $\sim 0.01$ cm range \cite{mitrofanov}.
Thus, the view that $M_{\rm Pl}=10^{19}\,{\rm GeV}$ is a fundamental energy
scale is based solely on the belief that gravity stays unmodified over
the 31 orders of magnitude going from $10^{-33} \,{\rm cm}$ to $\sim 0.01 \,{\rm cm}$,
where it is measured.  One can thus conceive ways to solve the
hierarchy problem, if gravity gets indeed modified at scales smaller
than $1 \,{\rm mm}$.  As a simple realization of this, suppose our world is
$(4+n)$-dimensional. The $n$ extra dimensions must be compact,
otherwise gravity would not behave as $\frac{1}{r}$ at large
distances, as we know it does. Consider therefore that each of these
$n$ extra dimensions have typical length $\sim R$.  Two test masses
$m_1$, $m_2$ placed within a distance $r \ll R$ will feel a
gravitational potential dictated by Gauss's law in $(4+n)$ dimensions,
\begin{equation}
V(r)=G_{4+n}\frac{m_1 m_2}{r^{n+1}}\,\,,\,\,\,\,r \ll R.
\label{gaussndim}
\end{equation}
Here, $G_{4+n}$ is Newton's constant in the $(4+n)$ dimensional world.
On the other hand, if the masses are placed at distances $r \gg R$, their gravitational
flux lines can not continue to penetrate in the extra dimensions, and the usual $\frac{1}{r}$
potential is recovered \cite{hamed}
\begin{equation}
V(r)=\frac{G_{4+n}}{R^n}\frac{m_1 m_2}{r}\,\,,\,\,\,\,r \gg R.
\label{gauss2ndim}
\end{equation}
In order to have the usual strength at large distances, one must impose 
\begin{equation}
\frac{G_{4+n}}{R^n}=G_4\,, 
\label{grel}
\end{equation}
where $G_4$ is our usual four dimensional Newton's
constant.
Let's now see what happens to the Planck scale.
The Planck mass can be obtained by equating the Schwarzschild radius
and the Compton wavelength of an object of mass $m$.
In general $(4+n)$ dimensions, the Schwarzschild radius $r_+$ of that object is 
\begin{equation}
r_+=\left[\frac{16\pi G_{4+n} m}{2(n+2)\pi^{(n+2)/2}}\Gamma[(n+3)/2]\right]^{1/(n+1)}\sim 
 (G_{4+n} m)^{1/(n+1)}\,,
\label{Sradiusobj}
\end{equation}
and the Compton wavelength is 
\begin{equation}
\lambda=\frac{1}{m}\,,
\label{compt}
\end{equation}
where we have set both Planck's constant $h$ and the velocity of light $c$ equal to 1.
Equating (\ref{Sradiusobj}) with (\ref{compt}) for a mass $m$ equal to the Planck mass
$M_{\rm Pl}$ we get 
\begin{equation}
M_{\rm Pl}^{2+n}=\frac{1}{G_{4+n}}\,.
\label{compt1}
\end{equation}
But in view of (\ref{grel}), and keeping in mind that the four dimensional Planck scale
is simply $M_{\rm 4Pl}^2=\frac{1}{G_4}$, it means that the fundamental $(4+n)$ Planck scale 
$M_{\rm Pl}$ 
is related to the effective $4$ dimensional Planck scale $M_{\rm 4 Pl}$ as
\begin{equation}
M_{\rm 4 Pl}^2 \sim M_{\rm Pl}^{2+n}R^n.
\label{relationPS}
\end{equation}
It is now very easy to solve the hierarchy problem: just make it go away,
by imposing equal scales, i.e., $ m_{\rm EW}= M_{\rm Pl}$.
This constraints the typical sizes $R$ of the extra dimensions to be
\begin{equation}
R \sim 10^{30/n-17} {\rm cm} (\frac{1 {\rm TeV}}{m_{\rm EW}})^{1+2/n}.
\label{sizesextradim}
\end{equation}
If $n=1$ this gives $R\sim 10^{13} \,{\rm cm}$, which is excluded empirically,
since this would imply deviations from Newtonian gravity over solar
system distances.  However any $n \geq 2$ gives modifications of
Newtonian gravity at distances smaller than those currently probed by
experiment.

On the other hand, Standard Model gauge forces have been
accurately measured at weak scale distances, as we remarked
earlier. This means that Standard Model particles cannot freely
propagate in the $n$ extra dimensions, but must, somehow, be localized
to a $4$ dimensional submanifold.  An important question is the
mechanism by which the Standard Model fields are localized to the
brane, which has several resolutions, see \cite{hamed}. 

An exciting consequence of TeV-scale gravity is the possibility of
production of black holes at the LHC and beyond \cite{bhprod}.  To understand how
this comes about, let's consider the formation of a black hole with
horizon radius $r_+$ much smaller than the typical size $R$ of the
extra dimensions. When the Schwarzschild radius of a black hole is
much smaller than the radius $R$ of the compactified dimensions, it
should be insensitive to the brane and the boundary conditions in the
$n$ transverse directions, and so is well approximated by a
$(4+n)$-dimensional Schwarzschild black hole
\cite{tangherlini,myersperry}.  The Schwarzschild radius of a
non-rotating $(4+n)$-dimensional black hole is
\begin{equation}
r_+=\frac{1}{\sqrt{\pi}M_{\rm Pl}}\left [\frac{M}{M_{\rm
Pl}}\frac{8\Gamma{\frac{(n+3)}{2}}}{n+2} \right]^{1/(n+1)}\,,
\label{SradiusBHD}
\end{equation}
where $M$ is the black hole mass.  To estimate the parton-level cross
section, consider partons scattering at center of mass energy
$\sqrt{s}$ and impact parameter $b$. A conjecture in general
relativity is Thorne's hoop conjecture \cite{thornehoop}, which states
that horizons form when and only when a mass $M$ is compacted into a
region whose circumference in every direction is less than $2\pi r_+$.
This conjecture implies that the cross section for black hole
production is
\begin{equation}
\sigma \sim \pi r_+^2=\frac{1}{M_{\rm Pl}^2}\left [\frac{M}{M_{\rm Pl}}
\frac{8\Gamma{\frac{n+3}{2}}}{n+2} \right]^{2/(n+1)}\,.
\label{crosssection bhd}
\end{equation}
It is important to note that this is only correct if the black hole 
is larger than the colliding particles. Since these partons have no internal structure,
at least up to $\sim 1\,{\rm TeV}$, then we can assume they are indeed pointlike
in such calculation.
A simple estimate gives an impressive result: for $M_{\rm Pl} \sim 1\,{\rm TeV}$, 
the LHC will produce one black hole per second with mass $> 5 M_{\rm Pl}$.
If this is true it will turn LHC into a black hole factory.

After the black hole forms, it will start to decay. First, it will
radiate all the excess multipole moments, because when the black hole is born,
it has an highly asymmetric shape and thus will settle down to a stationary
shape (which in four dimensions must be a Kerr-Newman black hole, by the uniqueness
theorem). One calls this the balding stage.  
After this phase we are left, in general, with a spinning black hole.
It will then start to loose angular momentum through the Hawking process:
as Page \cite{page} as shown, a spinning black hole Hawking radiates mainly along the 
equatorial plane, and the Hawking radiation carries away most of the black hole's
angular momentum. This is the spin-down phase.
After this we have a non-spinning black hole Hawking radiating. This is the 
Schwarzschild phase. When the black hole finally reaches the Planck size,
we loose track of what will happen: current physics is not able to go that deep.
This second part of the thesis will try to give an account of the balding phase,
as we shall explain below.

\section{Outline of Part II}
In this second part of the thesis we will begin, 
in Chapter \ref{chap:rads} by refining the calculation of the
gravitational radiation released when a high energy particle collides
with a black hole.  This study shows three highly interesting aspects:
first, it is in amazingly agreement with a similar, but flat space
calculation by Weinberg.  Second, the total result for the radiation,
when one takes the limit of equal mass particles, gives an answer
quite similar to that coming from a completely different formalism, but
also studying the high energy collision of two black holes.
Finally this study shows clearly that the fundamental QN frequencies
act as a cutoff in the energy spectra.
The main
advantage of the technique used here is that it allows for an important
generalization, which does not seem feasible using the other
formalism: it is easy to study the colliosion between rotating black
holes, as i shall also do in Chapter \ref{chap:radkerr}, and allows to study the
collision of higher dimensional black holes, a work which is now being
carried by some authors.

In Chapter \ref{chap:radads} we shall begin the
study of radiation in non-asymptotically flat spacetimes. We shall
focus on scalar radiation in asympotically anti-de Sitter background,
and consider the radial infall of massive particles into
Schwarzschild-anti-de Sitter black holes.  As expected, and as very
often remarked in Chapter
\ref{chap:Intro}, the waveform is dominated by the quasinormal
ringing. The work presented in this chapter is the first showing that
in fact QNMs dominate the answer even for non-asymptotically flat
spacetimes.  The infalling particle and consequent scalar radiation
have an interpretation in terms of the AdS/CFT: if one recognizes that
a black hole corresponds to a thermal state.  then the infalling
particle corresponds to some perturbation of this thermal state.
Again, the typical timescale of approach to equilibrium is therefore
dictated by the lowest QN frequency, since the signal is dominated by
QN ringing.

\thispagestyle{empty} \setcounter{minitocdepth}{1}
\chapter[Gravitational radiation in a curved asymptotically flat four dimensional background
: encounters of non-rotating black holes with particles]
{Gravitational radiation in a curved asymptotically flat four dimensional background: 
high energy encounters of non-rotating black holes with particles
} \label{chap:rads}
\lhead[]{\fancyplain{}{\bfseries Chapter \thechapter. \leftmark}}
\rhead[\fancyplain{}{\bfseries \rightmark}]{}
\minitoc \thispagestyle{empty}
\renewcommand{\thepage}{\arabic{page}}
\section{Introduction}

The study of gravitational wave emission by astrophysical objects has
been for the last decades one of the most fascinating topics in General
Relativity. This enthusiasm is of course partly due to the possibility
of detecting gravitational waves by projects such as GEO600 \cite{geo}, 
LIGO \cite{ligo} or VIRGO \cite{virgo}, already
operating.  Since gravity couples very weakly to matter, one needs to
have powerful sources of in order to hope for the detection of the
gravitational waves.  Of the candidate sources, black holes stand out
naturally, as they provide huge warehouses of energy, a fraction of
which may be converted into gravitational waves, by processes such as
collisions between two black holes.  As it often happens, the most
interesting processes are the most difficult to handle, and events such
as black hole-black hole collisions are no exception. An efficient
description of such events requires the use of the full non linear
Einstein's equations, which only begin to be manageable by numerical
methods, and state-of-the-art computing.  In recent years we have
witnessed serious progress in this field \cite{anninos}, and
we are now able to evolve numerically the collision of two black holes,
provided their initial separation is not much larger than a few
Schwarzschild radius. At the same time these numerical results have
been supplemented with results from first and second order perturbation
theory \cite{gleiser}, which simultaneously served as guidance into the
numerical codes. The agreement between the two methods is not only
reassuring, but it is also in fact impressive that a linearization of
Einstein's equations yield such good results (as Smarr \cite{smarrl}
puts it, ``the agreement is so remarkable that something deep must be
at work'').  In connection with this kind of events, the use of
perturbation methods goes back as far as 1970, when Zerilli
\cite{zerilli} and Davis et al \cite{davis} first computed the
gravitational energy radiated away during the infall from rest at
infinity of a small test particle of mass $m_0$, into a Schwarzschild
black hole with mass $M$.  Later, Ruffini \cite{ruffini} generalized
these results to allow for an initial velocity of the test particle 
(this problem has recently been the subject of further study \cite{lousto}, 
in order to investigate the question of choosing appropriate initial data
for black hole collisions).  Soon after, one began to realize that the
limit $m_0 \rightarrow M$ describing the collision of two black holes
did predict reasonable results, still within perturbation theory,
thereby making perturbation theory an inexpensive tool to study
important phenomena.

In this paper we shall extend the results of Davis et al \cite{davis}
and Ruffini \cite{ruffini} by considering a massless test particle
falling in from infinity through a radial geodesic. This process
describes the collision of an infalling test particle in the limit that
the initial velocity goes to the speed of light, thereby extending the
range of Ruffini's results into larger Lorentz boost parameters 
$\gamma$s.
If then one continues to rely on the agreement between
perturbation theory and the fully numerical outputs, these results
presumably describe the collision of two black holes near the speed of
light, these events have been extensively studied through matching
techniques by D'Eath (a good review can be found in his book
\cite{D'Eath}), and have also been studied in \cite{smarr2}. 
The extension is straightforward, the mathematics involved are quite
standard, but the process has never been studied. Again supposing that
these results hold for the head on collision of two black holes
travelling towards each other at the speed of light, we have a very
simple and tractable problem which can serve as a guide and supplement
the results obtained by Smarr and by D'Eath.  Another strong
motivation for this work comes from the possibility of black hole
formation in TeV-scale gravity \cite{hamed}.  Previous estimates
on how this process develops, in particular the final mass of the
black hole formed by the collision of relativistic particles have
relied heavily upon the computations of D'Eath and Payne \cite{payne}.
A fresher look at the problem is therefore recommended, and a
comparation between our results with results obtained years ago
\cite{D'Eath,smarr2} and with recent results \cite{bhprod}
are in order.

The fully relativistic results we will present here show an impressive agreement with
results by Smarr \cite{smarr2} for collisions of massive particles near
the speed of light, namely a flat spectrum, with a zero frequency limit
(ZFL) $(\frac{dE}{d\omega})_{\omega = 0}= 0.4244 m_0^2 \gamma^2$.  We
also show that Smarr underestimated the total energy radiated to
infinity, which we estimate to be $\Delta E=0.26 m_0^2 \gamma^2/M$,
with $M$ the black hole mass.  The quadrupole part of the perturbation
carries less than 50\% of this energy.  When applied to the head on
collision of two black holes moving at the speed of light, we obtain an
efficiency for gravitational wave generation of 13\%, quite close to
D'Eath and Payne's result of 16\% \cite{D'Eath,payne}.
This chapter will follow closely the analysis in \cite{cardosorads}

\noindent
\section{Basic Formalism}
\vskip 3mm
Since the mathematical formalism for this problem has been thoroughly
exploited over the years, we will just outline the procedure. Treating the
massless particle as a perturbation, we write the metric functions for
this spacetime, black hole + infalling particle,  as
\begin{equation}
g_{ab}(x^\nu)= g^{(0)}_{ab}(x^\nu)+h_{ab}(x^\nu)\,,
\label{perturbation}
\end{equation}
where the metric $g^{(0)}_{ab}(x^\nu)$ is the background metric,
(given by some known solution of Einstein's equations), which we now specialize
to the Schwarzschild metric
\begin{equation}
ds^{2}=-f(r)dt^{2}+\frac{dr^{2}}{f(r)}+
r^{2}(d\theta^{2}+\sin^2\theta d\phi^{2})\,,
\label{lineelement}
\end{equation}
where $f(r)=1-2M/r$.
Also, $ h_{ab}(x^\nu)$ is a small perturbation, induced by the
massless test particle, which is described by the stress energy tensor
\begin{equation}
T^{\mu \nu}= -\frac{p_0}{(-g)^{1/2}} \int d \lambda 
\delta^4 (x-z(\lambda))\dot{z}^{\mu} \dot{z}^{\nu}\,. 
\label{stresstensor}
\end{equation}
Here, $z^{\nu}$ is the trajectory of the particle along the world-line,
parametrized by an affine parameter $\lambda$ (the proper time in the case 
of a massive particle), 
and $p_0$ is the momentum of the particle. 
To proceed, we decompose Einstein's equations $G_{ab}=8\pi T_{ab}$
in tensorial spherical harmonics and
specialize to the Regge-Wheeler \cite{regge} gauge. 
For our case, in which the particle falls
straight in, only even parity perturbations survive. Finally,
following Zerilli's \cite{zerilli2} prescription, we arrive at a
wavefunction (a function of the time $t$ and radial $r$ coordinates only)
whose evolution can be followed by the wave equation
\begin{equation}
\frac{\partial^{2} {\bf \tilde Z}(\omega,r)}{\partial r_*^{2}} +
\left\lbrack\omega^2-V(r)\right\rbrack{\bf \tilde Z}(\omega,r)=(1-2M/r)S \,,
\label{waveequationrads}
\end{equation}
Here, the $l-$dependent potential $V$ is given by
\begin{equation}
V(r)=\frac{f(r)\left[2\sigma^2(\sigma+1)r^3+6\sigma^2 r^2M +18 
\sigma rM^2 +18M^3\right]}
{r^3(3M+\sigma r)^2}\,,
\label{potentialmaxwell}
\end{equation}
where $\sigma=\frac{(l-1)(l+2)}{2}$ and the tortoise 
coordinate $r_*$ is defined as
$\frac{\partial r}{\partial r_*}= f(r)\,$. 
The passage from the time variable $t$ to the frequency $\omega$ has been achieved
through a Fourier transform, 
${\bf \tilde  Z}(\omega,r)=
\frac{1}{(2\pi)^{1/2}}\int_{-\infty}^{\infty}e^{i \omega t}Z(t,r)dt$.
The source $S$ depends entirely on the stress energy tensor of the 
particle and on whether or not it is massive.
The difference between massive and massless particles
lies on the geodesics they follow.
The radial geodesics for massive particles are:
\begin{equation}
\frac{dT}{dr}=-\frac{E}{f(r)(E^2-1+2M/r)^{1/2}}\,;
\frac{dt}{d\tau} =\frac{E}{f(r)}\,,
\label{massgeo}
\end{equation}
where $E$ is a conserved energy parameter: For example, if the particle
has velocity $v_{\infty}$ at infinity then
$E=\frac{1}{(1-v_{\infty}^2)^{1/2}}\equiv\gamma$.  On the other hand,
the radial geodesics for massless particles are described by
\begin{equation}
\frac{dT}{dr}=-\frac{1}{f(r)}\,\,;\, 
\frac{dt}{d\tau} =\frac{\epsilon_0}{f(r)}\,,
\label{masslessgeo}
\end{equation}
where again $\epsilon_0$ is a conserved energy parameter, which in relativistic
units is simply $p_0$. We shall however keep $\epsilon_0$ for future use, to
see more directly the connection between massless particles and massive ones
traveling close to the speed of light.
One can see that, on putting $p_0 \rightarrow m_0$,  
$\epsilon_0 \rightarrow \gamma$ and  $\gamma \rightarrow \infty$, the radial
null geodesics reduce to radial timelike geodesics, so that all the results
we shall obtain in this paper can be carried over to the case of ultrarelativistic
(massive) test particles falling into a Schwarzschild black hole.
For massless particles the source term $S$ is 
\begin{equation}
S=\frac{4ip_0 e^{-i\omega r_*}\epsilon_0 (4l+2)^{1/2}\sigma }
{w(3M+ \sigma r)^2}\,.
\label{source}
\end{equation}
To get the energy spectra, we use
\begin{equation}
\frac{dE}{d\omega}=\frac{1}{32 \pi}\frac{(l+2)!}{(l-2)!} \omega^2 |{\bf \tilde Z}(\omega,r)|^2\,,
\label{spectra}
\end{equation}
and to reconstructe the wavefunction $Z(t,r)$ one uses the inverse Fourier transform
\begin{equation}
Z(t,r)=
\frac{1}{(2\pi)^{1/2}}\int_{-\infty}^{\infty}e^{-i \omega t}{\bf \tilde Z}(\omega,r)d\omega.
\label{invfourier}
\end{equation}
Now we have to find ${\bf \tilde Z}(\omega,r)$ from the differential
equation (\ref{waveequationrads}).  This is accomplished by a Green's
function technique. Imposing the usual boundary conditions, i.e., only
ingoing waves at the horizon and outgoing waves at infinity, we get
that, near infinity,
\begin{equation}
{\bf \tilde Z}=\frac{1}{W}\int_{r_+}^{\infty}z_L S dr.
\label{solutionrads}
\end{equation}
Here, $z_L$ is a homogeneous solution of (\ref{waveequationrads}) which
asymptotically behaves as
\begin{eqnarray}
z_L \sim e^{-i\omega r_*}\,,r_* \rightarrow -\infty \\
z_L \sim B(\omega)e^{i\omega r_*}+C(\omega)e^{-i\omega r_*}\,,r_* \rightarrow +\infty. 
\label{behavior1rads}
\end{eqnarray}
$W$ is the wronskian of the homogeneous solutions of (\ref{waveequationrads}). 
These solutios are, 
$z_L$ which has just been defined, and $z_R$ which behaves as $z_R \sim
e^{i\omega r_*}\,,r_* \rightarrow +\infty $. From this follows that
$W=2i\omega C(\omega)$.  We find $C(\omega)$ by solving
(\ref{waveequationrads}) with the right hand side set to zero, and with the
starting condition $z_L= e^{-i\omega r_*}$ imposed at a large negative
value of $r_*$. For computational purposes good accuracy is hard to
achieve with the form (\ref{behavior1rads}), so we used an asymptotic
solution one order higher in $1/(\omega r)$:
\begin{eqnarray}
&z_L = B(\omega)(1+\frac{i(\sigma+1)}{\omega r})e^{i\omega r_*}+
\nonumber\\ &
+C(\omega)(1-\frac{i(\sigma+1)}{\omega r})e^{-i\omega r_*}\,,r_* \rightarrow +\infty\,.&
\label{behavior2}
\end{eqnarray}
In the numerical work, we chose to adopt $r$ as the independent
variable, thereby avoiding the numerical inversion of $r_*(r)$.
A fourth order Runge-Kutta routine started the integration
of $z_L$ near the horizon, at $r=r_i=2M+2M\epsilon$, with tipically 
$\epsilon=10^{-5}$. It then integrated out to large values
of $r$, where one matches $z_L$ extracted numerically with the
asymptotic solution (\ref{behavior2}), in order to find $C(\omega)$.
To find $Z(t,r)$ the integral in (\ref{invfourier}) is
done by Simpson's rule. For both routines Richardson
extrapolation is used.
\noindent
\section{Numerical Results}
\begin{figure}
\centerline{\includegraphics[width=10 cm,height=10 cm]
{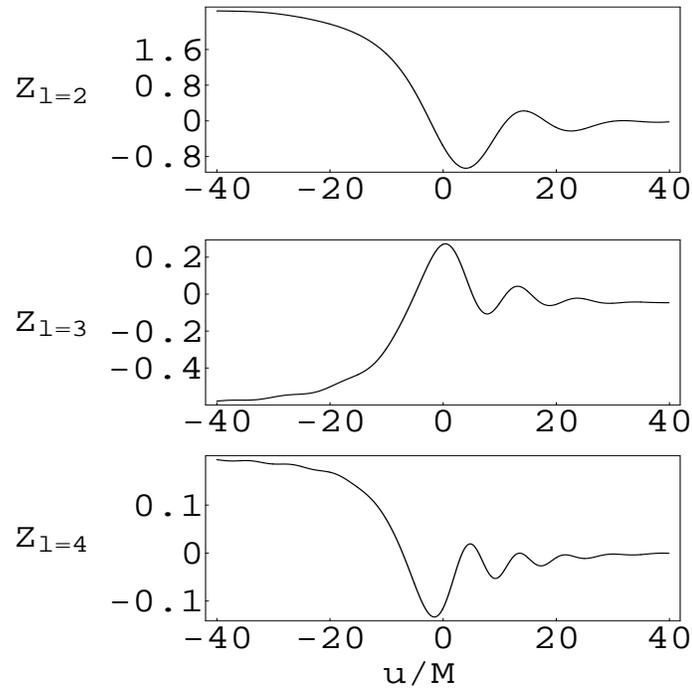}}
\caption{Waveforms for the three lowest
radiatable multipoles, for a massless particle falling from 
infinity into a Schwarzschild black hole. Here, the wavefunction 
$Z$ is measured in units of $\epsilon_0p_0$.}
\label{waveformgraph}
\end{figure}

\begin{figure}
\centerline{\includegraphics[width=12 cm,height=7 cm]
{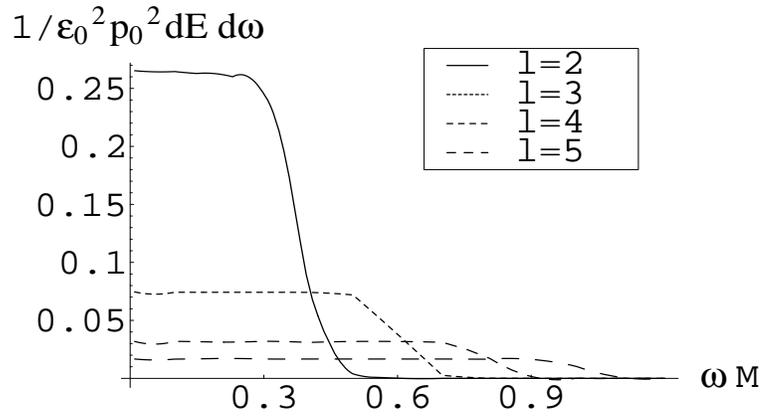}}
\caption{The energy spectra for the four
lowest radiatable multipoles, for a massless particle falling
from infinity into a Schwarzschild black hole. }
\label{spectragraph}
\end{figure}

\vskip 3mm
The results for the wavefunction $Z(t,r)$ as a function of the
retarded time $u\equiv t-r_*$ are shown in Figure \ref{waveformgraph}, for the
three lowest radiatable multipoles, $l=2, 3$ and $4$.  
As expected from the work of Ruffini \cite{ruffini} and Ferrari and
Ruffini \cite{ferrari}, the wavefunction is not zero at very early
times, reflecting the fact that the particle begins to fall with non
zero velocity. At late times, the $l=2$ (for example) signal is 
dominated by quasinormal ringing with 
frequency $\omega \sim 0.35/M$, the lowest quasinormal
frequency for this spacetime \cite{chandra}.  The energy
spectra is shown in Figure \ref{spectragraph}, for the four lowest radiatable
multipoles. First, as expected from Smarr's work \cite{smarr2}, the
spectra is flat, up to a certain critical value of the frequency, after
which it rapidly decreases to zero. This ($l$-dependent) critical
frequency is well approximated, for each $l$-pole, by the fundamental
quasinormal frequency. In Table \ref{tab:zflrads}, we list the zero frequency limit
(ZFL) for the first ten lowest radiatable multipoles.
\begin{table}
\caption{\label{tab:zflrads}  The zero frequency limit (ZFL) 
for the ten lowest radiatable multipoles.}
\begin{tabular}{llll}  \hline
$l$ & ZFL($\times\frac{1}{\epsilon_0^2p_0^2}$)&$l$& ZFL($\times \frac{1}{\epsilon_0^2p_0^2}$)\\ \hline
2    &  0.265 &  7 &  0.0068 \\ \hline 
3    &  0.075 &   8 &  0.0043  \\ \hline 
4    &  0.032 &   9 &  0.003 \\ \hline 
5    &  0.0166 &   10 &  0.0023  \\ \hline 
6    &  0.01  &   11 &  0.0017 \\ \hline 
\end{tabular}
\end{table}

For high
values of the angular quantum number $l$, a good fit to our numerical
data is
\begin{equation}
\left(\frac{dE_l}{d\omega}\right)_{\omega = 0} = 
\frac{2.25}{l^3} \epsilon_0^2 p_0^2
\label{fit}
\end{equation}
We therefore estimate the zero ZFL as
\begin{eqnarray}
&\left(\frac{dE_l}{d\omega}\right)_{\omega = 0}=
\left[\sum_{l=2}^{l=11}
\left(\frac{dE_l}{d\omega}\right)_{\omega = 0}\right] +
\frac12 \frac{2.25}{12^2} \epsilon_0^2p_0^2
\nonumber\\ &
=0.4244 \epsilon_0^2p_0^2\,.&
\label{ZFL}
\end{eqnarray}
To calculate the total energy radiated to infinity, we proceed as
follows: as we said, the spectra goes as $2.25/l^3$ as long as $\omega
< \omega_{lQN}$, where $\omega_{lQN}$ is the lowest quasinormal
frequency for that $l$-pole.  For $\omega > \omega_{lQN}$, $dE/d\omega
\sim 0$ (In fact, our numerical data shows that $dE/d\omega \sim
e^{-27\alpha\omega M}$, with $\alpha$ a factor of order unity, for
$\omega > \omega_{lQN}$).  Now, from the work of Ferrari and Mashhoon
\cite{ferrari2} and Schutz and Will \cite{schutz}, one knows that for
large $l$, $\omega_{lQN} \sim \frac{l+1/2}{3^{3/2}M}$.  Therefore, for
large $l$ the energy radiated to infinity in each multipole is
\begin{equation}
\Delta E_l=\frac{2.25(l+1/2)}{3^{3/2}l^3}\frac{\epsilon_0^2p_0^2}{M}\,, 
\label{totalEl}
\end{equation}
and an estimate to the total energy radiated is then
\begin{equation}
\Delta E=\sum_l \Delta E_l = 0.262\frac{\epsilon_0^2p_0^2}{M}
\label{totalE}
\end{equation}

\begin{figure}
\centerline{\includegraphics[width=10 cm,height=6.5 cm]
{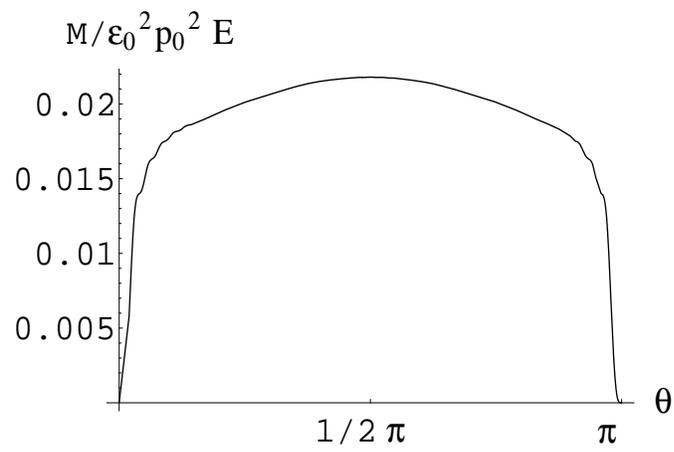}}
\caption{The energy radiated per solid angle as
a function of $\theta$. }
\label{angulardist}
\end{figure}

Let us now make the bridge between these results and previous results
on collisions between massive particles at nearly the speed of light
\cite{ruffini,D'Eath,smarr2}. As we mentioned, putting
$p_0 \rightarrow m_0$ and $\epsilon_0 \rightarrow \gamma$ does the trick.
So for ultrarelativistic test particles with mass $m_0$ falling into a 
Schwarzschild black hole, one should have 
$(dE/d\omega)_{\omega=0}= 0.4244 m_0^2 \gamma^2$ and 
$\Delta E=  0.262\frac{ m_0^2 \gamma^2}{M}$.
Smarr \cite{smarr2} obtains
\begin{eqnarray}
&\left(\frac{dE}{d\omega}\right)^{\rm Smarr}_{\omega=0}=
\frac{8}{6\pi}m_0^2 \gamma^2\sim 0.4244 m_0^2 \gamma^2\,,
\nonumber\\ &
\Delta E_{\rm Smarr}= 0.2 \frac{m_0^2 \gamma^2}{M}\,.&
\label{smarrresult}
\end{eqnarray}
So Smarr's result for the ZFL is in excellent agreement with ours,
while his result for the total energy radiated is seen to be an
underestimate. As we know from the work of Davis et al \cite{davis}
for a particle falling from infinity with $v_{\infty}=0$ most of the
radiation is carried by the $l=2$ mode.  Not so here, in fact in our
case less than 50\% is carried in the quadrupole mode (we obtain
$\Delta E_{l=2} = 0.1\frac{\epsilon_0^2p_0^2}{M}$, $\Delta E_{l=3} =
0.0447\frac{\epsilon_0^2p_0^2}{M}$ ).
This is reflected in the angular distribution of the radiated energy
(power per solid angle)
\begin{equation}
\frac{dE}{d\Omega}=\Delta E_l \frac{(l-2)!}{(l+2)!}
\left[2 \frac{\partial ^2}{\partial \theta ^2} Y_{l0} +
l(l+1)Y_{l0}\right]^2,
\label{powerangle}
\end{equation}
which we plot in Figure \ref{angulardist}. Compare with figure 5 of \cite{smarr2}.
If, as Smarr, we continue to assume that something deep is at work, and
that these results can be carried over to the case of two equal mass
black holes flying towards each other at (close to) the speed of light,
we obtain a wave generation efficiency of 13 \%. This is in close
agreement with results by D'Eath and Payne \cite{D'Eath,payne}, who
obtain a 16\% efficiency (Smarr's results cannot be trusted in this
regime, as shown by Payne \cite{payne2}). Now, D'Eath and Payne's
results were achieved by cutting an infinite series for the news
function at the second term, so one has to take those results with some
care.  However, the agreement we find between ours and their results
lead us to believe that once again perturbation theory has a much more
wider realm of validity.  To our knowledge, this is the first
alternative to D'Eath and Payne's method of computing the energy
release in such events.

\thispagestyle{empty} \setcounter{minitocdepth}{1}
\chapter[Gravitational radiation in a curved asymptotically flat four dimensional background: encounters of Kerr black holes with particles]{Gravitational radiation in a curved asymptotically flat four dimensional background: high energy encounters of Kerr black holes with particles} \label{chap:radkerr}
\lhead[]{\fancyplain{}{\bfseries Chapter \thechapter. \leftmark}}
\rhead[\fancyplain{}{\bfseries \rightmark}]{}
\minitoc \thispagestyle{empty}
\renewcommand{\thepage}{\arabic{page}}
\section{Introduction}
In the previous chapter we have studied the collision at the speed of light
between a point particle and a Schwarzschild black hole
\cite{cardosorads}.  This analyses can describe
several phenomena, such as, the collision between small and massive
black holes \cite{schutz}, the collision between stars and massive
black holes, or the collision between highly relativistic particles
like cosmic and gamma rays colliding with black holes \cite{piran}, to
name a few.  We have argued in \cite{cardosorads,cardosoradkerrs}
that these studies could give valuable
quantitative answers for the collision at the speed of light between
equal mass black holes, although we only work with perturbation theory
{\it {\`a} la} Regge-Wheeler-Zerilli-Teukolsky-Sasaki-Nakamura, which
formally does not describe this process.  In fact, extrapolating for
two equal mass, Schwarzschild black holes we obtained
\cite{cardosorads} results which were in very good agreement with
results \cite{payne} obtained through different methods.  For equal
mass black holes, we found that the collision at the speed of light
between a Schwarzschild black hole and a Kerr black hole along the
symmetry axis gave similar results to those in \cite{cardosorads} . In
particular, we found that $15.5 \%$ of the total energy gets converted
into gravitational waves (for two non-rotating holes the amount is
slightly less, $13 \%$). There are as yet no numerical results for
rotating black holes, so there is no clue as to the correctness of our
results for rotating holes.  However, should these results give
radiated energies larger than that allowed by the cosmic censorship,
and therefore by the area theorem, we would run into problems, and we
would know that the extrapolation to equal masses is strictly
forbidden.  Our previous results do not violate the area theorem, but
the mere possibility brings us to the study of the collision of a
non-rotating black hole with a rotating one, along the equatorial
plane, both approaching each other at nearly the speed of light.
Sasaki, Nakamura and co-workers \cite{nmothers} have studied the
infall of particles, at rest at infinity into Kerr black holes, along
the equatorial plane and along the axis.  They found that the energy
radiated for radial infall along the equatorial plane was about $2.6$
times larger than for radial along the axis, for extreme holes.
Accordingly, we expect the energy for highly relativistic collisions
along the equatorial plane to be larger than that along the symmetry
axis, but how larger?  If the ratio $\frac{\Delta E_{{\rm
equator}}}{\Delta E_{{\rm axis}}}$ should still be the same, $2.6$,
then the energy released in the collision along the equatorial plane
of a Kerr hole and a non-rotating hole at the speed of light would be
$\sim 40 \%$ of the total energy.  But this is slightly more than the
upper bound on the efficiency allowed by the area theorem, which is
$\sim 38.7 \%$.  Motivated mainly by this scenario, we shall study
here the collision at nearly the speed of light between a point
particle and a Kerr black hole, along its equatorial plane.  We will
then perform a boost in the Kerr black hole, and extrapolate for two
equal mass objects, one rotating, the other non-rotating.

Another point of interest to study this process is the hypothesis that
black holes could be produced at the Large Hadron Collider (LHC) at
CERN, which has recently been put forward \cite{bhprod} in the so called
TeV-scale gravity scenarios.  In such scenarios, the hierarchy problem
is solved by postulating the existence of $n$ extra dimensions,
sub-millimeter sized, such that the $4+n$ Planck scale is equal to the
weak scale $\sim$ 1TeV. The Standard Model lives on a 4-dimensional
sub-manifold, the brane, whereas gravity propagates in all
dimensions. If TeV-scale gravity is correct, then one could
manufacture black holes at the LHC.  The ability to produce black
holes would change the status of black hole collisions from a rare
event into a human controlled one.  This calls for accurate
predictions of gravitational wave spectra and gravitational energy
emitted during black hole formation from the high speed encounter of
two particles.  This investigation was first carried out by D'Eath and
Payne \cite{payne} by doing a perturbation expansion around the
Aichelburg-Sexl metric, describing a boosted Schwarzschild black
hole. Their computation was only valid for non-rotating black holes
and it seems quite difficult to extend their methods to include for
spinning black holes.  Our approach however allows one to also study
spinning black holes, which is of great importance, since if one forms
black holes at the LHC, they will most probably be rotating ones, the
chance for having a zero impact parameter being vanishingly small.

Suppose therefore one can produce black holes at LHC. The first thing
that should happen is a release of the hole's hair, in a phase termed
``balding'' \cite{bhprod}. The total amount of energy released in such
a phase is not well known, and is based mostly in D'Eath and Payne's
results.  With our results one can predict the total gravitational
energy radiated in the balding phase when the resulting holes are
rotating, a process which, as we shall see, radiates a tremendous
amount of energy.  This means that there will be a ``missing'' energy
during the formation of a black hole, this energy being carried away
by gravitational waves, and undetected, at least by any realistic
present technology.  Our results suggest that at most 35\% of the
center of mass energy can be leaking away as gravitational waves, and
therefore one should have at most 35\% missing energy.  Strictly speaking,
these values are only valid for a head-on collision, i.e. a collision
with zero impact parameter, along the equatorial plane. One expects
the total energy to decrease if the collision is taken along a
different plane, or specially with non-zero impact parameter.  
\section{The Teukolsky and Sasaki-Nakamura formalism}

In this section, we give a very brief account of the Teukolsky equation,
and of the Sasaki-Nakamura equation. Details about the Teukolsky formalism may
be found in the original literature \cite{teukolsky}, and also in 
\cite{breuerbook}. For a good account of the Sasaki-Nakamura 
formalism we refer the reader
to \cite{nmothers,nakamurasasaki,hughes1}.

We start from the Kerr background geometry, written in Boyer-Lindquist 
$(t, r, \theta, \phi)$ coordinates:

\begin{equation}
ds^{2}=-(1-\frac{2Mr}{\Sigma})dt^{2}-\frac{4Mar}{\Sigma}\sin^2\theta dt d\phi+
\frac{\Sigma}{\Delta} dr^{2}+\Sigma d\theta^{2}+
\frac{A\sin^2\theta}{\Sigma} d\phi^{2}\,,
\label{lineelementKerrBoyer}
\end{equation}
Here, $M$ is the mass of the black hole, and $a$ its angular momentum per
unit mass.
Also $\Sigma= r^2+a^2\cos^2\theta \,$ ; $\Delta=r^2+a^2-2Mr \,$ ; 
$A=(r^2+a^2)^2-\Delta a^2\sin^2\theta$.

Working with Kinnersley's null tetrad, one can show \cite{teukolsky}
that the equations for the Newman-Penrose quantities decouple and
separate, giving rise to the Teukolsky equation,

\begin{equation}
\Delta^{-s}\frac{d}{dr}(\Delta^{s+1}\frac{d}{dr} {_sR})-\, _sV \, _sR 
=-\, _sT \,.
\label{teukolskyequation}
\end{equation}
Here,
\begin{eqnarray}
 _sV=\frac{-K^2}{\Delta}+
\frac{isK \Delta'}{\Delta}-\frac{2isK'}{\Delta}-2isK'+\, _s\lambda \\
K=(r^2+a^2)\omega - am \,,
\label{teukolskyequationexplanation}
\end{eqnarray}
and a prime denotes derivative with respect to $r$.
The quantity $s$ denotes the spin weight (or helicity) 
of the field under consideration
(see for example \cite{newman}) and $m$ is an azimuthal quantum number. 
We are interested in gravitational perturbations, which have spin-weight
$s=-2,+2$. Solutions with $s=-2$ are related to those with $s=2$ via
the Teukolsky-Starobinsky identities, so we need only worry about a
specific one. For definiteness, and because that was the choice adopted
by Sasaki and others \cite{nmothers}, we work with $s=-2$.  The constants
$m$, $_s\lambda$ ($_s\lambda$ depends non-linearly on $\omega$) are
separation constants arising from the azimuthal function $e^{im\phi}$
and the angular eigenfunction $\,_sZ_{lm}^{a\omega}(\theta,\phi)$,
respectively. 
$\,_sZ_{lm}^{a\omega}(\theta,\phi)$ is the spin-weighted spheroidal
harmonic,
$_sZ_{lm}^{a\omega}(\theta,\phi)=
\frac{1}{(2\pi)^{1/2}}\,_sS_{lm}^{a\omega}(\theta)e^{im\phi}$, 
and in turn, $_sS_{lm}^{a\omega}(\theta)$ satisfies
\begin{equation}
\frac{1}{\sin\theta}\frac{d}{d\theta}(\sin \theta \frac{d}{d\theta} {S})+
(a^2\cos^2\theta-\frac{m^2}{\sin^2\theta}+4a
\omega\cos\theta+\frac{4m\cos\theta}{\sin^2\theta}
-4\cot^2\theta+c)S=0\,, 
\label{spinspheroidal}
\end{equation}
where $c=\lambda-2-a^2\omega^2+2am\omega$. The spheroidal functions
are normalized according to $\int d\Omega
|_sZ_{lm}^{a\omega}(\theta,\phi)|^2=1$.  These functions have not been
thoroughly exploited as far as we know, but the essentials, together
with tables for the eigenvalues $\lambda$ for $s=-2$ can be found in
\cite{pressteu,breuer}.  The source term $_sT$ appearing in
(\ref{teukolskyequation}) may be found in \cite{breuerbook}, and
depends on the stress-energy tensor of the perturbation under
consideration.  We note that, for a test particle falling in from
infinity with zero velocity, for example, $T \sim r^{7/2}$ and $V$ is
always long-range. This means that when we try to solve the Teukolsky
equation (\ref{teukolskyequation}) numerically we will run into
problems, since usually the numerical solution is accomplished by
integrating the source term times certain homogeneous solutions. At
infinity this integral will not be well defined since the source term
explodes there, and also because since the potential is long range,
the homogeneous solution will also explode at infinity.  We can remedy
this long-range nature of the potential and at the same time
regularize the source term in the Sasaki-Nakamura formalism. After a set of
transformations, the Teukolsky equation (\ref{teukolskyequation}) may
be brought to the Sasaki-Nakamura form
\cite{nakamurasasaki}:
\begin{equation}
\frac{d^2}{dr_*^2} {X(\omega,r)}- {\cal F}\frac{d}{dr_*}{X(\omega,r)} - 
{\cal U} X(\omega,r) = {\cal L} \,.
\label{sn}
\end{equation}
The functions ${\cal F}$ and ${\cal U}$ can be found in the original
literature \cite{nmothers,nakamurasasaki}. 
The source term ${\cal L}$ is given by
\begin{equation}
{\cal L}=\frac{\gamma_0\Delta}{r^2(r^2+a^2)^{3/2}}W e^{-i\int K/\Delta dr_*}\,,
\label{sourceterm}
\end{equation}
and $W$ satisfies
\begin{equation}
W''=-\frac{r^2}{\Delta}T e^{i\int K/\Delta dr_*}.
\label{W}
\end{equation}
The tortoise $r_*$ coordinate is defined by
$dr_*/dr=(r^2+a^2)/\Delta$, and ranges from $-\infty$ at the horizon
to $+\infty$ at spatial infinity.
The Sasaki-Nakamura equation (\ref{sn}) is to be solved under the
``only outgoing radiation at infinity'' boundary condition (see
section 4.2), meaning
\begin{equation}
X(\omega,r)=X^{{\rm out}}e^{i\omega r_*} \,, r_* \rightarrow \infty.
\label{xout}
\end{equation}
The two independent polarization modes of the metric, 
$h_+$ and $h_{\times}$, are given by
\begin{equation}
h_+ + ih_{\times}=\frac{8}{r\sqrt{2\pi}}\int_{-\infty}^{+\infty}d\omega 
\sum_{l,m} e^{i\omega(r_*-t)}   
\frac{X^{{\rm out}}S_{lm}^{a\omega}(\theta)}{\lambda(\lambda+2)-
12i\omega -12a^2\omega^2}e^{im\phi},
\label{definitionh}
\end{equation}
and the power per frequency per unit solid angle is
\begin{equation}
\frac{d^2E}{d\Omega d\omega}=
\frac{4\omega^2}{\pi}|\sum_{l,m}\frac{X^{{\rm out}}}{\lambda(\lambda+2)-
12i\omega -12a^2\omega^2} 
S_{lm}^{a\omega}|^2.
\label{power}
\end{equation}
Following Nakamura and Sasaki \cite{nmothers} the multipolar 
structure is given by 
\begin{equation}
h_+ + ih_{\times}=-\frac{(2\pi)^{1/2}}{r}\int_{-\infty}^{+\infty}d\omega 
\sum_{l,m}e^{i\omega(r_*-t)}\left[h^{lm}(\omega) _{-2}Y_{lm}\right]e^{im\phi},
\label{definitionhmult}
\end{equation}
which defines $h^{lm}$ implicitly. Moreover, 
\begin{equation}
\Delta E=\frac{\pi}{4}\int_{0}^{\infty}\sum_{lm}|h^{lm}(\omega)|^2+
|h^{lm}(-\omega)|^2.
\label{power2}
\end{equation}
Here, $_{-2}Y_{lm}(\theta)$ are spin-weighted spherical harmonics (basically
$_{2}S_{lm}^{a\omega=0}(\theta)$), whose properties are described by 
Newman and Penrose \cite{newman} and Goldberg et al 
\cite{goldberg}.

For the befenit of comparison, 
the general relation between Sasaki and Nakamura's $X(\omega,r)$ function
and Teukolsky's $R(\omega,r)$ function is given in \cite{nmothers} is 
terms of some complicated expressions. One finds however that near infinity, which is
the region we are interested in, this relation simplifies enormously.
In fact, when $r_* \rightarrow \infty$, the relation is
\begin{equation}
R= R^{{\rm out}} r^3 e^{i\omega r_*}\, 
\label{relationRX1}
\end{equation}
where, 
\begin{equation}
R^{{\rm out}}\equiv
-\frac{4\omega ^2 X^{{\rm out}}}{\lambda(\lambda+2)-12i
\omega -12a^2\omega^2}\,.
\label{relationRX2}
\end{equation}

\section{The Sasaki-Nakamura equations for a point particle
falling along a geodesic of the Kerr black hole spacetime}
The dependence of the emitted gravitational wave power on the
trajectory of the particle comes entirely from the source ${\cal L}$
in eqs. (\ref{sn})-(\ref{sourceterm}), which in turn through (\ref{W})
depends on the source term $T$ in Teukolsky's equation
(\ref{teukolskyequation}).  To compute the explicit dependence, one
would have to consider the general expression for the geodesics in the
Kerr geometry \cite{chandra}, which in general gives an end result not
very amenable to work with. However we find that by working in the
highly relativistic regime, the equations simplify enormously, and in
particular for radial infall (along the symmetry axis or along the
equator) it is possible to find a closed analytic and simple
expression for $W$ in (\ref{sourceterm})-(\ref{W}).  Accordingly, we
shall in the following give the closed expressions for ${\cal L}$ in
these two special situations: radiall infall along the symmetry axis
and radial infall along the equator of an highly energetic particle.

\subsection{The Sasaki-Nakamura equations for an highly relativistic
point particle falling along the symmetry axis of the Kerr black hole}

We now specialize then all the previous equations to the case under study.
We suppose the particle to be falling radially along the 
symmetry axis of the Kerr hole,
in which case all the equations simplify enormously. 
In this case the geodesics, written in Boyer-Lindquist coordinates, are
\begin{eqnarray}
\frac{dt}{d\tau}= \frac{\epsilon_0(r^2+a^2)}{\Delta}\,;\quad
\,(\frac{dr}{d\tau})^2=\epsilon_0^2-\frac{\Delta}{r^2+a^2}\,;\quad
\frac{d\phi}{d\tau}= \frac{a\epsilon_0}{\Delta}\,,
\label{geodesics1}
\end{eqnarray}
where the parameter $\epsilon_0$ is the energy
per unit rest mass of the infalling particle.  On considering highly
relativistic particles, $\epsilon_0 \rightarrow \infty$, we find that
the term ${\cal L}$ takes the simple form (compare with the source
term in \cite{cardosorads,cardosoradkerrs}) :
\begin{equation}
{\cal L}=-\frac{\mu C \epsilon_0 \gamma_0 \Delta}{2\omega^2 
r^2 (r^2+a^2)^{3/2}}e^{-i\omega r_*}.
\label{explicitS1}
\end{equation}
Here,
$C=\left[\frac{8Z_{l0}^{a\omega}}{\sin^2\theta}\right]_{\theta=0}$,
the function $\gamma_0=\gamma_0(r)$ can be found in
\cite{nmothers}, while $\mu$ is the mass of the particle.

\subsection{The Sasaki-Nakamura equations for an highly relativistic
point particle falling along the equatorial plane of the Kerr black
hole}

We now suppose the particle is falling radially along the
equator of the Kerr hole, where radially is defined in the sense of
Chandrasekhar \cite{chandra}, meaning that $L=a\epsilon_0$, where $L$
is the conserved angular momentum. In this case the geodesics, written
in Boyer-Lindquist coordinates, are
\begin{eqnarray}
\frac{dt}{d\tau}= \frac{\epsilon_0(r^2+a^2)}{\Delta}\,;\quad
\,r^2(\frac{dr}{d\tau})^2=\epsilon_0^2r^2-\Delta\,;\quad
\frac{d\phi}{d\tau}= \frac{a\epsilon_0}{\Delta}\,.
\label{geodesics2}
\end{eqnarray}
On considering highly relativistic particles, $\epsilon_0 \rightarrow
\infty$, we find that the source term takes again a very simple form
(compare with the source term in \cite{cardosorads,cardosoradkerrs}):
\begin{equation}
{\cal L}=-\frac{\mu \hat{S}\epsilon_0 \gamma_0 
\Delta}{\omega^2 r^2 (r^2+a^2)^{3/2}}
e^{-i\int\frac{K}{\Delta}dr}\,,
\label{explicitS2}
\end{equation}
where
\begin{equation}
\hat{S}=\left[\frac{\lambda}{2}-(m-a\omega)^2\right]Z_{lm}
({\theta=\frac{\pi}{2}})
+\left[(m-a\omega)\right]Z'_{lm}({\theta=\frac{\pi}{2}})
\label{termofontespin}
\end{equation}

Once $X^{{\rm out}}$ is known, we can get the Teukolsky 
wavefunction near infinity 
from (\ref{relationRX1}), and the waveform and energy spectra from 
(\ref{definitionh})-(\ref{power2}).

\section{Implementing a numerical solution}
To compute the spin-weighted spheroidal functions, one would have to
compute the eigenvalues $\lambda$, for each $\omega$, using a shooting
method \cite{numrecipes} or any other numerical scheme one chooses
\cite{nmothers,hughes1}.  We have chosen to lean ourselves on work
already done by Press and Teukolsky \cite{pressteu} and to verify at
each step the correctness of their results (we found no
discrepancies).  Press and Teukolsky have found, for each ($l$, $m$),
a sixth-order polynomial (in $a \omega$) which approximates the
eigenvalue $\lambda$, to whithin five decimal places (see their Table
1).  This is a good approximation as long as $a\omega<3$, which is
within our demands for $l<6$.  The computation of the spheroidal
function then follows trivially, by numerically integrating
(\ref{spinspheroidal}).

The next step is to determine $X(\omega,r)$ from the Sasaki-Nakamura
differential equation (\ref{sn}). This is accomplished by a Green's
function technique, constructed so as to satisfy the usual boundary
conditions, i.e., only ingoing waves at the horizon ($X \sim
e^{-i\omega r_*}\,,r_*\rightarrow -\infty$) and outgoing waves at
infinity($X \sim e^{i\omega r_*}\,,r_*\rightarrow \infty$).  We get
that, near infinity, and for infall along the equator (we are
interested in knowing the wavefunction in this region),
\begin{equation}
X=X^{\infty}\int \frac{X^{H}{\cal S}}{{\rm Wr}} dr_*\,;\quad
X^{\rm out}=\int \frac{X^{H}{\cal S}}{{\rm Wr}} dr_*.
\label{solutionradkerr}
\end{equation}
The similar expression for infall along the symmetry axis is given in 
\cite{cardosoradkerrs}.
Here, $X^{\infty}$ and $X^{H}$ are two linearly independent 
homogeneous solutions of (\ref{sn}) which
asymptotically behave as
\begin{eqnarray}
X^{H} \sim A(\omega)e^{i\omega r_*}+B(\omega)e^{-i\omega r_*}\,,r_* 
\rightarrow \infty \label{behavior1radkerr}\\
X^{H} \sim e^{-i\omega r_*}\,,r_* \rightarrow -\infty \label{behavior12}\\ 
X^{\infty} \sim e^{i\omega r_*}\,,r_* \rightarrow \infty 
\label{behavior13}\\
X^{\infty} \sim C(\omega)e^{i\omega r_*}+D(\omega)e^{-i\omega r_*}
\,,r_* \rightarrow -\infty\,, 
\label{behavior14}
\end{eqnarray}
and ${\rm Wr}$ is the wronskian of these two solutions.  Expression
(\ref{solutionradkerr}) can be further simplified in the case of infall
along the equator to
\begin{equation}
X^{{\rm out}}=-\frac{\mu \epsilon_0 c_0 \hat{S} }{2i\omega^3 B}\int 
\frac{e^{-i\int \frac{K}{\Delta}dr} X^{H}}{r^2(r^2+a^2)^{1/2}} dr.
\label{solution2}
\end{equation}

To implement a numerical solution, we first have to determine $B(\omega)$.
We find $B(\omega)$ by solving (\ref{sn}) with the right hand side set 
to zero, and with the
starting condition $X^H= e^{-i\omega r_*}$ imposed at a large negative
value of $r_*$. For computational purposes good accuracy is hard to
achieve with the form (\ref{behavior1radkerr}), so we used an asymptotic
solution one order higher in $1/(\omega r)$:
\begin{eqnarray}
&X^H = A(\omega)(1+\frac{i(\lambda+2+2am\omega -2i\omega P)}
{2\omega r})e^{i\omega r_*}+
\nonumber\\ &
+B(\omega)(1-\frac{i(\lambda+2+2am\omega -2i\omega P)}
{2 \omega r})e^{-i\omega r_*}\,,r_* \rightarrow +\infty\,.&
\label{behavior2radkerr}
\end{eqnarray}
Here, $P=\frac{8ai(-m\lambda +a\omega(\lambda-3))}
{12a^2\omega^2+12iM\omega-\lambda^2-2\lambda-12am\omega}$.
These expressions all refer to the case of infall along 
the equator. For analogous
expressions for infall along the symmetry axis see \cite{cardosoradkerrs}. 

In the numerical work, we chose to adopt $r$ as the independent
variable, thereby avoiding the numerical inversion of $r_*(r)$.
A fourth order Runge-Kutta routine started the integration
of $X^H$ near the horizon, at $r=r_+ + r_+\epsilon$
(the horizon radius $r_+$ is $r_+=M+(M^2-a^2)^{1/2}$), with tipically 
$\epsilon=10^{-5}$. It then integrated out to large values
of $r$, where one matches $X^H$ extracted numerically with the
asymptotic solution (\ref{behavior2radkerr}), in order to find $B(\omega)$.
To find $\Delta E$ the integral in (\ref{power2}) is
done by Simpson's rule. For both routines Richardson
extrapolation is used.

\section{Numerical Results and Conclusions}

Recent studies \cite{cardosorads,cardosoradkerrs} on high energy
collisions of point particles with black holes point to the existence
of some characteristic features of these processes, namely: (i) the
spectrum and waveform largely depend upon the lowest quasinormal
frequency of the spacetime under consideration;  (ii) there is a
non-vanishing zero frequency limit (ZFL) for the spectra,
$\frac{dE}{d\omega}_{\omega\rightarrow 0}$, and it seems to be
independent of the spin of the colliding particles whereas for
low-energy collisions the ZFL is zero;  (iii) the energy radiated in
each multipole has a power-law dependence rather than exponential for
low-energy collisions.

\begin{figure}
\centerline{\includegraphics[width=10 cm,height=6.5 cm]
{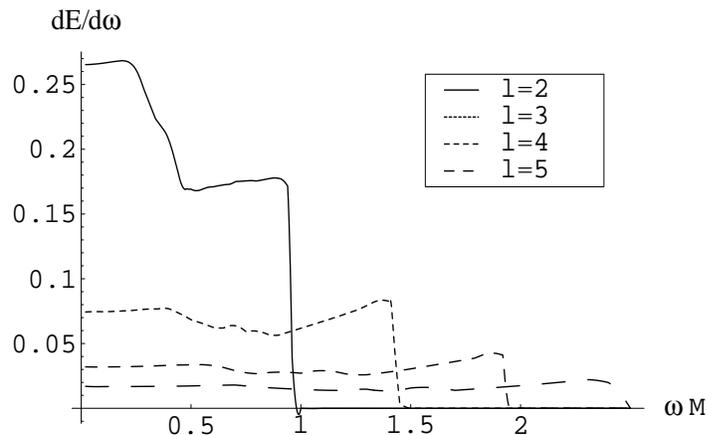}}
\caption{The energy spectra for a point particle moving at nearly the
speed of light and colliding, along the equatorial plane, with an
extreme (a=0.999M) Kerr black hole. The energy is normalized in units
of $\mu^2 \epsilon_0^2$.  Notice that the spectra is almost flat (for
large $l$), the ZFL is non-vanishing and that the quadrupole carries
less than half of the total radiated energy.}
\label{fig:1radkerr}
\end{figure}

\begin{figure}
\centerline{\includegraphics[width=10 cm,height=6.5 cm]
{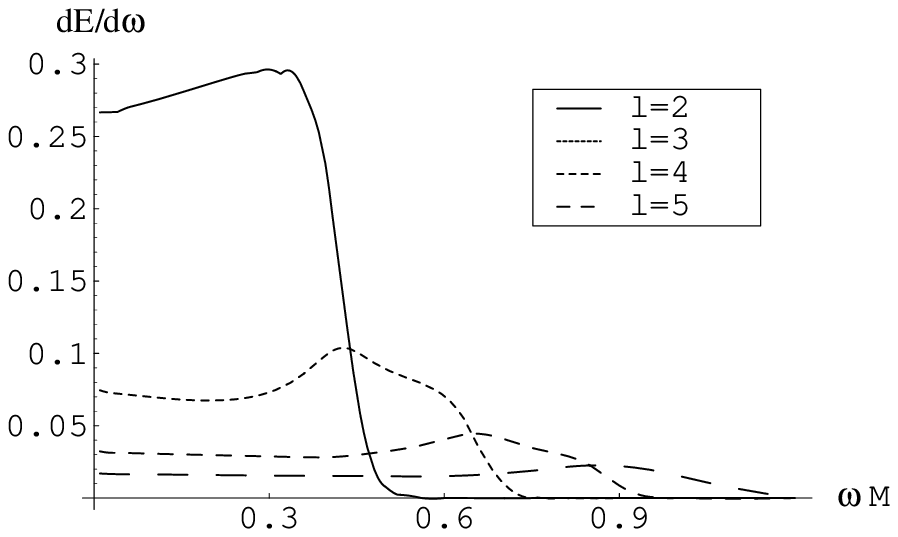}}
\caption{The energy spectra for a point particle moving at nearly the
speed of light and colliding, along the symmetry axis, with an extreme
(a=0.999M) Kerr black hole. The energy is normalized in units of
$\mu^2 \epsilon_0^2$.  Notice that the spectra is almost flat (for
large $l$), the ZFL is non-vanishing and that the quadrupole carries
less than half of the total radiated energy.}
\label{fig:2}
\end{figure}

The present study reinforces all these aspects.  In Fig. \ref{fig:1radkerr} we show,
for an almost extreme Kerr hole with $a=0.999M$, the energy spectra
for the four lowest values of $l$, when the particle collides along
the equatorial plane.  
In Fig. \ref{fig:2} we show the same values but for a
collision along the symmetry axis.  The existence of a non-vanishing
ZFL is evident, but the most important in this regard is that the ZFL
is exactly the same, whether the black hole is spinning or not, or
whether the particle is falling along the equator or along the
symmetry axis.  
In fact, our numerical results show that, up to the
numerical error of about $1\%$ the ZFL is given by Table \ref{tab:zflradkerr}
(see also \cite{cardosorads} and the exact value given by Smarr \cite{smarr2}),
and this holds for highly relativistic particles falling along the
equator (present work), along the symmetry axis \cite{cardosorads,cardosoradkerrs}, or
falling into a Schwarzschild black hole \cite{cardosorads}.

\begin{table}
\caption{\label{tab:zflradkerr}  The zero frequency limit (ZFL) 
for the ten lowest radiatable multipoles.}
\begin{tabular}{llll}  \hline
$l$ & ZFL($\times\frac{1}{\mu^2 \epsilon_0^2}$)&$l$& ZFL($\times 
\frac{1}{\mu^2 \epsilon_0^2}$)\\ \hline
2    &  0.265 &  7 &  0.0068 \\ \hline 
3    &  0.075 &   8 &  0.0043  \\ \hline 
4    &  0.032 &   9 &  0.003 \\ \hline 
5    &  0.0166 &   10 &  0.0023  \\ \hline 
6    &  0.01  &   11 &  0.0017 \\ \hline 
\end{tabular}
\end{table}
\vskip 1mm

The $l$-dependence of the energy radiated is a power-law; in fact for large $l$
we find for infall along the equator
\begin{equation}
\Delta E_l=0.61\frac{\mu^2 \epsilon_0^2}{M l^{1.666}}\;\;,\quad\quad a=0.999M.
\label{lbehav}
\end{equation} 
Such a power-law dependence seems to be universal for high energy
collisions.  Together with the universality of the ZFL this is one of
the most important results borne out of our numerical studies.  The
exponent of $l$ in (\ref{lbehav}) depends on the rotation
parameter. As $a$ decreases, the exponent increases monotonically, until
it reaches the Schwarzschild value of 2 ($\Delta E_l \sim
\frac{1}{l^{2}}$) which was also found for particles falling along the
symmetry axis of a Kerr hole. In Table \ref{tab:powerradkerr} we show the values of the
exponent, as well as the total energy radiated, for some values of the
rotation parameter $a$.
\vskip 1cm
\begin{table}
\caption{\label{tab:powerradkerr} Power-law dependence of the energy radiated in
each multipole $l$, here shown for some values of $a$, the rotation
parameter. We write $\Delta E_l=\frac{c}{l^b}$ for the energy emitted
for each $l$ and $\Delta E_{{\rm tot}}$ for the total energy radiated
away. The collision happens along the equatorial plane. }
\begin{tabular}{|l|l|l|l|}  \hline 
\multicolumn{4}{|c|}{ $\Delta E_l=cl^{-b}$} \\ \hline
$\frac{a}{M}$&  $c$  &  $b$  & $\Delta E_{{\rm tot}}\frac{M}{\mu^2 \epsilon_0^2}$ \\ \hline
0.999        &  0.61 & 1.666 & 0.69   \\ \hline
0.8          &  0.446& 1.856 & 0.36   \\ \hline
0.5          & 0.375 & 1.88  & 0.29   \\ \hline
0            & 0.4   & 2     & 0.26   \\ \hline
\end{tabular}
\end{table}
This power-law dependence and our numerical results allow us to infer
that the total energy radiated for a collision along the equator is
\begin{equation}
\Delta E_{{\rm tot}}=0.69 \frac{ \mu^2 \epsilon_0^2}{M}\;\;,\quad\quad 
a=0.999M.
\label{Etot}
\end{equation} 
This represents a considerable enhancement of the total radiated
energy, in relation to the Schwarzschild case \cite{cardosorads} or
even to the infall along the symmetry axis \cite{cardosoradkerrs}.  Again,
the energy carried by the $l=2$ mode ($\Delta E_{l=2}=0.2\frac{\mu^2
\epsilon_0^2}{M}\;,\;\; a=0.999M$) is much less than the total radiated
energy.  
\begin{figure}
\centerline{\includegraphics[width=10 cm,height=6.5 cm]
{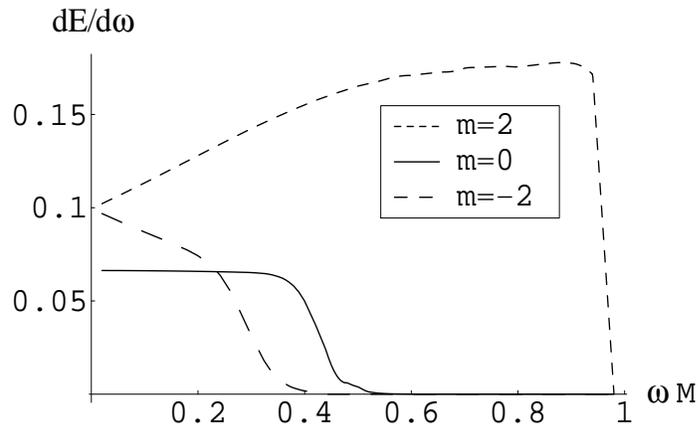}}
\caption{The energy spectra as a function of $m$ for $l=2$ 
and for an highly
relativistic particle falling along the equatorial plane 
of a Kerr black hole.
The energy is normalized in units of $\mu^2 \epsilon_0^2$.}
\label{fig:3}
\end{figure}
\begin{figure}
\centerline{\includegraphics[width=10 cm,height=6.5 cm]
{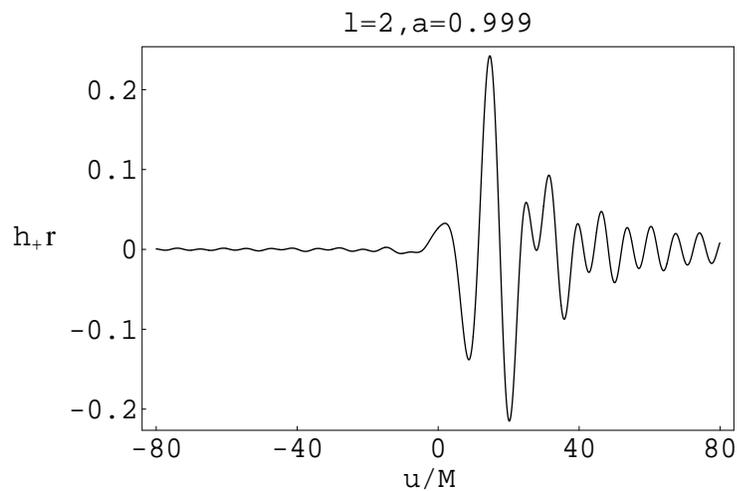}}
\caption{The $l=2$ waveform as defined by (\ref{definitionhmult}) 
for an highly
relativistic particle falling along the equatorial plane of 
an extreme
(a=0.999) Kerr black hole.The waveform is normalized in 
units of $\mu \epsilon_0$.}
\label{fig:4}
\end{figure}
In Fig. \ref{fig:3} we show, for $l=2$, the energy spectra as a
function of $m$, which allows us to see clearly the influence of the
quasinormal modes.  Indeed, one can see that the energy radiated for
$m=2$ is much higher than for $m=-2$, and this is due to the behaviour
of the quasinormal frequency for different azimuthal numbers $m$
\cite{QNKerr}, as emphasized by different authors \cite{nmothers}.  A
peculiar aspect is that the energy radiated for $m+l=$ odd is about
two orders of magnitude lower than for $m+l=$ even, and that is why
the spectra for $m=-1,1$ is not plotted in Fig. \ref{fig:3}.  Notice that the ZFL
is the same for $m=2$ and for $m=-2$ both conspiring to make the ZFL
universal.  In Fig. \ref{fig:4} we show the $l=2$ waveform for $a=0.999M$ as
seen in $\phi=0\,, \theta=\pi /2$.  By symmetry, $h_{\times}=0$.

Up to now, we dealt only with the head-on collision between a particle
and a black hole. What can we say about collisions with non-zero
impact parameter, and along the equator ?  As is clear from our
results, the fact that the quasinormal modes are excited is
fundamental in obtaining those high values for the total energy
radiated.  Previous studies \cite{QNKerr} show that the quasinormal
modes are still strongly excited if the impact parameter is less than
$2M$. We are therefore tempted to speculate that as long as the impact
parameter is less than $2M$ the total energy radiated is still given
by Table \ref{tab:powerradkerr}. For larger values of the impact parameter, one expects the
total energy to decrease rapidly. Moreover, for large impact
parameters, Smarr's calculation \cite{smarr2} should apply (but see
also other work cited in \cite{smarr2}).
On the other hand, If the
collision is not along the equatorial plane, we do expect the total
energy to decrease. For example, if the collision is along the
symmetry axis, we know \cite{cardosoradkerrs} that the total energy is
$0.31 \frac{ \mu^2 \epsilon_0^2}{M}$ for $a=0.999M$.  So we expect
that as the angle between the collision axis and the equator is varied
between $0$ and $\pi/2$ the total energy will be a monotonic function
decreasing from $0.69 \frac{ \mu^2 \epsilon_0^2}{M}$ to $0.31 \frac{
\mu^2 \epsilon_0^2}{M}$.  Still, more work is necessary to confirm
this.  Let us now consider, using these results, the collision at
nearly the speed of light between a Schwarzschild and a Kerr black
hole, along the equatorial plane.  We have argued in previous papers
\cite{cardosorads,cardosoradkerrs} that the naive extrapolation $\mu
\rightarrow M$ may give sensible results, so let's pursue that idea
here.  We obtain an efficiency of $34.5 \%$ for gravitational wave
generation, a remarkable increase relative to the
Schwarzschild-Schwarzschild collision.  Now, the area theorem gives an
upper limit of $38.7 \%$ so we may conclude with two remarks: (a)
these perturbative methods pass the area theorem test; (b) we are
facing the most energetic events in the Universe, with the amazing
fraction of $34.5 \%$ of the rest mass being converted into
gravitational waves.

\thispagestyle{empty} \setcounter{minitocdepth}{1}
\chapter[Scalar radiation in asymptotically anti-de Sitter in three, four and higher dimensional spacetimes]{Radiation in asymptotically anti-de Sitter in three, four and higher dimensional spacetimes} \label{chap:radads}
\lhead[]{\fancyplain{}{\bfseries Chapter \thechapter. \leftmark}}
\rhead[\fancyplain{}{\bfseries \rightmark}]{}
\minitoc \thispagestyle{empty}
\renewcommand{\thepage}{\arabic{page}}
\noindent
\section{ Introduction}
\vskip 3mm

Anti-de Sitter (AdS) spacetime has been considered of fundamental
meaning within high energy elementary particle physics, specially in
supersymmetric theories of gravity such as 11-dimensional supergravity
and M-theory (or string theory).  The dimension $D$ of AdS spacetime
is a parameter which can have values from
two to eleven, and where the other spare
dimensions either are joined as a
compact manifold $\cal M$ into the whole spacetime to yield ${\rm
AdS}_d\times {\cal M}^{11-D}$ or 
receive a Kaluza-Klein treatment.  AdS spacetime appears as the background
for black holes solutions and it also plays a
further crucial role since it is the near-horizon geometry, separated
by a (soft) boundary from an otherwise asymptotically flat spacetime,
of some black solutions \cite{horostrometal}.  In addition, by taking
low energy limits at strong coupling and through group theoretic
analysis, Maldacena conjectured a correspondence between the bulk of
AdS spacetime and a dual conformal field gauge theory (CFT) on the
spacetime boundary itself \cite{maldacena}.  A concrete
method to implement this correspondence is to identify the extremum of
the classical string theory action $I$ for the dilaton field $\varphi$,
say, at the boundary of AdS, with the generating functional $W$ of the
Green's correlation functions in the CFT for the operator $\cal O$
that corresponds to ${\varphi}$
\cite{gubserklebanovpolyakov}, 
$
I_{\varphi_0(x^\mu)}\,=
\, W[\varphi_0(x^\mu)]\,, 
$
where $\varphi_0$ is the value of $\varphi$ at the AdS boundary and the
$x^\mu$ label the coordinates of the boundary. The motivation for this
proposal can be seen in the reviews
\cite{aharonyetalklebanovhorowitzreview,maldacenareview}.  In its strongest form the
conjecture requires that the spacetime be asymptotically AdS, the
interior could be full of gravitons or containing a black hole.  The
correspondence realizes the holographic principle, since the bulk is
effectively encoded in the boundary, and is also a strong/weak
duality, it can be used to study issues of strong gravity using weak
CFT or CFT issues at strong coupling using classical gravity in the
bulk.

A particular important AdS dimension is three.  In AdS$_3$ Einstein
gravity is simple, the group of isometries is given by two copies of
$SL(2,R)$, it has no propagating degrees of freedom, is
renormalizable, it allows for the analytical computation of many
physical processes extremely difficult or even impossible in higher
dimensions, it belongs to the full string theory compactification
scheme \cite{aharonyetalklebanovhorowitzreview,maldacenareview}, 
the dual CFT$_2$ is
the low-energy field theory of a D1-D5-brane system which can be
thought of as living on a cylinder (the boundary of AdS$_3$)
\cite{maldacenastrom}, and it contains the BTZ black hole.  The BTZ
black hole is of considerable interest, not only because it can yield
exact results, but also because one hopes that the results can
qualitatively be carried through to higher dimensions.  Several
results related to the BTZ black hole itself and to the AdS/CFT
correspondence have been obtained
\cite{keski,danielsson,mat,balasu}.  The AdS/CFT
mapping implies that a black hole in the bulk corresponds to a thermal
state in the gauge theory \cite{wittenbanksetal}. Perturbing the black
hole corresponds to perturbing the thermal state and the decaying of
the perturbation is equivalent to the return to the thermal state.
Particles initially far from the black hole correspond to a blob (a
localized excitation) in the CFT, as the IR/UV duality teaches
\cite{susskindwitten}.  The evolution towards the black hole represents
a growing size of the blob with the blob turning into a bubble
traveling close to the speed of light \cite{danielsson}.

In this chapter we will extend some of the previous results and we begin by  studying in
detail the collision between a BTZ black hole and a scalar particle.
This will follow closely \cite{cardosoradbtz} (see also \cite{cardosogul}).
Generically, a charged particle falling towards a black hole emits
radiation of the corresponding field. In higher dimensions it also
emits gravitational waves, but since in three dimensions there is no
gravitational propagation in the BTZ case there is no emission. 

Thus, a scalar particle falling into a BTZ
black hole emits scalar waves. This collision process is important
from the points of view of three-dimensional dynamics and 
of the AdS/CFT conjecture. Furthermore, one can
compare this process with previous works, since there are exact
results for the quasinormal mode (QNM) spectrum of scalar 
perturbations which are known to govern their decay 
at intermediate and late times
\cite{horowitz,cardosoqnmbtz,cardosogul,cardosoqnmsads2,cardosoqnmtoro,cardosoqnmsads}.

The phenomenon of radiation emission generated from an infalling
particle in asymptotically flat spacetimes has been studied by several
authors \cite{davis,davis2,anninos,gleiser} and most recently in \cite{lousto},
where the results are to be compared to full scale numerical
computations for strong gravitational wave emission of astrophysical
events \cite{baker} which will be observed by the GEO600, 
LIGO and VIRGO projects. A scalar infalling particle as a model for
calculating radiation reaction in flat spacetimes has been considered in 
\cite{burko}.
Many of the
techniques  have been developed in connection
to such spacetimes.  Such an
analysis has not been carried to non-asymptotically flat spacetimes,
which could deepen our understanding of these kind of events, and of
Einstein's equations.  In this respect, for the mentioned reasons, 
AdS spacetimes are the most promising candidates.  
As asymptotically flat spacetimes
they provide well defined conserved charges and positive 
energy theorems, which makes them a good
testing ground if one wants to go beyond flatness. However, 
due to different boundary conditions it raises new
problems. The first one is that since the natural boundary conditions are 
boxy-like all of the generated radiation will eventually fall into
the black hole, thus infinity has no special meaning in
this problem, it is as good a place as any other, i.e., one can
calculate the radiation passing at any radius $r$, for
instance near the horizon.  Second, in contrast to asymptotically flat
spacetimes, here one
cannot put a particle at infinity (it needs an infinite amount of
energy) and thus the particle has to start from finite $r$. This has
been posed in \cite{lousto} but was not fully solved when applied
to AdS spacetimes.

We shall then extend the analysis, and compute
in detail the collision between a black hole and a scalar particle in general
dimensions \cite{cardosoradadsddim}. 
This collision process is interesting
from the point of view of the dynamics itself in relation to the
possibility of manufacturing black holes at LHC within the brane world
scenario \cite{hamed,bhprod}, and from the point of view of the AdS/CFT
conjecture, since the scalar field can represent the string theory
dilaton, and  $4$, $5$, $7$ (besides $3$) are  dimensions of interest for the
AdS/CFT correspondence \cite{maldacenareview,horowitz}.  In addition, one can
compare this process with previous works, since there are results for
the quasinormal modes of scalar, and electromagnetic perturbations
which are known to govern the decay of the perturbations, at
intermediate and late times \cite{cardosoqnmsads,cardosoqnmsads2,horowitz,cardosoradbtz,qnmads}.

Some general comments can be made about the mapping AdS/CFT 
when it envolves a black hole. 
A black hole in the bulk corresponds to a thermal 
state in the gauge theory \cite{wittenbanksetal}. Perturbing the 
black hole corresponds to perturbing the thermal state and the 
decaying of the perturbation is equivalent to the return 
to the thermal state. So one obtains a prediction for the 
thermal time scale in the strongly coupled CFT. 
Particles initially far from the black hole correspond to a 
blob (a localized excitation) in the CFT, as the IR/UV 
duality teaches (a  position in the bulk is equivalent 
to size of an object) \cite{susskindwitten}.
The evolution toward the black hole represents a growing 
size of the blob with the blob turning into a bubble travelling 
close to the speed of light \cite{danielsson}. 
For other processes in anti-de Sitter spacetime, we refer the reader to
\cite{cardosoradsync} and references therein.

\noindent
\section{Scalar radiation from infall of a particle into a BTZ black hole}
\vskip 3mm

\noindent
\subsection{Formulation of the problem and basic equations}
\vskip 3mm

We consider a small test particle of mass $m_0$ and charge 
$q_0$, coupled to a massless scalar field $\varphi$, moving along 
a radial timelike
geodesic outside a BTZ black hole of mass $M$. The metric
outside the BTZ black hole is
\begin{equation}
ds^{2}= f(r) dt^{2}- \frac{dr^{2}}{f(r)}-
r^{2}d\theta^{2} \,,
\label{lineelementc3}
\end{equation}
where, $f(r)=(-M+\frac{r^2}{l^2})$, $l$ is the AdS radius 
(G=1/8; c=1). 
The horizon radius is given by
$r_+=M^{1/2}l$.
We treat the scalar field as a perturbation,
so we shall neglect the back reaction of the field's stress
tensor on the metric (this does not introduce large errors 
\cite{gleiser}).
If we represent the particle's worldline by
$x^{\mu}=x_p^{\mu}(\tau)$, with $\tau$ the proper
time along a geodesic, then the interaction action $\cal I$ is 
\begin{eqnarray}
&{\cal I}
=-\frac{1}{8 \pi} \int g^{1/2} \varphi _{;a} \varphi ^{;a} d^3y-
\nonumber\\
&
m_0 \int (1+q_0 \varphi)(-g_{ab}\dot{x}^a \dot{x}^b)^{\frac{1}{2}} 
d\tau\,, &
\label{action}
\end{eqnarray}
and thus the scalar field satisfies
the inhomogeneous wave equation
$
\Box \varphi= -4 \pi m_0q_0 \int
\delta^3(x^{\mu}-x_p^{\mu}(\tau))(-g)^{-1/2}d\tau\,,
$
where $g$ is the metric determinant and $\Box$ denotes
the covariant wave operator.
As the particle moves on a timelike geodesic, we have
\begin{equation}
\dot{t}_p=\frac{\cal E}{f(r_p)}\;\;,\dot{r}_p=-({\cal E}^2-f(r_p))^{1/2}\,,
\label{geodesicc3}
\end{equation}
where $\dot{ }\equiv d/d\tau$, 
and $\cal E$ is a conserved energy parameter.
We shall be considering the test particle initially at rest
at a distance $r_0$ (where ${\cal E}^2=f(r_0)$) and at $\theta_p=0$.
Expanding the field as 
\begin{equation}
\varphi(t,r,\theta)=\frac{1}{r^{1/2}}\phi(t,r)\sum_m e^{im\theta} \,,
\label{decomposition}
\end{equation}
where $m$ is the angular momentum 
quantum number, the wave equation is given by
(after an integration in $\theta$)
\begin{eqnarray}
&\frac{\partial^{2} \phi(t,r)}{\partial r_*^{2}} -
\frac{\partial^{2} \phi(t,r)}{\partial t^{2}}-V(r)\phi(t,r)=
\nonumber\\
&
-\frac{2q_0 m_0 f}{r^{1/2}}(\frac{dt}{d\tau})^{-1}\delta(r-r_p) \,,&
\label{waveeq2}
\end{eqnarray}
with
$
V(r)=\frac{3r^2}{4 l^4} - \frac{M}{2 l^2}-\frac{M^2}{4 r^2}+\frac{m^2}{l^2}
 - \frac{Mm^2}{r^2}\,,
$
and 
$
r_*=-M^{-1/2}{\rm arcoth}(r\,M^{-1/2}).
$
\noindent
\subsection{ The initial data and boundary conditions}
\vskip 3mm

\begin{figure}
\centerline{\includegraphics[width=10 cm,height=6.5 cm]
{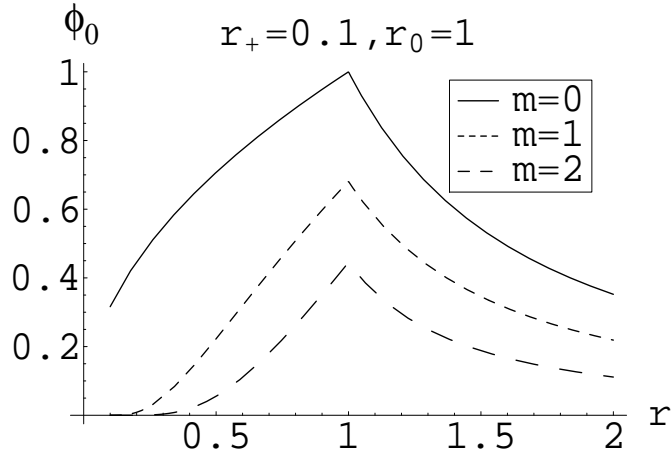}}
\caption{Initial data $\phi_0$ for a BTZ black hole with $r_+=0.1$, 
and with the particle at $r_0=1$, for several values of $m$, 
the angular quantum number. }
\label{staticbtz}
\end{figure}

In the case we study, and in contrast to asymptotically flat 
spacetimes where initial data can be pushed to infinity \cite{davis,davis2,anninos,gleiser}, 
initial data must be provided. 
Accordingly, we take the Laplace transform $\Phi(\omega,r)$ of
$\phi(t,r)$ to be 
\begin{equation}
\Phi(\omega,r)=\frac{1}{(2\pi)^{1/2}}\int_{0}^{\infty}
e^{i \omega t}\phi(t,r)dt.
\label{laplace}
\end{equation}   
Then, equation (\ref{waveeq2}) may be written as 
\begin{equation}
\frac{\partial^{2} \Phi(r)}{\partial r_*^{2}} +
\left\lbrack\omega^2-V(r)\right\rbrack\Phi(r)=S
+ \frac{i\omega \phi_0}{(2 \pi)^{1/2}}\,,
\label{waveeq3}
\end{equation}
with,
$
S=-\frac{f}{r^{1/2}}(\frac{2}{\pi})^{1/2}\frac{1}{\dot{r}_p} 
e^{i\omega t} \,
$
being the source function, 
and $\phi_0$ is the initial value of $\phi(t,r)$ satisfying
$
\frac{\partial^{2} \phi_0(r,m)}{\partial r_*^{2}} 
-V(r)\phi_0(r,m)=
-\frac{2f}{r^{1/2}}(\frac{dt}{d\tau})_{r_0}^{-1}\delta(r-r_p) \,.
$
We have rescaled $r$, $r\rightarrow\frac{r}{l}$, and measure
everything in terms of $l$, i.e., $\phi$, $r_+$ and $\omega$
are to be read, $\frac{1}{l^{1/2}q_0 m_0}\phi$, 
$\frac{r_+}{l}$ and $\omega l$, respectively.
One can numerically solve the equation for the initial
data $\phi_0$ by demanding regularity at both the horizon
and infinity (for a similar problem see \cite{wald}).
In Fig. \ref{staticbtz}, we show the form of $\phi_0$ for a typical case
$r_+=0.1$,  
$r_0=1$, and for three different values of $m$, 
$m=0,\,1,\,2$. 
Other cases like $r_+=1,\,10,\,...$ and several values of $r_0$ can be
computed.  Large black holes have a direct interpretation in the
AdS/CFT conjecture. The results for large or small black holes are
nevertheless similar, as we have checked.  As a test for the
numerical evaluation of $\phi_0$, we have checked that as $r_0
\rightarrow r_+$, all the multipoles fade away, i.e., $\phi_0
\rightarrow 0$, supporting the no-hair conjecture (that all the
multipoles go to zero).

To solve equation (\ref{waveeq3}) one has to impose physically
sensible boundary conditions, appropriate to AdS spacetimes.
In our case the potential diverges at infinity,
where $\phi_0$ vanishes, so we impose reflective boundary conditions
\cite{avis} there, i.e., $\Phi=0$ at infinity.  It has been common
practice to set $\Phi\sim F(\omega) e^{-iwr_*} $ near the horizon,
meaning ingoing waves there. This is an allowed boundary condition as
long as $\phi_0$ vanishes there. However, if $\phi_0$ does not vanish
there, one has to be careful in defining boundary conditions at the horizon.
When dealing with AdS spacetimes this detail is crucial to extract 
the correct information, and it has been overlooked when one deals 
with asymptotically flat spacetimes as in \cite{lousto}. 
Equation (\ref{waveeq3})
together with the source term $S$ 
allow us to conclude that near the horizon
$
\Phi \sim G(\omega)e^{i\omega r_*}+F(\omega) e^{-i\omega r_*}+
\frac{i \phi_0}{(2\pi)^{1/2}\omega }\,.
$
Since we want waves going down the black hole,  we shall require
\begin{equation}
\Phi \sim F(\omega) e^{-i\omega r_*}+
\frac{i \phi_0}{(2\pi)^{1/2}\omega }\,\,\,, r\rightarrow r_+
\label{boundarybehavior2}
\end{equation}
\noindent
\subsection{ Green's function solution}
\vskip 3mm

To proceed we must find a solution to equation (\ref{waveeq3}) through
a Green's function analysis.  A standard treatment \cite{lousto}
invokes contour integration to calculate the integrals near the
horizon.  There is no need for this here, by demanding regularized
integrals the correct boundary conditions appear in a natural way (see
\cite{poisson} for a regularization of the Teukolsky equation).  Let
$\Phi^{\infty}$ and $\Phi^{H}$ be two independent solutions of the
homogeneous form of (\ref{waveeq3}), satisfying:
$\Phi^H \sim e^{-i\omega r_*}\,,r \rightarrow r_+ \,$;
$\Phi^H \sim A(\omega) r^{1/2}+B(\omega)r^{-3/2}\,,r \rightarrow \infty \,$; 
$\Phi^{\infty} 
\sim C(\omega)e^{i\omega r_*}+D(\omega)e^{-i\omega r_*}\,,r 
\rightarrow r_+  \,$;
$\Phi^{\infty} \sim 1/r^{3/2}\,,r \rightarrow \infty\,$.
Define $h^H$ through $dh^H/dr_*=-\Phi^H$ and $h^{\infty}$ 
through $dh^{\infty}/dr_*=
-\Phi^{\infty}$.
We can then show that $\Phi$ given by 
\begin{eqnarray}
&
\Phi=\frac{1}{W} \left[ \Phi^{\infty} \int _{-\infty}^{r}\Phi^H S dr_* +
\Phi^H \int _{r}^{\infty}\Phi^{\infty} S dr_* \right]+
\nonumber\\
&\frac{i\omega}{(2\pi)^{1/2} W } \left[ \Phi^{\infty} 
\int _{-\infty}^{r}h^H \frac{d\phi_0}{dr_*} dr_* +\right.
\nonumber \\
&
\Phi^H \int _{r}^{\infty}h^{\infty} \frac{d\phi_0}{dr_*} dr_* + 
\nonumber \\
& 
\left. (h^{\infty}\phi_0 \Phi^H-h^H\phi_0\Phi^{\infty})(r)
\right]\,, 
\label{Phi}
\end{eqnarray}
is a solution to (\ref{waveeq3}) 
and satisfies
the boundary conditions.
The Wronskian $W= 2i\omega C(\omega)$ is a constant.
Near infinity, we get from (\ref{Phi}) that
\begin{eqnarray}
&
\Phi(r\rightarrow\infty)=\frac{1}{W} 
\left[ \Phi^{\infty}_{(r\rightarrow\infty)} \int _{-\infty}^{\infty}\Phi^H S dr_* \right]+
\nonumber\\
&\frac{i\omega}{(2\pi)^{1/2} W } \left[ \Phi^{\infty}_{(r\rightarrow\infty)} 
\int _{-\infty}^{\infty}h^H \frac{d\phi_0}{dr_*} dr_* +\right.
\nonumber \\
& 
\left. (h^{\infty}\phi_0 \Phi^H-h^H\phi_0\Phi^{\infty})(r\rightarrow\infty)
\right]\,.
\label{Phiinf}
\end{eqnarray}
Now, in our case, this is just zero, as it should, because both $\Phi^{\infty},
\phi_0 \rightarrow 0$, as $r\rightarrow \infty$.
However, if one is working with asymptotically flat space, as in \cite{lousto},
where $\Phi^{\infty}\rightarrow e^{i\omega r_*}$ at infinity, we get (recalling
that $\phi_0 \rightarrow 0$): 
\begin{eqnarray}
&
\Phi(r\rightarrow\infty)=\frac{1}{W} 
\left[ \Phi^{\infty}_{(r\rightarrow\infty)} \int _{-\infty}^{\infty}\Phi^H S dr_* \right]+
\nonumber\\
&\frac{i\omega}{(2\pi)^{1/2} W }  \Phi^{\infty}_{(r\rightarrow\infty)} 
\int _{-\infty}^{\infty}h^H \frac{d\phi_0}{dr_*} dr_* \,,
\label{Phiflat}
\end{eqnarray}
and where each integral is well defined.
In particular, integrating by parts the second integral can be put 
in the form
\begin{eqnarray}
&
\int _{-\infty}^{\infty}h^H \frac{d\phi_0}{dr_*} dr_* =
\left[h^H \phi_0 \right]_{-\infty}^{\infty}+\int _{-\infty}^{\infty}\Phi^H \phi_0{dr_*}\nonumber\\
&=\frac{i\phi_0 e^{-i\omega r_*} }{\omega}(r\rightarrow -\infty)+\int _{-\infty}^{\infty}\Phi^H \phi_0{dr_*}.
\label{Phiflat2}
\end{eqnarray}
Here, the final sum converges, but not each term in it.
Expression (\ref{Phiflat2}) is just expression (3.15) in \cite{lousto}, although
it was obtained imposing incorrect boundary conditions and not well defined regularization
schemes. Due to the fact that the initial data vanishes at infinity,
the results in \cite{lousto} are left unchanged.
In this work, we are interested in computing the wavefunction $\Phi(r,\omega)$ 
near the horizon ($r\rightarrow r_+$). In this 
limit we have
\begin{eqnarray}
&
\Phi(r\sim r_+)=\frac{1}{W} \left[ \Phi^H 
\int _{r_+}^{\infty}\Phi^{\infty} S dr_* \right]+&
\nonumber \\
&
\frac{i\omega}{(2\pi)^{1/2}W }
\Phi^H\left[ \int _{r_+}^{\infty}\Phi^{\infty} \phi_0 dr_*-
(h^{\infty}\phi_0)(r_+) \right]+
\nonumber \\
& 
+\frac{i\phi_0(r_+)}{(2\pi)^{1/2}\omega }\,,&
\label{Phi22}
\end{eqnarray}
where an integration by parts has been used.
Fortunately, one can obtain an exact expression for $\Phi^{\infty}$ 
in terms 
of hypergeometric functions \cite{cardosoradbtz}.
The results for $\Phi^{\infty}$ and $W$ are 
\begin{equation}
\Phi^{\infty}=
\frac{1}{r^{3/2}(1-M/r^2)^{\frac{i\omega}{2M^{1/2}}}}
F(a,b,2,\frac{M}{r^2})\,,
\label{solutionbtz}
\end{equation}
\begin{equation}
W=2\,i\,\omega\,
\frac{2^{\frac{i\omega}{M^{1/2}}}\Gamma(2)\Gamma(-\frac{i\omega}{M^{1/2}})}
{M^{3/4}\Gamma(1+i \frac{m-\omega}{2\sqrt{M}})
\Gamma(1-i \frac{m+\omega}{2\sqrt{M}})}.
\label{solutionwronskian}
\end{equation}
Here, $a=1+i \frac{m-\omega}{2\sqrt{M}}$ and $b=1-i \frac{m+\omega}{2\sqrt{M}}$.
So, to find $\Phi$ we only have to numerically integrate (\ref{Phi22}).
We have also determined $\Phi^{\infty}$ numerically by 
imposing the boundary conditions above. 
The agreement between the numerical 
computed $\Phi^{\infty}$ and
(\ref{solutionbtz}) was excellent. To find $\phi(t,r)$ one must 
apply the inverse Laplace transformation to $\Phi(\omega,r)$. 
Integrating the wave equation in this spacetime is simpler
than in asymptotically flat space, in the sense that, due to the
boundary conditions at infinity the solution is more stable,
and less effort is needed to achieve the same accuracy.
Using a similar method to that in \cite{lousto}, we estimate
the error in our results to be limited from above by 0.5\%.
\noindent
\subsection{ Numerical results for the waveforms and spectra}
\vskip 3mm

\begin{figure}
\centerline{\includegraphics[width=10 cm,height=9 cm]
{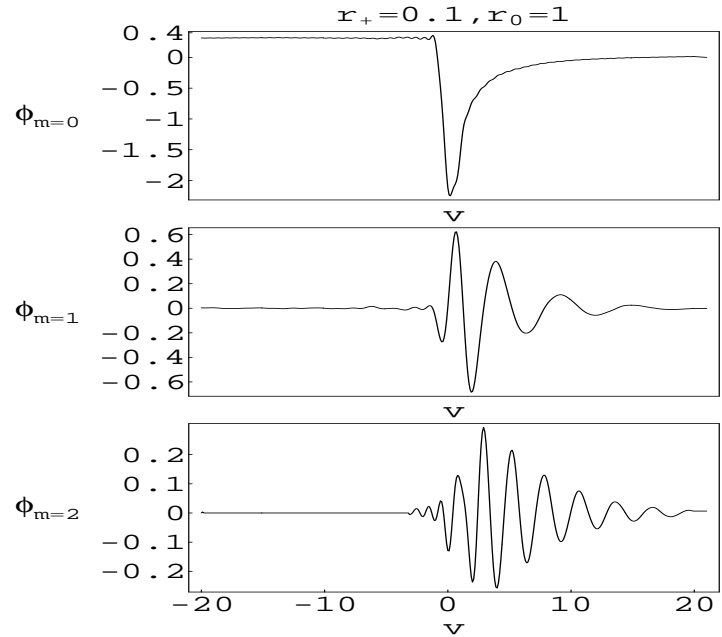}}
\caption{Waveforms $\phi(v)$ 
for a  $r_+=0.1\,,\,\,r_0=1$ BTZ black hole, 
for the three
lowest values of $m$. }
\label{wavebtz}
\end{figure}

\begin{figure}
\centerline{\includegraphics[width=10 cm,height=6.5 cm]
{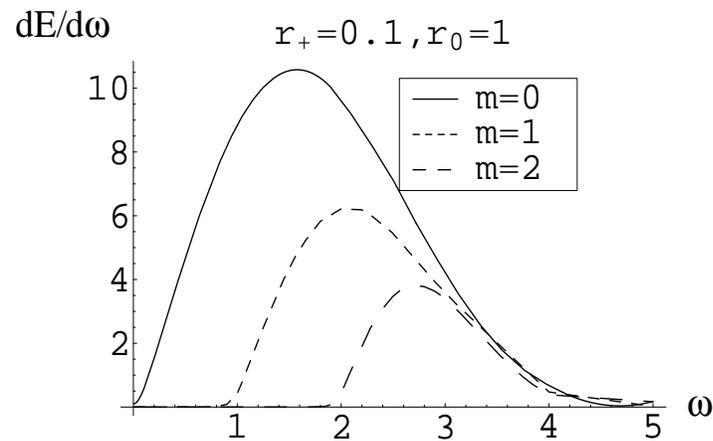}}
\caption{ Typical energy spectra, here
shown for $r_+=0.1$, and $r_0=1$.}
\label{spectra1btz}
\end{figure}

To better understand the numerical results, we first point out that
the QNM frequencies for this geometry, calculated by
Cardoso and Lemos \cite{cardosoqnmbtz} (see also \cite{birmingham2}
for a precise relation between these QNM frequencies and the poles of the 
correlation functions on the CFT side) are 
\begin{equation}
\omega_{\rm QNM}=\pm m -2iM^{1/2}(n+1).
\label{clequation}
\end{equation}
In Fig. \ref{wavebtz} we show the waveforms for the $r_+=0.1\,,\,\,r_0=1$ black hole, 
as a function of the advanced null-coordinate $v=t+r_*$. 
This illustrates in a beautiful way 
that QNMs govern the late time behavior of the waveform. 
\vskip 3mm
For example, for $m=0$, $\omega_{\rm QNM}=-0.2i(n+1)$, one expects to find
a purely decaying perturbation. This is evident from Fig. \ref{wavebtz}.  For
$m=1$, $\omega_{\rm QNM}= 1 -0.2i(n+1)$, so the signal should ring (at
late times) with frequency one. This is also clearly seen from Fig \ref{wavebtz}.
For $m=2$ we have the same kind of behavior. 
For large negative $v$ and fixed $t$ one has large negative $r_*$, 
so one is near the horizon. Thus $\phi(v\rightarrow-\infty)$ 
in Fig. \ref{wavebtz} should give the 
same values  as $\phi_0$ at $r_+$ in Fig. \ref{staticbtz}, which is the case.
The energy spectra peaks at higher $\omega$ when compared to the
fundamental $\omega_{\rm QNM}$ as is evident from Fig. \ref{spectra1btz}, which means
that higher modes are excited.  The total radiated energy as a
function of $m$ goes to zero slower than $1/m$ implying that the total
radiated energy diverges.  However, this divergence can be normalized
by taking a minimum size $L$ for the particle with a cut off given by
$m_{\rm max}\sim \frac{\pi}{2}\frac{r_+}{L}$ \cite{davis,davis2,anninos,gleiser}. We
have calculated for $r_+=0.1$ the total energy for the cases
$m=0,1,2$, yielding $E_{m=0}\simeq26$, $E_{m=1}\simeq12$,
$E_{m=2}\simeq 6$. An estimation of the total energy for a
particle with $m_{\rm max}\simeq 1000$ yields $E_{\rm total}\simeq
80$ (the energy is measured in units of $q_0^2m_0^2$).
\vskip 3mm

\vskip 3mm

We have also computed the radiated energy for several values of $r_0$
and verified that is not a monotonic function of $r_0$. For small
values of $r_0$ the energy radiated is a linear function of $(r_0-r_+)$,
for intermediate $r_0$ it has several peaks, and  it grows monotonically 
for large $r_0$.
The zero frequency
limit (ZFL), depends only on the initial data and one can prove that
it is given by $(\frac{dE}{d\omega})_{\omega\rightarrow 0}=
\phi_{0}^2$.  This is to be contrasted to the ZFL for outgoing
gravitational radiation in asymptotically flat spacetimes \cite{smarr2}
where it depends only on the initial velocity of the test particle.

\newpage

\noindent
\section{Scalar radiation from infall of a particle into a $D$-dimensional Schwarzschild-anti-de Sitter black hole}
\vskip 3mm

\noindent
\subsection{The Problem, the Equations and the Laplace transform, and 
the initial and boundary conditions for $D$-dimensions}
\vskip 3mm

\noindent
\subsubsection{The problem}
\vskip 3mm
In this paper we shall present the results of the following process:
the radial infall of a small particle coupled to a massless scalar
field, into a $d$-dimensional Schwarzschild-AdS black hole.
We will consider that both the mass $m_0$ and the scalar charge $q_s$ of the
particle are a perturbation on the background spacetime, i.e.,
$m_0\,,q_s\,<< M\,,R$, where $M$ is the mas of the black hole 
and $R$ is the AdS radius. In this approximation the background
metric is not affected by the scalar field, and is given by
\begin{equation}
ds^{2}= f(r) \,dt^{2}- \frac{dr^{2}}{f(r)}-
r^{2}d\Omega_{D-2}^{2}\,,
\label{lineelementddim}
\end{equation}
where, $f(r)=
(\frac{ r^{2}}{R^2}+1-\frac{16\pi M}{(D-2)A_{D-2}}\frac{1}{r^{D-3}})$,
$A_{D-2}$ is the area of a unit $(D-2)$ sphere, 
$A_{D-2}=\frac{2 \pi^{\frac{D-1}{2}}}{\Gamma(\frac{D-1}{2})}$, and
$d\Omega_{D-2}^{2}$ is the line element on the unit sphere $S^{D-2}$.
The action for the scalar field $\phi$ and particle is given by 
a sum of three parts, the action for the scalar field itself, 
the action for the particle and an interaction piece, 
\begin{equation}
{\cal I}=-\frac{1}{8 \pi} \int \phi _{;a} \phi ^{;a} \sqrt{-g} \,d^Dx-
 m_0 \int (1+q_s \phi)(-g_{ab}\dot{z}^a \dot{z}^b)^{\frac{1}{2}} 
d\lambda \,,
\label{actiond}
\end{equation}
where $g_{ab}$ is the background metric, $g$ its determinant, and 
$z^a(\lambda)$ represents the worldline of the particle as a function
of an affine parameter $\lambda$.

\noindent
\subsubsection{The equations and the Laplace transform}
\vskip 3mm
We now specialize to the radial infall case. In the usual
(asymptotically flat) Schwarzschild geometry, one can for example let
a particle fall in from infinity with zero velocity there
\cite{davis,davis2,anninos,gleiser}. The peculiar properties of AdS spacetime do not
allow  a particle at rest at infinity \cite{cardosoradbtz} (we
would need an infinite amount of energy for that) so we consider the
mass $m_0$ to be held at rest at a given distance $r_0$ in
Schwarzschild coordinates. At $t=0$ the particle starts falling into
the black hole.  As the background is spherically symmetric, Laplace's
equation separates into the usual spherical harmonics
$Y(\theta,\varphi_1,..,\varphi_{D-3})$ defined over the $(D-2)$ unit
sphere \cite{bateman}, where $\theta$ is the polar angle and
$\varphi_1,..,\varphi_{D-3}$ are going to be considered azimuthal
angles of the problem. In fact, since we are considering radial
infall, the situation is symmetric with respect to a $(d-3)$
sphere. We can thus decompose the scalar field as
\begin{equation}
\phi(t,r,\theta,\varphi_1,..,\varphi_{D-3})=  \frac{1}{r^{\frac{D-2}{2}}}
\sum_{l}Z_l(t,r) Y_{l0..0}(\theta)\,.
\label{sphericalharmonics1}
\end{equation}
The polar angle $\theta$ carries all the angular information, and $l$ is the
angular quantum number associated with $\theta$. From now on, instead 
of   $Y_{l0..0}(\theta)$, we shall simply write 
$Y_{l}(\theta)$ for the spherical harmonics over the $(D-2)$ unit sphere.
In fact $Y_{l}(\theta)$ is, apart from normalizations, just a Gegenbauer
polynomial $C_{l}^{\frac{D-3}{2}}(\cos\theta)$ \cite{bateman}. 
Upon varying the action (\ref{actiond}), integrating over the $(D-2)$ sphere
and using the orthonormality properties of the spherical harmonics we
obtain the following equation for $Z_l(t,r)$
\begin{equation}
\frac{\partial^{2} Z_l(t,r)}{\partial r_*^{2}}-
\frac{\partial^{2} Z_l(t,r)}{\partial t^{2}}-
V(r)Z_l(t,r)=
\frac{4\pi q_s m_0 f}{r^{\frac{D-2}{2}}}(\frac{dt}{d\tau})^{-1}
\delta(r-r_p(t))Y_{l}(0) \,.
\label{waveequation1}
\end{equation}
The potential $V(r)$ appearing in equation (\ref{waveequation1}) is given by
\begin{equation}
V(r)=
f(r)\left\lbrack\frac{a}{r^2}+
\frac{(D-2)(D-4)f(r)}{4r^2}+\frac{(D-2)f'(r)}{2r}\right\rbrack \,,
\label{potentiald}
\end{equation}
where $a=l(l+D-3)$ is the eigenvalue of the Laplacian on $S^{D-2}$,
and the tortoise coordinate $r_*$ is defined as
$\frac{\partial r}{\partial r_*}= f(r)$.
By defining the Laplace transform ${\bf \tilde Z_l}(\omega,r)$ of
$Z_l(t,r)$ as 
\begin{equation}
{\bf \tilde Z}(\omega,r)=
\frac{1}{(2\pi)^{1/2}}\int_{0}^{\infty}e^{i \omega t}Z_l(t,r)dt.
\label{laplaced}
\end{equation}   
Then, equation (\ref{waveequation1}) transforms into
\begin{equation}
\frac{\partial^{2} {\bf \tilde Z}(\omega,r)}{\partial r_*^{2}} +
\left\lbrack\omega^2-V(r)\right\rbrack
{\bf \tilde Z}(\omega,r)=S_l(\omega,r)+
\frac{i\omega}{(2\pi)^{1/2}}{Z_0}_l(r) \,,
\label{waveequation2d}
\end{equation}
with the source term $S_l(\omega,r)$ given by,
\begin{equation}
S_l(\omega,r)=\frac{2 (2\pi)^{1/2} 
q_s m_0 f Y_{l}(0)}{r^{(D-2)/2}(E^2-f)^{1/2}} 
e^{i \omega T(r)}\,.
\label{sourced}
\end{equation}
Note that ${Z_0}_l(r)$ is the initial value of $Z(t,r)$, i.e., 
${Z_0}_l(r)=Z(t=0,r)$, satisfying
\begin{equation}
\frac{\partial^{2} {Z_0}_l(r)}{\partial r_*^{2}} 
-V(r){Z_0}_l(r)=
-\frac{4 \pi q_s m_0 f(r)
Y_{l}(0)}{r^{(D-2)/2}}(\frac{dt}{d\tau})_{r_0}^{-1}\delta(r-r_0) \,,
\label{initial}
\end{equation}
where $r_0=r_p(t=0)$. 
We have represented the particle's worldline by
$z^{\mu}=z_p^{\mu}(\tau)$, with $\tau$ the proper
time along a geodesic.
Here, $t=T(r)$ describes the particle's radial trajectory giving the time
as a function of radius along the geodesic
\begin{equation}
\frac{dT(r)}{dr}=-\frac{E}{f(E^2-f)^{1/2}} \,\,
\label{time}
\end{equation}
with initial conditions $T(r_0)=0 $, and $E^2=f(r_0)$.

We have rescaled $r$, $r\rightarrow\frac{r}{R}$, and measure
everything in terms of $R$, i.e., $\omega$ is to be read $\omega R$,
$\Psi$ is to be read $\frac{R}{q_s m_0}\Psi$ and $r_+$, the horizon
radius is to be read $\frac{r_+}{R}$.

\subsubsection{The initial data}
We can obtain ${Z_0}_l(r)$, the initial value of $Z(t,r)$, by
solving numerically equation (\ref{initial}), 
demanding regularity at both the horizon and infinity
(for a similar problem, see for example \cite{wald,burko,ruffini}).
To present the initial data and the results we divide the problem 
into two categories: (i) small black holes with $r_+<<1$, and (ii) 
intermediate and large black holes with 
$r_+\buildrel>\over\sim 1$.

\vskip .5cm

{\bf (i): Initial data for small black holes, $r_+=0.1$}

\begin{figure}
\centerline{\includegraphics[width=18 cm,height=5 cm]
{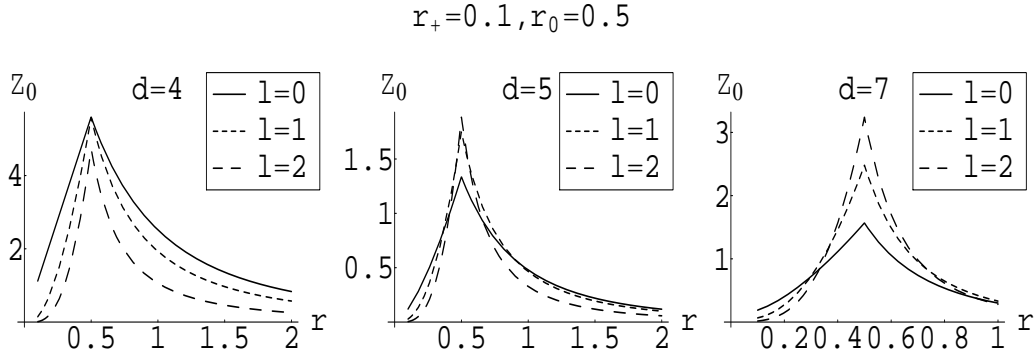}}
\caption{Initial data ${Z_0}_l$ for a small black hole with $r_+=0.1$, for
$D=4,\, 5$ and $7$ (from left to right respectively). The small scalar particle
is located at $r_0=0.5$. The results are 
shown for the lowest values of the angular quantum number $l$. }
\label{static1}
\end{figure}

\begin{figure}
\centerline{\includegraphics[width=18 cm,height=5 cm]
{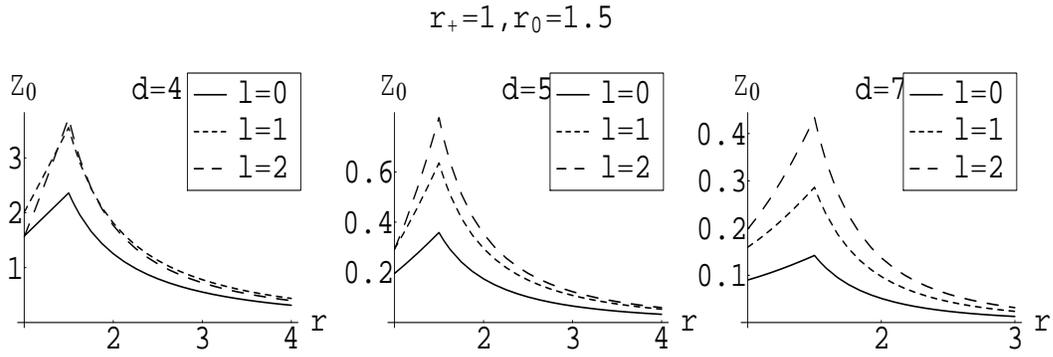}}
\caption{Initial data ${Z_0}_l$ for a black hole 
with $r_+=1$, and with the
particle at $r_0=1.5$, for some values of $l$ the angular quantum
number. Again, we show the results for $D=4,\,5$ and $7$ 
from left to right respectively.  }
\label{static2}
\end{figure}

\begin{figure}
\centerline{\includegraphics[width=18 cm,height=5 cm]
{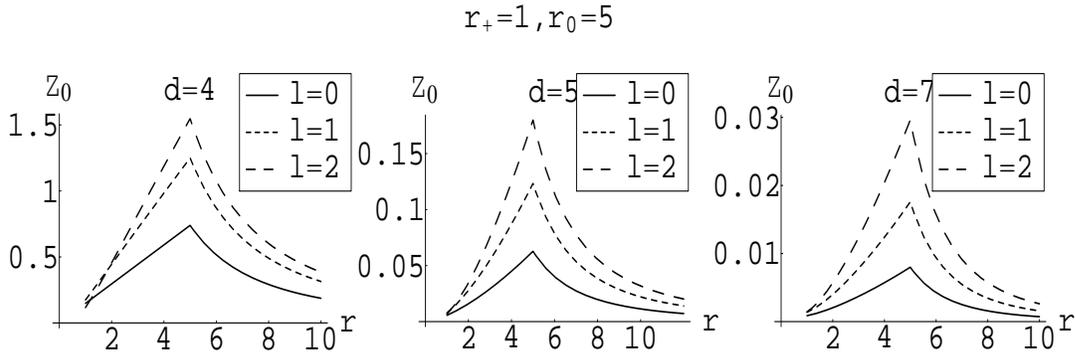}}
\caption{Initial data ${Z_0}_l$ for a black hole 
with $r_+=1$, and with the
particle at $r_0=5$, for some values of $l$ the angular quantum
number. Again, we show the results for $D=4,\,5$ and $7$ 
from left to right respectively.  }
\label{static3}
\end{figure}

In Fig. \ref{static1} we present initial data for small black holes with $r_+=0.1$
in the dimensions of interest ($d=4,\,5$ and $7$).  In this case, the
fall starts at $r_0=0.5$.  Results referring to initial data in $d=3$
(BTZ black hole) are given in \cite{cardosoradbtz}.
We show a typical form of ${Z_0}_l$ for $r_+=0.1$ and
$r_0=0.5$, and for different values of $l$.  As a test for the
numerical evaluation of ${Z_0}_l$, we have checked that as $r_0
\rightarrow r_+$, all the multipoles fade away, i.e., ${Z_0}_l
\rightarrow 0$, supporting the No Hair Conjecture.
Note that ${Z_0}_l$ has to be small. 
We are plotting $\frac{{Z_0}_l}{q_sm_0/R}$. 
Since $q_sm_0/R<<1$ in our approximation one has from Figs. \ref{static1}-\ref{static3} 
that indeed ${Z_0}_l<<1$. 
\vskip .5cm

{\bf (ii): Initial data for intermediate and large black holes, $r_+=1$}
\vskip5mm

In Figs. \ref{static2} and \ref{static3} we show initial data for an
intermediate to large black hole, $r_+=1$.  In Fig.\ref{static2}  the fall starts at
$r_0=5$. We show a typical form of ${Z_0}_l$ for $r_+=1$ and
$r_0=5$, and for different values of $l$.  
In Fig. \ref{static3} it starts further down at $r_0=1.5$. We 
show a typical form of ${Z_0}_l$ for $r_+=1$ and
$r_0=1.5$, and for different values of $l$.  Again, 
we have checked that as $r_0 \rightarrow r_+$, 
all the multipoles fade away, i.e., ${Z_0}_l
\rightarrow 0$, supporting the No Hair Conjecture.

Two important remarks are in order:
first, it is apparent from Figs. \ref{static1}-\ref{static3} that the field (sum over the
multipoles) is divergent at the particle's position $r_0$. This is to
be expected, as the particle is assumed to be point-like;  second, one
is led to believe from Figs. \ref{static1}-\ref{static3} 
(but especially from Figs.\ref{static2} -\ref{static3})
that ${Z_0}_l$ increases with $l$.  This is not true however, as this
behavior is only valid for small values of the angular quantum number
$l$. For large $l$, ${Z_0}_l$ decreases with $l$, in such a manner as
to make $\phi(t,r)$ in (\ref{sphericalharmonics1}) convergent and
finite. For example, for $r_+=1, r_0=5$ and $D=4$, we have at $r=6$,
${Z_0}_{l=20}=0.781$ and ${Z_0}_{l=40}=0.3118$.

\subsubsection{Boundary conditions and the Green's function}

Equation (\ref{waveequation2d}) is to be solved with the boundary
conditions appropriate to AdS spacetimes, but special attention must
be paid to the initial data \cite{cardosoradbtz}: ingoing waves at the
horizon,
\begin{equation}
{\bf \tilde Z} \sim F(\omega) e^{-i\omega r_*}+
\frac{i {Z_0}_l}{(2\pi)^{1/2}\omega }\,\,\,, r\rightarrow r_+\,,
\label{boundarybehavior2d}
\end{equation}
and since the potential diverges at infinity we 
impose reflective boundary conditions
(${\bf \tilde Z} = 0$) there \cite{avis}.
Naturally, given these boundary conditions, 
all the energy eventually
sinks into the black hole. 
To implement a numerical solution, we note that two independent
solutions ${\bf \tilde Z}^H$ and ${\bf \tilde Z^{\infty}}$ 
of (\ref{waveequation2d}),
with the source term set to zero, have the behavior:
\begin{eqnarray}
{\bf \tilde Z}^H \sim e^{-i\omega r_*}\quad\quad r \rightarrow r_+ \,,\\
{\bf \tilde Z}^H \sim A r^{D/2-1} +Br^{-D/2}\quad r \rightarrow \infty\,, \\
{\bf \tilde Z^{\infty}} \sim Ce^{i\omega r_*}+
De^{-i\omega r_*}\quad r \rightarrow r_+ \,,\\
{\bf \tilde Z^{\infty}} \sim r^{-D/2} \quad r \rightarrow \infty \,, \\
\label{behavior}
\end{eqnarray}
Here, 
the wronskian $W$ of these two solutions is a constant, $W=2Ci\omega$. 
Define as in \cite{cardosoradbtz} $h^H$ through $dh^H/dr_*
=-{\bf \tilde Z}^H$ and $h^{\infty}$ 
through $dh^{\infty}/dr_*=
-{\bf \tilde Z^{\infty}}$.
We can then show that ${\bf \tilde Z}$ given by 
\begin{eqnarray}
&
{\bf \tilde Z}=\frac{1}{W} \left[ {\bf \tilde Z^{\infty}} 
\int _{-\infty}^{r}{\bf \tilde Z}^H S dr_* +
{\bf \tilde Z}^H\int _{r}^{\infty}{\bf \tilde Z^{\infty}} S dr_* \right]+
\frac{i\omega}{(2\pi)^{1/2} W } \left[ {\bf \tilde Z^{\infty}}
\int _{-\infty}^{r}h^H \frac{d{Z_0}_l}{dr_*} dr_* +\right.
\nonumber \\
&
{\bf \tilde Z}^H 
\int _{r}^{\infty}h^{\infty} \frac{d{Z_0}_l}{dr_*} dr_* 
+  
\left. (h^{\infty}{Z_0}_l{\bf \tilde Z}^H -
h^H{Z_0}_l{\bf \tilde Z^{\infty}})(r)
\right]\,, 
\label{Z}
\end{eqnarray}
is a solution to (\ref{waveequation2d}) 
and satisfies
the boundary conditions.
In this work, we are interested in computing 
the wavefunction ${\bf \tilde Z}(\omega,r)$ 
near the horizon ($r\rightarrow r_+$). In this 
limit we have
\begin{eqnarray}
&
{\bf \tilde Z}(r\sim r_+)=\frac{1}{W} \left[ {\bf \tilde Z}^H
\int _{r_+}^{\infty}{\bf \tilde Z^{\infty}} S dr_* \right]+&
\nonumber \\
&
\frac{i\omega}{(2\pi)^{1/2}W }
{\bf \tilde Z}^H\left[ \int _{r_+}^{\infty}{\bf \tilde Z^{\infty}} {Z_0}_l dr_*-
(h^{\infty}{Z_0}_l)(r_+) \right]
+\frac{i{Z_0}_l(r_+)}{(2\pi)^{1/2}\omega }\,,&
\label{Z22}
\end{eqnarray}
where an integration by parts has been used.

All we need to do is to find a solution ${\bf \tilde Z_2}$ of the
corresponding homogeneous equation satisfying the above mentioned
boundary conditions (\ref{behavior}), and then numerically integrate
it in (\ref{Z22}).  In the numerical work, we chose to adopt $r$ as
the independent variable, therefore avoiding the numerical inversion
of $r_*(r)$.  To find ${\bf \tilde Z_2}$,the integration (of the
homogeneous form of (\ref{waveequation2d})) was started at a large
value of $r=r_i$, which was $r_i=10^5$ typically.  Equation
(\ref{behavior}) was used to infer the boundary conditions ${\bf
\tilde Z_2}(r_i)$ and ${\bf \tilde Z_2}'(r_i)$.  We then integrated
inward from $r=r_i$ in to typically $r=r_++10^{-6}r_+$.  Equation
(\ref{behavior}) was then used to get $C$.

\noindent
\subsection{Results}
\vskip 3mm

\subsubsection{Numerical results}
Our numerical evolution for the field showed that some drastic changes
occur when the size of the black hole varies, so we have chosen to
divide the results in (i) small black holes and (ii) intermediate and
large black holes . We will see that the behavior of these two classes
is indeed strikingly different.

\bigskip

{\bf (i): Wave forms and spectra for small black holes, $r_+=0.1$}

\begin{figure}
\centerline{\includegraphics[width=18 cm,height=5 cm]
{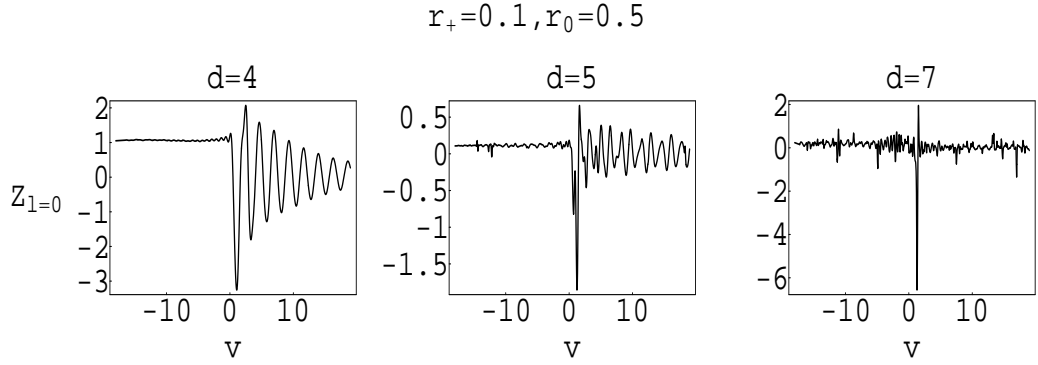}}
\caption{The spherically symmetric ($l=0$)
waveform for the case of a particle falling from $r_0=0.5$ into a 
$r_+=0.1$ black hole. The results are displayed for $D=4,\,5$ and $7$
from left to right respectively. The coordinate
$v=t+r_*$ is the usual Eddington coordinate.}
\label{waveformsmall1}
\end{figure}

\begin{figure}
\centerline{\includegraphics[width=18 cm,height=5 cm]
{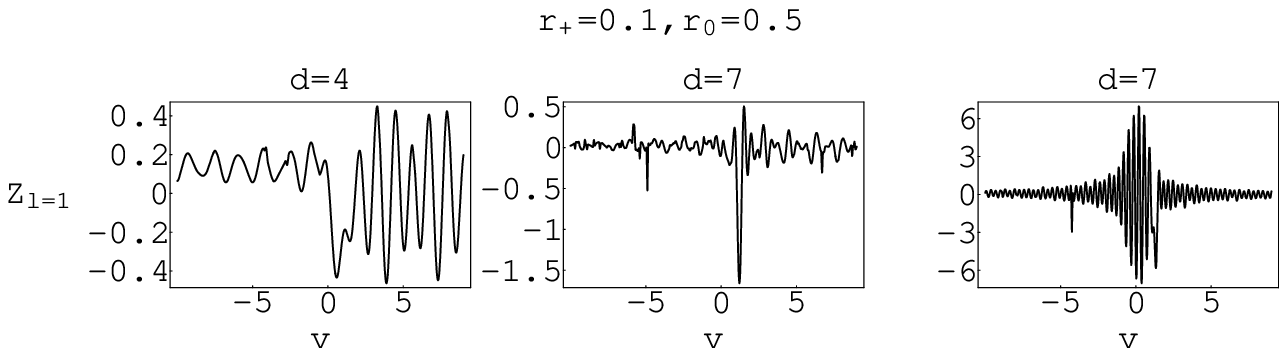}}
\caption{The $l=1$ waveform for the case of a
particle falling from $r_0=0.5$ into a $r_+=0.1$ black hole.The
results are displayed for $D=4,\,5$ and $7$ from left to right
respectively.}
\label{waveformsmall2}
\end{figure}

\begin{figure}
\centerline{\includegraphics[width=18 cm,height=5 cm]
{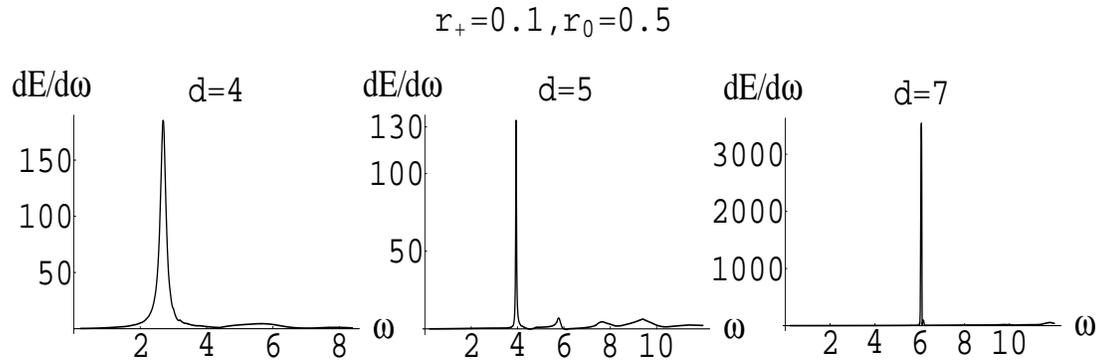}}
\caption{Typical energy spectra for the spherically
symmetric part of the perturbation ($l=0$), here
shown for $r_+=0.1$, and $r_0=0.5$, and for $D=4,\, 5$ and $7$. 
Total energy in this mode: for d=4 we have $E_{l=0, d=4} \sim 75$. 
For $D=5$, we have 
$E_{l=0, D=5} \sim 34$. For $D=7$, we have 
$E_{l=0, D=7} \sim 1500$.}
\label{spectrasmall1}
\end{figure}

\begin{figure}
\centerline{\includegraphics[width=18 cm,height=5 cm]
{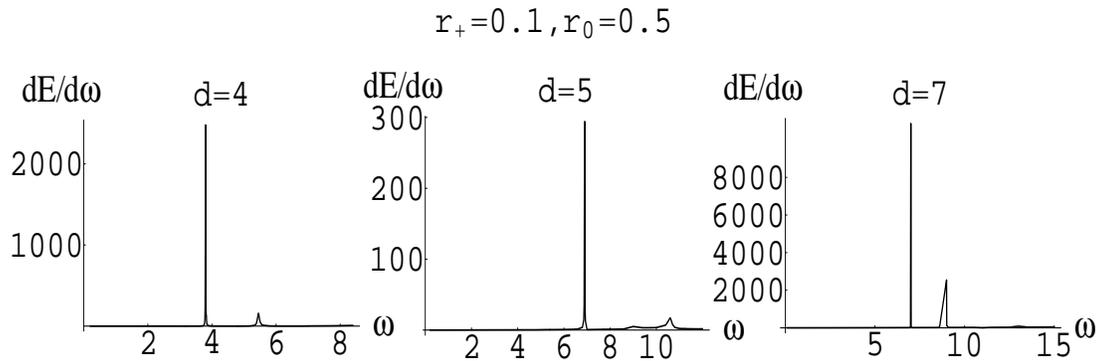}}
\caption{Typical energy spectra (for $l=1$), here
shown for $r_+=0.1$, and $r_0=0.5$.}
\label{spectrasmall2}
\end{figure}

\medskip
We plot the waveforms and the spectra.
Figs. \ref{waveformsmall1}-\ref{spectrasmall2} are typical plots for small black holes of waveforms 
and spectra for $l=0$ and $l=1$ (for $l=2$ and higher the conclusions 
are not altered).  
They show the first interesting aspect of our
numerical results: for small black holes the $l=0$ signal is clearly
dominated by quasinormal, exponentially decaying, ringing modes with a
frequency $\omega \sim D-1$ (scalar quasinormal frequencies of
Schwarzschild-AdS black holes can be found in
\cite{horowitz,cardosoqnmsads2}).  This particular limit is a pure AdS mode
\cite{burgess,konoplyasmall}.  For example, Fig. \ref{waveformsmall1}  gives, for $D=4$,
$\omega=2\pi/T \sim 2\pi/(10/4.5)\sim 2.7$.  This yields a value near
the pure AdS mode for $d= 4$, $\omega= 3$. Likewise, Fig. \ref{waveformsmall1} gives 
$\omega \sim 4$ when $d=5$, the pure AdS mode for $D=5$.  All these
features can be more clearly seen in the energy spectra plots, Fig.\ref{spectrasmall1} ,
where one can observe the intense peak at $\omega \sim D-1$.  The
conclusion is straightforward: spacetimes with small black holes
behave as if the black hole was not there at all. This can be checked
in yet another way by lowering the mass of the black hole. We have
done that, and the results we have obtained show that as one lowers
the mass of the hole, the ringing frequency goes to $\omega \sim 3$
(for $D=4$) and the imaginary part of the frequency, which gives us
the damping scale for the mode, decreases as $r_+$ decreases. In this
limit, the spacetime effectively behaves as a bounding box in which
the modes propagate ``freely'', and are not absorbed by the black
hole.

Not shown is the spectra for higher values of the angular quantum
number $l$. The total energy going down the hole increases slightly
with $l$. This would lead us to believe that an infinite amount of
energy goes down the hole. However, as first noted in \cite{davis2},
this divergence results from treating the incoming object as a point
particle. Taking a minimum size $L$ for the particle implies a cutoff
in $l$ given by $l_{\rm max} \sim \frac{\pi}{2}\frac{r_+}{L}$, and this
problem is solved.

\bigskip

{\bf (ii): Wave forms and spectra for intermediate and large black holes, 
$r_+=1$}
 
\medskip

We plot the waveforms and the spectra.  As
we mentioned, intermediate and 
large black holes (which are of more direct interest to
the AdS/CFT) behave differently. The signal is dominated by a sharp
precursor near $v=r_{*0}$ and there is no ringing: the waveform
quickly settles down to the final zero value in a pure decaying
fashion.  The timescale of this exponential decay is, to high
accuracy, given the inverse of the imaginary part of the quasinormal
frequency for the mode.  The total energy is not a monotonic function
of $r_0$ and still diverges if one naively sums over all the
multipoles.  In either case, there seem to be no power-law tails, as
was expected from the work of Ching et. al. \cite{ching2}.  Note that
$E$ is given in terms of $E/(q_sm_0/r)^2$. Since $q_sm_0/r<<1$ the
total energy radiated is small in accord with our approximation.
\begin{figure}
\centerline{\includegraphics[width=18 cm,height=5 cm]
{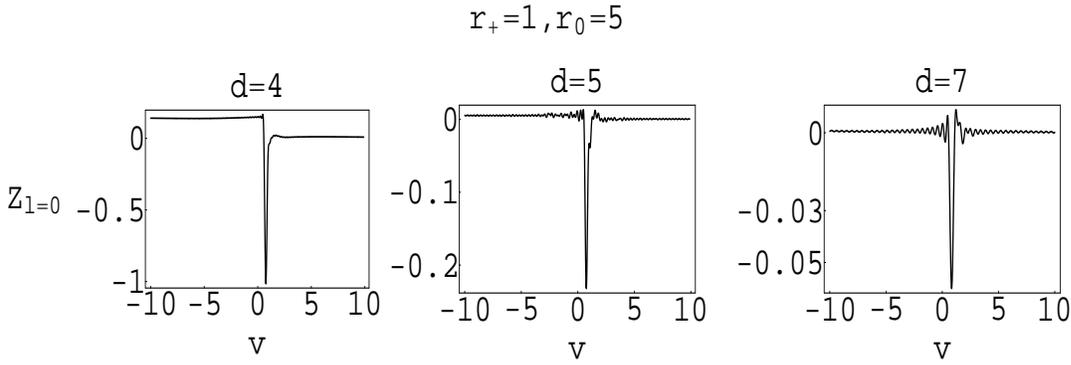}}
\caption{The spherically symmetric ($l=0$)
waveform for the case of a particle falling from $r_0=5$ into a 
$r_+=1$ black hole. The results are displayed for $D=4,\,5$ and $7$
from left to right respectively.}
\label{waveformlarge1}
\end{figure}
\begin{figure}
\centerline{\includegraphics[width=18 cm,height=5 cm]
{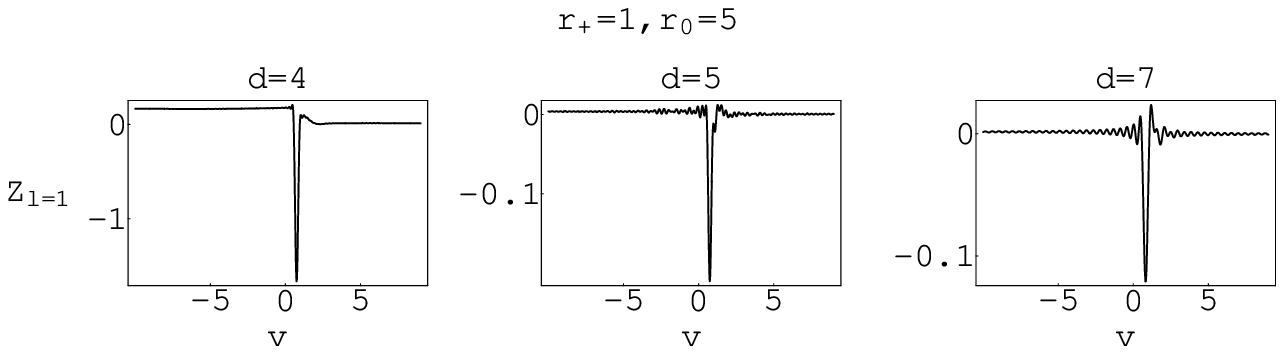}}
\caption{The $l=1$ waveform for the case of a particle
falling from $r_0=5$ into a $r_+=1$ black hole.}
\label{waveformlarge2}
\end{figure}
\begin{figure}
\centerline{\includegraphics[width=18 cm,height=5 cm]
{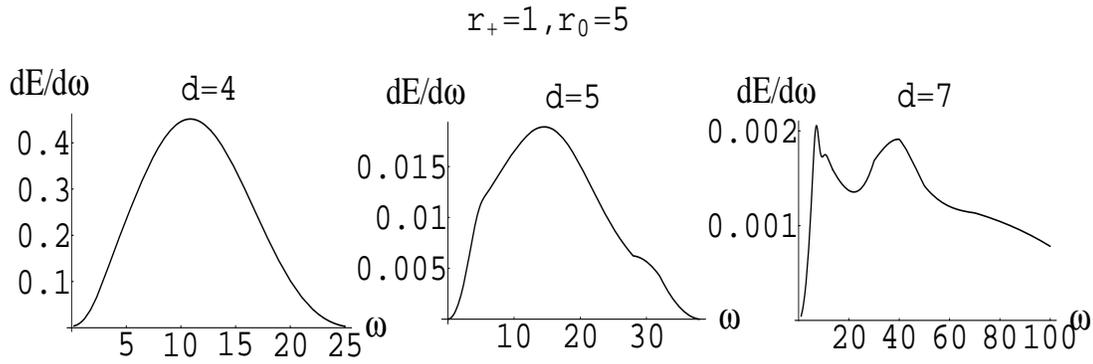}}
\caption{Typical energy spectra for the spherically
symmetric part of the perturbation ($l=0$), here
shown for $r_+=1$, and $r_0=5$, and for $D=4,\, 5$ and $7$.}
\label{spectralarge1}
\end{figure}
\begin{figure}
\centerline{\includegraphics[width=18 cm,height=5 cm]
{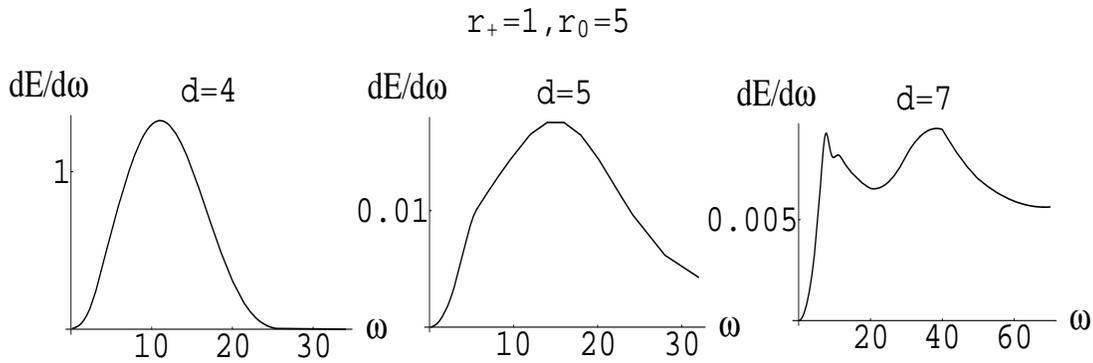}}
\caption{Typical energy spectra (for $l=1$), here
shown for $r_+=1$, and $r_0=5$.}
\label{spectralarge2}
\end{figure}

\begin{figure}
\centerline{\includegraphics[width=18 cm,height=5 cm]
{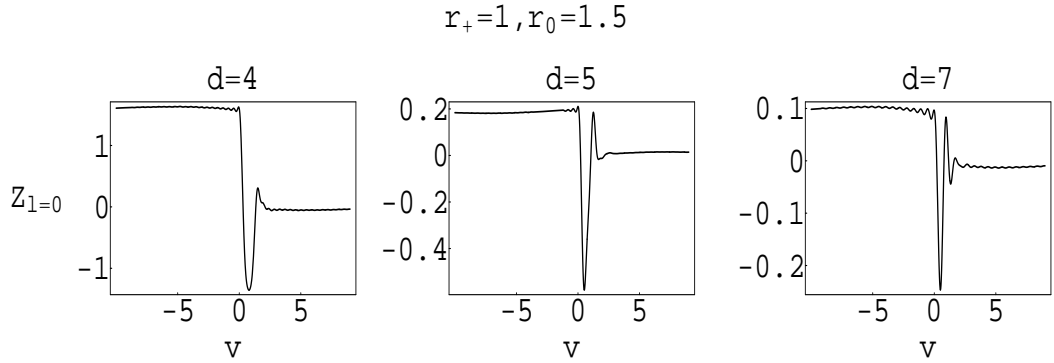}}
\caption{The $l=0$ waveform for the case of a particle
falling from $r_0=1.5$ into a $r_+=1$ black hole.}
\label{waveformlarge3}
\end{figure}

\begin{figure}
\centerline{\includegraphics[width=18 cm,height=5 cm]
{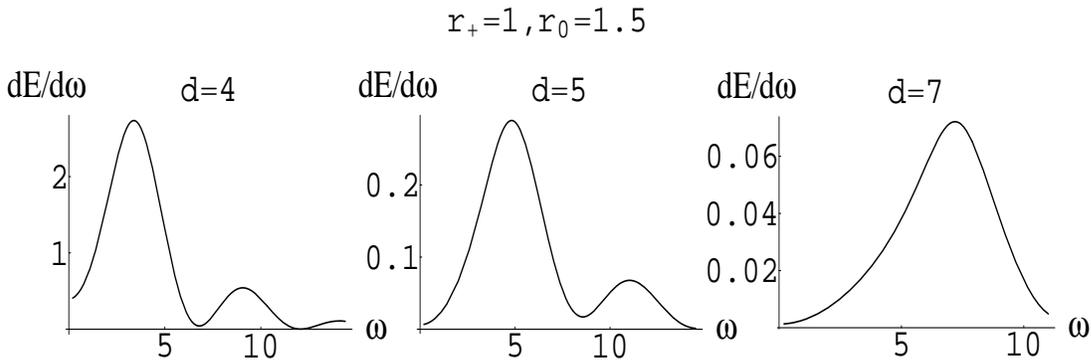}}
\caption{Typical energy spectra (for $l=0$), here
shown for $r_+=1$, and $r_0=1.5$.}
\label{spectralarge3}
\end{figure}

The value attained by ${\bf \tilde Z}$ for large negative $v$ -
Fig. \ref{waveformlarge3} - is the initial data, and this can be most easily seen by
looking at the value of $Z_0$ near the horizon in Fig. \ref{static3} (see also
\cite{cardosoradbtz}).  This happens for small black holes also, which is
only natural, since large negative $v$ means very early times, and at
early times one can only see the initial data, since no information
has arrived to tell that the particle has started to fall.  The
spectra in general does not peak at the lowest quasinormal frequency
(cf Figs. \ref{spectralarge1}-\ref{spectralarge3}), as it did in flat spacetime
\cite{davis}. (Scalar quasinormal frequencies of Schwarzschild-AdS
 black holes can be found in \cite{horowitz,cardosoqnmsads2}). Most
importantly, the location of the peak seems to have a strong
dependence on $r_0$ (compare Figs. \ref{spectralarge1} and \ref{spectralarge3}).  This discrepancy has
its roots in the behavior of the quasinormal frequencies.  In fact,
whereas in (asymptotically) flat spacetime the real part of the
frequency is bounded and seems to go to a constant \cite{nollert,nollert2,andersson1},
in AdS spacetime it grows without bound as a function of the principal
quantum number $n$ \cite{horowitz,cardosoqnmsads2}. Increasing the distance
$r_0$ at which the particle begins to fall has the effect of
increasing this effect, so higher modes seem to be excited at larger
distances.

\noindent
\subsubsection{Discussion of results}
\vskip 3mm

Two important remarks regarding these results can be made:

(i) the total energy radiated depends on the size of the infalling object,
and the smaller the object is, the more energy it will radiate.
This is a kind of a scalar analog in AdS space of a well known result
for gravitational radiation in flat space \cite{naka}.

(ii) the fact that the radiation emitted in each multipole is high
even for high multipoles leads us to another important point, first
posed by Horowitz and Hubeny \cite{horowitz}.  While we are not able to
garantee that the damping time scale stays bounded away from infinity
(as it seems), it is apparent from the numerical data that the damping
time scale increases with increasing $l$. Thus it looks like the late
time behavior of these kind of perturbations will be dominated by the
largest $l$-mode ($L_{\rm max} \sim \frac{r_+}{{\rm 
size}\, {\rm of}\, {\rm object}}$), and
this answers the question posed in \cite{horowitz}. Thus a
perturbation in $<F^2>$ in the CFT ]
 with given angular dependence $Y_l$ on $S^3$
will decay exponentially with a time scale given by the imaginary part
of the lowest quasinormal mode with {\it that} value of $l$.

\noindent
\section{ Conclusions}
\vskip 3mm
We have computed the scalar energy emitted by a point test particle
falling from rest into a Schwarzschild-AdS black hole.
From the point of view of the AdS/CFT conjecture, where the (large)
black hole corresponds on the CFT side to a thermal state, the
infalling scalar particle corresponds to a specific perturbation of
this state (an expanding bubble), while the scalar radiation is
interpreted as particles decaying into bosons of the associated
operator of the gauge theory.  Previous works \cite{horowitz,cardosoqnmsads2}
have shown that a general perturbation should have a timescale
directly related to the inverse of the imaginary part of the
quasinormal frequency, which means that the approach to thermal
equilibrium on the CFT should be governed by this timescale. We have
shown through a specific important problem that this is in fact
correct, but that it is not the whole story, since some important
features of the waveforms highly depend on $r_0$.

Overall, we expect to find the same type of features, at least 
qualitatively, in the
gravitational or electromagnetic radiation by test particles falling
into a Schwarzschild-AdS black hole.  For example, if the
black hole is small, we expect to find in the gravitational radiation
spectra a strong peak located at $ \omega^2=4n^2+l(l+1)\;, \quad
n=1,2,...$.
Moreover, some major results in perturbation theory and numerical
relativity \cite{anninos,gleiser}, studying the collision of two black
holes, with masses of the same order of magnitude, allow us to infer
that evolving the collision of two black holes in AdS spacetime, should
not bring major differences in relation to our results (though it is of
course a much more difficult task, even in perturbation theory).  In
particular, in the small black hole regime, the spectra and waveforms
should be dominated by quasinormal ringing.

\part{Gravitational waves in higher dimensions}

\thispagestyle{empty} \setcounter{minitocdepth}{1}
\chapter[Gravitational waves in higher dimensions]
{Gravitational waves in higher dimensions} \label{chap:Intro3}
\lhead[]{\fancyplain{}{\bfseries Chapter \thechapter. \leftmark}}
\rhead[\fancyplain{}{\bfseries \rightmark}]{}
\minitoc \thispagestyle{empty}
\renewcommand{\thepage}{\arabic{page}}
\section{Introduction}

One expects to finally detect gravitational waves in the forthcoming
years. If this happens, and if the observed waveforms match the
predicted templates, General Relativity will have pass a crucial test.
Moreover, if one manages to cleanly separate gravitational waveforms,
we will open a new and exciting window to the Universe, a window from
which one can look directly into the heart of matter, as gravitational
waves are weakly scattered by matter.  A
big effort has been spent in the last years trying to build
gravitational wave detectors, and a new era will begin with
gravitational wave astronomy \cite{schutz1,hughes}.  What makes
gravitational wave astronomy attractive, the weakness with which
gravitational waves are scattered by matter, is also the major source
of technical difficulties when trying to develop an apparatus which
interacts with them. Nevertheless, some of these highly non-trivial
technical difficulties have been surmounted, and we have detectors
already operating \cite{geo,ligo,virgo}. Another effort is being
dedicated by theoreticians trying to obtain accurate templates for the
various physical processes that may give rise to the waves impinging
on the detector. We now have a well established theory of wave generation
and propagation, which started with Einstein and his quadrupole formula.  The
quadrupole formula expresses the energy lost to gravitational waves by
a system moving at low velocities, in terms of its energy
content.  The quadrupole formalism is the
most famous example of slow motion techniques to compute wave
generation. All these techniques break Einstein's equations
non-linearity by imposing a power series in some small quantity and
keeping only the lowest or the lowest few order terms. The quadrupole
formalism  starts from a flat background and 
expands the relevant quantities 
in $R/ \lambda$, where $R$ is the size of source and $\lambda$ the 
wavelength of waves.
Perturbation formalisms on the other hand, start from some
non-radiative background, whose metric is known exactly, for example
the Schwarzschild metric, and expand in deviations from that
background metric. For a catalog of the various methods and their
description we refer the reader to the review works by Thorne
\cite{thorne} and Damour \cite{damour}. The necessity to develop all
such methods was driven of course by the lack of exact radiative
solutions to Einstein's equations (although there are some worthy
exceptions, like the C-metric \cite{kinnersley}), 
and by the fact that even nowadays solving the full set of
Einstein's equations numerically is a monumental task, and has been
done only for the more tractable physical situations. All the
existing methods seem to agree with each other when it comes down to
the computation of waveforms and energies radiated during physical
situations, and also agree with the few available results from a fully
numerical evolution of Einstein's equations.

We extend some of these
results to higher dimensional spacetimes.
There are several reasons why one should now try to do it. 
It seems impossible to
formulate in four dimensions a consistent theory which unifies gravity
with the other forces in nature. Thus, most efforts in this direction
have considered a higher dimensional arena for our universe, one
example being string theories which have recently made some remarkable
achievements. Moreover, recent investigations \cite{hamed} propose the
existence of extra dimensions in our Universe in order to solve the
hierarchy problem, i.e., the huge difference between the electroweak
and the Planck scale, $m_{\rm EW}/M_{\rm Pl}\sim 10^{-17}$. 
The fields of standard model would inhabit a 4-dimensional
sub-manifold, the brane, whereas the gravitational degrees of freedom
would propagate throughout all dimensions. One of the most spectacular
consequences of this scenario would be the production of black holes
at the Large Hadron Collider at CERN \cite{bhprod} (for recent
relevant work related to this topic we refer the reader to
\cite{decay,cardosorads,cardosoradkerrs,cardosoradkerr,cardosoradel}).
Now, one of the experimental
signatures of black hole production will be a missing energy, perhaps
a large fraction of the center of mass energy \cite{cardosoradkerr}. 
This will happen because when the partons collide to form a black
hole, some of the initial energy will be converted to gravitational
waves, and due to the small amplitudes involved, there is no
gravitational wave detector capable of detecting them, so they will
appear as missing. Thus, the collider could in fact indirectly serve
as a gravitational wave detector.  This calls for the calculation of
the energy given away as gravitational waves when two high energy
particles collide to form a black hole, which lives in all the
dimensions. The work done so far on this subject
\cite{eardley,yoshino} in higher dimensions, is mostly geometric, and
generalizes a construction by Penrose to find trapped surfaces on the
union of two shock waves, describing boosted Schwarzschild black
holes. On the other hand, there are clues
\cite{cardosorads,cardosoradkerrs,cardosoradkerr} indicating that a formalism
described by Weinberg \cite{weinberg} to compute the gravitational
energy radiated in the collision of two point particles, gives results
correct to a order of magnitude when applied to the collision of two
black holes.  This formalism assumes a hard collision, i.e., a
collision lasting zero seconds.  It would be important to apply this
formalism in higher dimensions, trying to see if there is agreement
between both results. This will be one of the topics discussed here. 
The other topic we study is the quadrupole 
formula in higher dimensions. Due to the difficulties in handling the 
wave tails in odd dimensions we concentrate our study in even 
dimensions.
We shall also see, in Chapter \ref{chap:Tails} that in odd dimensional
spacetimes there are tails associated with wave propagation, even if the spacetime
is flat. On the other hand, for even dimensional spacetimes, this tail only
appears if the spacetime is curved, by including for example a black hole.

\section{Outline of Part III}
In this third part of the thesis we will begin, in Chapter
\ref{chap:radddim}, by reviewing our understanding of linearized
gravitational waves in a flat background.  As there is now a great
interest in physics in higher dimensions, due to the TeV-scale gravity
scenario, I shall extend all of the results we have in four
dimensional spacetimes to $D$ dimensional spacetimes.  This is a
non-trivial extension, mainly because the Green's function for the
wave equation in $D$-dimensions no longer has the simple structure
that we know so well when $D=4$. In fact, whereas for even $D$ there
is still propagation on the light cone (this means Huygens principle still
holds), for odd $D$ this propagation on the light cone is lost (this
shows up cleanly in the Green's function expression).  Due to
difficulties in handling the wave tails that arise in odd dimensional
spacetimes, we will therefore focus on even-dimensional spacetimes
(even $D$).  We will compute the $D$-dimensional quadrupole formula,
thereby extending Einstein's formula to general dimension, and we shall
also compute the gravitational radiation when a high energy particle
collides with a black hole. It is unnecessary to write the importance
this may have, should the TeV-scale scenario be correct, but we shall
do it nonetheless.  If this scenario is correct, we may be lucky
enough to produce black holes at the LHC, as we remarked earlier. If
this happens, one needs desperately some estimates for the
gravitational radiation released during the balding phase, the first
stage of black hole formation.  So, the computation of gravitational
radiation when black holes collide in higher dimensions is extremely
important.  Finally i shall compute, for the first time, the
gravitational radiation released during the quantum process of black
hole pair creation.  The quantum process of black hole pair creation
as a process for gravitational wave generation seems to have been
overlooked over the years. We show it may well be a strong generator of
gravitational waves.

In Chapter \ref{chap:Tails} we shall conclude this thesis by continuing 
the program of gravitational wave
physics in higher dimensional spacetimes. We consider higher dimensional
Schwarzschild black holes.  Supporting what has been said about (flat
space) Green's function in higher dimensions, we show that if the
spacetime has odd number of dimensions then the appearance of tails is
inevitable, even if there is no black hole. We also show that the late
time behaviour of any field is dictated by $\Psi \sim t^{-(2l+D-2)}$,
where $l$ is the angular index determining the angular dependence of
the field.
One can show directly from the flat space Green's function that such a
power-law is indeed expected in flat odd dimensional spacetimes, and
it is amusing to note that the same conclusion can be reached directly
from an indepent analysis, as i shall do.  For even dimensional
spacetimes we find also a power-law decay at late times, but with a
much more rapid decay, $\Psi \sim t^{-(2l+3D-8)}$.  For even $D$, this
power-law tail is entirely due to the black hole, as opposed to the
situation in odd $D$.  These results are strictly valid for
$D>4$. Four dimensional Schwarzschild geometry is special, having the
well known power-law tail $\Psi \sim t^{-(2l+3)}$.

\thispagestyle{empty} \setcounter{minitocdepth}{1}
\chapter[Gravitational waves in a flat higher dimensional background]{Gravitational waves in a flat higher dimensional background} \label{chap:radddim}
\lhead[]{\fancyplain{}{\bfseries Chapter \thechapter. \leftmark}}
\rhead[\fancyplain{}{\bfseries \rightmark}]{}
\minitoc \thispagestyle{empty}
\renewcommand{\thepage}{\arabic{page}}
\section{Introduction}
This chapter is organized as follows, and is based on \cite{vitoroscarjose}: 
In section II we linearize
Einstein's equations in a flat $D$-dimensional background and arrive at
an inhomogeneous wave equation for the metric perturbations. The
source free equations are analyzed in terms of plane waves, and then
the general solution to the homogeneous equation is deduced in terms
of the $D$-dimensional retarded Green's function.  In section III we
compute the $D$-dimensional quadrupole formula (assuming slowly moving
sources), expressing the metric and the radiated energy in terms of
the time-time component of the energy-momentum tensor. We then apply
the quadrupole formula to two cases: a particle in circular motion in a 
generic background, and
a particle falling into a $D$-dimensional Schwarzschild black hole.
In section IV we consider the hard collision between two particles, i.e., 
the collision takes zero seconds, and introduce a cutoff frequency
necessary to have meaningful results. We then apply to the case where 
one of the colliding
particles is a black hole. We propose that this cutoff should be
related to the gravitational quasinormal frequency of the black hole, 
and compute some values of the scalar quasinormal frequencies for
higher dimensional Schwarzschild black holes, expecting that the
gravitational quasinormal frequencies will behave in the same manner.  
Finally, we apply this formalism to
compute the generation of gravitational radiation during black hole
pair creation in four and higher dimensions, a result that has never been 
worked out, even for $D=4$.
In our presentation we shall mostly follow Weinberg's \cite{weinberg}
exposition.

\section{Linearized $D$-dimensional Einstein's equations}

Due to the non-linearity of Einstein's equations, the treatment of
the gravitational radiation problem is not an easy one since the
energy-momentum tensor of the gravitational wave contributes to
its own gravitational field. To overcome this difficulty it is a
standard procedure to work only with the weak radiative solution,
in the sense that the energy-momentum content of the gravitational
wave is small enough in order to allow us to neglect its
contribution to its own propagation. This approach is justified in
practice since we expect the detected gravitational radiation to
be of low intensity. For some non-linear effects in gravitational
wave physics, we refer the reader to \cite{cardosononlinear} and references therein.
\subsection{The inhomogeneous wave equation}
We begin this subsection by introducing the general background
formalism (whose details can be found, e.g., in \cite{weinberg}) that
will be needed in later sections. Then we obtain the linearized
inhomogeneous wave equation.

Greek indices vary as $0,1,\cdots, D-1$ and latin indices as
$1,\cdots, D-1$ and our units are such that $c \equiv 1$.
We work on a $D$-dimensional spacetime described by a
metric $g_{\mu\nu}$ that approaches asymptotically the
$D$-dimensional Minkowski metric $\eta_{\mu\nu}={\rm
diag}(-1,+1,\cdots,+1)$, and thus we can write
\begin{equation}
g_{\mu\nu}=\eta_{\mu\nu}+h_{\mu\nu}\,
\hspace{1cm}\mu,\nu=0,1,\cdots,D-1 \,,
 \label{g}
\end{equation}
where $h_{\mu\nu}$ is small, i.e., $|h_{\mu\nu}|<<1$, so that it
represents small corrections to the flat background.
The exact Einstein field equations,
 $G_{\mu\nu}=8\pi {\cal G} T_{\mu\nu}$ (with ${\cal G}$ being the usual Newton
constant), can then be written as
\begin{equation}
R^{(1)}_{\;\;\;\;\mu\nu}-\frac{1}{2}\eta_{\mu\nu}
 R^{(1)\,\alpha}_{\;\;\;\;\;\;\;\alpha}=8\pi {\cal G}\,\, \tau_{\mu\nu} \,,
 \label{R1}
\end{equation}
with
\begin{equation}
\tau^{\mu\nu} \equiv \eta^{\mu \alpha} \eta^{\nu\beta}
\,(T_{\alpha\beta}+t_{\alpha\beta}).
 \label{tau}
\end{equation}
Here $R^{(1)}_{\;\;\;\;\mu\nu}$ is the part of the Ricci tensor
linear in $h_{\mu\nu}$,
 $R^{(1)\,\alpha}_{\;\;\;\;\;\;\;\alpha}=\eta^{\alpha\beta}
 R^{(1)}_{\;\;\;\;\beta\alpha}$,
and $\tau_{\mu\nu}$ is the effective energy-momentum tensor, containing
contributions from $T_{\mu\nu}$, the energy-momentum tensor of the matter source,
and $t_{\mu\nu}$ which represents the gravitational contribution.
The pseudo-tensor $t_{\mu\nu}$ contains the difference between
the exact Ricci terms and the Ricci terms linear in $h_{\mu\nu}$,
\begin{equation}
t_{\mu\nu}=\frac{1}{8\pi {\cal G}}\left [ R_{\mu\nu}
-\frac{1}{2}g_{\mu\nu} R^{\alpha}_{\;\;\;\alpha}
 -R^{(1)}_{\;\;\;\;\mu\nu}+\frac{1}{2}\eta_{\mu\nu}
 R^{(1)\,\alpha}_{\;\;\;\;\;\;\;\alpha} \right ] \,.
 \label{t}
\end{equation}
The Bianchi identities imply that $\tau_{\mu\nu}$ is locally conserved,
\begin{equation}
\partial_{\mu} \tau^{\mu\nu}=0 \,.
 \label{constau}
\end{equation}
Introducing the cartesian coordinates $x^{\alpha}=(t,{\bf x})$ with ${\bf
x}=x^i$, and considering a $D-1$ volume $V$ with a boundary spacelike
surface $S$ with dimension $D-2$ whose unit exterior normal is ${\bf
n}$, eq. (\ref{constau}) yields
\begin{equation}
\frac{d}{dt}\int_{V} d^{D-1}{\bf x} \;\tau^{0\nu}
=-\int_{S}
 d^{D-2}{\bf x} \;n_i \tau^{i\nu} \,.
 \label{constau1}
\end{equation}
This means that one may interpret
\begin{equation}
p^{\nu} \equiv \int_{V} d^{D-1}{\bf x} \;\tau^{0\nu}
 \label{p}
\end{equation}
as the total energy-momentum (pseudo)vector of the system, including
matter and gravitation, and $\tau^{i\nu}$ as the corresponding flux.
Since the matter contribution is contained in $t^{\mu\nu}$,
the flux of gravitational radiation is
\begin{equation}
{\rm Flux}=\int_{S} d^{D-2}{\bf x} \;n_i t^{i \nu} \,.
 \label{fluxgrav}
\end{equation}

In this context of linearized general
relativity, we neglect terms of order higher than the first in
$h_{\mu\nu}$ and all the indices are raised and lowered using
$\eta^{\mu\nu}$. We also neglect the contribution of the
gravitational energy-momentum tensor $t_{\mu\nu}$ (i.e.,
$|t_{\mu\nu}|<<|T_{\mu\nu}|$) since from (\ref{t}) we see that
$t_{\mu\nu}$ is of higher order in $h_{\mu\nu}$. Then, the
conservation equations (\ref{constau}) yield
\begin{equation}
\partial_{\mu}T^{\mu\nu}=0 \,.
 \label{consT}
\end{equation}
In this setting and choosing the convenient coordinate system
that obeys the harmonic (also called Lorentz) gauge conditions,
\begin{equation}
2\partial_{\mu}h^{\mu}_{\;\;\;\nu}=\partial_{\nu}h^{\alpha}_{\;\;\;\alpha}
 \label{hargauge}
\end{equation}
(where $\partial_{\mu}=\partial/\partial x^{\mu}$), the first
order Einstein field equations (\ref{R1}) yield
\begin{equation}
\square h_{\mu\nu}=-16\pi {\cal G} S_{\mu\nu} \,,
 \label{ineq}
\end{equation}
\begin{equation}
S_{\mu\nu}=T_{\mu\nu}-\frac{1}{D-2}\,\eta_{\mu\nu}\,T^{\alpha}_{\;\;\;\alpha}
\,,
 \label{S}
\end{equation}
where $\square=\eta^{\mu\nu}\partial_{\mu}\partial_{\nu}$ is the
$D$-dimensional Laplacian, and $S_{\mu\nu}$ will be called the
modified energy-momentum tensor of the matter source. Eqs.
(\ref{ineq}) and (\ref{S}) subject to (\ref{hargauge}) allow us to
find the gravitational radiation produced by a matter source
$S_{\mu\nu}$.
\subsection{The plane wave solutions}

In vacuum, the linearized equations for the gravitational field are
$R^{(1)}_{\;\;\;\;\mu\nu}=0$ or, equivalently, the homogeneous
equations $\square h_{\mu\nu}=0$, subjected to the harmonic gauge
conditions (\ref{hargauge}). The solutions of these equations, the
plane wave solutions, are important since the general solutions of the
inhomogeneous equations (\ref{hargauge}) and (\ref{ineq}) approach the
plane wave solutions at large distances from the source.  Setting
$k_{\alpha}=(-\omega,{\bf k})$ with $\omega$ and ${\bf k}$ being
respectively the frequency and wave vector, the plane wave solutions
can be written as a linear superposition of solutions of the kind
\begin{equation}
 h_{\mu\nu}(t,{\bf x})=e_{\mu\nu}\,e^{i k_{\alpha} x^{\alpha}}+
 e_{\mu\nu}^*\,e^{-i k_{\alpha} x^{\alpha}} \,,
  \label{homsol}
\end{equation}
where $e_{\mu\nu}=e_{\nu\mu}$ is called the polarization tensor
and $^*$ means the complex conjugate. These solutions satisfy eq.
(\ref{ineq}) with $S_{\mu\nu}=0$ if $k_{\alpha} k^{\alpha}=0$, and
obey the harmonic gauge conditions (\ref{hargauge}) if
$2\,k_{\mu}\,e^{\mu}_{\;\;\;\nu}=k_{\nu}\,e^{\mu}_{\;\;\;\mu}$.

An important issue that must be addressed is the number of
different polarizations that a gravitational wave in $D$
dimensions can have. The polarization tensor $e_{\mu\nu}$, being
symmetric, has in general $D(D+1)/2$ independent components.
However, these components are subjected to the $D$ harmonic gauge
conditions that reduce by $D$ the number of independent
components. In addition, under the infinitesimal change of
coordinates $x'^{\mu}=x^{\mu}+\xi^{\mu}(x)$, the
polarization tensor transforms into
$e'_{\mu\nu}=e_{\mu\nu}-\partial_{\nu}\xi_{\mu}
-\partial_{\mu}\xi_{\nu}$. Now, $e'_{\mu\nu}$ and
$e_{\mu\nu}$ describe the same physical system for arbitrary
values of the $D$ parameters $\xi^{\mu}(x)$. Therefore, the number
of independent components of $e_{\mu\nu}$, i.e., the number of
polarization states of a gravitational wave in $D$ dimensions is
$D(D+1)/2-D-D=D(D-3)/2$. From this computation we can also see
that gravitational waves are present only when $D>3$. Therefore,
from now on we assume ${\rm D}>3$ whenever we refer to $D$. In what concerns
the helicity of the gravitational waves, for arbitrary $D$ the
gravitons are always spin $2$ particles.

To end this subsection on gravitational plane wave solutions, we
present the average gravitational energy-momentum tensor of a
plane wave, a quantity that will be needed later. Notice that
in vacuum, since the matter contribution is zero
($T_{\mu\nu}=0$), we cannot neglect the contribution of the
gravitational energy-momentum tensor $t_{\mu\nu}$. From eq.
(\ref{t}), and neglecting terms of order higher than $h^2$, the
gravitational energy-momentum tensor of a plane wave is given by
\begin{equation}
t_{\mu\nu}\simeq \frac{1}{8\pi {\cal G} }\left [
R^{(2)}_{\;\;\;\;\mu\nu}-\frac{1}{2}\eta_{\mu\nu}
 R^{(2)\,\alpha}_{\;\;\;\;\;\;\;\alpha} \right ] \,,
 \label{tplane}
\end{equation}
and through a straightforward calculation (see e.g.
\cite{weinberg} for details) we get the average gravitational
energy-momentum tensor of a plane wave,
\begin{equation}
\langle t_{\mu\nu} \rangle=\frac{k_{\mu}k_{\nu}}{16\pi {\cal G}}\left [
e^{\alpha\beta}e_{\alpha\beta}^*
-\frac{1}{2}|e^{\alpha}_{\;\;\;\alpha}|^2\right ] \,.
 \label{taverage}
\end{equation}

\subsection{The $D$-dimensional retarded Green's function}
The general solution to the inhomogeneous differential 
equation (\ref{ineq}) may be found in the usual way in terms of
a Green's function as 
\begin{equation}
h_{\mu\nu}(t,{\bf x})=-16 \pi {\cal G}\int dt'\int d^{D-1}{\bf x'} 
S_{\mu \nu}(t',{\bf x'})
G(t-t',{\bf x - x'})+{\rm homogeneous \,solutions}\,,
\label{insol}
\end{equation}
where the Green's function $G(t-t',{\bf x - x'})$ satisfies
\begin{equation}
\eta^{\mu\nu}\partial_{\mu}\partial_{\nu} 
G(t-t',{\bf x - x'})=\delta(t-t')\delta({\bf x - x'})\,,
\label{greendef}
\end{equation}
where $\delta(z)$ is the Dirac delta function.
In the momentum representation this reads 
\begin{equation}
G(t,{\bf x})=-\frac{1}{(2\pi)^{D}} \int d^{D-1} {\bf k} 
e^{i {\bf k}\cdot{\bf x}}\int d\omega \frac{e^{-i\omega t}}{\omega^2-k^2}\,,
\label{greenmomentum}
\end{equation}
where $k^2=k_{1}^2+k_{2}^2+...+k_{D-1}^2$.
To evaluate this, it is convenient to perform the $k$-integral by
using spherical coordinates in the ($D-1$)-dimensional $k$-space.  The
required transformation, along with some useful formulas which shall
be used later on, is given in Appendix \ref{apendiceradddim}.  The result for the retarded
Green's function in those spherical coordinates is
\begin{equation}
G^{\rm ret}(t,{\bf x})=-\frac{\Theta(t)}{(2\pi)^{(D-1)/2}}\times\frac{1}{r^{(D-3)/2}}
\int k^{(D-3)/2} J_{(D-3)/2}(kr)\sin(kt)dk \,,
\label{greenretgeral}
\end{equation}
where $r^2=x_{1}^2+x_{2}^2+...+x_{D-1}^2$, and  $\Theta(t)$ is the Heaviside 
function defined as
\begin{equation}
\Theta(t)=\left\{ \begin{array}{ll}
             1   & \mbox{if $t>0$}\\
             0    & \mbox{if $t<0$}\,.
\end{array}\right.
\label{Heaviside}
\end{equation}
The function $J_{({\rm D}-3)/2}(kr)$ is a Bessel function
\cite{watson,stegun}.  The structure of the retarded Green's function
will depend on the parity of $D$, as we shall see. This dependence on
the parity, which implies major differences between even and odd
spacetime dimensions, is connected to the structure of the Bessel
function. For even $D$, the index of the Bessel function is
semi-integer and then the Bessel function is expressible in terms of
elementary functions, while for odd $D$ this does not happen. A
concise explanation of the difference between retarded Green's
function in even and odd $D$, and the physical consequences that
entails is presented in \cite{courant} (see also \cite{hadamard,barrow,galtsov}).
 A complete derivation of the Green's function in
higher dimensional spaces may be found in Hassani \cite{hassani}.
The result is
\begin{equation}
G^{\rm ret}(t,{\bf x})=\frac{1}{4\pi}\left[-\frac{\partial}{2\pi r \partial r} \right]^{(D-4)/2}
\left[\frac{\delta(t-r)}{r}\right]\,,\,\,\,\,D \,\,{\rm even}.
\label{greenfinaleven}
\end{equation}
\begin{equation}
G^{\rm ret}(t,{\bf x})=\frac{\Theta(t)}{2\pi}\left[-\frac{\partial}{2\pi r \partial r} \right]^{(D-3)/2}
\left[\frac{1}{\sqrt{t^2-r^2}}\right]\,,\,\,\,\,D \,\,{\rm odd}.
\label{greenfinalodd}
\end{equation}
It is sometimes convenient to work with the Fourier transform (in the time
coordinate) of the Green's function. One finds \cite{hassani} an
analytical result independent of the parity of $D$
\begin{equation}
G^{\rm ret}(\omega,{\bf x})=\frac{i^D \pi}{2(2\pi)^{(D-1)/2}} 
\left(\frac{\omega}{r}\right)^{(D-3)/2} H^1_{(D-3)/2}(\omega r)\,,
\label{greenfinalevenfourier}
\end{equation}
where $H^1_{\nu}(z)$ is a modified Bessel function
\cite{stegun,watson}.  Of course, the different structure of the
Green's function for different $D$ is again embodied in these Bessel
functions.  Equations (\ref{greenfinaleven}) and
(\ref{greenfinalevenfourier}), are one of the most important results
we shall use in this paper.  For $D=4$ (\ref{greenfinaleven})
obviously reproduce well known results \cite{hassani}.  Now, one sees
from eq. (\ref{greenfinaleven}) that although there are delta function
derivatives on the even-$D$ Green's function, the localization of the
Green's function on the light cone is preserved.
However, eq. (\ref{greenfinalodd}) tells us that the retarded Green's
function for odd dimensions is non-zero inside the light
cone. The consequence, as has been emphasized by different authors
\cite{courant,galtsov, kazinski}, is that for odd $D$ the Huygens principle
does not hold: the fact that the retarded Green's function support
extends to the interior of the light cone implies the appearance
of radiative tails in (\ref{insol}). 
In other words, we still have a propagation
phenomenum for the wave equation in odd dimensional spacetimes, in so
far as a localized initial state requires a certain time to reach a
point in space. Huygens principle no longer holds, because
the effect of the initial state is not sharply limited in time: once
the signal has reached a point in space, it persists there
indefinitely as a reverberation.

This fact coupled to the analytic structure of the Green's function in
odd dimensions make it hard to get a grip on radiation generation in
odd dimensional spacetimes. Therefore, from now on we shall focus on
even dimensions, for which the retarded Green's function is given by
eq. (\ref{greenfinaleven}).

\subsection{The even $D$-dimensional retarded solution in the wave zone}
The retarded solution for the metric perturbation $h_{\mu\nu}$,
obtained by using the retarded Green's function (\ref{greenfinaleven})
and discarding the homogeneous solution in (\ref{insol})
will be given by
\begin{equation}
h_{\mu\nu}(t,{\bf x})= 16 \pi {\cal G}\int dt'\int d^{D-1}{\bf x'} 
S_{\mu \nu}(t',{\bf x'})
G^{\rm ret}(t-t', {\bf x - x'})\,,
\label{retsol}
\end{equation}
with $G^{\rm ret}(t-t',{\bf x - x'})$ as in eq. (\ref{greenfinaleven}).
For $D=4$ for example one has
\begin{equation}
G^{\rm ret}(t,{\bf x})=\frac{1}{4\pi}\frac{\delta(t-r)}{r}\,\,,\,\,\,\,\,D=4\,,
\label{green4}
\end{equation}
which is the well known result.
For $D=6$, we have
\begin{equation}
G^{\rm ret}(t,{\bf x})=\frac{1}{8\pi^2}\left(\frac{\delta'(t-r)}{r^2}+\frac{\delta(t-r)}{r^3}\right)
\,\,,\,\,\,\,\,D=6\,,
\label{green6}
\end{equation}
where the $\delta'(t-r)$ means derivative of the Dirac delta function with respect to its argument.
For $D=8$, we have
\begin{equation}
G^{\rm ret}(t,{\bf x})=\frac{1}{16\pi^3}\left(\frac{\delta''(t-r)}{r^3}+
3\frac{\delta'(t-r)}{r^4}+3\frac{\delta(t-r)}{r^5}\right)
\,\,,\,\,\,\,\,D=8\,.
\label{green8}
\end{equation}
We see that in general even-$D$ dimensions the Green's function
consists of inverse integer powers in r, spanning all values between
$\frac{1}{r^{(D-2)/2}}$ and $\frac{1}{r^{D-3}}$, including these ones.
Now, the retarded solution is given by eq. (\ref{retsol}) as a product
of the Green's function times the modified energy-momentum tensor
$S_{\mu \nu}$. The net result of having derivatives on the delta
functions is to transfer these derivatives to the energy-momentum
tensor as time derivatives (this can be seen by integrating
(\ref{retsol}) by parts in the $t$-integral).

A close inspection then shows that the retarded field possesses a kind
of peeling property in that it consists of terms with different
fall off at infinity. Explicitly, this means that the retarded field
will consist of a sum of terms possessing all integer inverse powers
in $r$ between $\frac{D-2}{2}$ and $D-3$.  The term that dies off more
quickly at infinity is the $\frac{1}{r^{D-3}}$, typically a
static term, since it comes from the Laplacian.  As a matter of fact
this term was already observed in the higher dimensional black hole
by Tangherlini \cite{tangherlini} (see also Myers and Perry 
\cite{myersperry}). 
We will see that the
term falling more slowly, the one that goes like
$\frac{1}{r^{(D-2)/2}}$, gives rise to gravitational radiation. It is 
well defined, in the sense that the power crossing
sufficiently large hyperspheres with different radius is the same, because
the volume element goes as $r^{D-2}$ and the energy as $|h|^2
\sim \frac{1}{r^{D-2}}$.

In radiation problems, one is interested in finding out the field at
large distances from the source, $r>>\lambda$, where $\lambda$ is the 
wavelength of the waves, and also much larger
than the source's dimensions $R$. This is defined as the wave zone.  In
the wave zone, one may neglect all terms in the Green's function that
decay faster than $\frac{1}{r^{(D-2)/2}}$. So, in the wave zone, we
find
\begin{equation}
h_{\mu\nu}(t,{\bf x})= -8 \pi {\cal G} \frac{1}{(2\pi r)^{(D-2)/2}}
\partial_{t}^{(\frac{D-4}{2})}\left[ \int d^{D-1}{\bf x'} 
 S_{\mu \nu}(t-|{\bf x-x'}|,{\bf x'})\right]\,,
\label{retsolwavezone}
\end{equation}
where $\partial_{t}^{(\frac{D-4}{2})}$ stands for the $\frac{D-4}{2}$th derivative
with respect to time.
For $D=4$ eq. (\ref{retsolwavezone}) yields the standard result
\cite{weinberg}:
\begin{equation}
h_{\mu\nu}(t,{\bf x})= -\frac{4 {\cal G}}{r} \int d^{D-1}{\bf x'} 
 S_{\mu \nu}(t-|{\bf x-x'}|,{\bf x'})\,,\,\,\,\,\,D=4.
\label{retsolwavezoned4}
\end{equation}
To find the Fourier transform of the metric, one uses the
representation (\ref{greenfinalevenfourier}) for the Green's
function. Now, in the wave zone, the Green's function may be
simplified using the asymptotic expansion for the Bessel function
\cite{stegun}
\begin{equation}
H^1_{(D-3)/2}(\omega r) \sim \sqrt{\frac{2}{\pi(\omega r)}} 
e^{i\left[\omega r-\frac{\pi}{4}(D-2)\right]}\,\,,\,\,\,\,\,
\omega r \rightarrow \infty.
\label{asymbessel}
\end{equation}
This yields 
\begin{equation}
h_{\mu\nu}(\omega,{\bf x})= -\frac{8 \pi {\cal G}}{(2\pi r)^{(D-2)/2}}\omega^{(D-4)/2}e^{i\omega r}
\int d^{D-1}{\bf x'} S_{\mu \nu}(\omega,{\bf x'})\,.
\label{retsolwavezonefourier}
\end{equation}
This could also have been arrived at directly from
(\ref{retsolwavezone}), using the rule time derivative $\rightarrow
-i\omega $ for Fourier transforms.  Equations (\ref{retsolwavezone})
and (\ref{retsolwavezonefourier}) are one of the most important
results derived in this paper, and will be the basis for all the
subsequent section. Similar equations, but not as general as the ones 
presented here, were given by Chen, 
Li and Lin \cite{lin} in the context of gravitational radiation by 
a rolling tachyon. 
 
To get the energy spectrum, we use (\ref{S})
yielding
\begin{equation}
\frac{d^2E}{d\omega d\Omega}= 2 {\cal G} \frac{\omega^{D-2}}{(2\pi)^{D-4}} 
\left( T^{\mu\nu}(\omega,{\bf k})T_{\mu\nu}^*(\omega,{\bf k})-\frac{1}{D-2}
|T^{\lambda}_{\:\:\:\lambda}(\omega, {\bf k})|^2\right)\,.
\label{powerwavezone}
\end{equation}

\section{The even $D$-dimensional quadrupole formula}
\subsection{Derivation of the even $D$-dimensional quadrupole formula}
When the velocities of the sources that generate the gravitational
waves are small, it is sufficient to know the $T^{00}$ component of
the gravitational energy-momentum tensor in order to have a good
estimate of the energy they radiate. In this subsection, we will
deduce the $D$-dimensional quadrupole formula and in the next
subsection we will apply it to (1) a particle in circular orbit and
(2) a particle in free fall into a $D$-dimensional Schwarzschild black
hole.

We start by recalling that the Fourier transform of the
energy-momentum tensor is
\begin{equation}
T_{\mu\nu}(\omega,{\bf k})=\int d^{D-1}{\bf x'} e^{-i {\bf k}\cdot
{\bf x'}} \int dt\, e^{i \omega t}\,T^{\mu\nu}(t,{\bf x})+ {\rm c.c.} \,,
\label{fourierT}
\end{equation}
where ${\rm c.c.}$ means the complex conjugate of the preceding term.
Then, the conservation equations (\ref{consT}) for
$T^{\mu\nu}(t,{\bf x})$ applied to eq. (\ref{fourierT}) yield
$k^{\mu}\,T_{\mu\nu}(\omega,{\bf k})=0$. Using this last result we
obtain $T_{00}(\omega,{\bf k})=\hat{k}^j \,\hat{k}^i
\,T_{ji}(\omega,{\bf k})$ and $T_{0i}(\omega,{\bf k})=-\hat{k}^j\,
T_{ji}(\omega,{\bf k})$, where ${\bf \hat{k}}={\bf k}/\omega$. 
We can then write the energy spectrum, eq.
(\ref{powerwavezone}), as a function only of the spacelike
components of $T^{\mu\nu}(\omega,{\bf k})$,
\begin{equation}
\frac{d^2E}{d\omega d\Omega}= 2 {\cal G}
\frac{\omega^{D-2}}{(2\pi)^{D-4}}\, \Lambda_{ij,\,lm}(\hat{k})\,
T^{*\,ij}(\omega,{\bf k})\, T^{\,ij}(\omega,{\bf k}) \,,
\label{powerwavezone2}
\end{equation}
where
\begin{equation}
 \Lambda_{ij,\,lm}(\hat{k})=\delta_{il}\delta_{jm}
 -2\hat{k}_j \hat{k}_m \delta_{il}
 +\frac{1}{D-2}\left (-\delta_{ij}\delta_{lm}
 + \hat{k}_l \hat{k}_m \delta_{ij}
 +\hat{k}_i \hat{k}_j \delta_{lm}\right )
 +\frac{D-3}{D-2}\hat{k}_i \hat{k}_j \hat{k}_l \hat{k}_m \,.
\label{Lambda}
\end{equation}
At this point, we make a new approximation (in addition to the
wave zone approximation) and assume that $\omega R<<1$, where $R$
is the source's radius. In other words, we assume that the internal 
velocities of the sources are small and thus the source's radius is much smaller
than the characteristic wavelength $\sim 1/\omega$ of the emitted gravitational
waves. Within this approximation, one can set $e^{-i {\bf k}\cdot
{\bf x'}} \sim 1$  in eq. (\ref{fourierT}) (since $R=|{\bf
x'}|_{\rm max}$). Moreover, after a straightforward calculation,
one can also set in eq. (\ref{powerwavezone2}) the approximation
$T^{\,ij}(\omega,{\bf k})\simeq -(\omega^2/2)D_{ij}(\omega)$,
where
\begin{equation}
D_{ij}(\omega)=
 \int d^{D-1} {\bf x}\, x^i \,x^j \,T^{00}(\omega,{\bf x})\,.
 \label{Dij}
\end{equation}
Finally, using
\begin{eqnarray}
\int d\Omega_{D-2}\hat{k}_i
\hat{k}_j=\frac{\Omega_{D-2}}{D-1}\delta_{ij}\,, \nonumber \\
\int d\Omega_{D-2}\hat{k}_i \hat{k}_j \hat{k}_l \hat{k}_m=
 \frac{3\Omega_{D-2}}{D^2-1} (\delta_{ij}\delta_{lm}
  +\delta_{il}\delta_{jm}+\delta_{im}\delta_{jl})\,,
\label{intk}
\end{eqnarray}
where $\Omega_{D-2}$ is the $(D-2)$-dimensional solid angle
defined in (\ref{integratedsolidangle}), we obtain the
$D$-dimensional quadrupole formula
\begin{equation}
\frac{dE}{d\omega}=\frac{2^{2-D}\pi^{-(D-5)/2}{\cal G}\,(D-3)D}
{\Gamma[(D-1)/2](D^2-1)(D-2)}\, \omega^{D+2} {\biggl [}
(D-1)D^*_{ij}(\omega)D_{ij}(\omega)-|D_{ii}(\omega)|^2 {\biggr ]}
 \,,
\label{quadw}
\end{equation}
where the Gamma function $\Gamma[z]$ is defined in Appendix
\ref{apendiceradddim}.  As the dimension $D$ grows it is seen that the rate
of gravitational energy radiated increases as $\omega^{D+2}$.
Sometimes it will be more useful to have the time rate of emitted
energy
\begin{equation}
\frac{dE}{dt}=\frac{2^{2-D}\pi^{-(D-5)/2} {\cal G}\,(D-3)D}
{\Gamma[(D-1)/2](D^2-1)(D-2)}\,{\biggl [}
(D-1)\partial_t^{(D+2)/2}D^*_{ij}(t)\partial_t^{(D+2)/2}D_{ij}(t)-
|\partial_t^{(D+2)/2}D_{ii}(t)|^2 {\biggr ]}
 \,.
\label{quad}
\end{equation}
For $D=4$, eq. (\ref{quad}) yields the well known result \cite{weinberg}
\begin{equation}
\frac{dE}{dt}=\frac{{\cal G}}{5}\, {\biggl [}
\partial_t^{3}D^*_{ij}(t)\partial_t^{3}D_{ij}(t)-
\frac{1}{3}|\partial_t^{3}D_{ii}(t)|^2
{\biggr ]}
 \,.
\label{quad4}
\end{equation}

\subsection{Applications of the quadrupole formula: test particles in a background geometry}
The quadrupole formula has been used successfully
in almost all kind of problems involving gravitational wave
generation. By successful we mean that it agrees with other
more accurate methods. Its simplicity and the fact that it gives
results correct to within a few percent, 
makes it an invaluable tool in estimating
gravitational radiation emission. We shall in the following present
two important examples of the application of the quadrupole formula.
\subsubsection{A particle in circular orbit}
The radiation generated by particles in circular motion was perhaps
the first situation to be considered in the analysis of gravitational
wave generation. For orbits with low frequency, the quadrupole formula
yields excellent results. As expected it is difficult to find in
nature a system with perfect circular orbits, they will in general be
elliptic. In this case the agreement is also remarkable, and one finds
that the quadrupole formalism can account with precision for the
increase in period of the pulsar PSR 1913$+$16, due to gravitational
wave emission \cite{pulsar}. In four dimensions the full treatment of
elliptic orbital motion is discussed by Peters \cite{peters}.  In
dimensions higher than four, it has been shown \cite{tangherlini} that
there are no stable geodesic circular orbits, and so geodesic circular
motion is not as interesting for higher $D$. For this reason, and also
because we only want to put in evidence the differences that arise in
gravitational wave emission as one varies the spacetime dimension $D$, we
will just analyze the simple circular, not necessarily geodesic
motion, to see whether the results are non-trivially changed as one
increases $D$.  Consider then two bodies of equal mass $m$ in circular
orbits a distance $l$ apart.  Suppose they revolve around the center
of mass, which is at $l/2$ from both masses, and that they orbit with
frequency $\omega$ in the $x-y$ plane.  A simple calculation
\cite{peters,schutzlivro} yields
\begin{eqnarray}
D_{xx}=\frac{ml^2}{4}\cos({2\omega t}) \,+\,{\rm const}\,\,, \\
D_{yy}=-D_{xx}\,\,, \\
D_{xy}=\frac{ml^2}{4}\sin({2\omega t}) \,+\,{\rm const}\,\,,
\label{moments}
\end{eqnarray}
independently of the dimension in which they are imbedded and with all
other components being zero.  We therefore get from eq. (\ref{quad})

\begin{equation}
\frac{dE}{dt}=\frac{2{\cal G}D(D-3)}{\pi^{(D-5)/2}\Gamma[(D-1)/2](D+1)(D-2)}m^2 l^4
 \omega^{D+2}.
\label{totalenecircular}
\end{equation}
For $D=4$ one gets
\begin{equation}
\frac{dE}{dt}=\frac{8{\cal G}}{5}m^2 l^4 \omega^{6}\,,
\label{totalenecircularD4}
\end{equation}
which agrees with known results \cite{peters,schutzlivro}.
Eq. (\ref{totalenecircular}) is telling us that as one climbs up
in dimension number $D$, the frequency effects gets more pronounced.

\subsubsection{A particle falling radially into a higher dimensional
Schwarzschild black hole}
As yet another example of the use of the quadrupole formula
eq. (\ref{quad}) we now calculate the energy given away as
gravitational waves when a point particle, with mass $m$ falls into a
$D$-dimensional Schwarzschild black hole, a metric first given in
\cite{tangherlini}.  Historically, the case of a particle falling into
a $D=4$ Schwarzschild black hole was one of the first to be studied
\cite{zerilli,davis} in connection with gravitational wave generation,
and later served as a model calculation when one wanted to evolve
Einstein's equations fully numerically \cite{smarrl,gleiser}.  This
process was first studied \cite{davis} by solving numerically
Zerilli's \cite{zerilli} wave equation for a particle at rest at
infinity and then falling into a Schwarzschild black hole. Davis et al
\cite{davis} found numerically that the amount of energy radiated to
infinity as gravitational waves was $\Delta E_{\rm num} =0.01
\frac{m^2}{M}$, where $m$ is the mass of the particle falling in and
$M$ is the mass of the black hole.
It is found that the $D=4$ quadrupole formula yields \cite{quem}
$\Delta E_{\rm quad} =0.019 \frac{m^2}{M}$, so it is of the order of
magnitude as that given by fully relativistic numerical results.
Despite the fact that the quadrupole formula fails somewhere near the
black hole (the motion is not slow, and the background is certainly
not flat), it looks like one can get an idea of how much radiation
will be released with the help of this formula.  Based on this good
agreement, we shall now consider this process but for higher
dimensional spacetimes.  The metric for the $D$-dimensional
Schwarzschild black hole in ($t,r,\theta_1,\theta_2,..,\theta_{D-2}$)
coordinates (see Appendix \ref{apendiceradddim}) is
\begin{equation}
ds^2= -\left(1-\frac{16\pi{\cal G} M}{(D-2)\Omega_{D-2}}\frac{1}{r^{D-3}}\right)dt^2+
\left(1-\frac{16\pi{\cal G} M}{(D-2)\Omega_{D-2}}\frac{1}{r^{D-3}}\right)^{-1}dr^2
+r^{D-2}d\Omega_{D-2}^2.
\label{metricmyersperry} 
\end{equation}
Consider a particle falling along a radial geodesic, and at rest at infinity.
Then, the geodesic equations give
\begin{equation}
\frac{dr}{dt} \sim \frac{16\pi{\cal G} M}{(D-2)\Omega_{D-2}}\frac{1}{r^{D-3}}\,,
\label{geodesic}
\end{equation}
where we make the flat space approximation $t=\tau$.  We then have, in
these coordinates, $D_{11}=r^2$, and all other components vanish.
From (\ref{quad}) we get the energy radiated per second, which yields
\begin{equation}
\frac{dE}{dt}= \frac{2^{2-D}\pi^{-(D-5)/2}{\cal G}\,(D-3)}
{\Gamma[(D-1)/2](D^2-1)}D|\partial_{t}^{(\frac{D+2}{2})}
D_{11}|^2\,,
\label{enerpersecinfall}
\end{equation}
We can perform the derivatives and integrate to get the total energy
radiated.  There is a slight problem though, where do we stop the
integration?  The expression for the energy diverges at $r=0$ but this
is no problem, as we know that as the particle approaches the horizon,
the radiation will be infinitely red-shifted. Moreover, the standard
picture \cite{quem} is that of a particle falling in, and in the last
stages being frozen near the horizon.  With this in mind we integrate
from $r=\infty$ to some point near the horizon, say $r=b\times r_+$,
where $r_+$ is the horizon radius and $b$ is some number larger than
unit, and we get
\begin{equation}
\Delta E=
A \frac{D(D-2)\pi}{2^{2D-4}} \times b^{(9-D^2)/2} \times \frac{m^2}{M}\,,
\label{totalenergyinfall}
\end{equation}
where
\begin{equation}
A=\frac{(3-D)^2(5-D)^2(7-3D)^2(8-4D)^2(9-5D)^2...
(D/2+4-D^2/2)^2}{\Gamma[(D-1)/2]^2(D-1)(D+1)(D+3)}
\label{A}
\end{equation}
To understand the effect of both the dimension number $D$ and the
parameter $b$ on the total energy radiated according to the quadrupole
formula, we list in Table \ref{tab:zzz} some values $\Delta E$ for different
dimensions, and $b$ between $1$ and $1.3$.

\vskip 1mm
\begin{table}
\caption{\label{tab:zzz} The energy radiated by a particle falling
from rest into a higher dimensional Schwarzschild black hole, as a
function of dimension.  The integration is stopped at $b\times r_+$
where $r_+$ is the horizon radius.}
\begin{tabular}{llll}  \hline
\multicolumn{1}{c}{} &
\multicolumn{3}{c}{ $\Delta E \times \frac{M}{m^2}$}\\ \hline
$D$ & $b=1$:  &     $b=1.2$:   & $b=1.3$:\\ \hline
4   &  0.019  &  0.01 &  0.0076 \\ \hline 
6   &  0.576  &  0.05 &  0.0167 \\ \hline 
8   &  180    &  1.19 &  0.13   \\ \hline 
10  &  24567  &  6.13 &  0.16   \\ \hline 
12  &$3.3\times 10^6$ & 14.77 & 0.0665 \\ \hline 
\end{tabular}
\end{table}
\vskip 1mm
The parameter $b$ is in fact a measure of our ignorance of what goes
on near the black hole horizon, so if the energy radiated doesn't vary
much with $b$ it means that our lack of knowledge doesn't affect the
results very much.  For $D=4$ that happens indeed. Putting $b=1$ gives
only an energy $2.6$ times larger than with $b=1.3$, and still very
close to the fully relativistic numerical result of $0.01
\frac{m^2}{M}$.  However as we increase $D$, the effect of $b$
increases dramatically. For $D=12$ for example, we can see that a
change in $b$ from $1$ to $1.3$ gives a corresponding change in
$\Delta E$ of $3\times 10^6$ to $0.0665$. This is $8$ orders of
magnitude lower!  Since there is as yet no 
Regge-Wheeler-Zerilli \cite{zerilli,regge} treatment of this process for higher
dimensional Schwarzschild black holes, there are no fully
relativistic numerical results to compare our results with. 
Thus $D=4$ is just the perfect dimension to predict, through the
quadrupole formula, the gravitational energy coming from collisions
between particles and black holes, or between small and massive black
holes.  It is not a problem related to the quadrupole formalism, but
rather one related to $D$. A small change in parameters translates
itself, for high $D$, in a large variation in the final result. Thus,
as the dimension $D$ grows, the knowledge of the cutoff radius
$b\times r_+$ becomes essential to compute accurately the energy
released.
We note however that a decomposition of gravitational perturbations
in higher dimensions has recently become available, with the work of
Kodama and Ishibashi \cite{kodama}. We hope their work allows the computation
in a perturbation framework of the gravitational radiation generated by the infall
of particles into higher dimensional Schwarzschild black holes.

\section{Instantaneous collisions in even $D$-dimensions}
In general, whenever two bodies collide or scatter there will be
gravitational energy released due to the changes in momentum involved
in the process. If the collision is hard meaning that the incoming
and outgoing trajectories have constant velocities, there is a method
first envisaged by Weinberg \cite{weinberg,wein1}, later explored
in \cite{smarr2} by Smarr to compute exactly the metric perturbation and energy
released.The method is valid for arbitrary velocities 
(one will still be working in the linear approximation, so energies have to be
low). Basically, it assumes a collision lasting for zero
seconds. It was found that in this case the resulting spectra were flat,
precisely what one would expect based on one's experience with
electromagnetism \cite{jackson}, and so to give a meaning to the total
energy, a cutoff frequency is needed. This cutoff frequency depends
upon some physical cutoff in the particular problem.
We shall now generalize this construction for arbitrary dimensions.
\subsection{Derivation of the radiation formula 
in terms of a cutoff for a head-on collision}
\label{HeadonCollision}
Consider therefore a system of freely moving particles with
$D$-momenta $P_{i}^{\mu}$, energies $E_i$ and ($D-1$)-velocities
${\bf v}$, which due to the collision change abruptly at $t=0$, to
corresponding primed quantities. For such a system, the
energy-momentum tensor is
\begin{equation}
T^{\mu\nu}(t, {\bf v})=  
\sum \frac{P_{i}^{\mu}P_{i}^{\nu}}{E_i} 
\delta^{D-1}({\bf x}-{\bf v}t)\Theta(-t)+
\frac{{P'}_{i}^{\mu}{P'}_{i}^{\nu}}{E'_i} 
\delta^{D-1}({\bf x'}-{\bf v'}t)\Theta(t)\,,
\label{enmomtenpointpctles}
\end{equation}
from which, using eqs. (\ref{retsolwavezonefourier}) and
(\ref{powerwavezone}) one can get the quantities $h_{\mu\nu}$ and also
the radiation emitted. Let us consider the particular case in which
one has a head-on collision of two particles, particle $1$ with mass
$m_1$ and Lorentz factor $\gamma_1$, and particle $2$ with mass
$m=m_2$ with Lorentz factor $\gamma_2$, colliding to form a particle
at rest.  Without loss of generality, one may orient the axis so that
the motion is in the $(x_{D-1},x_D)$ plane, and the $x_D$ axis is the
radiation direction (see Appendix \ref{apendiceradddim}). We then have
\begin{eqnarray}
P_{1}=
\gamma_1m_1 (1,0,0,...,v_1\sin\theta_1,v_1\cos\theta_1)\,\,;\,\,\,\,\,
P'_1=(E'_{1},0,0,...,0,0)
\label{momenta1}
\\
P_{2}=
\gamma_2m_2 (1,0,0,...,-v_2\sin\theta_1,-v_2\cos\theta_1)\,\,;\,\,\,\,\,
P'_2=(E'_{2},0,0,...,0,0).
\label{momenta2}
\end{eqnarray}
Momentum conservation leads to the additional relation
$\gamma_1m_1v_1=\gamma_2m_2v_2$.  Replacing (\ref{momenta1}) and
(\ref{momenta2}) in the energy-momentum tensor
(\ref{enmomtenpointpctles}) and using (\ref{powerwavezone}) we find
\begin{equation}
\frac{d^2E}{d\omega d\Omega}=\frac{2{\cal G}}{(2\pi)^{D-2}}
\frac{D-3}{D-2}\frac{\gamma_{1}^2m_{1}^2v_{1}^2(v_1+v_2)^2
\sin{\theta_1}^4}{(1-v_1\cos\theta_1)^2(1+v_2\cos\theta_1)^2}\times \omega^{D-4}\,.
\label{energypersolidanglefreqinstcol}
\end{equation}
We see that the for arbitrary (even) $D$ the spectrum is not
flat. Flatness happens only for $D=4$. For any $D$ the total energy,
integrated over all frequencies would diverge so one needs a cutoff
frequency which shall depend on the particular problem under
consideration.  Integrating (\ref{energypersolidanglefreqinstcol})
from $\omega=0$ to the cutoff frequency $\omega_c$ we have
\begin{equation}
\frac{dE}{d\Omega}=\frac{2{\cal G}}{(2\pi)^{D-2}}
\frac{1}{D-2}\frac{\gamma_{1}^2m_{1}^2v_{1}^2(v_1+v_2)^2
\sin{\theta_1}^4}{(1-v_1\cos\theta_1)^2(1+v_2\cos\theta_1)^2}\times \omega_{c}^{D-3}\,.
\label{energypersolidangleinstcol}
\end{equation}
Two limiting cases are of interest here, namely (i) the collision
between identical particles and (ii) the collision between a light
particle and a very massive one.  In case (i) replacing $m_1=m_2=m$,
$v_1=v_2=v$, eq. (\ref{energypersolidangleinstcol}) gives
\begin{equation}
\frac{dE}{d\Omega}=\frac{8{\cal G}}{(2\pi)^{D-2}}
\frac{1}{D-2}\frac{\gamma^2m^2v^4
\sin{\theta_1}^4}{(1-v^2\cos^2\theta_1)^2}\times \omega_{c}^{D-3}\,.
\label{energypersolidangleinstcolident}
\end{equation}
In case (ii) considering $m_1\gamma_1 \equiv m\gamma <<m_2\gamma_2$, $v_1\equiv v>>v_2$,
eq. (\ref{energypersolidangleinstcol}) yields
\begin{equation}
\frac{dE}{d\Omega}=\frac{2{\cal G}}{(2\pi)^{D-2}}
\frac{1}{D-2}\frac{\gamma^2m^2v^4
\sin{\theta_1}^4}{(1-v\cos\theta_1)^2}\times \omega_{c}^{D-3}\,.
\label{energypersolidangleinstcoliheavylight}
\end{equation}
Notice that the technique just described is expected to break down if
the velocities involved are very low, since then the collision would not
be instantaneous. In fact a condition for this method to work would can be stated

 Indeed, one can see from eq.
(\ref{energypersolidangleinstcol}) that if $v\rightarrow 0$,
$\frac{dE}{d\omega}\rightarrow 0$, even though we know (see Subsection
(\ref{HeadonCollision})) that $\Delta E \neq 0$. In any case, if the
velocities are small one can use the quadrupole formula instead.

\subsection{Applications:
the cutoff frequency when one of the particles is
a black hole and radiation from black hole pair creation}

\subsubsection{The cutoff frequency when one of the head-on colliding 
particles is a black hole}

\begin{figure}
\centerline{\includegraphics[width=10 cm,height=6.5 cm]
{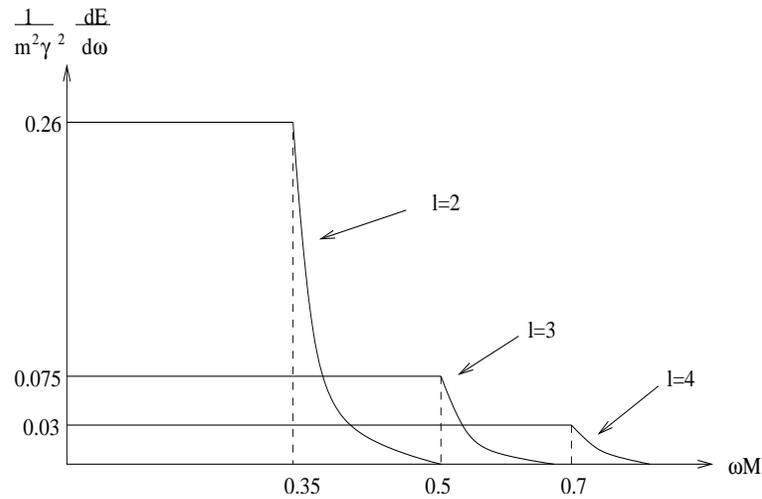}}
\caption{The energy spectra as a function of the angular number $l$,
for a highly relativistic particle falling into a $D=4$ Schwarzschild
black hole \cite{cardosorads}. The particle begins to fall with a
Lorentz factor $\gamma$. Notice that for each $l$ there is a cutoff
frequency $\omega_{c_l}$ which is equal to the quasinormal frequency
$\omega_{\rm QN}$ after which the spectrum decays exponentially. So it
is clearly seen that $\omega_{\rm QN}$ works as a cutoff
frequency. The total energy radiated is a given by a sum over $l$,
which is the same as saying that the effective cutoff frequency is
given by a weighted average of the various $\omega_{c_l}$.}
\label{fig:1}
\end{figure}

We shall now restrict ourselves to the case (ii) of last subsection, 
in which at least one of the particles participating in the collision is a 
massive black hole, with mass $M>>m$ (where we have put $m_1=m$ and $m_2=M$). 
 Formulas
(\ref{energypersolidangleinstcol})-
(\ref{energypersolidangleinstcoliheavylight})
are useless unless one is able to determine the cutoff frequency
$\omega_c$ present in the particular problem under consideration.  In
the situation where one has a small particle colliding at high
velocities with a black hole, it has been suggested by Smarr
\cite{smarr2} that the cutoff frequency should be $\omega_c \sim 1/2M$,
presumably because the characteristic collision time is dictated by
the large black hole whose radius is $2M$.  Using this cutoff he finds
\begin{equation}
\Delta E_{\rm Smarr}\sim 0.2 \gamma^2 \frac{m^2}{M}. 
\label{smarr}
\end{equation}
The exact result, using a relativistic perturbation approach which
reduces to the numerical integration of a second order differential
equation (the Zerilli wavefunction), has been given by Cardoso and
Lemos \cite{cardosorads}, as
\begin{equation}
\Delta E_{\rm exact} = 0.26 \gamma^2 \frac{m^2}{M}. 
\label{exact}
\end{equation}
This is equivalent to saying that $\omega_c =\frac{0.613}{M} \sim \frac{1}{1.63
M}$, and so it looks like the cutoff is indeed the inverse of the
horizon radius.  However, in the numerical work by Cardoso and Lemos,
it was found that it was not the presence of an horizon that
contributed to this cutoff, but the presence of a potential barrier $V$
outside the horizon. By decomposing the field in tensorial spherical
harmonics with index $l$ standing for the angular quantum number, we
found that for each $l$, the spectrum is indeed flat (as predicted by
eq.  (\ref{energypersolidangleinstcol}) for $D=4$), until a cutoff
frequency $\omega_{c_l}$ which was numerically equal to the lowest
gravitational quasinormal frequency $\omega_{\rm QN}$. For $\omega>
\omega_{c_l}$ the spectrum decays exponentially. This behavior is
illustrated in Fig. \ref{fig:1}.  
The quasinormal frequencies \cite{kokkotas} are
those frequencies that correspond to only outgoing waves at infinity
and only ingoing waves near the horizon.  As such the gravitational
quasinormal frequencies will in general have a real and an imaginary
part, the latter denoting gravitational wave emission and therefore a
decay in the perturbation.  There have been a wealth of works dwelling
on quasinormal modes on asymptotically flat spacetimes
\cite{kokkotas}, due to its close connection with gravitational wave
emission, and also on non-asymptotically flat spacetimes, like
asymptotically anti-de Sitter \cite{qnmads} or asymptotically de
Sitter \cite{mellor,moss,cardosoqnmds,molina,brinkds,suneeta,zhidenko,shijunds} spacetimes, mainly due to the AdS/CFT and dS/CFT
\cite{maldacena} correspondence conjecture.
We argue here that it is indeed the quasinormal frequency that
dictates the cutoff, and not the horizon radius.  For $D=4$ it so
happens that the weighted average of $\omega_{c_l}$ is
$\frac{0.613}{M}$, which, as we said, is quite similar to
$r_+=\frac{1}{2M}$.  The reason for the cutoff being dictated by the
quasinormal frequency can be understood using some WKB
intuition. The presence of a potential barrier outside the horizon
means that waves with some frequencies get reflected back on the
barrier while others can cross. Frequencies such that $\omega^2$ is
lower than the maximum barrier height $V_{\rm max}$ will be reflected
back to infinity where they will be detected. However, frequencies
$\omega^2$ larger than the maximum barrier height cross the barrier
and enter the black hole, thereby being absorbed and not contributing
to the energy detected at infinity.  So only frequencies $\omega^2$
lower than this maximum barrier height are detected at infinity.  It
has been shown \cite{schutz} that the gravitational quasinormal
frequencies are to first order equal to the square root of the maximum
barrier height.  In view of this picture, and considering the physical
meaning of the cutoff frequency, it seems quite natural to say that
the cutoff frequency is equal to the quasinormal frequency. If the
frequencies are higher than the barrier height, they don't get
reflected back to infinity.
This discussion is very important to understand how the total energy
varies with the number $D$ of dimensions. In fact, if we set $\omega_c
\sim \frac{1}{r_+}$, we find that the total energy radiated decreases
rapidly with the dimension number, because $r_+$ increases rapidly
with the dimension. This conflicts with recent results
\cite{eardley,yoshino}, which using shock waves that describe boosted
Schwarzschild black holes, and searching for apparent horizons,
indicate an increase with $D$.
So, we need the gravitational quasinormal frequencies for higher
dimensional Schwarzschild black holes. 
To arrive at an wave equation for gravitational perturbations of
higher dimensional Schwarzschild black holes, and therefore to compute
its gravitational quasinormal frequencies, one needs to decompose
Einstein's equations in D-dimensional tensorial harmonics, which would
lead to some quite complex expressions (this has recently been carried
through in \cite{kodama}).  It is not necessary to go
that far though, because one can get an idea of how the gravitational
quasinormal frequencies vary by searching for the quasinormal
frequencies of scalar perturbations, and scalar quasinormal
frequencies are a lot easier to find.  One hopes that the scalar
frequencies will behave with $D$ in the same manner as do the
gravitational ones.  Scalar perturbations in $D$-dimensional
Schwarzschild spacetimes obey the wave equation 
(consult \cite{cardosoradadsddim} for details)

\begin{equation}
\frac{\partial^{2} \phi(\omega,r)}{\partial r_*^{2}} +
\left\lbrack\omega^2-V(r)\right\rbrack
\phi(\omega,r)=0 \,.
\label{scalwavequat}
\end{equation}
The potential $V(r)$ appearing in equation (\ref{scalwavequat}) is given by
\begin{equation}
V(r)=
f(r)\left\lbrack\frac{a}{r^2}+
\frac{(D-2)(D-4)f(r)}{4r^2}+\frac{(D-2)f'(r)}{2r}\right\rbrack \,,
\label{potential}
\end{equation}
where $a=l(l+D-3)$ is the eigenvalue of the Laplacian on the
hypersphere $S^{D-2}$, the tortoise coordinate $r_*$ is defined as
$\frac{\partial r}{\partial r_*}=f(r)=\left(1-\frac{16\pi {\cal G}
M}{(D-2)\Omega_{D-2}}\frac{1}{r^{D-3}}\right)$, and $f'(r)=\frac{df(r)}{dr}$.
We have found the quasinormal frequencies of spherically symmetric
($l=0$) scalar perturbations, by using a WKB approach developed by
Schutz, Will and collaborators \cite{schutz,will}.  The results are
presented in Table \ref{tab:zfl}, where we also show the maximum barrier height of
the potential in eq. (\ref{potential}), as well as the horizon radius.
(A more complete treatment using a novel sixth order WKB approach has been
given by Konoplya \cite{konoplyawkb}).
\vskip 1mm
\begin{table}
\caption{\label{tab:zfl} The lowest scalar quasinormal frequencies for
spherically symmetric ($l=0$) scalar perturbations of higher dimensional
Schwarzschild black holes, obtained using a WKB method \cite{schutz,will}.
Notice that the real part of the quasinormal frequency is always the
same order of magnitude as the square root of the maximum barrier height. 
We show also the maximum barrier height as well as the horizon radius as a
function of dimension $D$. The mass $M$ of the black hole has been set
to 1.}
\begin{tabular}{lllll}  \hline
$D$&${\rm Re}[\omega_{QN}]$:&${\rm Im}[\omega_{QN}]:$&$\sqrt{\rm V_{\rm max}}$:&$1/r_+$:\\ \hline
4   &  0.10      & -0.12         &  0.16 &  0.5 \\ \hline 
6   &  1.033     &  -0.713      &  1.441   &  1.28  \\ \hline 
8   &  1.969      &  -1.023       &  2.637      &  1.32   \\ \hline 
10  &  2.779     &   -1.158     &  3.64   &  1.25   \\ \hline 
12  &  3.49     &    -1.202     &  4.503   &  1.17 \\ \hline 
\end{tabular}
\end{table}
\vskip 1mm
The first thing worth noticing is that the real part of the scalar
quasinormal frequency is to first order reasonably close to the square root
of the maximum barrier height $\sqrt{V_{\rm max}}$, supporting the
previous discussion.  Furthermore, the scalar quasinormal frequency
grows more rapidly than the inverse of the horizon radius
$\frac{1}{r_+}$ as one increases $D$. In fact, the scalar quasinormal
frequency grows with $D$ while the horizon radius $r_+$ gets smaller.
Note that from pure dimensional arguments, for fixed $D$,
$\omega \propto \frac{1}{r_+}$.
The statement here is that the constant of proportionality depends on
the dimension $D$, more explicitly it grows with $D$,  and can be found 
from Table \ref{tab:zfl}.
Assuming that the gravitational quasinormal frequencies will have the
same behavior (and some very recent studies \cite{konoplyawkb} seem to point that way
), the total energy radiated will during high-energy
collisions does indeed increase with $D$, as some studies
\cite{eardley,yoshino} seem to indicate.
\subsubsection{The gravitational 
energy radiated during black hole pair creation}
As a new application of this instantaneous collision formalism, we
will now consider the gravitational energy released during the quantum
creation of pairs of black holes, a process which as far as we know
has not been analyzed in the context of gravitational wave emission,
even for $D=4$.  It is well known that vacuum quantum fluctuations
produce virtual electron-positron pairs. These pairs can become real
\cite{schwinger} if they are pulled apart by an external electric
field, in which case the energy for the pair materialization and
acceleration comes from the external electric field energy.  Likewise,
a black hole pair can be created in the presence of an external field
whenever the energy pumped into the system is enough in order to make
the pair of virtual black holes real (see Dias
\cite{diaslemos2} for a review on black hole pair creation). If one
tries to predict the spectrum of radiation coming from pair creation,
one expects of course a spectrum characteristic of accelerated masses
but one also expects that this follows some kind of signal indicating
pair creation. In other words, the process of pair creation itself,
which involves the sudden creation of particles, must imply
emission of radiation. It is this phase we shall focus on, forgetting
the subsequent emission of radiation caused by the acceleration.

Pair creation is a pure quantum-mechanical process in nature, with no
classical explanation. But given that the process does occur, one may
ask about the spectrum and intensity of the radiation accompanying
it. The sudden creation of pairs can be viewed for our purposes as an
instantaneous creation of particles (i.e., the time reverse
process of instantaneous collisions), the violent acceleration of
particles initially at rest to some final velocity in a very short
time, and the technique described at the beginning of this section
applies. This is quite similar to another pure quantum-mechanical
process, the beta decay.  
The electromagnetic radiation emitted during beta decay has been
computed classically by Chang and Falkoff \cite{chang} and is also
presented in Jackson \cite{jackson}.  The classical calculation is
similar in all aspects to the one described in this section (the
instantaneous collision formalism) assuming the sudden
acceleration to energies $E$ of a charge initially at rest, and
requires also a cutoff in the frequency, which has been assumed to be
given by the uncertainty principle $\omega_c \sim \frac{E}{\hbar}$.
Assuming this cutoff one finds that the agreement between the
classical calculation and the quantum calculation \cite{chang} is
extremely good (specially in the low frequency regime), and more
important, was verified experimentally.
Summarizing, formula (\ref{energypersolidangleinstcolident}) also describes
the gravitational energy radiated when two black holes, each with mass
$m$ and energy $E$ form through quantum pair creation.  The typical
pair creation time can be estimated by the uncertainty principle
$\tau_{\rm creation}\sim \hbar/E \sim \frac{\hbar}{m\gamma}$, and thus
we find the cutoff frequency as 
\begin{equation}
\omega_c\sim \frac{1}{\tau_{\rm creation}}\sim \frac{m\gamma}{\hbar}\,.  
\label{cutofffrequency}
\end{equation}
Here we would
like to draw the reader's attention to the fact that the units of
Planck's constant $\hbar$ change with dimension number $D$: according
to our convention of setting $c=1$ the units of $\hbar$ are
$[M]^{\frac{D-2}{D-3}}$.  With this cutoff, we find the spectrum of the
gravitational radiation emitted during pair creation to be given by
(\ref{energypersolidanglefreqinstcol}) with $m_1=m_2$ and $v_1=v_2$
(we are considering the pair creation of two identical black holes):
\begin{equation}
\frac{d^2E}{d\omega d\Omega}=\frac{8{\cal G}}{(2\pi)^{D-2}}
\frac{D-3}{D-2}\frac{\gamma^2m^2v^4
\sin{\theta_1}^4}{(1-v^2\cos^2\theta_1)^2}\times \omega^{D-4}\,,
\label{specpaircreat}
\end{equation}
and the total frequency integrated energy per solid angle is
\begin{equation}
\frac{dE}{d\Omega}=\frac{8{\cal G}}{(2\pi)^{D-2}(D-2)}\frac{v^4
\sin{\theta_1}^4}{(1-v^2\cos^2\theta_1)^2}\times \frac{(m \gamma)^{D-1}}{\hbar^{D-3}}\,.
\label{specpaircreatintfreq}
\end{equation}
For example, in four dimensions and for pairs with $v \sim 1$ one
obtains 
\begin{equation}
\frac{dE}{d\omega} =\frac{4{\cal G}}{\pi} \gamma^2m^2 \,,
\label{specpair4d}
\end{equation}
and will have for the total energy radiated during production itself,
using the cutoff frequency (\ref{cutofffrequency})
\begin{equation}
\Delta E =\frac{4{\cal G}}{\pi} 
\frac{\gamma^3m^3}{\hbar} \,.
\label{totenergpair4d}
\end{equation}
This could lead, under appropriate numbers of $m$ and $\gamma$ to huge
quantities. Although one cannot be sure as to the cutoff frequency,
and therefore the total energy (\ref{totenergpair4d}), it is extremely
likely that, as was verified experimentally in beta decay, the zero
frequency limit, eq. (\ref{specpair4d}), is exact.

\subsection{Spherical coordinates in ($D-1$)-dimensions}
\label{apendiceradddim}
In this appendix we list some important formulas and results used
throughout this paper.  We shall first present the transformation
mapping a ($D-1$) cartesian coordinates, ($x_1,x_2,x_3,...,x_{D-1}$)
onto ($D-1$) spherical coordinates,
($r,\theta_1,\theta_2,...,\theta_{D-2}$).
The transformation reads
\begin{eqnarray*}
x_1=r\sin\theta_1\sin\theta_2...\sin\theta_{D-2}\\
x_2=r\sin\theta_1\sin\theta_2...\sin\theta_{D-3}\cos\theta_{D-2}\\
:\\
x_i=r\sin\theta_1\sin\theta_2...\sin\theta_{D-i-1}\cos\theta_{D-i}\\
:\\
x_{D-1}=r\cos\theta_1
\label{transformation}
\end{eqnarray*}
The Jacobian of this transformation is 
\begin{equation}
J=r^{D-2}\sin\theta_1^{D-3}
\sin\theta_2^{D-4}...\sin\theta_i^{D-i-2}...\sin\theta_{D-3}\,,
\label{jacobian}
\end{equation}
and the volume element becomes
\begin{equation}
d^{D-1}{\bf x}= 
J dr d\theta_1 d\theta_2...d\theta_{D-2}=r^{D-2}dr d\Omega_{D-2}\,,
\label{volumeelement}
\end{equation}
where 
\begin{equation} 
d\Omega_{D-2}=\sin\theta_1^{D-3}\sin\theta_2^{D-4}...
\sin\theta_{D-3}d\theta_1d\theta_2...d\theta_{D-2}.
\label{solidangle}
\end{equation}
is the element of the ($D-1$) dimensional solid angle.
Finally, using \cite{grad}
\begin{equation}
\int_{0}^{\pi} \sin\theta^n=\sqrt{\pi} \frac{\Gamma[(n+1)/2]}{\Gamma[(n+2)/2]}\,,
\label{sinintegral}
\end{equation}
this yields
\begin{equation}
\Omega_{D-2}=\frac{2\pi^{(D-1)/2}}{\Gamma[(D-1)/2]}.
\label{integratedsolidangle}
\end{equation}
Here, $\Gamma[z]$ is the Gamma function, whose definition and
properties are listed in \cite{stegun}. In this work the main
properties of the Gamma function which were used are
$\Gamma[z+1]=z\Gamma[z]$ and $\Gamma[1/2]=\sqrt{\pi}$.
\section{Summary and discussion}
We have developed the formalism to compute gravitational wave
generation in higher $D$ dimensional spacetimes, with $D$
even. Several examples have been worked out, and one cannot help the
feeling that our apparently four dimensional world is the best one to
make predictions about the intensity of gravitational waves in
concrete situations, in the sense that a small variation of parameters
leads in high $D$ to a huge variation of the energy radiated.  A lot
more work is still needed if one wants to make precise predictions
about gravitational wave generation in $D$ dimensional spacetimes.
Since a decomposition of gravitational perturbations in higher
dimensional spacetimes has recently become available \cite{kodama} it
is now urgent to use this formalism to compute the gravitational
radiation generated in typical processes, in a perturbation manner as
described in the following chapter, i.e., assuming a curved background
(for example, the background of a higher dimensional Schwarzschild
black hole).  The results can then be compared with the estimates
presented here in a more complete fashion.
One of the examples worked out, the gravitational radiation
emitted during black hole pair creation, had not been previously
considered in the literature, and it seems to be a good candidate,
even in $D=4$, to radiate intensely through gravitational waves.

\thispagestyle{empty} \setcounter{minitocdepth}{1}
\chapter[Late-time tails in higher dimensional spacetimes]{Late-time tails in higher dimensional spacetimes} \label{chap:Tails}
\lhead[]{\fancyplain{}{\bfseries Chapter \thechapter. \leftmark}}
\rhead[\fancyplain{}{\bfseries \rightmark}]{}
\minitoc \thispagestyle{empty}
\renewcommand{\thepage}{\arabic{page}}

\section{Introduction}
It is an everyday life experience that light rays and waves in general
travel along a null cone. For example, if one lights a candle or a
lighter for five seconds and then turns it off, any observer (at rest
relative to the object) will see the light for exactly five seconds
and then suddenly fade out completely.  Mathematically this is due to
the well known fact that the flat space 4-dimensional Green's
function has a delta function character and therefore has support only on
the light cone.  There are however situations where this only-on-the
light cone propagation is lost. For instance, in a curved spacetime a
propagating wave leaves a ``tail'' behind, as shown by DeWitt and
Brehme's seminal work \cite{dewitt}.  This means that a pulse of
gravitational waves (or any massless field for that matter) travels
not only along the light cone but also spreads out behind it, and
slowly dies off in tails. Put it another way, even after the candle is
turned off in a curved spacetime, one will still see its shinning
light, slowly fading away, but never completely. This is due to
backscattering off the potential \cite{price} at very large spatial
distances.

The existence of late-time tails in black hole spacetimes is by now
well established, both analytically and numerically, in linearized
perturbations and even in a non-linear evolution, for massless or 
massive fields
\cite{price,price2,price3,leaver,ching1,ching2,tom,hod}.  This is a problem of
more than academic interest: one knows that a black hole radiates away
everything that it can, by the so called no hair theorem (see
\cite{bek} for a nice review), but how does this hair loss proceed
dynamically?
A more or less complete picture is now available. The study of a
fairly general class of initial data evolution shows that the signal
can roughly be divided in three parts:
(i) the first part is the prompt response, at very early times, and
the form depends strongly on the initial conditions. This is the most
intuitive phase, being the obvious counterpart of the light cone
propagation.
(ii) at intermediate times the signal is dominated by an exponentially
decaying ringing phase, and its form depends entirely on the black
hole characteristics, through its associated quasinormal modes
\cite{kokkotas,nollert,hod,dreyer,dreyerall,cardosoqnmbtz,cardosoqnmsads2,cardosoqnmds,qnmads,motl1,motl2}.
(iii) a late-time tail, usually a power law falloff of the field. This
power law seems to be highly independent of the initial data, and
seems to persist even if there is no black hole horizon. In fact it
depends only on the asymptotic far region.  Mathematically each of
these stages has been associated as arising from different
contributions to the Green's function. The late-time tail is due to a
branch cut \cite{leaver}.  The study of linearized (we note that
non-linear numerical evolution also displays these tails, but here we
shall work at the linearized level) perturbations in the black hole
exterior can usually be reduced to the simple equation
\begin{equation}
[\partial_t^2-\partial_{x}^2+V(x)]\Psi=0\,,
\label{eveq}
\end{equation}
where the potential $V(x)$ depends on what kind of field one is
considering and also, of course, on the spacetime.  A detailed study of
the branch cut contribution by Ching, Leung, Suen and Young 
\cite{ching1,ching2} has
provided analytical results for some specific but quite broad class of
potentials.  These analytical results concerning the late time tails
were confirmed numerically.

It is not generally appreciated that there is another case in which
wave propagation develops tails: wave propagation in odd dimensional
{\it flat} spacetimes.  In fact, the Green's function in a
$D$-dimensional spacetime \cite{greend,vitoroscarjose,amj} have a
completely different structure depending on whether $D$ is even or
odd. For even $D$ it still has support only on the light cone, but for
odd $D$ the support of the Green's function extends to the interior of
the light cone, and leads to the appearance of tails. It is hard to
find good literature on this subject, but a complete and pedagogical
discussion of tails in flat $D$-dimensional backgrounds can be found
in \cite{amj}.

A study of wave physics in higher dimensions is now, more than ever, needed.
It seems impossible to
formulate in four dimensions a consistent theory which unifies gravity
with the other forces in nature. Thus, most efforts in this direction
have considered a higher dimensional arena for our universe, one
example being string theories which have recently made some remarkable
achievements. Moreover, recent investigations \cite{hamed} propose the
existence of extra dimensions in our Universe in order to solve the
hierarchy problem, i.e., the huge difference between the electroweak
and the Planck scale, $m_{\rm EW}/M_{\rm Pl}\sim 10^{-17}$. 
The fields of standard model would inhabit a 4-dimensional
sub-manifold, the brane, whereas the gravitational degrees of freedom
would propagate throughout all dimensions. 

A first step towards the understanding of gravitational wave physics
in higher dimensions was given by Cardoso, Dias and Lemos
\cite{vitoroscarjose}, by studying wave generation and propagation
in generic $D$-dimensional flat spacetimes.  Here we shall take a step
further, by studying wave tails in higher dimensional black hole
spacetimes. We will restrict the analysis to higher dimensional 
Schwarzschild black holes. As expected, if one now considers tails in higher
dimensional black hole spacetimes two aspects should emerge: in odd
dimensional spacetimes one expects the black hole contribution to the
tail to be smaller than that of the background itself. Therefore for
odd dimensions the tail should basically be due to the flat space
Green's function.  However, for even $D$-dimensional black hole
spacetimes there is no background contribution, and one expects to see
only the black hole contribution to the tail.
A recent study by Barvinsky and Solodukhin \cite{barv} has showed that
such tails may not be impossible to detect. Unfortunately, the weakness 
of gravitational waves impinging on Earth, make this an unlikely event. 
We note however that they worked with small length, compact extra-dimensions, 
whereas we shall consider large extra dimensions. Our results will be strictly
correct if the extra dimensions are infinite, but also allow us to determine
the correct answer if the large extra dimensions are large enough that
the timescale for wave reflection at the boundaries is larger than the timescales
at which the tail begins to dominate.
We shall follow closely the analysis in \cite{cardosotails}
   
The evolution problem in a $D$-dimensional Schwarzschild background
can be cast in the form (\ref{eveq}), and we will show also that the
potential can be worked out in such a way as to belong to the class of
potentials studied in \cite{ching1,ching2}.  Therefore,
their analytical results carry over to the $D$-dimensional Schwarzschild
black holes as well. We will verify this by a direct numerical
evolution.  The main results are:
the late-time behavior is dominated by a tail, and this is a
power-law falloff.  For odd dimensions the power-law is determined not
by the presence of the black hole, but by the fact that the spacetime
is odd dimensional.  In this case the field decays as $\Psi \sim
t^{-(2l+D-2)}$, where $l$ is the angular index determining the angular
dependence of the field.  This is one of the most interesting results
obtained here. One can show directly from the flat space Green's
function that such a power-law is indeed expected in flat odd
dimensional spacetimes, and it is amusing to note that the same
conclusion can be reached directly from the analysis of \cite{ching1,ching2}.
For even dimensional spacetimes we find also a power-law decay at late
times, but with a much more rapid decay, $\Psi \sim t^{-(2l+3D-8)}$.
For even $D$, this power-law tail is entirely due to the black hole,
as opposed to the situation in odd $D$.  These results are strictly
valid for $D>4$. Four dimensional Schwarzschild geometry is special,
having the well known power-law tail $\Psi \sim t^{-(2l+3)}$.
\section{A brief summary of previous analytical results 
for a specific class of potentials}
In a very complete analysis, Ching, Leung, Suen and Young 
\cite{ching1,ching2} have studied
the late-time tails appearing when one deals with evolution equations 
of the form
(\ref{eveq}), and the potential $V$ is of the form
\begin{equation}
V(x) \sim \frac{\nu(\nu+1)}{x^2}+
\frac{c_1\log{x}+c_2}{x^{\alpha}}\,\,\,\,,x\rightarrow \infty.
\label{potching}
\end{equation}
By a careful study of the branch cut contribution to the associated
Green's function they concluded that in general the late-time behavior
is dictated by a power-law or by a power-law times a logarithm, and
the exponents of the power-law depend on the leading term at very
large spatial distances.  The case of interest for us here, as we
shall verify in the following section, is when $c_1=0$. Their
conclusions, which we will therefore restrict to the $c_1=0$ case, are
(see Table 1 in \cite{ching1} or \cite{ching2}): \newline (i) if $\nu$ is an
integer the term $\frac{\nu(\nu+1)}{x^2}$ does not contribute to the
late-time tail. We note this term represents just the pure centrifugal
barrier, characteristic of flat space, so one can expect that indeed
it does not contribute, at least in four-dimensional spacetime. We
also note that since even dimensional spacetimes have on-light cone
propagation, one may expect to reduce the evolution equation to a form
containing the term $\frac{\nu(\nu+1)}{x^2}$ with $\nu$ an integer. We
shall find this is indeed the case.  Therefore, for integer $\nu$, it
is the $\frac{c_2}{x^{\alpha}}$ term that contributes to the late-time
tail. In this case, the authors of \cite{ching1,ching2} find that the 
tail is given by a
power-law,
\begin{equation}
\Psi \sim t^{-\mu}\,\,,\,\,\mu>2\nu+\alpha\,\,,\,\, \alpha\, 
{\rm odd}\,\,{\rm integer}<2\nu+3.
\label{tailintnu1}
\end{equation}
For this case ($\alpha$ an odd integer smaller than $2\nu +3$) the
exponent $\mu$ was not determined analytically. However, they argue both
analytically and numerically, that
$\mu=2\nu+2\alpha-2$.  For all other real $\alpha$, the tail is
\begin{equation}
\Psi \sim t^{-(2\nu+\alpha)}\,\,,\,\,{\rm all}\,\,{\rm other}\,\, 
{\rm real}\,\,\alpha.
\label{tailintnu2}
\end{equation}
\newline (ii) if $\nu$ is not an integer, then the main contribution to 
the late-time tail
comes from the $\frac{\nu(\nu+1)}{x^2}$ term. In this case the tail is
\begin{equation}
\Psi \sim t^{-(2\nu+2)}\,\,,\,\,{\rm non-integer}\,\,\nu.
\label{tailnonintnu}
\end{equation}
We will now see that for a $D$-dimensional Schwarzschild geometry the
potential entering the evolution equations is asymptotically of the
form (\ref{potching}) and therefore the results
(\ref{tailintnu1})-(\ref{tailnonintnu}) can be used.

\section{The evolution equations and late-time tails in the 
$D$-dimensional Schwarzschild geometry}
Here, we shall consider the equations describing the evolution of
scalar, electromagnetic and gravitational weak fields outside the
$D$-dimensional Schwarzschild geometry. We shall then, based on the
results presented in the previous section, derive the late-time tails
form of the waves.  We will find they are always a power-law falloff.
\subsection{The evolution equations and the reduction of the potential 
to the standard form }
The metric of the $D$-dimensional
Schwarzschild black hole in ($t,r,\theta_1,\theta_2,..,\theta_{D-2}$)
coordinates is \cite{tangherlini}
\begin{equation}
ds^2= -fdt^2+
f^{-1}dr^2
+r^2d\Omega_{D-2}^2\,,
\label{metrictang} 
\end{equation}
with
\begin{equation}
f=1-\frac{M}{r^{D-3}}.
\label{fdef}
\end{equation}
The mass of the black hole is given by
$\frac{(D-2)\Omega_{D-2} M}{16\pi{\cal G}}$, where
$\Omega_{D-2}=\frac{2\pi^{(D-1)/2}}{\Gamma[(D-1)/2]}$ is the area of a
unit $(D-2)$ sphere, and $d\Omega_{D-2}^{2}$ is the line element on
the unit sphere $S^{D-2}$.  We will only consider the linearized
approximation, which means that we are considering wave fields outside
this geometry that are so weak they do not alter this
background. Technically this means that all covariant derivatives are
taken with respect to the background metric (\ref{metrictang}).  The
evolution equation for a massless scalar field follows directly from
the (relativistic) Klein-Gordon equation. After a separation of the
angular variables with Gegenbauer functions (see \cite{cardosoradadsddim} for
details) we get that the scalar field follows (\ref{eveq}) with a
potential
\begin{equation}
V_{\rm s}(r_*)=
f(r)\left\lbrack\frac{a}{r^2}+
\frac{(D-2)(D-4)f(r)}{4r^2}+\frac{(D-2)f'(r)}{2r}\right\rbrack \,,
\label{potentialscalar1}
\end{equation}
where $r$ is a function of the tortoise coordinate $r_*$ according to
$\frac{\partial r}{\partial
r_*}=f(r)$.
The constant $a=l(l+D-3)$ is the eigenvalue of the Laplacian on the
hypersphere $S^{D-2}$, and
$f'(r)=\frac{df(r)}{dr}$. $l$ can take any nonnegative integer value.
Of course the evolution equation is (\ref{eveq}) where the variable
$x$ is in this case the tortoise coordinate $r_*$.  This is the
standard form in which the potential is presented. However, one can
collect the different powers of $r$ and get
\begin{equation}
V_{\rm s}(r_*)=
f(r)\left\lbrack\frac{\nu (\nu+1)}{r^2}+
\frac{1}{r^{D-1}}\frac{(D-2)^2 M}{4}\right\rbrack \,,
\label{potentialscalar2}
\end{equation}
where
\begin{equation}
\nu=l-2+\frac{D}{2}.
\label{nudef}
\end{equation}
Asymptotically for large $r_*$ one can show that
\begin{equation}
V_{\rm s}(r_*) {\biggl |}_{r_* \rightarrow \infty}\!\!\!\!\!=\frac{\nu 
(\nu+1)}{r_{*}^2}+
\frac{1}{r_*^{D-1}}\frac{(D-2)Ml}{D-4}(3-l-D).
\label{potentialscalarasympt}
\end{equation}
This is strictly valid for $D>4$. In the $D=4$ case there is a logarithm
term \cite{price}.
Notice that the coefficient $\nu$ appearing in the centrifugal barrier
term $\frac{\nu (\nu+1)}{r_{*}^2}$ is, as promised, an integer for
even $D$, and a half-integer for odd $D$.  The gravitational evolution
equations have recently been derived by Kodama and Ishibashi
\cite{kodama}. There are three kinds of gravitational perturbations,
according to Kodama and Ishibashi's terminology: the scalar
gravitational, the vector gravitational and the tensor gravitational
perturbations.  The first two already have their counterparts in
$D=4$, which were first derived by Regge and Wheeler
\cite{regge} and by Zerilli \cite{zerilli}. The tensor type is a new
kind appearing in higher dimensions. However, it obeys exactly the
same equation as massless scalar fields, so the previous result
(\ref{potentialscalar2})-(\ref{potentialscalarasympt}) holds.  It can
be shown in fact that the scalar and vector type also obey the same
evolution equation with a potential that also has the form
(\ref{potentialscalarasympt}) with a slightly different coefficient
for the $1/r_*^{D-1}$ term. For example, for the vector type the
potential is
\begin{equation}
V_{\rm gv}(r_*)=
f(r)\left\lbrack\frac{\nu (\nu+1)}{r^2}-\frac{1}{r^{D-1}}
\frac{3(D-2)^2 M}{4}\right\rbrack \,,
\label{potentialgravvetor}
\end{equation}
where $\nu$ is defined in (\ref{nudef}).
Therefore asymptotically for large $r_*$,
\begin{eqnarray}
\!\!\!\!\!V_{\rm gv}(r_*){\biggl |}_{r_* \rightarrow \infty}\!\!\!\!\!&=&
\frac{\nu (\nu+1)}{r_{*}^2}-\frac{1}{r_*^{D-1}}\frac{(D-2)M}{D-4}\times
\nonumber \\
& & \left\lbrack8+D^2+D(l-6)+l(l-3)\right\rbrack.
\label{potentialgravvetorasympt}
\end{eqnarray}
which is of the same form as the scalar field potential.  The scalar
gravitational potential has a more complex form, but one can show that
asymptotically it has again the form (\ref{potentialscalarasympt}) or
(\ref{potentialgravvetorasympt}) with a different coefficient in the
$1/r_*^{D-1}$ term. Since the explicit form of this coefficient is not
important here, we shall not give it explicitly.  Electromagnetic
perturbations in higher dimensions were considered in \cite{higuchi}.
Again, asymptotically for large $r_*$ they can be reduced to the form
(\ref{potentialscalarasympt})(where again $\nu=l-2+\frac{D}{2}$ and
the $1/r_*^{D-1}$ coefficient is different) , so we shall not dwell on
them explicitly.

\subsection{Late-time tails}
Now that we have shown that the potentials appearing in evolution of
massless fields in the $D$-dimensional Schwarzschild geometry belong
to the class of potentials studied in \cite{ching1,ching2}
we can easily find the form of the late-time tail.  For odd
dimensional spacetimes, $\nu$ is not an integer, therefore the
centrifugal barrier gives the most important contribution to the
late-time tail.  According to (\ref{tailnonintnu}) we have that the
late-time tail is described by the power-law falloff
\begin{equation}
\Psi \sim t^{-(2l+D-2)}\,, {\rm odd}\, D.
\label{tailodd}
\end{equation}
According to the discussion in the introduction, this tail is
independent of the presence of the black hole, and should therefore
already appear in the flat space Green's function. Indeed it does
\cite{vitoroscarjose,amj}.  The flat, odd dimensional Green's
function has a tail term \cite{vitoroscarjose,amj} proportional to
$\frac{\Theta(t-r)}{(t^2-r^2)^{D/2-1}}$, where $\Theta$ is the
Heaviside step function. It is therefore immediate to conclude that,
for spherical perturbations, for example, the tail at very large times
should be $t^{-(D-2)}$ which is in agreement with (\ref{tailodd}) for
$l=0$ (which are the spherical perturbations).  It is amusing to note
that the analysis  of \cite{ching1,ching2} gives the correct behavior 
at once, simply
by looking at the centrifugal barrier!  
We have checked numerically
the result (\ref{tailodd}) for $D=5$, and the results are shown in
Figs. 1 and 2.

\begin{figure}
\centerline{\includegraphics[width=8 cm,height=8 cm]
{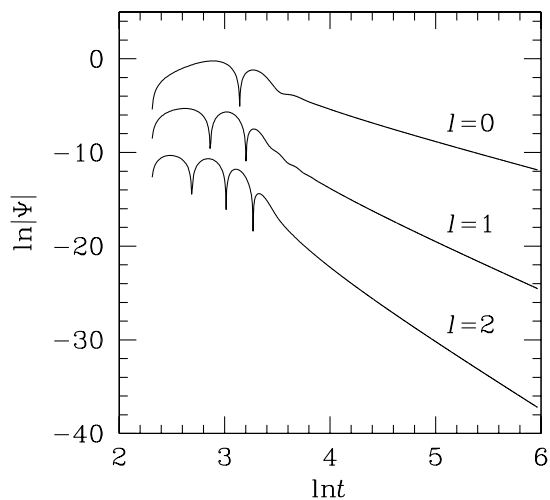}}
\caption{Generic time dependence of a scalar field $\Psi$ in a five
dimensional Schwarzschild geometry, at a fixed spatial position.  We
took as initial conditions a Gaussian wave packet with $\sigma=3$ and
$v_c=10$.  We have performed other numerical extractions for different
initial values.  The results for the late-time behavior are
independent of the initial data, as far as we can tell. For $l=0$ the
late-time behavior is a power-law with $\Psi \sim t^{-3.1}$, for
$l=1$, $\Psi \sim t^{-5.2}$ at late times and for $l=2$, $\Psi
\sim t^{-7.3}$. The predicted powers are $-3$, $-5$ and $-7$,
respectively.}
\label{fig:TailScalar5D}
\end{figure}

\begin{figure}
\centerline{\includegraphics[width=8 cm,height=8 cm]
{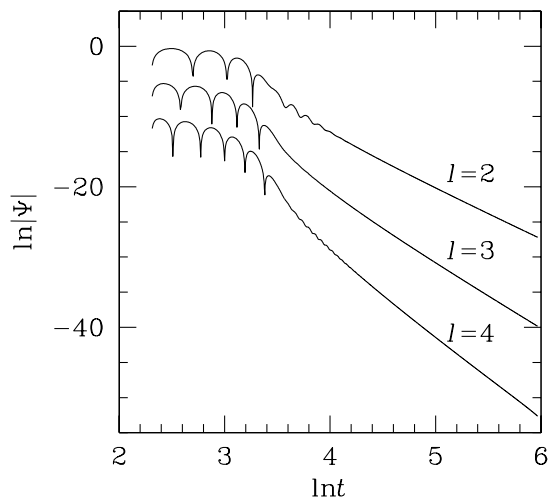}}
\caption{Generic time dependence of a gravitational Gaussian
wavepacket $\Psi$ in a five dimensional Schwarzschild geometry, at a
fixed spatial position.  The results for the late-time behavior are
independent of the initial data.  For $l=2$ the late-time behavior is
a power-law with $\Psi \sim t^{-7.1}$, for $l=3$ the falloff is
given by $\Psi \sim t^{-9.2}$ at late times, and for $l=4$ it is
$\Psi \sim t^{-11.4}$.  The predicted powers are
$-7$, $-9$ and $-11$, respectively.}
\label{fig:TailGrav5D}
\end{figure}
The numerical procedure followed the one outlined in \cite{price2},
with constant data on $v=v_0$, where $v$ is the advanced time coordinate,
$v=t+r_*$.
One final remark is in order here. When numerically evolving the
fields using the scheme in \cite{price2}, we have found that for $D>5$
the tail looked always like $t^{-(2\nu+4)}$. This is a fake behavior,
and as pointed out in \cite{ching2} (see in particular their Appendix
A) it is entirely due to the ghost potential, appearing for potentials
vanishing faster than $\frac{1}{r_*^4}$ for a second order scheme.
Technically the presence of ghost potentials can be detected by
changing the grid size \cite{ching2}. If the results with different grid
sizes are different, then the ghost potential is present.  Our
numerical results for $D=5$, presented in Figs. 1-3 are free from any
ghosts.

The numerical results are in excellent agreement with the analytical
predictions (\ref{tailodd}), and seems moreover to be quite
independent of the initial data. This also means that for odd
dimensional spacetimes the late time behavior is dictated not by the
black hole, but by the fact that spacetime has an odd number of
dimensions.  To further check that it is in fact the centrifugal
barrier term that is controlling the tail, we have performed numerical
evolutions with a five dimensional model potential.  To be concrete,
we have evolved a field subjected to the potential
(\ref{potentialscalar2}) (with $D=5$) but we have considered an
integer value for $\nu$, namely $\nu=1$.  The result is shown in
Fig. 3.
 Of course the true potential has a semi-integer value for $\nu$, but
this way one can verify the dependence of the tail on the centrifugal
term.  Indeed if $\nu=1$ then by (\ref{tailintnu2}), with $\nu=1$ and
$\alpha=4$, the late time tail should be $\Psi \sim t^{-6}$. The
agreement with the numerical evolution is great.  It is therefore the
centrifugal barrier that controls the tail in odd dimensional
spacetimes.
\begin{figure}
\centerline{\includegraphics[width=8 cm,height=8 cm]
{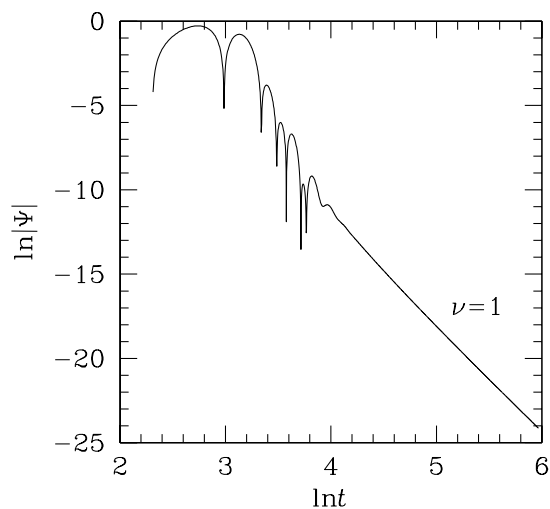}}
\caption{Evolution of a field subjected to a five-dimensional
model potential.  The potential is given by (\ref{potentialscalar2})
but here we take $\nu$ as an integer, $\nu=1$. We obtain at late times
$\Psi \sim t^{-6.1}$, whereas the analytical prediction is $t^{-6}$.
This shows nicely that for integer $\nu$ the centrifugal barrier
contribution vanishes, and it is the next term in the asymptotic
expansion of the potential that gives the most important contribution.
We have checked that this power law is indeed the correct one, and not
a numerical artifact of the ghost potential.
}
\label{fig:TailGrav5Dnu1}
\end{figure}
For even dimensional spacetimes, $\nu=l-2+\frac{D}{2}$ is an integer. 
Moreover, 
$\alpha=D-1<2\nu+3=2l+D-1$. Therefore we are in situation (\ref{tailintnu1}).
So the late time tail of wave propagation in an even $D$-dimensional 
Schwarzschild
spacetime is a power-law,
\begin{equation}
\Psi \sim t^{-(2l+3D-8)}\,, {\rm even}\, D.
\label{taileven}
\end{equation}

\section{Conclusions}
We have determined the late time behavior of massless fields (scalar,
electromagnetic and gravitational) outside a $D$-dimensional
Schwarzschild black hole. For odd $D$, the field at late times has a
power-law falloff, $\Psi \sim t^{-(2l+D-2)}$, and this tail is
independent of the presence of the black hole. It depends solely on
the flat spacetime background, through the properties of the flat
space odd dimensional Green's function.  For even $D$, the late time
behavior is again a power-law but this time it is due to the presence
of the black hole, and is given by $\Psi \sim t^{-(2l+3D-8)}$, at late
times for $D>4$.  We have focused on large extra dimensions
only. Recent investigations \cite{barv}, focusing on brane world
models, and therefore compact extra dimensions suggest that the tail
is more slowly damped if the extra dimensions are compact.  In fact,
for $D=5$ they obtain a power-law $\Psi \sim t^{-5/2}$ whereas we have
$\Psi \sim t^{-3}$ for spherically symmetric perturbations.  This may
be due to the reflection of the field at the boundaries of the extra
dimension.

\clearemptydoublepage
\newpage
\setcounter{page}{195}  
\pagestyle{fancyplain}
\thispagestyle{empty}
\lhead[]
      {\fancyplain{}{\bfseries Bibliography}}
\rhead[\fancyplain{}{\bfseries Bibliography}]
      {}
\nocite{*}
\thispagestyle{empty}
\addcontentsline{toc}{part}{\numberline{}Bibliography}
\bibliographystyle{acm}
\bibliography{BibThese}

\begin{thebibliography}{99}


\bibitem{vish} C. V. Vishveshwara, 
Nature {\bf 227}, 936 (1970).

\bibitem{davis} M. Davis, R. Ruffini, W. H. Press, and
R. H. Price, Phys. Rev. Lett.  {\bf 27}, 1466 (1971).

\bibitem{davis2} M. Davis, R. Ruffini, and J. Tiomno,
 Phys. Rev. D {\bf 5}, 2932 (1971);
 R. Gleiser, C. Nicasio, R. Price, J. Pullin
Phys. Rept. {\bf 325}, 41  (2000).

\bibitem{anninos} P. Anninos, D. Hobill, E. Seidel,
L. Smarr, and W. M. Suen, 
Phys. Rev. Lett. {\bf 71}, 2851(1993).  

\bibitem{gleiser} R. J. Gleiser, C. O. Nicasio, R. H. Price,
and J. Pullin, Phys. Rev. Lett. {\bf 77}, 4483 (1996). 

\bibitem{seidell} E. Seidel,
{\it Relativistic Astrophysics}, ed. H. Riffert et al
(Braunschweig: Vieweg).

\bibitem{smarrl} S. L. Smarr (ed.), 
{\it Sources of Gravitational Radiation},
(Cambridge University Press, 1979). 

\bibitem{press} W. H. Press, 
Astrophys. J. {\bf 170}, L105 (1971).

\bibitem{kokkotas} K.D. Kokkotas, B.G. Schmidt,
Living Rev. Relativ. {\bf 2}, 2 (1999);

\bibitem{nollert} H.-P. Nollert, 
CQG {\bf 16}, R159 (1999).

\bibitem{vish2} C. V. Vishveshwara, 
Phys. Rev. D {\bf 1}, 2870 (1970).

\bibitem{chandradet} S. Chandrasekhar, and S. Detweiler,
Proc. R. Soc. London A {\bf 344}, 441 (1975).

\bibitem{schutz} B. F. Schutz and C. M. Will,
Astrophys. Journal {\bf 291}, L33 (1985);

\bibitem{will} C. M. Will and S. Iyer,
Phys. Rev. D {\bf 35}, 3621 (1987);
S. Iyer,
Phys. Rev. D {\bf 35}, 3632 (1987).

\bibitem{konoplyawkb} R. A. Konoplya,
Phys. Rev. D {\bf 68}, 024018 (2003).

\bibitem{ferrari} V. Ferrari and B. Mashhoon,
Phys. Rev. D {\bf 30}, 295 (1984).

\bibitem{L} E. W. Leaver, 
Proc. Roy. Soc. Lon. {\bf A402}, 285 (1985).

\bibitem{nollert2} H.-P. Nollert, 
Phys. Rev. D {\bf 47}, 5253 (1993).

\bibitem{strominger} A. Strominger, C. Vafa,
Phys. Lett. B {\bf 379}, 99 (1996).

\bibitem{geo} K. Danzmann et al., in 
{\it Gravitational Wave Experiments}, eds. E. Coccia, G. Pizzella and
F. Ronga (World Scientific, Singapore, 1995).

\bibitem{ligo}A. Abramovici et al., Science {\bf 256}, 325 (1992).

\bibitem{virgo} C. Bradaschia et al., in {\it Gravitation} 1990,
Proceedings of the Banff Summer Institute, Banff, Alberta, 1990,
edited by R. Mann and P. Wesson (World Scientific, Singapore, 1991).

\bibitem{echeverria} F. Echeverria,
Phys. Rev. D {\bf 40}, 3194 (1989);
L. S. Finn, 
Phys. Rev. D {\bf 46}, 5236 (1992).

\bibitem{nakanoringing} H. Nakano, H. Takahashi, H. Tagoshi, M. Sasaki,
gr-qc/0306082.


\bibitem{green} M. B. Green, J. H. Schwarz, E. Witten,
{\it Superstring theory},
(Cambridge University Press, Cambridge, 1987);
J. Polchinski, 
{\it String theory},
(Cambridge University Press, Cambridge, 1998).

\bibitem{hooft} G. 't Hooft,
Nucl. Phys. B {\bf 72}, 461 (1974).

\bibitem{maldacena} J. M. Maldacena, 
Adv. Theor. Math. Phys. {\bf 2}, 253 (1998);
E. Witten, Adv. Theor. Math. Phys. 
{\bf 2}, 505 (1998).

\bibitem{maldacenareview} O. Aharony, S. S. Gubser, J. Maldacena,
H. Ooguri, Y. Oz, Physics Reports {\bf 323}, 183 (2000).

\bibitem{bek0} J. Bekenstein,
Lett. Nuovo Cimento {\bf 11}, 467 (1974).

\bibitem{bek} J. Bekenstein, in
{\it Cosmology and Gravitation}, edited by M. Novello
(Atlasciences, France 2000), pp. 1-85, gr-qc/9808028 (1998).

\bibitem{hod} S. Hod, 
Phys. Rev. Lett. {\bf 81}, 4293 (1998).

\bibitem{bek2} J. Bekenstein and V. F. Mukhanov, 
Phys. Lett. B {\bf 360}, 7 (1995).

\bibitem{dreyer} O. Dreyer,
Phys. Rev. Lett. {\bf 90}, 081301 (2003). 

\bibitem{motl1} L. Motl,
Adv. Theor. Math. Phys. {\bf 6}, 1135 (2003).

\bibitem{motl2} L. Motl, A. Neitzke,
Adv. Theor. Math. Phys. {\bf 7}, 2 (2003).

\bibitem{banados} M. Ba\~nados, C. Teitelboim, and 
J. Zanelli, 
Phys. Rev. Lett. {\bf 69}, 1849 (1992).  

\bibitem{cardosoqnmbtz} V. Cardoso and J. P. S. Lemos, 
Phys. Rev. D {\bf 63}, 124015 (2001).

\bibitem{regge} T. Regge, J. A. Wheeler, Phys. Rev. {\bf 108},
 1063 (1957).  

\bibitem{zerilli2}
F. Zerilli,
 Phys. Rev. Lett.  {\bf 24}, 737 (1970).

\bibitem{chandra} S. Chandrasekhar, 
in {\it The Mathematical Theory of Black Holes}
(Oxford University, New York, 1983).

\bibitem{lemos1} J. P. S. Lemos, Class. Quantum Grav. {\bf 12}, 1081
(1995); Phys. Lett. B {\bf 353}, 46 (1995); J. P. S. Lemos and
V. T. Zanchin, Phys. Rev. D {\bf 54}, 3840 (1996); see also the review
paper: J. P. S. Lemos, {\it Black holes with toroidal, cylindrical and
planar horizons in anti-de Sitter spacetimes in general relativity and
their properties}, ``Recent developments in astronomy and 
astropysics'', Proceedings of the 10th Portuguese Meeting on
Astronomy and Astrophysics, edited by J. P. S. Lemos et al (World
Scientific, 2001, gr-qc/0011092.

\bibitem{cardosoqnmtoro} V. Cardoso and J. P. S. Lemos,
Class. Quant. Grav. {\bf 18}, 5257 (2001).

\bibitem{cardosoqnmkerr} E. Berti, V. Cardoso, K. D. Kokkotas, H. Onozawa
hep-th/0307013.

\bibitem{cardosokerr2} E. Berti, V. Cardoso and S. Yoshida; gr-qc/0401052. 
\bibitem{beyer} H. R. Beyer, Commun. Math. Phys. {\bf 204}, 397(1999).

\bibitem{Ching} E. S. C. Ching, P. T. Leung,
W. M. Suen, S. S. Tong, and K. Young, Rev. Mod. Phys. {\bf 70}, 1545(1998).

\bibitem{Detweiler} S.L. Detweiler, and E. Szedenits, 
Astrophys. J. {\bf 231}, 211(1979).  

\bibitem{horowitz} G. T. Horowitz, and V. Hubeny, 
Phys. Rev. D {\bf 62}, 024027(2000).  

\bibitem{birmingham} D. Birmingham,
Phys. Rev. D {\bf 64}, 064024 (2001).

\bibitem{chop} M. W. Choptuik, 
Phys. Rev. Lett. {\bf 70}, 9(1993).

\bibitem{fernando} S. Fernando, hep-th/0306214.

\bibitem{shen} Z-X. Shen, B. Wang, R-K. Su (CCAST World Lab, Beijing),
gr-qc/0307097.

\bibitem{birmingham2} D. Birmingham, I. Sachs and S. N. Solodukhin,
Phys. Rev. Lett. {\bf 88}, 151301 (2002).

\bibitem{satoh} I. Ichinose, and K. Satoh,
Nucl. Phys. B {\bf 447}, 340(1995).

\bibitem{btz_PRD} M. Ba\~nados, M. Henneaux, C. Teitelboim, J. Zanelli,
Phys. Rev. D{\bf 48}, 1506 (1993).

\bibitem{CL1} G. Cl\'ement, Class. Quantum Grav. {\bf 10}, L49 (1993).

\bibitem{BTZ_Q} C. Mart\'{\i}nez, C. Teitelboim, J. Zanelli,
Phys. Rev. D{\bf 61}, 104013 (2000).

\bibitem{HW} E. W. Hirschmann, D. L. Welch, Phys. Rev. D{\bf 53},
5579 (1996).

\bibitem{Cat_Sal} M. Cataldo, P. Salgado, Phys. Rev. D{\bf 54}, 2971
(1996).

\bibitem{OscLemBTZ}  O. J. C. Dias, J. P. S. Lemos,
JHEP {\bf 0201}, 006 (2002).

\bibitem{edmonds} A. R. Edmonds,  
{\it Angular Momentum in Quantum Mechanics},
(Princeton U. P., Princeton, 1957). 

\bibitem{ruffini} R. Ruffini, 
{\it Black Holes: les Astres Occlus},
(Gordon and Breach Science Publishers, 1973).

\bibitem{uvarov} A. F. Nikiforov, V. B. Uvarov,  
{\it Special Functions of Mathematical Physics},
(Birkh\"{a}user, Boston, 1988).  

\bibitem{govinda} T. R. Govindarajan, and V. Suneeta, Class. Quantum
Grav. {\bf 18}, 265 (2001).

\bibitem{arfken} G. B. Arfken, H. J. Weber, 
{\it Mathematical Methods for Physicists},
(Academic Press, 1995).

\bibitem{cooper} F. Cooper, A. Khare, and U. Sukhatme,
 Phys. Rep. {\bf 251}, 267(1995).

\bibitem{birmingham3} D. Birmingham, S. Carlip and Y. Chen,
hep-th/0305113.
\bibitem{york} J. W. York, 
Phys. Rev. D {\bf 28}, 2929(1983).

\bibitem{baez} J. Baez, in {\it Matters of gravity},
p. 12, ed. J. Pullin, gr-qc/0303027.

\bibitem{dreyerall}
G. Kunstatter, 
Phys. Rev. Lett. {\bf 90}, 161301 (2003);
A. Corichi, 
Phys. Rev. D {\bf 67}, 087502 (2003);
J. Oppenheim, gr-qc/0307089.

\bibitem{bertikerr} E. Berti and K. D. Kokkotas, hep-th/0303029;

\bibitem{brinkasympt} A. M. van den Brink, gr-qc/0303095.

\bibitem{birminghamd} D. Birmingham,
hep-th/0306004 (2003).

\bibitem{konoplyasmall} R. A. Konoplya, 
Phys. Rev. D {\bf 66}, 044009 (2002);

\bibitem{qnmads} B. Wang, C. Y. Lin, and E. Abdalla, 
Phys. Lett. B {\bf481}, 79 (2000);  
B. Wang, C. M. Mendes, and E. Abdalla,
Phys. Rev. D {\bf 63}, 084001 (2001);
J. Zhu, B. Wang, and E. Abdalla,
Phys. Rev. D {\bf 63}, 124004(2001);
W. T. Kim, J. J. Oh, Phys. Lett. B {\bf 514}, 155 (2001);
S. F. J. Chan and R. B. Mann,
Phys. Rev. D {\bf 55}, 7546 (1997);
A. O. Starinets, Phys. Rev. D {\bf 66}, 124013 (2002);
Y. Kurita and M. Sakagami, Phys. Rev. D {\bf 67}, 024003 (2003);
R. Aros, C. Martinez, R. Troncoso and J. Zanelli,
Phys. Rev. D {\bf 67}, 044014 (2003);
D. T. Son, A. O. Starinets,
J.H.E.P. {\bf 0209}:042, (2002);
S. Musiri and G. Siopsis, 
Phys. Lett. B {\bf 563}, 102 (2003)..

\bibitem{moss} I. G. Moss and J. P. Norman,
Class. Quant. Grav. {\bf 19}, 2323-2332 (2002).

\bibitem{cardosoqnmsads} V. Cardoso and J. P. S. Lemos, 
Phys. Rev. D {\bf 64}, 084017 (2001).

\bibitem{cardosoqnmsads2} V. Cardoso, R. Konoplya, J. P. S. Lemos,
Phys. Rev. D {\bf 68}, 044024 (2003).

\bibitem{berti} E. Berti and K. D. Kokkotas, 
Phys. Rev. D {\bf 67}, 064020 (2003).


\bibitem{zerillimath}
F. Zerilli,  J. Math. Phys. {\bf 11}, 2203(1970).

\bibitem{mathews}
 J. Mathews, J. Soc. Ind. Appl. Math. {\bf 10}, 768(1962).

\bibitem{vish1} C. V. Vishveshwara, 
Phys. Rev.  D {\bf 1}, 2870(1970).

\bibitem{mellor}
F. Mellor, I. Moss ,
 Phys. Rev.  D {\bf 41}, 403(1990).

\bibitem{avis}
S. J. Avis, C. J. Isham and D. Storey,
 Phys. Rev. D {\bf 18}, 3565(1978).

\bibitem{breit}
P. Breitenlohner and D. Z. Freedman,
 Phys. Lett. B {\bf 115}, 197(1982).

\bibitem{burgess}
C. P. Burgess and C. A. Lutken,
Phys. Lett.  B{\bf 153}, 137(1985); 
see also  I. I. Cotaescu,  
Phys. Rev. D {\bf 60}, 107504 (1999).

\bibitem{dasgupta}
A. Dasgupta,
 Phys. Lett. B {\bf 445}, 279(1999).

\bibitem{ching2} E. S. C. Ching, P. T. Leung, W. M. Suen, and K. Young,
Phys. Rev. D {\bf 52}, 2118(1995). 

\bibitem{liu}
 H. Liu,  Class. Quantum Grav. {\bf 12}, 543(1995).

\bibitem{chandrass} S. Chandrasekhar,
 Proc. R. Soc. London, Ser. A {\bf 392}, 1(1984). 

\bibitem{hubeny} V. Hubeny, L. Susskind, and N. Toumbas, hep-th/0011164.


\bibitem{gibbons} S. R. Das, G. Gibbons and S. D. Mathur,
Phys. Rev. Lett. {\bf 78}, 417(1997).
       
\bibitem{futterman} J. A. H. Futterman, F. A. Handler and R. A. Matzner,
{\it Scattering from Black Holes},
(Cambridge University Press, Cambridge, 1988);
N. Sanchez, hep-th/9711068.



\bibitem{bcarter} B. Carter, ``General theory of stationary black hole 
states'', in {\it Black Holes}, eds B. DeWitt, C. DeWitt, 
(Gordon Breach, New York 1973).

\bibitem{win} E. Winstanley, gr-qc/0106032.

\bibitem{peca} C. S. Pe\c ca, J. P. S. Lemos, J. Math. Phys. 
{\bf 41}, 4783 (2000).

\bibitem{Liu2} H. Liu,  B. Mashhoon, Class. Quantum Grav. 
{\bf 13}, 233 (1996).

\bibitem{immirzi} G. Immirzi, 
Nucl. Phys. Proc. Suppl. {\bf 57}, 65 (1997).

\bibitem{vitoroscarjose} V. Cardoso, O. J. C. Dias and J. P. S. Lemos,
Phys. Rev. D {\bf 67}, 064026 (2003).

\bibitem{Nariai} H. Nariai, {\it On some static solutions of
Einstein's gravitational field equations in a spherically
symmetric case}, Sci. Rep. Tohoku Univ. {\bf 34}, 160 (1950); {\it
On a new cosmological solution of Einstein's field equations of
gravitation}, Sci. Rep. Tohoku Univ. {\bf 35}, 62 (1951).

\bibitem{GinsPerry} P. Ginsparg, M. J. Perry, {\it Semiclassical
perdurance of de Sitter space}, Nucl. Phys. B {\bf 222}, 245
(1983).

\bibitem{Bousso60y}
R. Bousso,  {\it Adventures in de Sitter space}, {\tt
hep-th/0205177}.

\bibitem{OscLemNariai} O. J. C. Dias and J. P. S. Lemos,
Phys. Rev. D. (2003, in press); hep-th/0306194.



\bibitem{brady} P. Brady, C. Chambers, W. Laarakkers and Eric Poisson,
Phys. Rev. D {\bf 60}, 064003 (1999).

\bibitem{teller} G. P\"oshl and E. Teller,
Z. Phys. {\bf 83}, 143 (1933).

\bibitem{cardosoqnmds}  V. Cardoso and J. P. S. Lemos,
Phys. Rev. D {\bf 67}, 084020 (2003).

\bibitem{molina} C. Molina, gr-qc/0304053.

\bibitem{brinkds} A. M. van den Brink, gr-qc/0304092.

\bibitem{suneeta} V. Suneeta, gr-qc/0303114.

\bibitem{zhidenko} A. Zhidenko, gr-qc/0307012.

\bibitem{shijunds} S. Yoshida and T. Futamase,
gr-qc/0308077.


\bibitem{kurita} Y. Kurita and M. Sakagami, 
Phys. Rev. D {\bf 67}, 024003 (2003);

\bibitem{ling} Y. Ling, H. Zhang, gr-qc/0309018.

\bibitem{musiri2} S. Musiri and G. Siopsis,
hep-th/0308168.

\bibitem{roman2} R. A. Konoplya, hep-th/0309030.

\bibitem{karlucio} K. H. C. Castello-Branco and E. Abdalla, gr-qc/0309090. 

\bibitem{ricardo} V. Cardoso, J. Nat\'ario and R. Schiappa, hep-th/0403132.

\bibitem{ida} D. Ida, Y. Uchida and Y. Morisawa,
Phys. Rev. D {\bf 67}, 084019 (2003).

\bibitem{guinn} J. W. Guinn, C. M. Will, Y. Kojima and B. F. Schutz,
Class. Quant. Grav. {\bf 7}, L47 (1990).

\bibitem{cardosoqnms} V. Cardoso, J. P. S. Lemos, S. Yoshida,
submitted (2003), gr-qc/0309112.

\bibitem{matsas} L. Crispino, A. Higuchi and G. Matsas,
Phys. Rev. D {\bf 63}, 124008 (2001).

\bibitem{bertihighd} E. Berti, M. Cavaglia and L. Gualtieri, hep-th/0309203.

\bibitem{Le90} E. W. Leaver, Phys. Rev. D {\bf 41}, 2986 (1990). 

\bibitem{Gu67} W. Gautschi, SIAM Rev. {\bf 9}, 24 (1967). 


\bibitem{andersson1} N. Andersson, 
CQG {\bf 10}, L61 (1993).

\bibitem{kodama} H. Kodama, A. Ishibashi,
hep-th/0305147 (2003); hep-th/0305185; hep-th/0308128.

\bibitem{neitzke} A. Neitzke,
hep-th/0304080 (2003).

\bibitem{siopsis} S. Musiri, G. Siopsis, hep-th/0308168.

\bibitem{abdallads} E. Abdalla, K. H. C. Castello-Branco, A. Lima-Santos,
gr-qc/0301130 (2003).

\bibitem{andersson2} N. Andersson, C. Howls,
private communication.

\bibitem{onozawa} H. Onozawa,
Phys. Rev. D {\bf 55}, 3593 (1997).

\bibitem{MVDBdoublet} 
P. T. Leung, A. Maassen van den Brink, K. W. Mak, K. Young,
gr-qc/0301018 (2003).

\bibitem{MVDBas} A. Maassen van den Brink, 
Phys. Rev. D {\bf 62}, 064009 (2000).

\bibitem{De} S. Detweiler,
Astrophy. J. {\bf 239}, 292 (1980).

\bibitem{KerrQNM} E. Seidel, S. Iyer,
PRD {\bf 41}, 374 (1990); 
K. D. Kokkotas, 
Class. Quant. Grav. {\bf 8}, 2217 (1991).

\bibitem{GA} K. Glampedakis, N. Andersson,
gr-qc/0304030 (2003).

\bibitem{H2} S. Hod,
gr-qc/0301122 (2003).

\bibitem{CR} D. Christodoulou, 
Phys. Rev. Lett. {\bf 25}, 1596 (1970);
D. Christodoulou, R. Ruffini, 
Phys. Rev. D {\bf 4}, 3552 (1971).

\bibitem{W} R. Wald, 
J. Math. Phys. (N.Y.) {\bf 14}, 1453 (1973).

\bibitem{Cas} S. Chandrasekhar, 
Proc. R. Soc. London {\bf A392}, 1 (1984).

\bibitem{A} N. Andersson,
Class. Quant. Grav. {\bf 11}, L39 (1994).

\bibitem{SepSca} M. Abramowitz, I. A. Stegun, 
{\it Handbook of Mathematical Functions},
(Dover, New York, 1970); 
T. Oguchi, 
Radio Sci. {\bf 5}, 1207 (1970).

\bibitem{bondi} H. Bondi, 
Nature {\bf 179}, 1072 (1957).

\bibitem{thorne} K. S. Thorne, Rev. Mod. Phys, {\bf 52}, 299 (1980);
K. S. Thorne, in {\it General Relativity: An Einstein Centenary Survey}, 
eds. S. W. Hawking and W. Israel (Cambridge University Press, 1989).

\bibitem{damour} T. Damour, in 
{\it Gravitational Radiation},  
eds Nathalie Deruelle and Tsvi Piran
(North Holland Publishing Company, New York, 1983).

\bibitem{weisbergtaylor} J. M. Weisberg and J. H. Taylor,
Phys. Rev. Lett. {\bf 52}, 1348 (1984).

\bibitem{weber} J. Weber,
Phys. Rev. {\bf 117}, 306 (1960).

\bibitem{schutz1} B. F. Schutz, Class. Quant. Grav. {\bf 16}, A131 (1999);
B. F. Schutz and F. Ricci, in {\it Gravitational Waves}, 
eds I. Ciufolini et al,
(Institute of Physics Publishing, Bristol, 2001).

\bibitem{hughes} S. A. Hughes,
Annals Phys. {\bf 303}, 142 (2003).

\bibitem{miller} M. C. Miller, E. J. M Colbert, astro-ph/0308402.
R. P. van der Marel, astro-ph/0302101.

\bibitem{hamed} N. Arkani-Hamed, S. Dimopoulos and G. Dvali, 
Phys. Lett. B {\bf 429}, 263 (1998); Phys. Rev. D {\bf 59}, 086004 (1999);
I. Antoniadis, N. Arkani-Hamed, S. Dimopoulos and G. Dvali,
Phys. Lett. B {\bf 436}, 257 (1998).

\bibitem{mitrofanov} for a review on short-ranges searches for non-Newtonian
gravity, see
M. Varney and J. Long, gr-qc/0309065.


\bibitem{polc} J. Polchinski, TASI lectures on D-branes,
hep-th/9611050.

\bibitem{tangherlini} F. R. Tangherlini,
Nuovo Cim. {\bf 27}, 636 (1963).

\bibitem{myersperry} R. C. Myers and M. J. Perry, 
Annals Phys. {\bf 172}, 304 (1986).

\bibitem{thornehoop} K. S. Thorne, in
{\it J. R. Klauder, Magic without Magic},
San Francisco, 1972.

\bibitem{page} D. N. Page, 
Phys. Rev. D {\bf 13}, 198 (1976);
Phys. Rev. D {\bf 14}, 3260 (1976).

\bibitem{bhprod} P. C. Argyres, S. Dimopoulos and J. March-Russell,
Phys. Lett. B {\bf 441}, 96 (1998);
S. Dimopoulos and G. Landsberg, 
Phys. Rev. Lett. {\bf 87}, 161602 (2001). 
S. B. Giddings and S. Thomas,
Phys. Rev. D {\bf 65}, 056010 (2002); 
S. D. H. Hsu, hep-ph/0203154;
H. Tu, hep-ph/0205024;
A. Jevicki and J. Thaler, Phys. Rev. D {\bf 66}, 024041 (2002); 
Y. Uehara, hep-ph/0205199;

\bibitem{cardosorads} V. Cardoso and J. P. S. Lemos, 
Phys. Lett. B {\bf 538}, 1 (2002);

\bibitem{cardosoradkerrs} V. Cardoso and J. P. S. Lemos, 
Gen. Rel. Grav. {\bf 35}, L327 (2003).
       
\bibitem{cardosoradkerr} V. Cardoso and J. P. S. Lemos, 
Phys. Rev. D {\bf 67 }, 084005 (2003).

\bibitem{cardosoradel} V. Cardoso, J. P. S. Lemos, S. Yoshida,
Phys. Rev. D (in press), gr-qc/0307104.

\bibitem{lousto} C. O. Lousto, and R. H. Price, 
Phys. Rev. D {\bf 55}, 2124 (1997); K. Martel, and E. Poisson, gr-qc/0107104.

\bibitem{D'Eath} P. D. D'Eath, in {\it Black Holes: gravitational interactions}, 
(Clarendon Press, Oxford, 1996).

\bibitem{payne} P. D. D'Eath and P. N. Payne,
Phys. Rev. D {\bf 46}, 658 (1992);
Phys. Rev. D {\bf 46}, 675 (1992);
Phys. Rev. D {\bf 46}, 694 (1992);

\bibitem{ferrariruf} V. Ferrari and R. Ruffini,
Phys. Lett. B {\bf 98}, 381 (1981).


\bibitem{ferrari2} V. Ferrari and B. Mashhoon,  
Phys. Rev. Lett. {\bf 52}, 1361 (1984).


\bibitem{payne2} P. N. Payne, Phys. Rev. D {\bf 28}, 1894 (1983);


\bibitem{piran} T. Piran, {\it Gravitational Waves: 
A Challenge to Theoretical Astrophysics}, 
edited by V. Ferrari, J.C. Miller and L. Rezzolla 
(ICTP, Lecture Notes Series).

\bibitem{nmothers} M. Sasaki and T. Nakamura, 
Prog. Theor. Phys. {\bf 67}, 1788 (1982);
T. Nakamura and M. Sasaki, Phys. Lett. A {\bf 89}, 185 (1982);
Y. Kojima and T. Nakamura, Phys. Lett. A {\bf 96}, 335 (1983);
Y. Kojima and T. Nakamura, Phys. Lett. A {\bf 99}, 37 (1983);
Y. Kojima and T. Nakamura, Prog. Theor. Phys. {\bf 71}, 79 (1984).


 
\bibitem{teukolsky} S. A. Teukolsky, Astrophys. J. {\bf 185}, 635 (1973).

\bibitem{breuerbook}  R. A. Breuer, in 
{\it Gravitational Perturbation Theory and Synchrotron Radiation}, 
(Lecture Notes in Physics, Vol. 44), (Springer, Berlin 1975).

\bibitem{nakamurasasaki} 
T. Nakamura and M. Sasaki, Phys. Lett. A {\bf 89}, 68 (1982);
M. Sasaki and T. Nakamura, Phys. Lett. A {\bf 87}, 85 (1981).

\bibitem{hughes1} S. A. Hughes, Phys. Rev. D {\bf 61}, 0804004 (2000);
D. Kennefick, Phys. Rev. D {\bf 58}, 064012 (1998); 

\bibitem{newman} E. Newman and R. Penrose,
J. Math. Phys. {\bf 3}, 566 (1966);

\bibitem{pressteu} W. H. Press and S. A. Teukolsky, 
Astrophys. J. {\bf 185}, 649 (1973).

\bibitem{breuer} R. A. Breuer, M. P. Ryan Jr, and S. Waller,
Proc. R. Soc. London A {\bf 358}, 71 (1977).

\bibitem{goldberg} J. N. Goldberg, A. J. MacFarlane, E. T. Newman,
F. Rohrlich and C. G. Sudarshan,
J. Math. Phys. {\bf 8}, 2155 (1967);

\bibitem{numrecipes} W. H. Press, B. P. Flannery, S. A. Teukolsky
and W. T. Vetterling, {\it Numerical Recipes}
(Cambridge University Press, Cambridge, England, 1986).

\bibitem{QNKerr} S. Detweiler, Astrophys. J. {\bf 239}, 292 (1980);
V. Ferrari and B. Mashhoon, Phys. Rev. D {\bf 30}, 295 (1984);
V. Ferrari and B. Mashhoon, Phys. Rev. Lett. {\bf 52}, 1361 (1984);
V. P. Frolov, and I. D. Novikov,
in {\it Black Hole Physics - Basic Concepts and New Developments}, 
(Kluwer Academic Publishers, Dordrecht, 1998).


\bibitem{horostrometal} G. Horowitz, A. Strominger, Nucl. Phys. B 
{\bf360}, 197 (1991).
M. J. Duff, ``TASI Lectures on Branes, Black
Holes and Anti-de Sitter Space'', hep-th/9912164; A. W. Peet, ``TASI
lectures on black holes in string theory'', hep-th/0008241; D. Marolf,
``String/M-branes for Relativists'', gr-qc/9908045.


\bibitem{gubserklebanovpolyakov} S. S. Gubser, I. R. Klebanov,
A. M. Polyakov, Phys. Lett. B {\bf 428} 105 (1998).

\bibitem{aharonyetalklebanovhorowitzreview} I. R. Klebanov, 
``TASI Lectures: Introduction to the AdS/CFT
Correspondence'', hep-th/0009139; G. T. Horowitz,
Class. Quant. Grav. {\bf 17}, 1107 (2000).

\bibitem{maldacenastrom} J. M. Maldacena, ``Black Holes in String
Theory'', Ph.D. Thesis, Princeton, hep-th/9607235; J. M. Maldacena,
A. Strominger.  JHEP {\bf 9812}, 005 (1998).

\bibitem{keski} E. Keski-Vakkuri, Phys. Rev D {\bf 59}, 104001
(1999);

\bibitem{danielsson} U. H. Danielsson, E. Keski-Vakkuri, M. Kruczenski, 
Nucl. Phys. B {\bf 563}, 279 (1999); JHEP {\bf 0002}, 039 (2000).

\bibitem{mat}  H. Matschull, Class. Quant. Grav. {\bf 16}, 1069 (1999).

\bibitem{balasu}  V. Balasubramanian, P. Krauss, 
A. Lawrence, S. Trivedi, 
Phys. Rev. D {\bf 59}, 104021 (1999); 
V. Balasubramanian, J. de Boer, E. Keski-Vakkuri, S. F. Ross, 
Phys. Rev. D {\bf 64} 064011 (2001); 
E. Kiritsis, T. R. Taylor, hep-th/9906048.

\bibitem{wittenbanksetal}
T. Banks, M. R. Douglas, 
G. T. Horowitz, E. Martinec, 
hep-th/9808016.

\bibitem{susskindwitten} L. Susskind, E. Witten, 
hep-th/9805114.

\bibitem{baker} J. Baker, B. Br\"ugmann, M. Campanelli, C. O. Lousto, 
R. Takahashi, Phys. Rev. Lett. {\bf 87}, 121103 (2001). 

\bibitem{burko} L.Barack and L. M. Burko, Phys. Rev. D {\bf 62},
084040 (2001).

\bibitem{wald} J. M. Cohen,  R. M. Wald,
J. Math. Phys. {\bf 12}, 1845 (1971).

\bibitem{poisson} E. Poisson, Phys. Rev. D {\bf 55}, 639 (1997).

\bibitem{cardosoradbtz} V. Cardoso, J. P. S. Lemos, 
Phys. Rev. D {\bf 65}, 104032 (2002).

\bibitem{cardosogul} V. Cardoso, ``Numerical Analysis of Partial Differential Equations
in General Relativity and String Theory: Applications to Gravitational Radiation
and to the AdS/CFT Conjecture'' 
(Programa Gulbenkian de Est\'{\i}mulo \`a Investiga\c c\~ao Cient\'{\i}fica, 2001).

\bibitem{cardosoradsync} V. Cardoso, J. P. S. Lemos, 
Phys. Rev. D {\bf 65}, 104033 (2002).


\bibitem{thooftsusskind} G. 't Hooft, gr-qc/9310026; 
L. Susskind, J. Math. Phys. {\bf 36}, 6377 (1995).


\bibitem{bateman} A. Erdelyi, W. Magnus, F. Oberlettinger,
 and F. Tricomi, 
{\it Higher Transcendental Functions}, 
(McGraw-Hill Book Co., Inc, New York, 1953);  
M. Abramowitz, and I. A. Stegun in 
{\it Handbook of Mathematical Functions},
(Dover, New York, 1970); A. F. Nikiforov, V. B. Uvarov,  
{\it Special Functions of Mathematical Physics},
(Birkh\"{a}user, Boston, 1988).

\bibitem{naka} M. Sasaki and T. Nakamura, Phys. Lett. B {\bf 89}, 68 (1982).
 




\bibitem{kinnersley} W. Kinnersley and M. Walker,
Phys. Rev. D {\bf 2}, 1359 (1970); 
O. J. C. Dias, J. P. S. Lemos,
Phys. Rev. D {\bf 67}, 064001 (2002);
Phys. Rev. D {\bf 67}, 084018 (2003);
hep-th/0306194.

\bibitem{decay} P. Kanti and J. March-Russell, 
Phys. Rev. D {\bf 66}, 024023 (2002). 
P. Kanti and J. March-Russell, hep-ph/0212199.
D. Ida, Kin-ya Oda and S. C. Park, hep-th/0212108;
V. Frolov and D. Stojkovic, 
Phys. Rev. Lett. {\bf 89}, 151302 (2002); 
gr-qc/0211055;

\bibitem{eardley} D. M. Eardley and S. B. Giddings, 
Phys. Rev. D {\bf 66}, 044011 (2002).

\bibitem{yoshino} H. Yoshino and Y. Nambu, gr-qc/0209003.

\bibitem{weinberg} S. Weinberg, {\it Gravitation and Cosmology} (Wiley, 
New York, 1972).
      
\bibitem{cardosononlinear} J. T. Mendonca, Vitor Cardoso,
Phys. Rev. D {\bf 66}, 104009 (2002);
M. Servin, M. Marklund, G. Brodin, J.T. Mendonca, V. Cardoso,
Phys. Rev. D {\bf 67}, 087501 (2003);
J.T. Mendonca, V. Cardoso, M. Marklund, M. Servin, G. Brodin,
Phys. Rev. D (in press), gr-qc/0307031.

\bibitem{watson} G. N. Watson, 
{\it A Treatise on the Theory of Bessel Functions}
(Cambridge University Press, 1995)

\bibitem{stegun} M. Abramowitz, I. A. Stegun, 
{\it Handbook of Mathematical Functions},
(Dover, New York, 1970). 


\bibitem{courant} R. Courant and D. Hilbert,
{\it Methods of Mathematical Physics},  (Interscience, 
New York, 1962).

\bibitem{hadamard} J. Hadamard,
{\it Lectures on Cauchy's Problem in Linear Partial Differential
Equations}
(Yale University Press, New Haven, 1923).

\bibitem{barrow} J. D. Barrow and F. J. Tipler,
{\it The Anthropic Cosmological Principle}
(Oxford University Press, Oxford, 1986).

\bibitem{galtsov} D. V. Gal'tsov, 
Phys. Rev. D {\bf 66}, 025016 (2002).

\bibitem{hassani} S. Hassani, 
{\it Mathematical Physics},
(Springer-Verlag, New York, 1998). 

\bibitem{kazinski} P. O. Kazinski, S. L. Lyakhovich and
A. A. Sharapov,
Phys. Rev. D {\bf 66}, 025017 (2002);
B. P. Kosyakov,
Theor. Math. Phys. {\bf 119}, 493 (1999).


\bibitem{lin} B. Chen, M. Li and Feng-Li Lin,
JHEP {\bf 0211}, 050 (2002).

\bibitem{pulsar} J. M. Weisberg and J. H. Taylor,
astro-ph/0211217.

\bibitem{peters} P. C. Peters, 
Phys. Rev. {\bf 136}, B1224 (1964);
P. C. Peters and J. Mathews, 
Phys. Rev. {\bf 131}, 435 (1963).

\bibitem{schutzlivro} B. F. Schutz, 
{\it A First Course in General Relativity},
(Cambridge University Press, 1985).

\bibitem{zerilli}
F. Zerilli, Phys. Rev. D {\bf 2}, 2141 (1970).

\bibitem{quem} R. Ruffini, Phys. Rev. D {\bf 7}, 972 (1973);
M. J. Fitchett, {\it The Gravitational Recoil Effect and its Astrophysical
Consequences}, (PhD Thesis, University of Cambridge, 1984).


\bibitem{wein1} S. Weinberg, Phys. Lett. {\bf 9}, 357 (1964);
Phys. Rev. {\bf 135}, B1049 (1964).

\bibitem{smarr2} L. Smarr, Phys. Rev. D {\bf 15},
2069 (1977); 

R. J. Adler and B. Zeks, 
Phys. Rev. D {\bf 12}, 3007 (1975). 

\bibitem{jackson} J. D. Jackson, 
{\it Classical Electrodynamics}, (J. Wiley, New York 1975).

\bibitem{cardosoradadsddim} V. Cardoso and J. P. S. Lemos, 
Phys. Rev. D {\bf 66}, 064006 (2002).

\bibitem{schwinger} J. Schwinger, 
Phys. Rev. {\bf 82}, 664 (1951).

\bibitem{diaslemos2} \'O. J. C. Dias, 
{\it Proceedings of the Xth Portuguese Meeting on
Astronomy and Astrophysics}, Lisbon, July 2000, eds J. P. S. Lemos et al, 
(World Scientific, 2001), p. 109; gr-qc/0106081.

\bibitem{chang} C. S. Wang Chang and D. L. Falkoff, 
Phys. Rev. {\bf 76}, 365 (1949).

\bibitem{grad} I. S. Gradshteyn and I. M. Ryzhik,
{\it Table of Integrals, Series and Products},
(Academic Press, New York, 1965). 

\bibitem{dewitt} B. S. DeWitt and R. W. Brehme, 
Ann. Phys. (NY){\bf 9} 220 (1965).

\bibitem{price} R. H. Price, 
Phys. Rev. D {\bf 5}, 2419 (1972). 

\bibitem{price2} C. Gundlach, R. H. Price and J. Pullin, 
Phys. Rev. D {\bf 49}, 883 (1994). 

\bibitem{price3} C. Gundlach, R. H. Price and J. Pullin, 
Phys. Rev. D {\bf 49}, 890 (1994). 

\bibitem{leaver} E. Leaver, 
Phys. Rev. D {\bf 34}, 384 (1986). 

\bibitem{ching1} E. S. C. Ching, P. T. Leung, W. M. Suen and K. Young,
Phys. Rev. Lett. {\bf 74}, 2414 (1995). 

\bibitem{tom} H. Koyama and A. Tomimatsu,
Phys. Rev. D {\bf 64}, 044014 (2001). 

\bibitem{hod1} S. Hod,
Class. Quant. Grav. {\bf 18}, 1311 (2001).
       
\bibitem{greend} S. Hassani, 
{\it Mathematical Physics},
(Springer-Verlag, New York, 1998);
R. Courant and D. Hilbert,
{\it Methods of Mathematical Physics}, Chapter VI (Interscience, 
New York, 1962).

\bibitem{amj} H. Soodak and M. S. Tiersten,
Am. J. Phys. {\bf 61}, 395 (1993).

\bibitem{barv} A. O. Barvinsky, S. N. Solodukhin, 
hep-th/0307011.

\bibitem{cardosotails} V. Cardoso, S. Yoshida, O. J. C. Dias, J. P. S. Lemos,
Phys. Rev. D [rapid communications] (in press), hep-th/0307122.

\bibitem{higuchi} L. C. B. Crispino, A. Higuchi, G. E. A. Matsas,
Phys. Rev. D {\bf 63}, 124008 (2001). 





 








\end{thebibliography}

\end{document}